\documentclass{aa}

\usepackage{graphicx}
\usepackage{subcaption}
\usepackage{cleveref}           % citing multiple references
\usepackage{natbib}
\usepackage[english]{babel}
\usepackage[varg]{txfonts}
\bibpunct{(}{)}{;}{a}{}{,} % to follow the A&A style
\usepackage{tabularx,stmaryrd,amsmath,multirow}
\SetSymbolFont{stmry}{bold}{U}{stmry}{m}{n}
\let\oldpageref\pageref
\renewcommand{\pageref}{\oldpageref*}

\usepackage{array}
\newcolumntype{L}[1]{>{\raggedright\let\newline\\\arraybackslash\hspace{0pt}}m{#1}}
\newcolumntype{C}[1]{>{\centering\let\newline\\\arraybackslash\hspace{0pt}}m{#1}}
\newcolumntype{R}[1]{>{\raggedleft\let\newline\\\arraybackslash\hspace{0pt}}m{#1}}

\begin{document}

   \title{Sensitivity of gas-grain chemical models to surface reaction barriers:\\}

   \subtitle{Effect from a key carbon-insertion reaction, C  + H$_2$  $\rightarrow$ CH$_2$ }

   \author{M. Simončič
          \inst{1}
          \and
          D. Semenov\inst{2,3}
          \and
          S. Krasnokutski\inst{4}
          \and
          Th. Henning\inst{2}
          \and
          C. J\"{a}ger\inst{4}
          }

   \institute{Faculty of Chemistry and Chemical Technology,
   University of Ljubljana,
              Večna pot 113, 1000 Ljubljana, Slovenia\\
              \email{matjaz.simoncic@fkkt.uni-lj.si}
         \and
             Max-Planck-Institut f\"{u}r Astronomie, K\"{o}nigstuhl 17, D-69117 Heidelberg, Germany\\
             \email{semenov@mpia.de}
         \and
         Department of Chemistry,
Ludwig Maximilian University,
Butenandtstr. 5-13,
81377 Munich,
Germany
         \and
                 Laboratory Astrophysics and Cluster Physics Group
                 of the Max Planck Institute for Astronomy
                         at the Friedrich Schiller University Jena,
                         Institute for Solid State Physics,
                         Helmholtzweg 3,
                     D-07743 Jena, Germany\\
                 \email{sergiy.krasnokutskiy@uni-jena.de}
%         \and
%                University of Michigan: Ann Arbor, MI, United States\\
%                \email{mail}
                 }

 \titlerunning{Sensitivity of chemical models to surface reaction barriers}
\authorrunning{M. Simončič et al.}

   \date{}

% \abstract{}{}{}{}{}
% 5 {} token are mandatory

  \abstract
   {The feasibility of contemporary gas-grain astrochemical models depends on the availability
   of accurate kinetics data, in particular, for surface processes.}
   {We study the sensitivity of gas-grain chemical models to 
the energy barrier $E_{\rm a}$ of the important surface reaction between some of the most abundant species: C and H$_2$ (surface C + surface H$_2$ $\rightarrow$ surface CH$_2$).}
   {We used the gas-grain code {\sc ALCHEMIC} to model the time-dependent chemical evolution over a
   2D grid of densities ($n_{\rm H} \in 10^{3}, 10^{12}~{\rm cm}^{-3}$) and temperatures ($T \in 10, 300~{\rm K}$),
   assuming UV-dark ($A_{\rm V}$ = 20 mag) and partly UV-irradiated
   ($A_{\rm V}$ = 3 mag) conditions that are typical of the dense interstellar medium.
   We considered two values for the energy barrier of the surface reaction,
   $E_{\rm a}=2\,500$~K (as originally implemented in the networks) and $E_{\rm a}=0$~K (as measured in the laboratory
   and computed by quantum chemistry simulations).}
   {We find that if the C $+$ H$_2$ $\rightarrow$ CH$_2$ surface reaction is barrierless,
   a more rapid conversion of the surface carbon atoms into methane ice occurs. Overproduction of the CH$_n$ hydrocarbon ices
   affects the surface formation of more complex hydrocarbons, cyanides and nitriles, and CS-bearing species at low temperatures $\lesssim 10-15$~K. The surface hydrogenation of CO and hence the synthesis of complex (organic) molecules become affected as well.
   As a result, important species whose abundances may change by more than a factor of two at $1$~Myr include atomic carbon, small mono-carbonic (C$_1$) and di-carbonic (C$_2$) hydrocarbons, CO$_{2}$, CN, HCN, HNC, HNCO, CS, H$_{2}$CO, H$_{2}$CS, CH$_{2}$CO, and CH$_{3}$OH (in either gas and/or ice).
   The abundances of key species, CO, H$_2$O, and N$_2$ as well as  O, HCO$^+$, N$_2$H$^+$, NH$_3$, NO, and most of the S-bearing molecules, remain almost unaffected.}
   {Further accurate laboratory measurements and quantum chemical calculations of the surface reaction barriers
   will be crucial to improve the accuracy of astrochemical models.}


   \maketitle
%
%-------------------------------------------------------------------
\section{Introduction}
%-------------------------------------------------------------------
To date, more than 200 interstellar molecular species have been
discovered\footnote{\url{http://www.astro.uni-koeln.de/cdms/molecules}}\citep{McGuire18}.
These molecules include terrestrial-like stable molecules,
more exotic unstable radicals, ions, complex species, macromolecules, and various ices.
%Among these interstellar molecules are unsaturated (long) carbon chains and
%fullerenes (C$_{60}$, C$_{60}^{+}$, and C$_{70}$;
%\citep{cami2010,campbell2015}.
Among these complex species, a variety of complex organic molecules
such as CH$_3$OH, CH$_3$CN, HCOOCH$_3$, CH$_3$CHO, and CH$_3$OCH$_3$ have been
found in the interstellar medium (ISM).

According to our current understanding of the chemical processes in space, molecules
can be produced and destroyed through gas-phase and/or dust grain surface chemistry, with
high-energy radiation and cosmic rays playing an important role \citep{Cuppen2017}.
For example, CO, CS, carbon chains, and cyanopolyynes are mainly formed through ion-molecule and
dissociative recombination reactions in the gas phase. In contrast, some simple key
molecules such as H$_2$, NH$_3$, H$_2$O, CH$_4$,  and most of the complex organics
are mainly synthesized through surface processes \citep{Hollenbach_Salpeter71,Herbst_vDishoeck09,Tielens_10}.
Among these processes, surface recombination of highly mobile atomic hydrogen with other species such as CO
plays a very important role in building the molecular complexity of ice mantles.
Similar reactions involving molecular hydrogen are usually considered to be of less importance
because of substantial energy barriers.
Modern public astrochemical databases such as KIDA\footnote{\url{http://kida.obs.u-bordeaux1.fr}} and
UDFA\footnote{\url{http://udfa.net}} include thousands of low-temperature reactions that can occur under
the ISM conditions. These reaction networks are  used in various astrochemical models that
simulate the chemical evolution of the ISM clouds, protoplanetary disks, circumstellar shells, and exoplanetary
atmospheres \citep{herbst2013,Agundez2013,2013ChRv..113.9016H,Rimmer_Helling16}.

Recently, a high reactivity of single carbon atoms toward different molecules has been demonstrated \citep{Chastaing00,Chastaing01,Kaiser02,Kaiser99b,Krasnokutski17,Shannon14}. In particular,
efficient barrierless reactions of C atoms toward saturated closed-shell molecules have been identified.
The high reactivity of carbon atoms persists over a broad temperature range starting from temperatures
slightly above absolute zero. The reactions of carbon atoms with small molecules in the gas phase can lead to
their efficient destruction \citep{Krasnokutski17,Shannon14}. At the same time, for
large molecules or C atoms reacting on dust surfaces, such reactions could lead to the growth of the molecular
size and the bottom-up formation of diverse complex molecules \citep{Krasnokutski2017,Krasnokutski17}.
%A high reactivity of C atoms toward different carbon bearing molecules is expected based on the fact that C atoms are often found to react barrierlessly with molecules via insertion of C atoms into an existing C-C bond \citep{Bettinger00,Kaiser99a,Kaiser97,Krasnokutski17}.
%All this implies a large impact of C atom reactions on the interstellar chemistry.

Until recently, no clean source of C atoms was available \citep{Krasnokutski14}, and thus only a limited number
of reactions of C atoms has been investigated in the laboratory. Moreover, most studies of these reactions
have been performed in the gas phase, and the information about similar surface processes remains limited.
As a result, the data about the surface reaction rates, barriers, and products have often been adopted from the
gas-phase studies. However, because the third body (a dust grain) enables association
reactions, surface reactions with C atoms might be more efficient than their gas-phase counterparts.
%Limited information on the surface reactivity of C atoms implies that many important reactions have not yet been fully taken into account.

One of the most important reactions involving carbon atoms is the reaction between C and H$_{2}$.
This reaction is very slow in the gas phase at low temperatures. The radiative association channel
leading to the formation of CH$_{2}$ has a low reaction constant $\sim 10^{-17}$~cm$^{3}$\,s$^{-1}$ \citep{Becker89,Dean1991,Husain1975},
while the neutral-neutral reaction channel leading to CH + H is highly
endothermic, with a barrier of $12\,000$~K \citep{KIDA}.
%However, the comparison of the rates of the C + H$_2$ $\rightarrow$ CH + H and CH + H $\rightarrow$ C + H$_2$ reactions, obtained at different high temperatures, suggested the presence of an energy barrier in reaction pathways \citep{Becker89,Dean1991,Husain1975}. The energy barrier could only be present for the transition between two states C$\cdot$H$_2$ and HCH. This implies the presence of an energy barrier for the C + H$_2$ + M $\rightarrow$ HCH + M reaction and consequently its low rate at low-temperatures. This uncertainty resulted in the fact that neither the gas-phase nor the surface reaction C + H$_2$ + M $\rightarrow$ HCH + M was considered in the reaction networks modeling the chemistry in the ISM.
%As it has already been mentioned, is not always possible to adopt the results of gas-phase studies to predict the outcome of surface reactions.

In contrast, as found by quantum chemical computations, the surface reaction C + H$_2$ + M $\rightarrow$ CH$_{2}$ + M
does not possess any significant energy barrier \citep{Bussery-Honvault2005,Harding1993,Lin2004}.
We recently performed a series of laboratory studies of this reaction. The superfluid helium nanodroplets
flying through a vacuum chamber were doped with carbon atoms and dihydrogen molecules. The outcome of the
reactions was monitored using mass spectrometry and calorimetric technique. The calorimetry technique allowed us
to estimate the amount of energy released in the reactions before the ionization point.
The liquid helium acts like a chemically inert dust surface because it absorbs the excess of
reaction energy. The reaction C + H$_2$ + M $\rightarrow$ CH$_{2}$ + M was found to be fast at T = 0.37~K \citep{Krasnokutski2016,Krasnokutski19b}.
This demonstrates the absence of the energy barrier in the surface reaction pathway and ensures
a high reaction rate at low temperatures in the ISM.
%Moreover, the addition of CO molecules to the C and H$_2$ reactants resulted in the formation of large variety of complex organic molecules \citep{Krasnokutski17}.
%The obtained results suggested that C + H$_2$ reaction is important for the formations of complex molecules in molecular regions of the ISM.

However, in many gas-grain astrochemical models the surface reaction of C with H$_{2}$ is
assumed to have a barrier of $2\,500$~K \citep{Garrod_Herbst06, Belloche_ea14, KIDA},
including our {\sc ALCHEMIC} model \citep{Semenov_ea10}.
In this paper, we investigate how strongly the results of astrochemical modelling of the ISM
are sensitive to the adopted value of this barrier. In Sec.~\ref{sec:model}, the adopted physical and
chemical model is described. Results and discussion are presented in Sec.~\ref{sec:results}.
The Conclusions follow.

%-------------------------------------------------------------------
\section{Physical and chemical model}
\label{sec:model}
%-------------------------------------------------------------------
The adopted chemical model is based on the public gas-grain {\sc ALCHEMIC}
code\footnote{\url{http://www.mpia.de/homes/semenov/disk_chemistry_OSU08ggs_UV.zip}}
\citep[see][]{Semenov_ea10}. The chemical network is based on the osu.2007 ratefile with the recent updates to
the reaction rates from the Kinetic Database for Astrochemistry (KIDA) \citep{KIDA} and the high-temperature
network of \citet{2010ApJ...721.1570H} and \citet{SW2011}. The model considers both gas-phase and grain-surface chemistry and consists of 653 atomic and molecular species made of 12 elements, and neutral, negatively, and positively charged grains,
which are involved in 7907 reactions.

The standard cosmic ray (CR) ionization rate was assumed, $\zeta_{\rm CR}=1.3\times10^{-17}$~s$^{-1}$.
The interstellar UV radiation field with the intensity of $1$ in the Draine units was used.
The self-shielding of H$_2$ from photodissociation was calculated by Eq.~(37) from \citet{DB96}.
The shielding of CO by dust grains, H$_2$ , and the CO self-shielding was calculated using a precomputed table of
\citet[][Table~11]{1996A&A...311..690L}.

The gas-grain interactions include sticking of neutral species and electrons to dust grains with 100\% probability and desorption of ices by
thermal, CRP-, and UV-driven processes. Uniform compact amorphous
silicate particles of olivine stoichiometry with a
density of $3$~g\,cm$^{-3}$ and a radius of
$0.1\,\mu$m were considered. Each grain provides $1.88\times10^6 $ surface sites
for surface recombinations \citep[][]{Bihamea01} through the Langmuir-Hinshelwood mechanism
\citep{HHL92}. A UV photodesorption yield of $10^{-5}$ was adopted \citep{Cruz_Diaz2016,Bertin2016}.
Photodissociation processes of solid species were taken from \citet{Garrod_Herbst06} and \citet{SW2011}.
A 1\% probability for nonthermal chemical desorption was assumed
\citep{2007A&A...467.1103G,2013ApJ...769...34V}. The ratio between diffusion and desorption barriers
for ices was taken to be 0.75 for all ices. The standard rate equation
approach to model the surface chemistry was used.
As initial abundances, the low metal elemental abundances of \citet{1982ApJS...48..321G}, \citet{Lea98}, and \citet{Agundez2013} were adopted, with hydrogen being mainly in a molecular form (see Table~\ref{tab:init_abunds}).

We considered two distinct scenarios in which our chemical network remained almost identical except for the energy barrier of
the single C $+$ H$_2$ $\rightarrow$ CH$_{2}$ surface reaction. In the first scenario, the original energy barrier $E_{\rm a} = 2\,500$~K
was used (ORG). In the second scenario, we assumed that this reaction has no barrier
(MOD). In what follows, for convenience we use the word ``ice'' when giving the names
of the surface species.

Using this chemical model and   the two distinct networks,
we computed the ISM chemical evolution over $10^{6}$~years on a 2D physical grid with 400 cells.
The gas volume densities range between $10^{3}$ and  $10^{12}$~cm$^{-3}$,
and the kinetic temperature varies between 10 and 300~K.
We considered dark and partly UV-illuminated ISM conditions with the visual extinction factors of $A_{\rm V}$ = 20 and 3~mag, respectively.
The chemical simulations took about two hours on
the Intel Core~i7 4-core 2.5~GHz CPU with the ALCHEMIC code compiled with gfortran~6.4.0.

\begin{table}
\caption{Initial abundances in the chemical model.}
\label{tab:init_abunds}
\centering
\begin{tabular}{ll}
\hline\hline
Species  & Relative abundances\\
\hline
H$_2$ &   $0.4999$     \\
H    &   $2.00 (-4)$  \\
He   &   $1.4 (-1)$  \\
C$^{+}$    &   $7.3 (-5)$  \\
N    &   $2.14 (-5)$  \\
O    &   $1.76 (-4)$  \\
S$^{+}$    &   $2.0 (-8)$  \\
Si$^{+}$   &   $3.0 (-9)$  \\
Na$^{+}$   &   $3.0 (-9)$  \\
Mg$^{+}$   &   $3.0 (-9)$  \\
Fe$^{+}$   &   $3.0 (-9)$  \\
P$^{+}$    &   $3.0 (-9)$ \\
Cl$^{+}$   &   $3.0 (-9)$  \\
e$^{-}$     & $7.3038 (-05)$ \\
\hline
\end{tabular}
\end{table}

To compare the computed time-dependent abundances between the two chemical networks,
we used the abundance ratio $\delta x$ and abundance difference $\Delta x$ as follows:
%The data analysis was continued with the comparison of the time-dependent abundances between the two calculated chemical networks considering the energy barrier of the aforementioned reaction (MOD; $\Delta E$=0 K and ORG; $\Delta E$=2.5$\times$10$^3$ K). The ratio was calculated as follows:
\begin{equation}
\label{eq:1}
\delta x_{i,k}(t_{j}) = \frac{x_{i,k}^{\rm MOD}(t_{j})}{x_{i,k}^{\rm ORG}(t_{j})},
\end{equation}
\begin{equation}
\label{eq:2}
\Delta x_{i,k}(t_{j}) = x_{i,k}^{\rm MOD}(t_{j}) - x_{i,k}^{\rm ORG}(t_{j}),
\end{equation}
where $x_{i,k}(t_{j})$ is the relative abundance of a species $i$ with respect to the total amount of hydrogen nuclei (n$\rm _H$ + 2n$\rm _{H_2}$) at a particular time moment $t_{j}$
in the grid cell number $k$ in the original (ORG) and modified (MOD) networks.

The abundance ratio $\delta x$ is useful to show the strongest deviations among the
considered molecular species, but it can be prone to large jumps in abundances of
minor species with low concentrations. To avoid this, we imposed a relative abundance threshold
of 10$^{-12}$, meaning that if the value of $x_{i,k}^{\rm ORG}$ or $x_{i,k}^{\rm MOD}$ was lower than this threshold,
their abundance ratio $\delta x$ was forced to be 1.
In contrast, the abundance difference $\Delta x$ is useful to
study how elemental atomic carbon is being redistributed among major chemical species
by the change in the energy barrier of the C + H$_{2}$ surface reaction. Below we use both of
these criteria to perform the chemical analysis of the differences between the two chemical networks.

\section{Results}
\label{sec:results}
%-------------------------------------------------------------------
To facilitate the chemical analysis, we split it into
several parts. First, we focus on investigating the overall effect of the barrier change in the single  surface reaction
on our gas-grain chemical model. For this, we used relative abundance ratios computed for all
time moments and the entire 2D physical grid and isolated sets of the most affected
species. Next, we restricted our analysis to the time moment $t_{j}= 10^{6}$~years as a
representative timescale for many interstellar environments such as low-mass molecular clouds,
and protoplanetary disks.

\subsection{Most affected species at all time moments}
\label{sec:top20}

In Table~\ref{tab:top_20_both} we list the 40 most affected species (with the highest increased or decreased abundances) over the entire
T-n$_{{\rm H}}$ grid for all computed time moments and for the two visual extinction factors.
We recall that the same molecule can fall into both categories of the species with
the highest increased or decreased abundances because it depends on physical conditions.
The majority of the most affected species for all computed time moments clearly are ices and radicals.
Only a few gaseous species appear in Table~\ref{tab:top_20_both}. For the $A_{\rm V}$ = 3~mag case, these include
C, CN, O$_{2}$ , and CH$_{2}$OH. In the $A_{\rm V}$ = 20~mag case, only C appears as a gaseous species in  Table~\ref{tab:top_20_both}.
The ice abundances that strongly increase when the surface reaction between C and H$_{2}$ becomes rapid include small hydrocarbons (e.g., CH$_{n}$ ($n=2-4$), C$_{2}$H$_{4}$, C$_{3}$), HCO, O$_{2}$H, N-bearing ices (NH$_{2}$, C$_3$N, H$_{2}$CN, HCCN, OCN), a few metal and Si-bearing ices, and many organic and complex organic species (CH$_2$CN, CH$_2$OH, etc.). The ice abundances that strongly decrease include atomic C, H$_{2}$, larger hydrocarbons (up to six C atoms), O$_{2}$H, HCO, HC$_{2}$O,
N-bearing ices (C$_{2}$N, HCCN), and a smaller number of COMs. The degree of abundance increase or decrease is much higher in the UV-dark
$A_{\rm V}$ = 20~mag case,
where the role of surface chemistry is stronger.

\begin{figure}
\centering
\setlength\tabcolsep{-0.5pt}
\renewcommand{\arraystretch}{0}
  \begin{tabular}{@{}cc@{}}
        \includegraphics[width=0.249\textwidth]{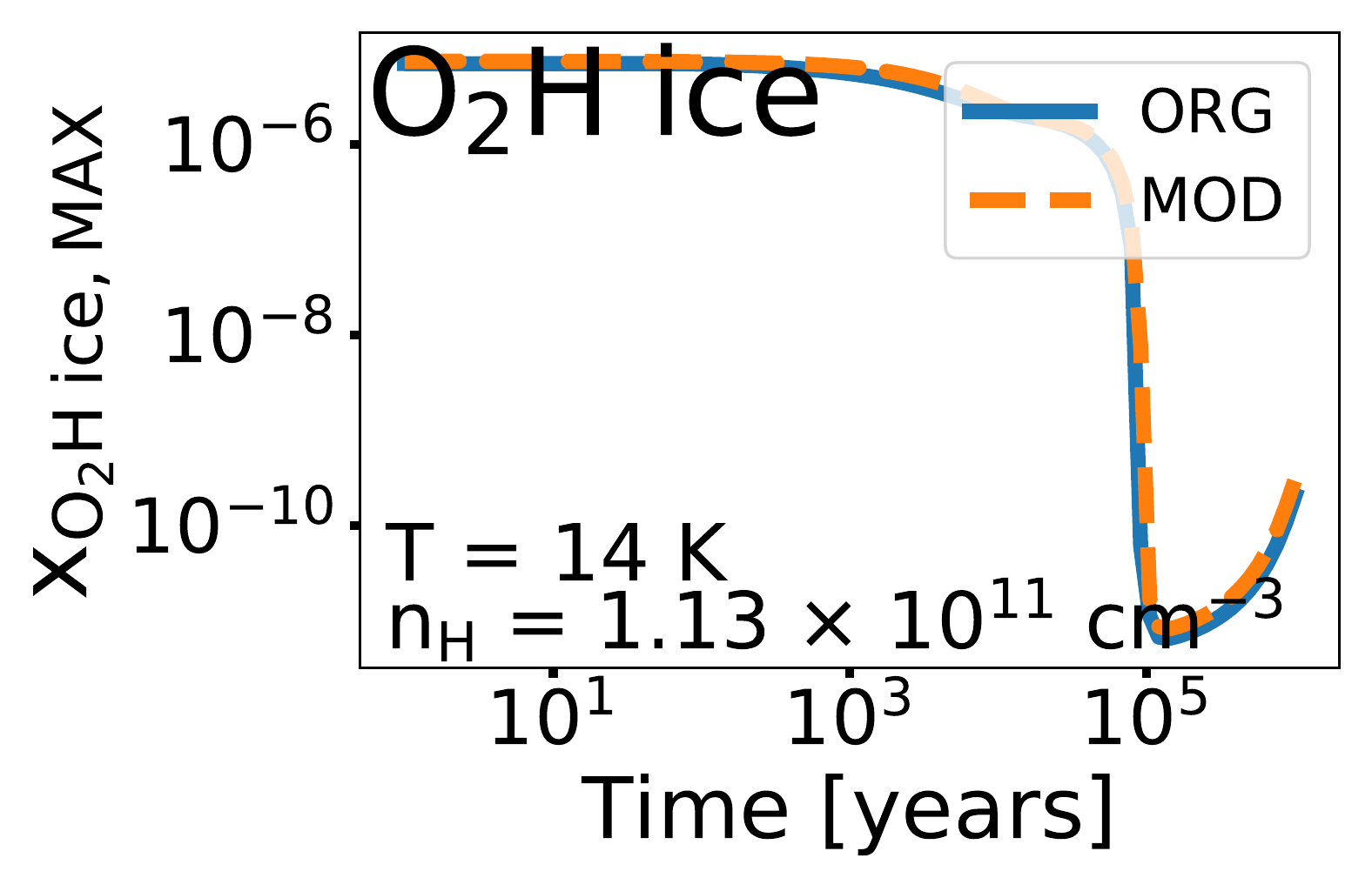} &
        \includegraphics[width=0.249\textwidth]{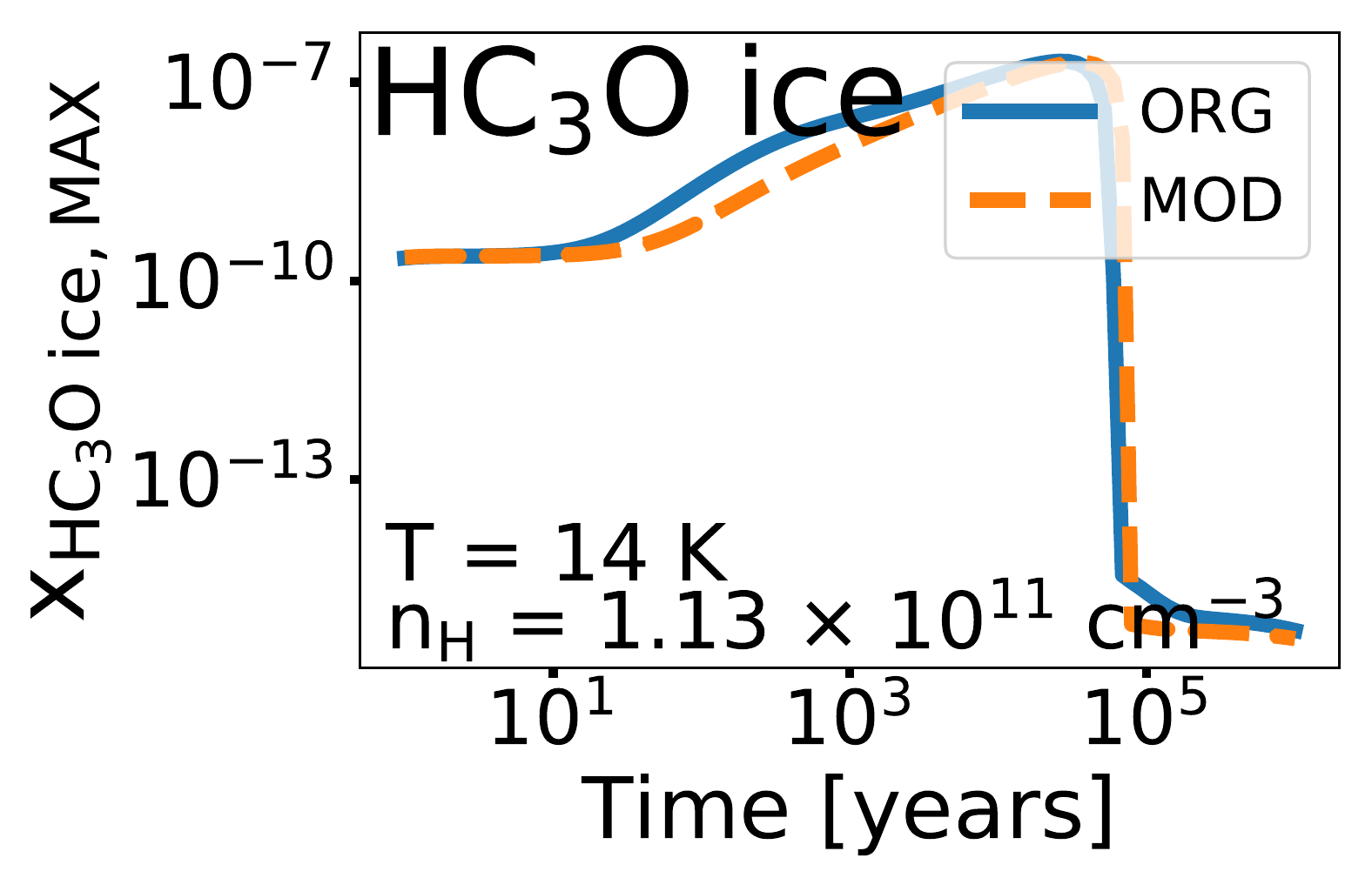} \\
        \includegraphics[width=0.249\textwidth]{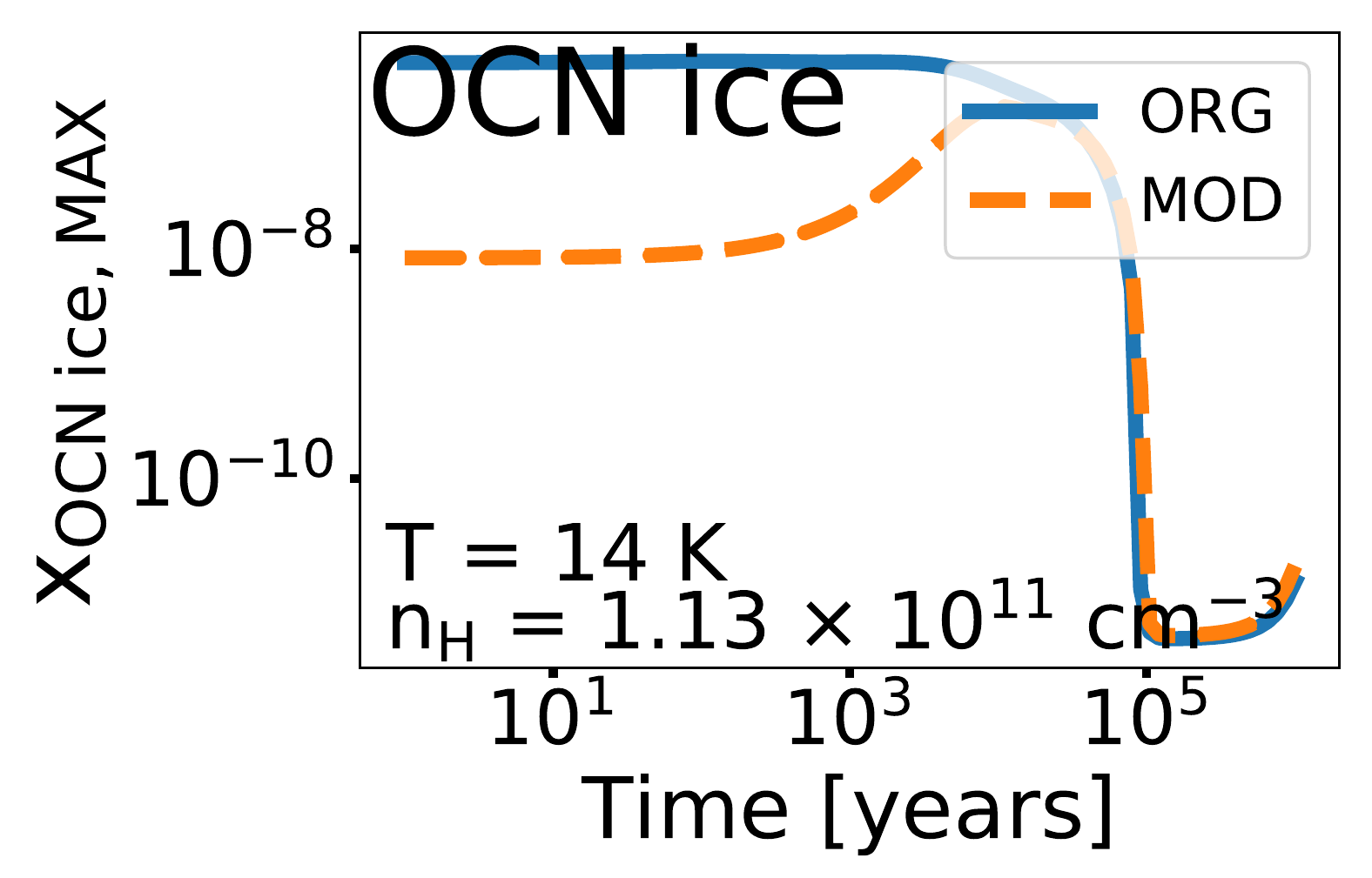} &
        \includegraphics[width=0.249\textwidth]{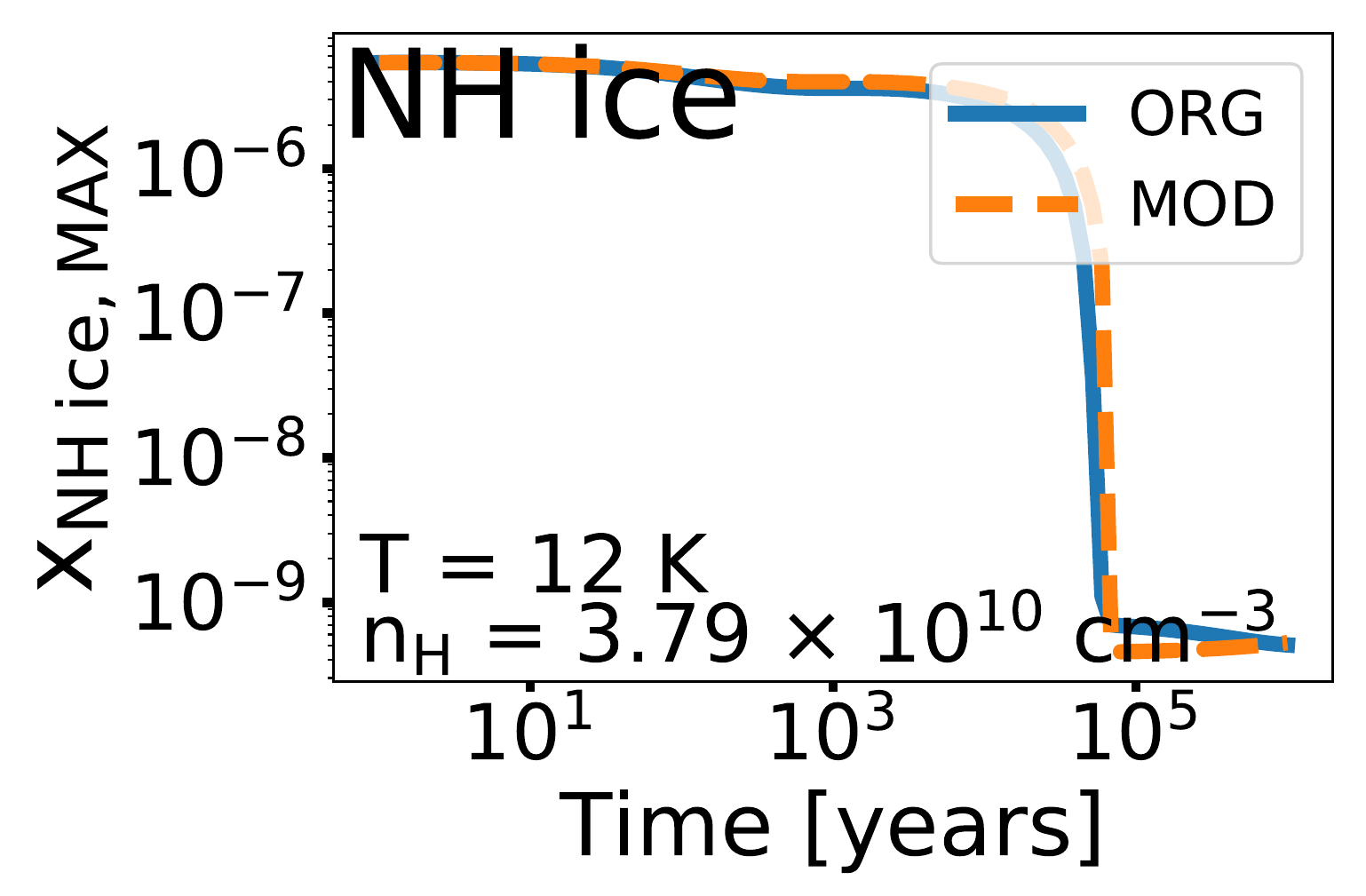} \\      
        \includegraphics[width=0.249\textwidth]{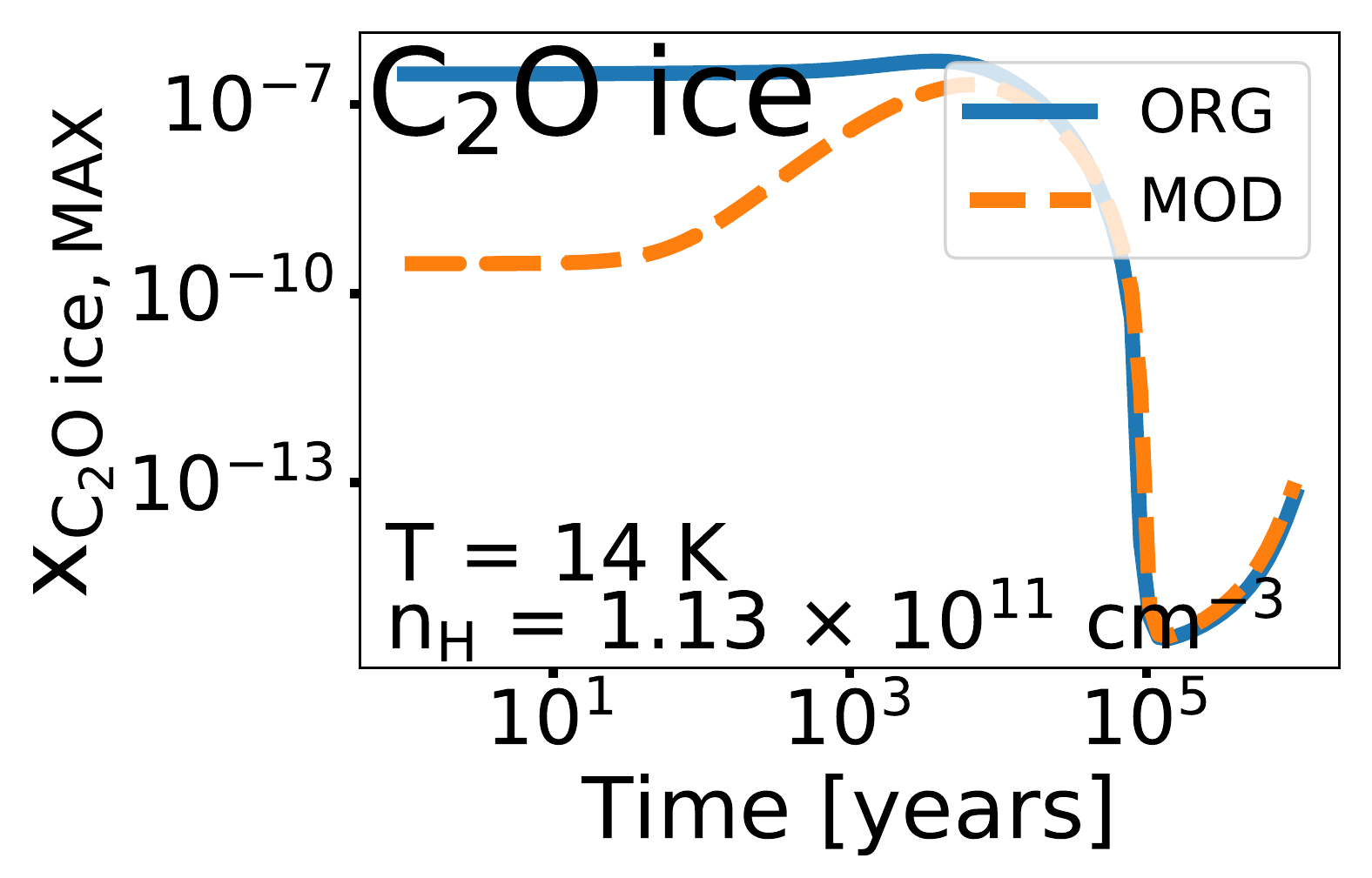} &
        \includegraphics[width=0.249\textwidth]{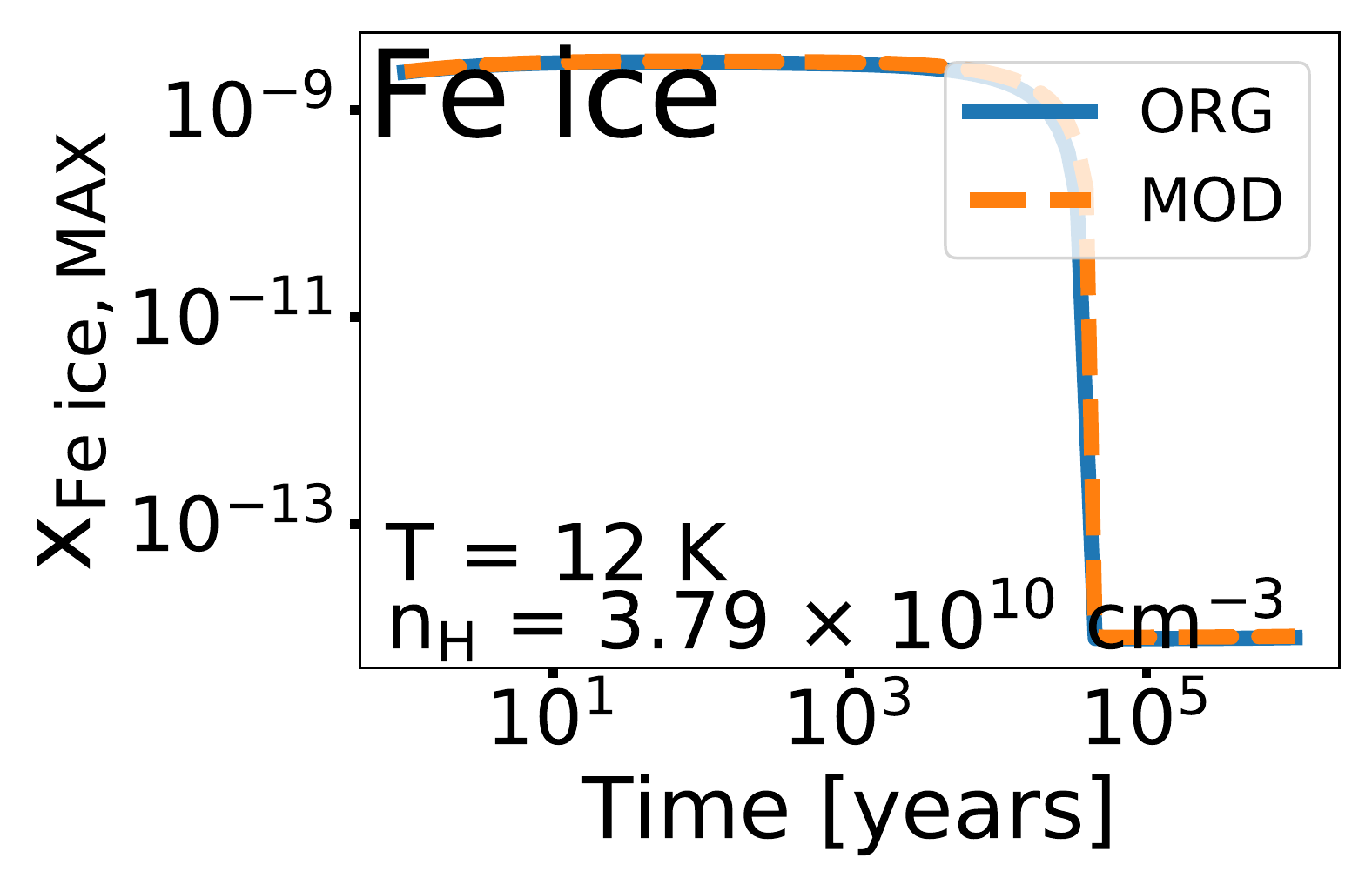} \\      
        \end{tabular}
        \caption{Time-dependent abundances of some surface species from Table \ref{tab:top_20_both} at physical conditions where the difference between both chemical models is most significant. The partly UV-irradiated model with $A_{\rm V}$ = 3~mag is shown at the left, and the ``dark'' model with $A_{\rm V}$ = 20~mag at the right.}
        \label{fig:gfe_1D}
\end{figure}

We used the abundance ratio $\delta x(i)$ to construct this table. These values are
prone to wide variations for the species whose chemical evolution undergoes a rapid transition
to another chemical state, for instance, as a result of freeze-out. This includes transient minor species such as Si- and metal-bearing ices. For example, for the O$_2$H, HC$_3$O, NH, and Fe ices in Figure \ref{fig:gfe_1D}, there is a slight shift in the time moment when their abundance drops rapidly at $\sim 10^{5}$~years, causing large abundance variations. However, the overall shape of their time-dependent abundances remains very similar in the original and modified chemical models. The chemical evolution for other species may start differently, as it does for OCN and C$_2$O ices in Figure \ref{fig:gfe_1D}, although large changes are still attributed to the same factor. 

The other most affected species include those that are chemically closely related to the C and CH$_{2}$ ices.
This is expected because the surface reaction C + H$_{2}$ $\rightarrow$ CH$_{2}$ is at the root of
their chemical evolution. The higher efficiency of this single reaction without the barrier
leads to a lower concentration of C atoms
on the dust surfaces and to more efficient production of CH$_{n}$ ices by surface hydrogenation.
Because there is less surface atomic carbon available for other reactions,
carbon insertion surface reactions leading to C$_{\rm n}$ ($n \geq 2$) and related species
become less effective and the abundances of these polyatomic
carbon species decrease in general.

However, the abundances of some C$_2$ and C$_3$-bearing species
do increase in the modified chemical model, for instance,
C$_2$H$_4$ ice, C$_2$N ice, C$_2$ and HCCN ices ($A_{\rm V}$ = 3~mag case), and C$_2$H$_3$ ice, HC$_3$O ice, C$_3$O ice, C$_3$N, C$_2$H$_5$, and
C$_3$ ($A_{\rm V}$ = 20~mag case). This is related to more efficient surface reactions
involving CH$_2$ and CH$_3$ radicals in the modified model, for example, CH$_2$ + CH$_2$ $\rightarrow$
C$_2$H$_4$ and CH$_2$ + CH$_3$ $\rightarrow$
C$_2$H$_5$. Further reactions of these hydrocarbon ices with
O- or N-bearing species and photoprocessing by UV lead to the synthesis of polycarbonaceous
N- and O-bearing ices.
%In addition, the surface concentration of H$_{2}$ consumed in the C + H$_{2}$ reactions
%also decreases, lowering the efficiency of other surface reactions involving H$_{2}$, including
%reactions forming complex organic molecules and metal-bearing ices.
%Please also note that in the partly UV-irradiated case, surface species can more efficiently desorb into the gas phase, leading to smaller changes for surface species as compared to the  ``UV-dark'' model (Table~\ref{tab:top_20_both}). Finally, among the most affected gas-phase species, CN, CCH and C$_{2}$O have been detected in the ISM.

\begin{table}
\caption{The 40 species that were most affected by the barrier change for all time moments and the entire 2D physical grid for both visual extinction factors. Species whose abundance changes are caused by rapid transitions to another chemical state are listed in boldface (see explanation in text).}
\label{tab:top_20_both}
\centering
\begin{tabular}{ll|ll}
\hline\hline
\multicolumn{2}{c}{$A_{\rm V}$ = 3~mag} & \multicolumn{2}{c}{$A_{\rm V}$ = 20~mag}\\
\hline
species & $\delta x_{i,k}(t_{j})$ & species & $\delta x_{i,k}(t_{j})$ \\
\hline
ice CH$_2$ & 3.52E+02 & \textbf{ice O$_2$H} & 1.38E+05 \\
ice CH$_3$ & 3.05E+02 & \textbf{ice CH$_2$OH} & 7.82E+04 \\
\textbf{ice O$_2$H} & 1.25E+02 & \textbf{ice CH$_3$NH} & 1.87E+04 \\
ice CH$_4$ & 1.16E+02 & \textbf{ice H$_2$CN} & 1.81E+04 \\
ice CH$_2$OH & 1.02E+02 & \textbf{ice CH$_2$NH} & 1.13E+04 \\
\textbf{ice CH$_2$NH$_2$} & 8.29E+01 & \textbf{ice C$_3$O} & 3.42E+03 \\
\textbf{ice CH$_3$NH} & 8.29E+01 & \textbf{ice C$_3$N} & 2.99E+03 \\
ice C$_2$H$_4$ & 8.21E+01 & \textbf{ice C$_3$} & 2.80E+03 \\
\textbf{ice CH$_2$NH} & 8.05E+01 & \textbf{ice C$_2$H$_5$} & 2.30E+03 \\
ice CH$_2$CN & 6.48E+01 & \textbf{ice HCO} & 1.75E+03 \\
ice HCCN & 6.47E+01 & \textbf{ice HC$_3$O} & 8.29E+02 \\
ice C$_2$N & 6.08E+01 & ice CH$_2$ & 3.52E+02 \\
\textbf{ice OCN} & 5.83E+01 & ice CH$_3$ & 3.05E+02 \\
\textbf{ice HCO} & 5.22E+01 & \textbf{ice NH$_2$} & 2.82E+02 \\
ice CH$_3$OH & 5.18E+01 & \textbf{ice MgH} & 2.50E+02 \\
ice H$_2$CN & 3.52E+01 & ice C$_2$H$_3$ & 2.21E+02 \\
ice NH$_2$CHO & 3.00E+01 & \textbf{ice NH} & 1.89E+02 \\
O$_2$ & 2.23E+01 & \textbf{ice Na} & 1.78E+02 \\
CH$_2$OH & 2.04E+01 & \textbf{ice SiH$_2$} & 1.63E+02 \\
ice C$_2$ & 1.84E+01 & \textbf{ice Fe} & 1.63E+02 \\
\hline
ice C & 1.59E-07 & ice C & 5.22E-08 \\
ice C$_3$H$_3$ & 1.65E-04 & \textbf{ice H$_2$} & 4.19E-05 \\
ice C$_3$H$_4$ & 4.29E-04 & \textbf{ice O$_2$H} & 9.81E-05 \\
C & 6.39E-04 & ice C$_3$H$_4$ & 1.49E-04 \\
ice H$_2$ & 6.79E-04 & \textbf{ice CH$_2$NH} & 1.66E-04 \\
\textbf{ice C$_2$O} & 9.77E-04 & ice C$_3$H$_3$ & 2.11E-04 \\
ice HCCN & 1.02E-03 & \textbf{ice HCO} & 2.19E-04 \\
ice C$_2$H$_3$ & 1.05E-03 & C & 3.05E-04 \\
\textbf{ice HC$_2$O} & 1.13E-03 & \textbf{ice C$_2$H$_3$} & 3.99E-04 \\
ice C$_2$H & 1.26E-03 & \textbf{ice CH$_2$OH} & 4.38E-04 \\
ice C$_2$ & 1.55E-03 & ice C$_2$ & 6.13E-04 \\
ice C$_2$N & 1.91E-03 & ice HCCN & 6.81E-04 \\
ice C$_6$H & 2.14E-03 & \textbf{ice CH$_2$CN} & 8.00E-04 \\
ice C$_5$H & 2.58E-03 & ice HC$_2$O & 8.90E-04 \\
ice C$_6$ & 3.31E-03 & ice C$_2$O & 8.96E-04 \\
ice C$_6$H$_2$ & 3.32E-03 & ice C$_2$H & 1.24E-03 \\
ice CH & 3.37E-03 & ice C$_6$H & 1.30E-03 \\
ice CH$_2$CN & 4.77E-03 & \textbf{ice C$_3$} & 1.56E-03 \\
ice C$_5$H$_2$ & 5.54E-03 & ice C$_5$H & 1.76E-03 \\
CN & 6.21E-03 & ice C$_2$N & 1.83E-03 \\

\hline
\end{tabular}
\end{table}

\subsection{Most affected species at $t_{j}=10^{6}$~years}
\label{sec:top_1Myr}

Because many low-mass prestellar and star-forming environments have a typical evolutionary timescale of about
1~Myr, we list in Table~\ref{tab:top_20_both_1Myr} the 40 species for which the abundance changes are the largest at the final evolutionary time moment in our simulations ($t_{j}= 10^{6}$~years). By that time,
many species that are produced mainly by gas-phase reactions attain a chemical quasi-steady-state, and the corresponding abundance ratios $\delta x_{i}$ are therefore much lower than in Table~\ref{tab:top_20_both}. In contrast,
complex organic species that are solely synthesized by slow surface processes do not reach steady state by
$t=1$~Myr. Table~\ref{tab:top_20_both_1Myr}
contains many (complex) organic species, including H$_2$CO,  HNCO, CH$_3$OH, CH$_{2}$CN, and CH$_3$OCH$_3$.
The majority of species in Table~\ref{tab:top_20_both_1Myr} are still hydrocarbon and
N-bearing carbonaceous ices,
as in Table~\ref{tab:top_20_both}.
The exotic metals, Si-bearing ices, and other transient radicals such as O$_{2}$H
are no longer present because strong variations in their abundances mainly occur during the build-up phase of the ice mantles.
Moreover, there are more gaseous species in the list of the most affected species at 1~Myr, such as
C, small hydrocarbons, O$_{2}$, H$_{2}$CO, NO, NH, and OCN. Most of these simple gaseous species have been detected in the diffuse and dense ISM.

\begin{table}
\caption{The 40 species that were most affected by the barrier change for the final time moment ($t_{j}= 10^{6}$~years) and the entire 2D physical grid for the two visual extinction factors.}
\label{tab:top_20_both_1Myr}
\centering
\begin{tabular}{ll|ll}
\hline\hline
\multicolumn{2}{c}{$A_{\rm V}$ = 3~mag} & \multicolumn{2}{c}{$A_{\rm V}$ = 20~mag}\\
\hline
species & $\delta x_{i,k}(10^{6}~yr)$ & species & $\delta x_{i,k}(10^{6}~yr)$ \\
\hline
C$_2$H$_4$ & 1.09E+01 & ice HCO & 1.48E+01 \\
ice CH$_2$ & 8.01E+00 & ice NH$_2$CHO & 1.47E+01 \\
ice CH$_2$OH & 7.99E+00 & ice CH$_3$OCH$_3$ & 1.31E+01 \\
ice CH$_3$ & 7.93E+00 & ice C$_2$H$_5$OH & 1.31E+01 \\
ice CH$_3$OH & 7.90E+00 & ice CH$_3$OH & 1.30E+01 \\
ice CH$_4$ & 7.60E+00 & ice C$_2$H$_6$ & 1.21E+01 \\
ice H$_2$CO & 7.02E+00 & ice CH$_2$NH & 7.42E+00 \\
ice C$_2$H$_4$ & 5.24E+00 & ice CH$_3$NH & 7.42E+00 \\
CH$_2$ & 4.38E+00 & ice CH$_5$N & 7.22E+00 \\
CH$_4$ & 4.15E+00 & ice OCN & 5.47E+00 \\
ice NH$_2$CHO & 3.97E+00 & H$_2$CO & 5.09E+00 \\
C$_2$H$_2$ & 3.46E+00 & ice O$_3$ & 4.74E+00 \\
CH$_3$ & 3.39E+00 & C$_2$H$_4$ & 4.70E+00 \\
H$_2$CO & 3.38E+00 & O$_2$ & 4.60E+00 \\
C$_2$H & 3.18E+00 & ice O$_2$ & 3.92E+00 \\
ice O$_3$ & 2.90E+00 & NO & 3.63E+00 \\
ice H$_2$CN & 2.83E+00 & CH$_4$ & 3.35E+00 \\
CH & 2.70E+00 & NH & 3.30E+00 \\
ice CH$_2$NH & 2.69E+00 & ice CH$_3$C$_3$N & 3.25E+00 \\
ice CH$_3$NH & 2.59E+00 & OCN & 3.23E+00 \\
\hline
ice C$_5$H$_2$ & 1.42E-02 & ice C$_3$H$_4$ & 3.21E-04 \\
ice C$_6$H$_2$ & 3.43E-02 & ice C$_6$H$_2$ & 3.28E-03 \\
ice C$_6$H & 4.25E-02 & ice CH$_3$CN & 4.71E-03 \\
ice C$_7$H$_2$ & 4.48E-02 & ice C$_5$H$_2$ & 6.20E-03 \\
ice C & 1.32E-01 & ice HNCO & 1.37E-02 \\
ice CH$_3$CN & 1.38E-01 & ice CH$_2$CO & 1.41E-02 \\
C & 1.44E-01 & ice C$_4$H$_4$ & 2.59E-02 \\
ice C$_2$H$_4$ & 1.46E-01 & ice C$_4$H$_2$ & 2.70E-02 \\
HCCN & 1.60E-01 & ice C$_4$H$_3$ & 3.14E-02 \\
ice CH$_2$CN & 1.67E-01 & ice C$_7$H$_2$ & 3.18E-02 \\
ice CS & 1.68E-01 & ice C$_4$H & 3.18E-02 \\
ice HCCN & 1.72E-01 & ice CH$_2$CN & 4.42E-02 \\
ice C$_2$N & 1.75E-01 & ice C$_2$H$_4$ & 5.34E-02 \\
ice C$_2$H$_2$ & 1.88E-01 & ice HNC$_3$ & 6.84E-02 \\
ice C$_2$H$_3$ & 1.96E-01 & ice HC$_5$N & 7.55E-02 \\
ice C$_4$H$_2$ & 1.98E-01 & ice C$_4$N & 9.01E-02 \\
C$_2$H & 2.13E-01 & ice HNC & 1.22E-01 \\
ice C$_2$O & 2.14E-01 & ice H$_2$CS & 1.31E-01 \\
C$_2$H$_2$ & 2.22E-01 & ice C & 2.00E-01 \\
ice HC$_2$O & 2.23E-01 & ice HCCN & 3.47E-01 \\
\hline
\end{tabular}
\end{table}

\begin{figure}
\centering
\setlength\tabcolsep{-0.5pt}
\renewcommand{\arraystretch}{0}
\begin{minipage}{0.49\textwidth}

  \begin{tabular}{@{}cc@{}}
        \includegraphics[width=0.49\textwidth]{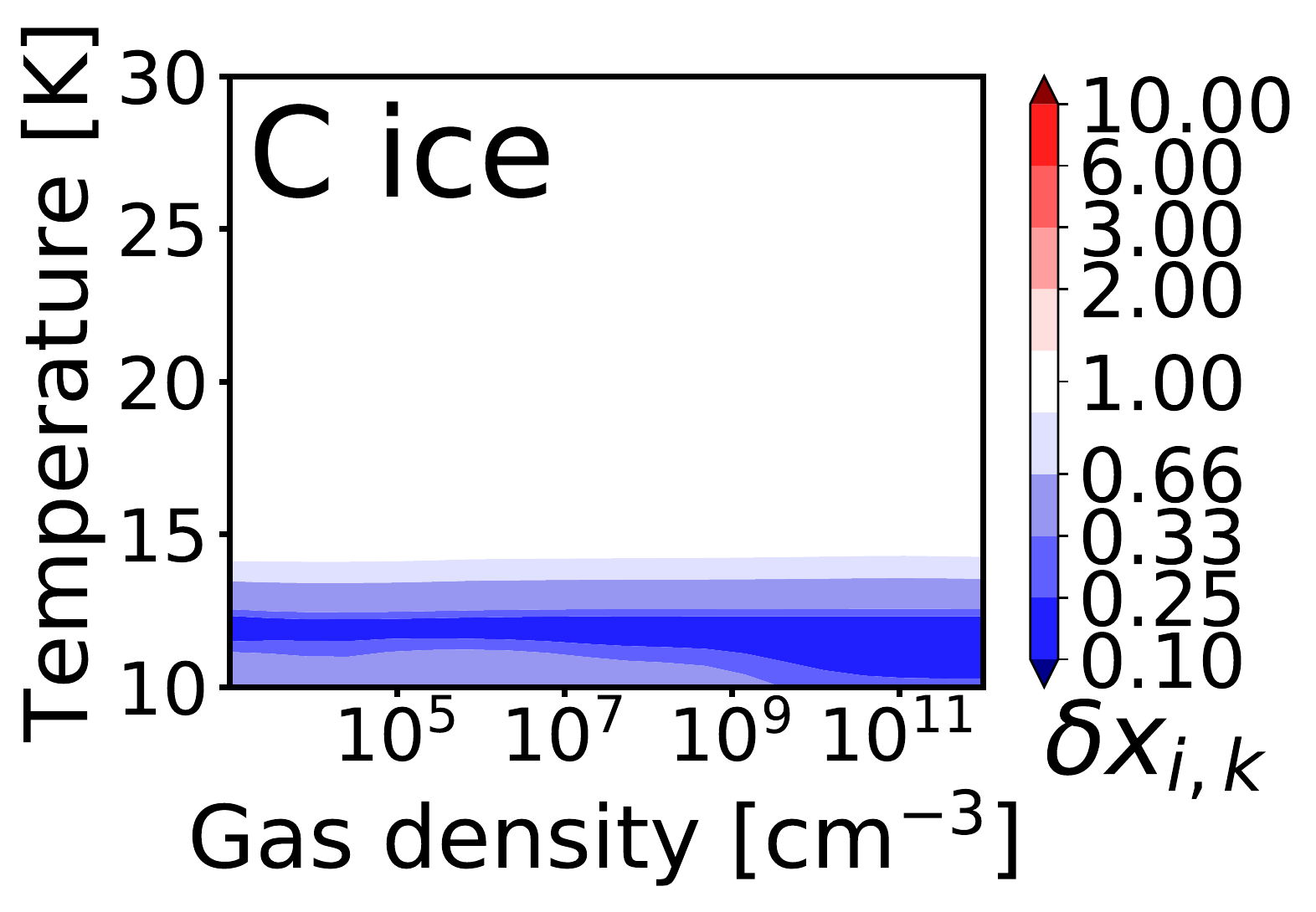} &
        \includegraphics[width=0.49\textwidth]{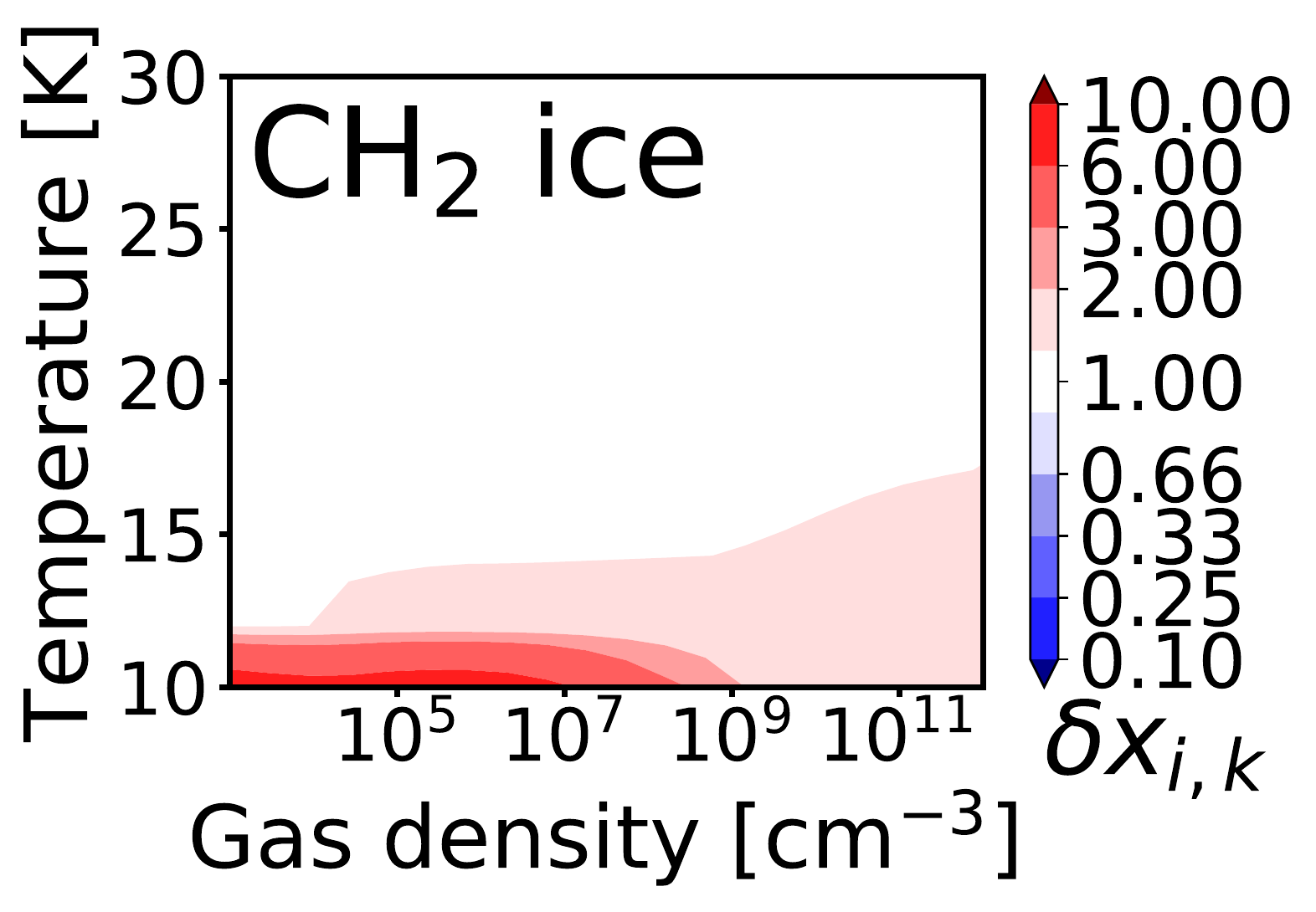} \\
        \includegraphics[width=0.49\textwidth]{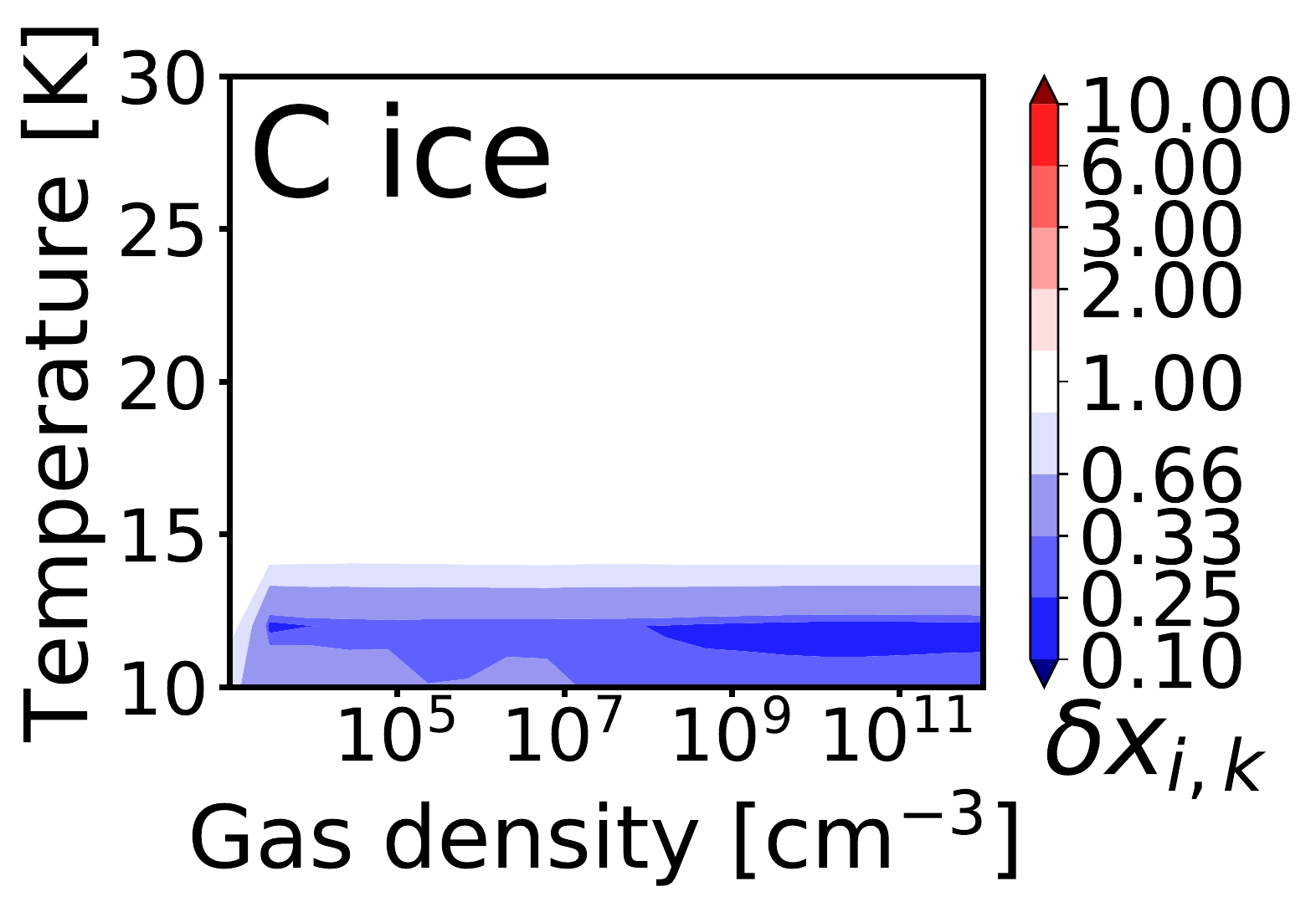} &
        \includegraphics[width=0.49\textwidth]{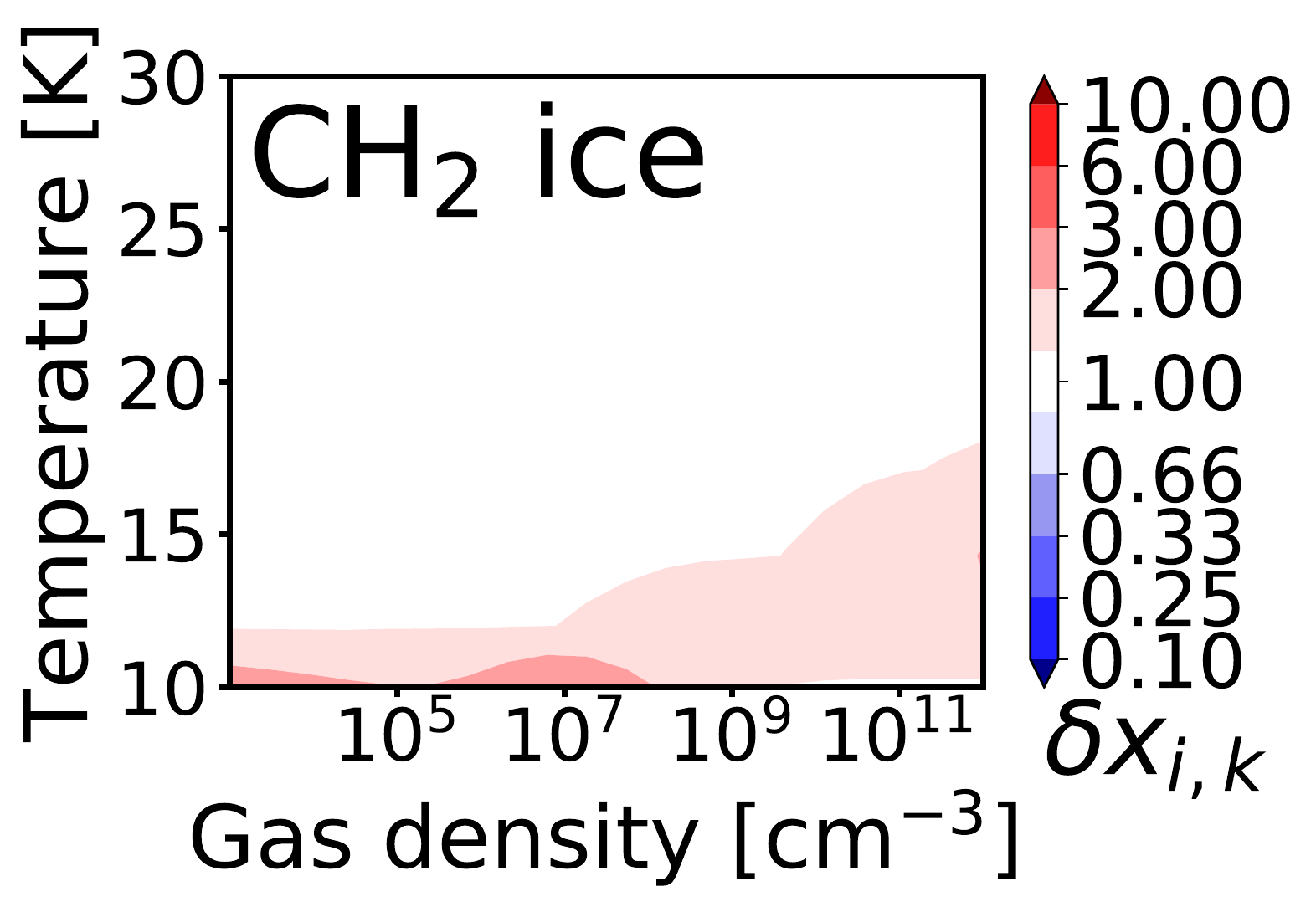}
        \end{tabular}
        \caption{Abundance ratios $\delta x$ for the surface C and CH$_{2}$ as a function of density and temperature at $t_{j}=10^{6}$~years.
        The partly UV-irradiated model with $A_{\rm V}$ = 3~mag is shown at the top, and the dark  model with $A_{\rm V}$ = 20~mag at the bottom.
        }
        \label{fig:gc_gch2_rel}
\end{minipage}
\hspace{0.10cm}
\begin{minipage}{0.49\textwidth}

        \begin{tabular}{@{}cc@{}}
        \includegraphics[width=0.49\textwidth]{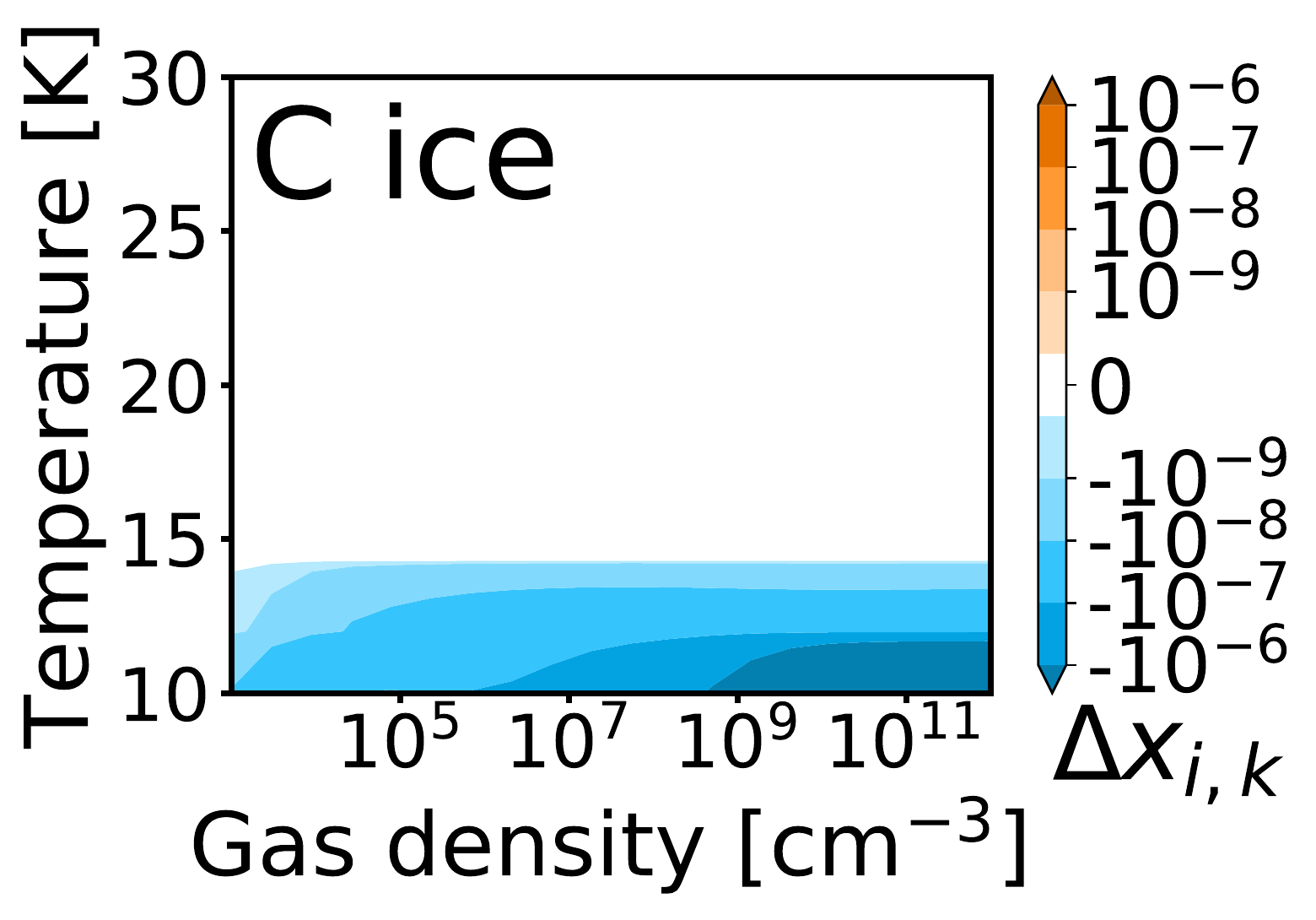} &
        \includegraphics[width=0.49\textwidth]{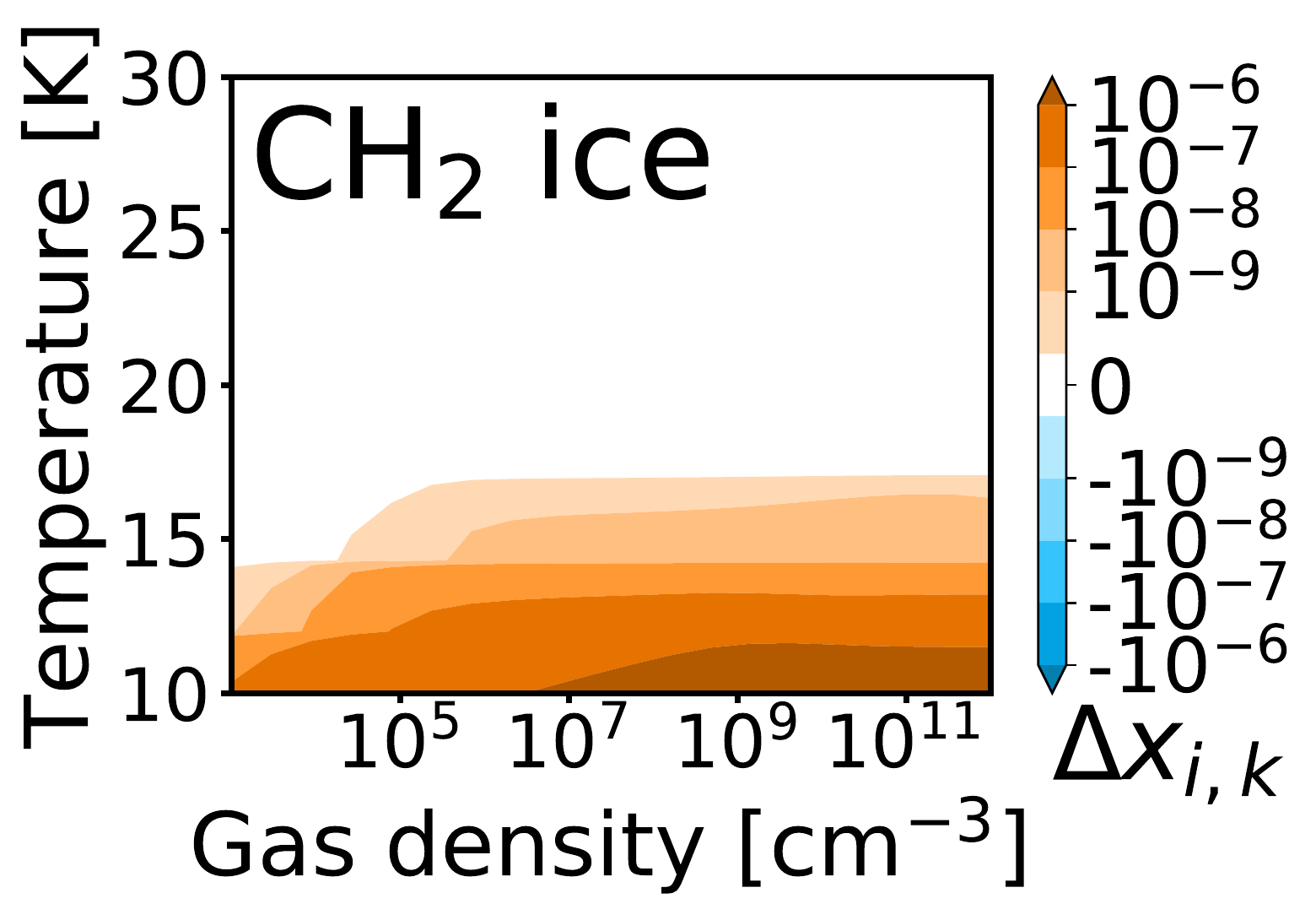} \\
        \includegraphics[width=0.49\textwidth]{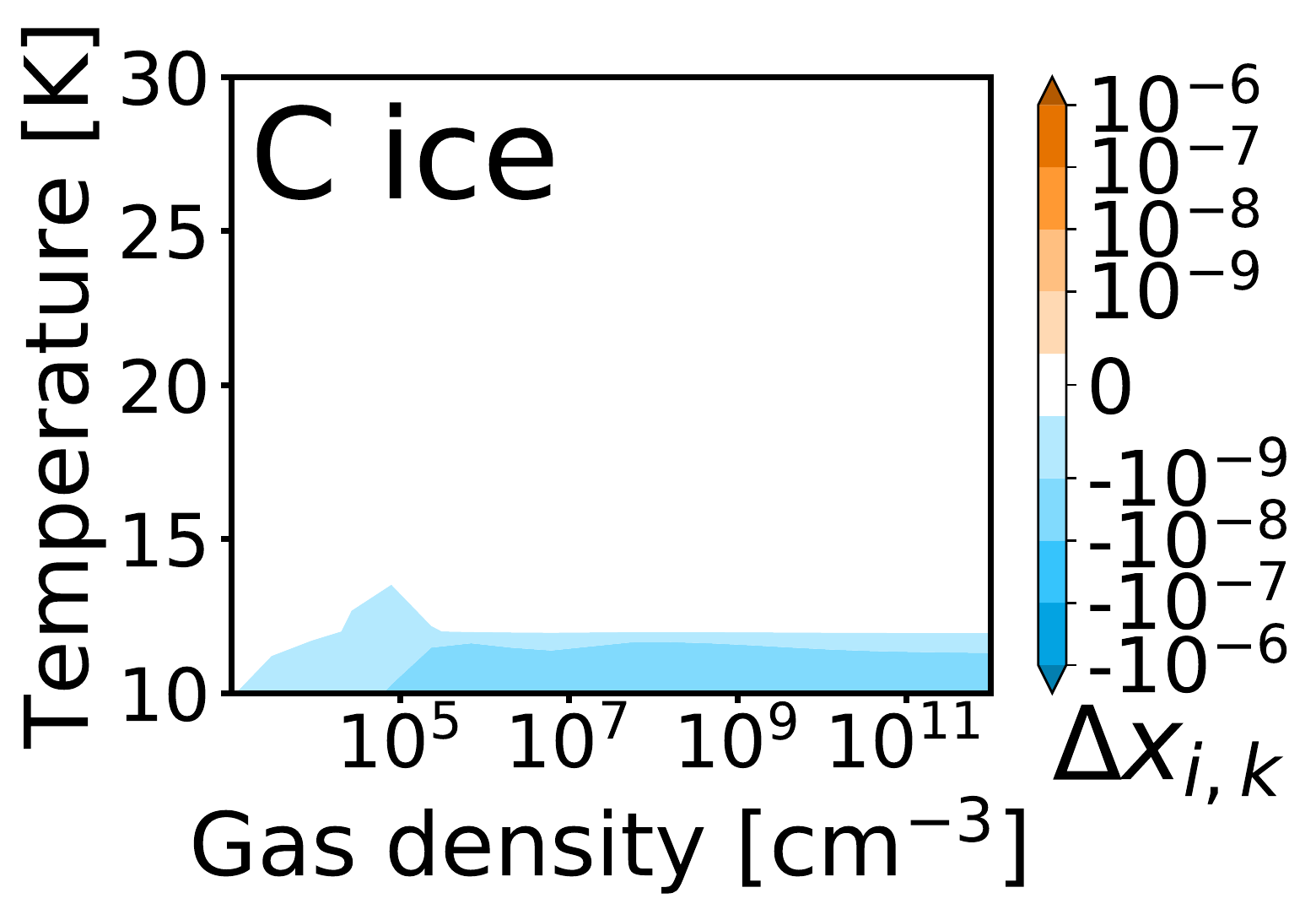} &
        \includegraphics[width=0.49\textwidth]{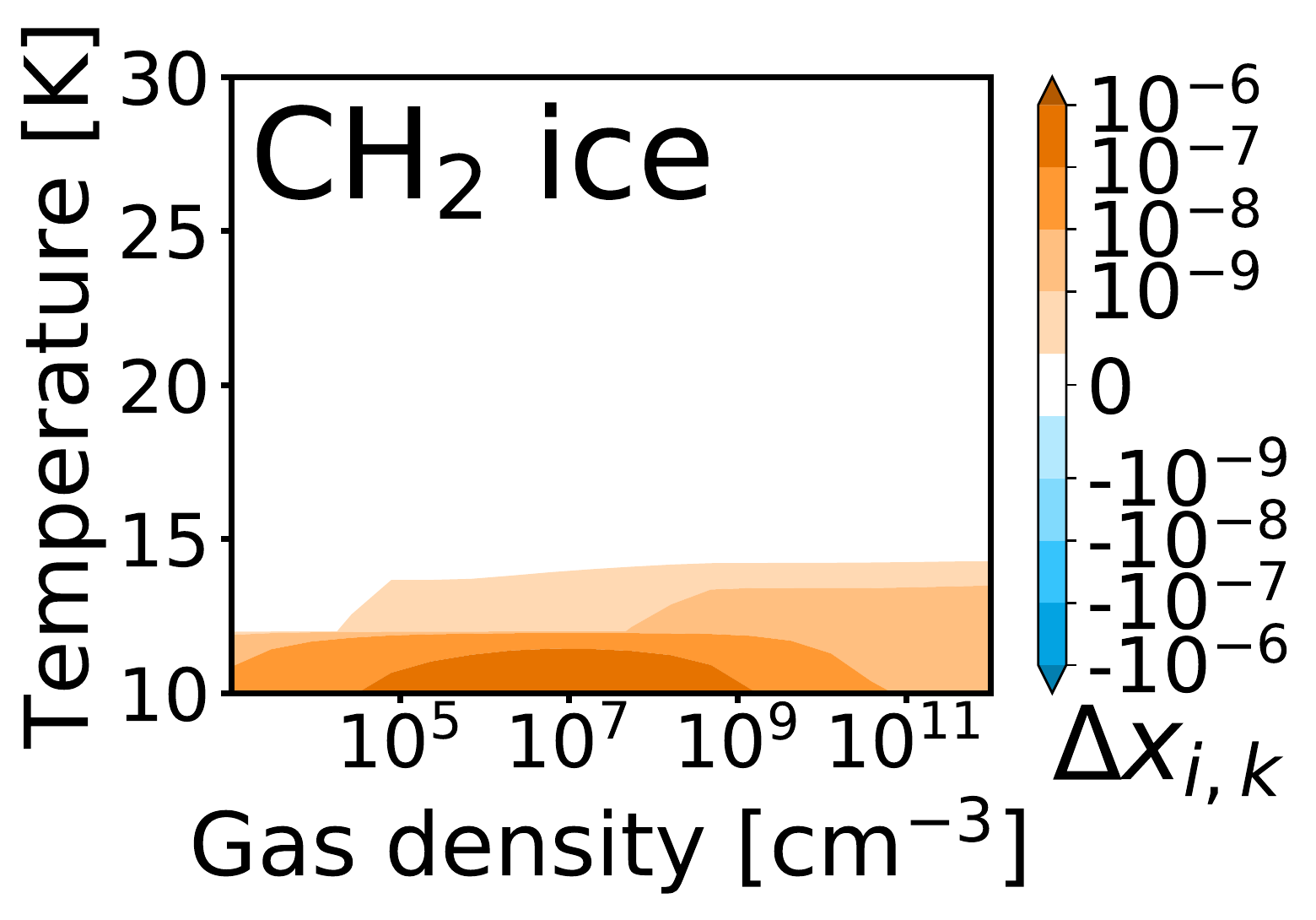}
        \end{tabular}
        \caption{Abundance difference $\Delta x$ for the surface C and CH$_{2}$ molecules
        as a function of density and temperature at $t_{j}=10^{6}$~years.
        The partly UV-irradiated model with $A_{\rm V}$ = 3~mag is shown at the top, and the dark model with $A_{\rm V}$ = 20~mag at the bottom.}
        \label{fig:gc_gch2_abs}
\end{minipage}
\end{figure}

Next, we aim to better understand how late-time chemistry of key molecules is affected  by the barrier change in the C + H$_{2}$ surface reaction.
While we computed our grid for a wide range of temperatures, we found that most
of the abundance variations occur at low temperatures of $\lesssim 10-30$~K, when most of the volatiles are frozen and where
surface reactions are most active. We therefore limit the analysis to this temperature range below.

\begin{figure}
\centering
\setlength\tabcolsep{-0.5pt}
\renewcommand{\arraystretch}{0}
        \begin{tabular}{@{}cc@{}}
        \includegraphics[width=0.245\textwidth]{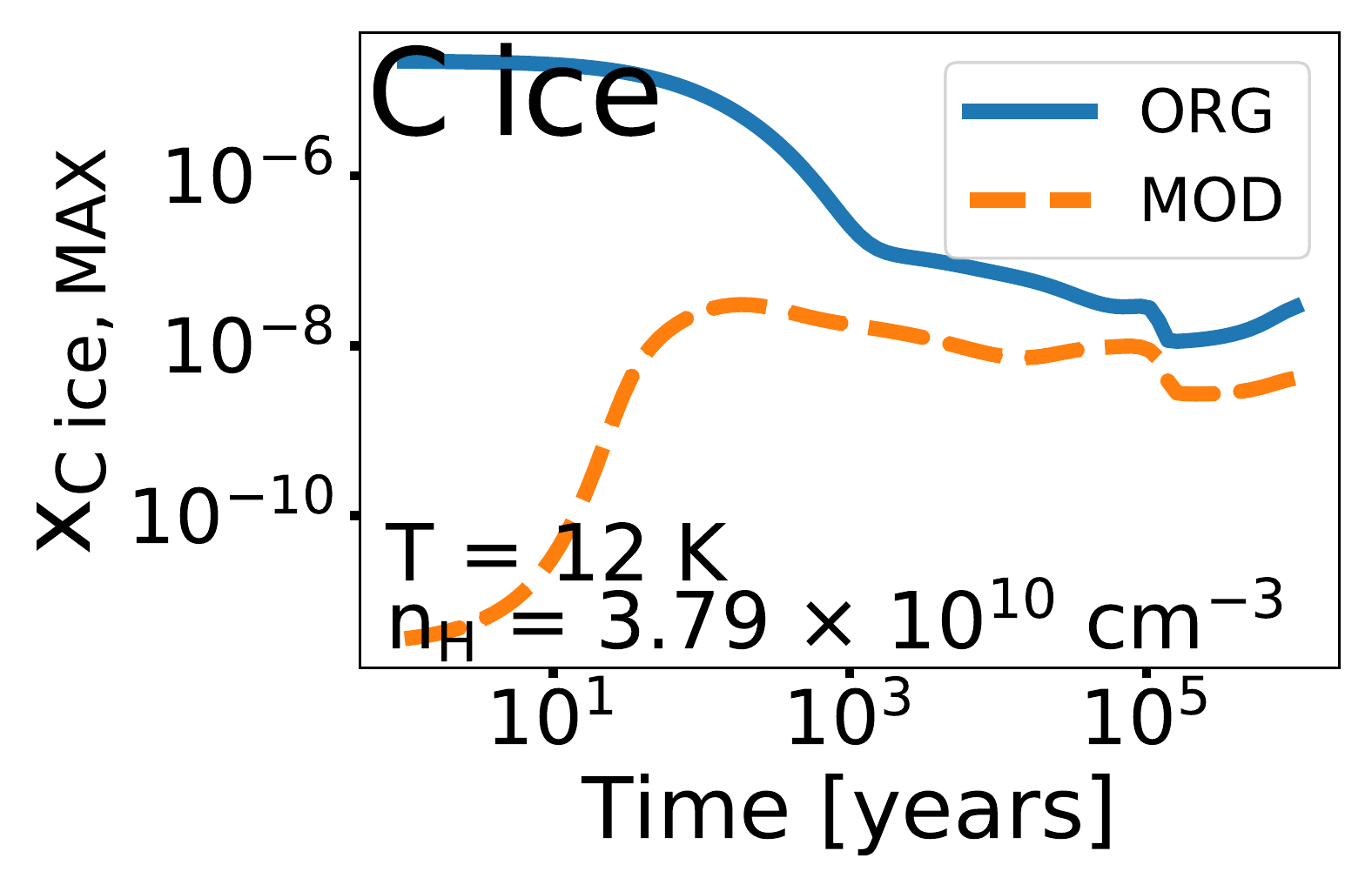} &
        \includegraphics[width=0.245\textwidth]{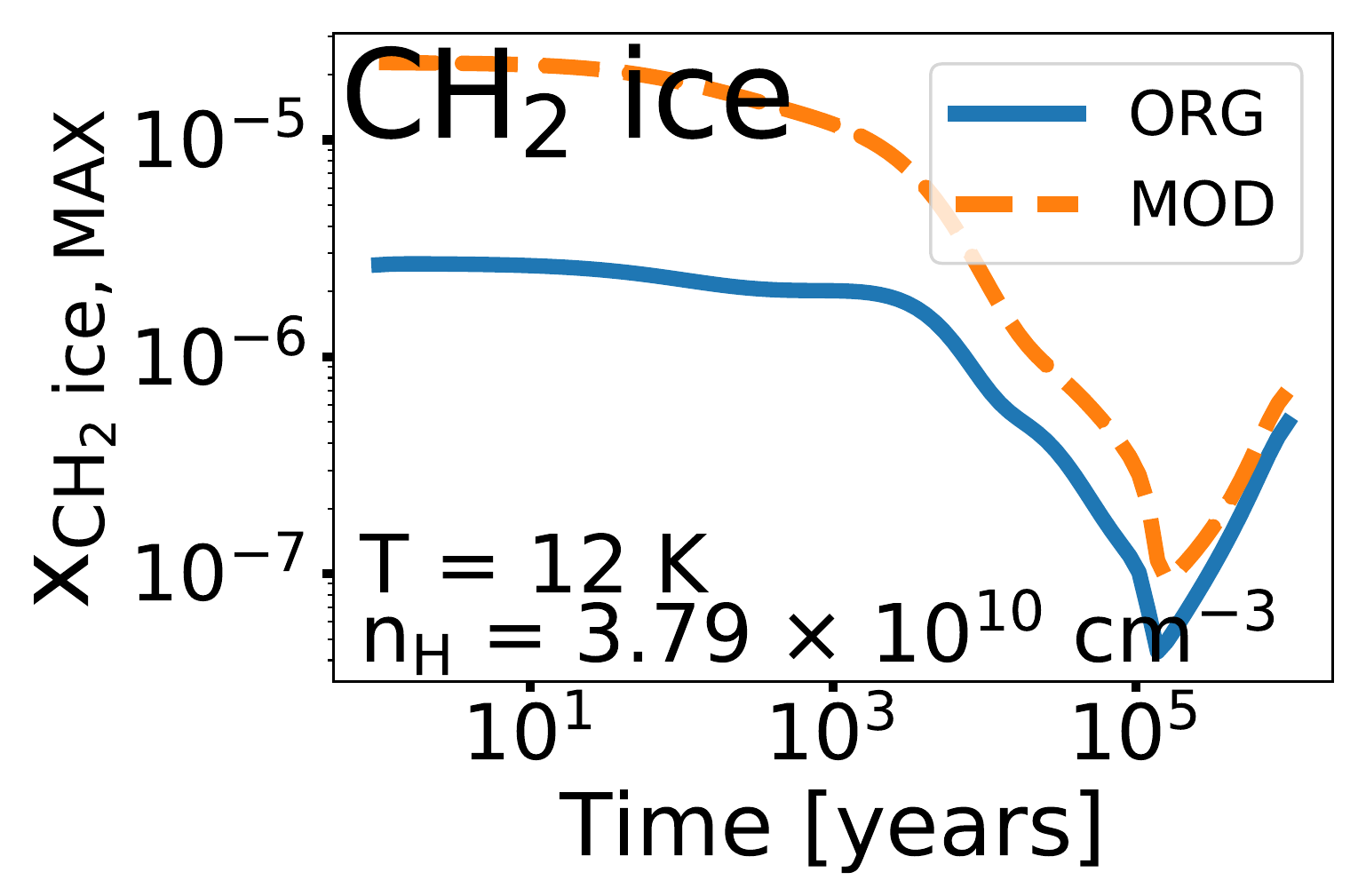} \\
        \includegraphics[width=0.245\textwidth]{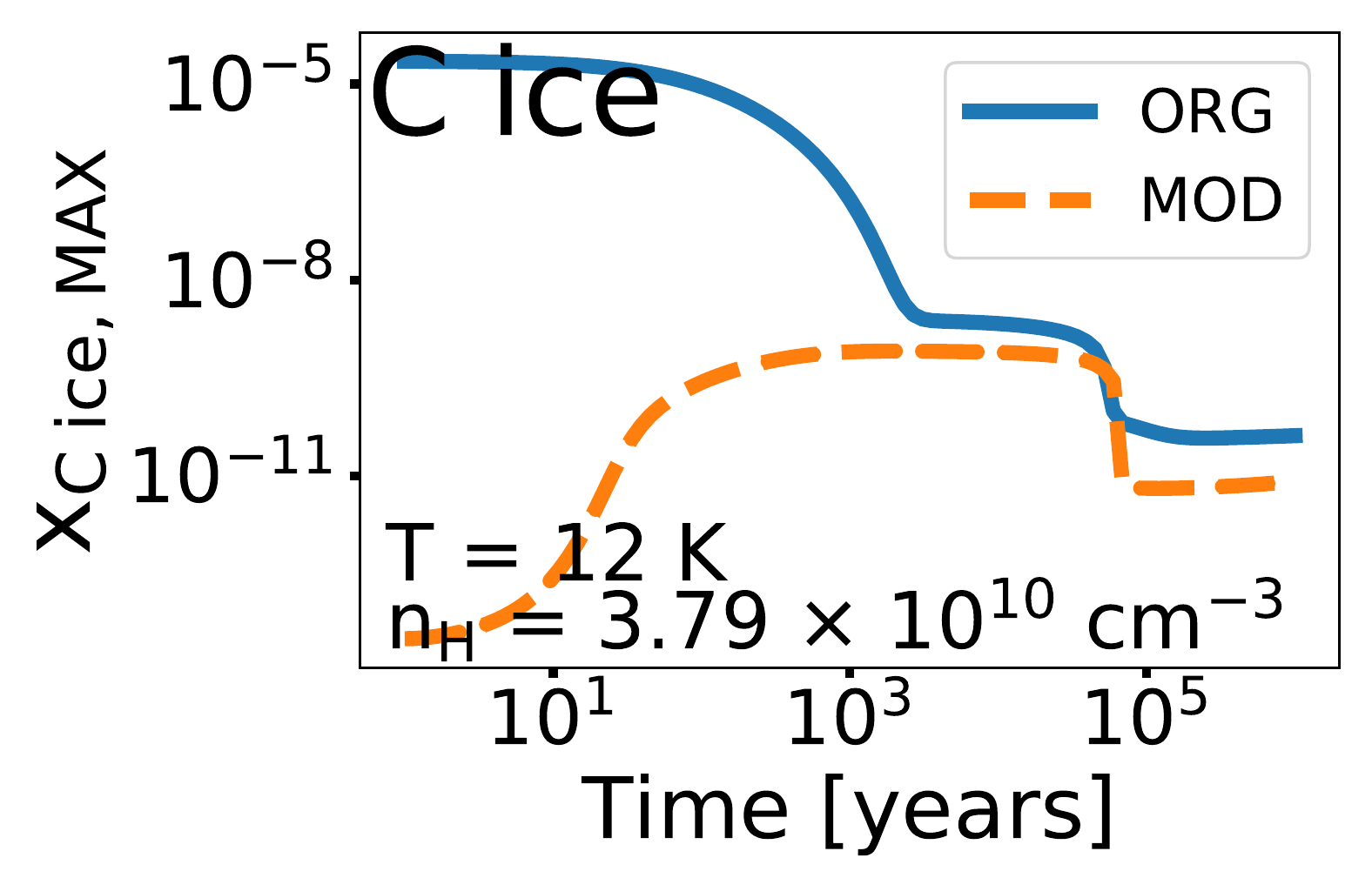} &
        \includegraphics[width=0.245\textwidth]{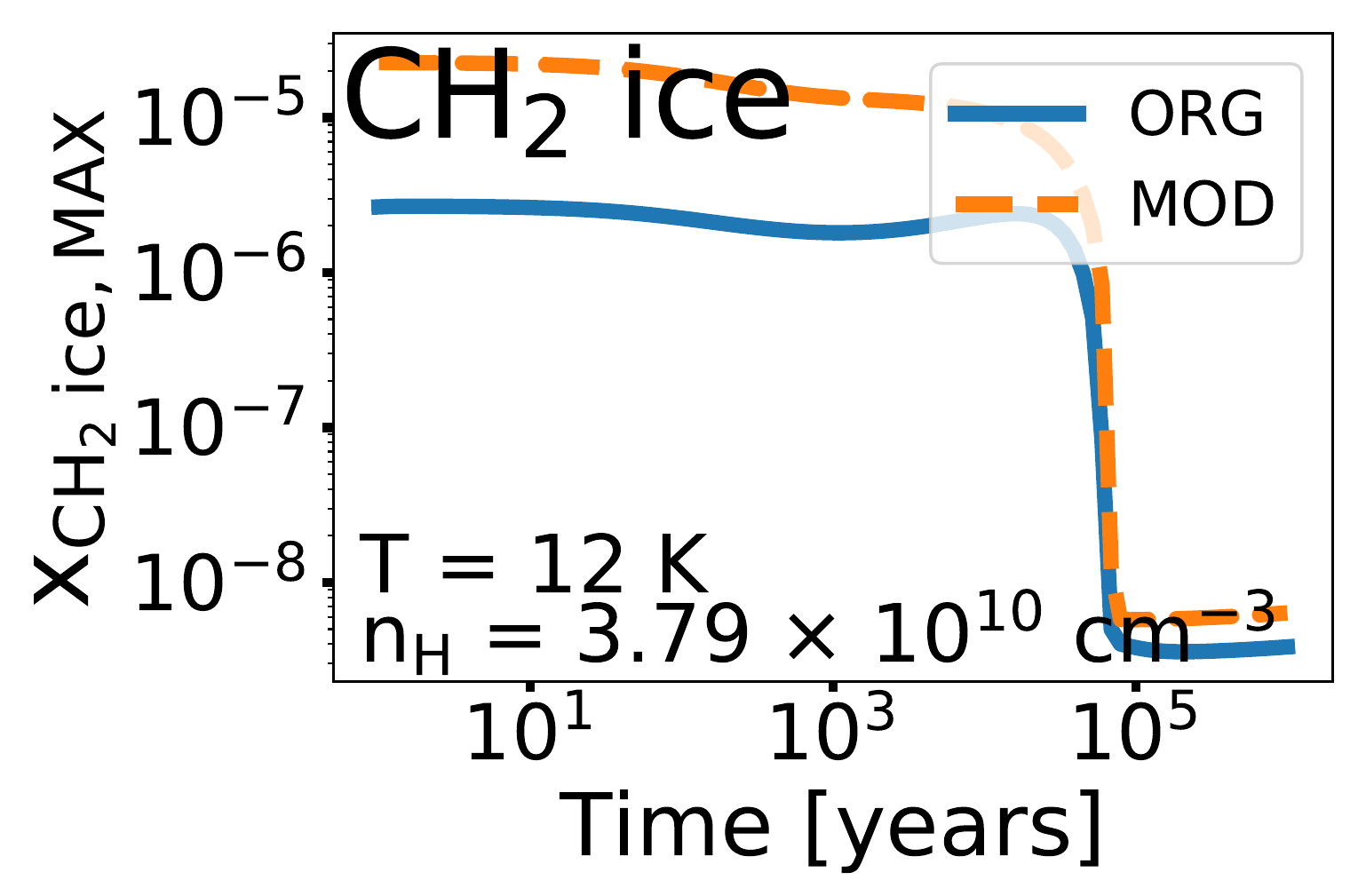}
        \end{tabular}
                \caption{Time-dependent abundances of surface species C and CH$_2$ at physical conditions where the difference between the two chemical models is most significant. The partly UV-irradiated model with $A_{\rm V}$ = 3~mag is shown at the top, and the dark model with $A_{\rm V}$ = 20~mag at the bottom. }
        \label{fig:gc_gch2_1D}
\end{figure}

The abundance ratios and differences for the C and CH$_{2}$ ices and their time-dependent abundances in a representative grid cell with $n_{\rm H} = 3.79 \times 10^{10}$~cm$^{-3}$ and $T=12$~K are shown in \Cref{fig:gc_gch2_rel,fig:gc_gch2_abs,fig:gc_gch2_1D}.
Figures~\ref{fig:gc_gch2_rel} and~\ref{fig:gc_gch2_abs} clearly show that the surface abundances of C atoms  decrease by a factor of $\sim 2-5$, while those of the surface CH$_{2}$ increase by a factor of $\gtrsim 2-6$ when the C + H$_{2}$ surface reaction proceeds without a barrier. The abundance variations for both species
occur at $T \lesssim 14-16$~K, and for the entire density range, they lie between $10^3$ and $10^{12}$~cm$^{-3}$.
Interestingly, the model with more interstellar UV photons shows stronger variations than the UV-dark model. Enhanced local UV radiation constantly liberates atomic carbon from C-bearing species, making the effect of the barrierless C + H$_{2}$ surface reaction on species abundances stronger (Fig.~\ref{fig:gc_gch2_abs}).

The abundance ratios of C and CH$_{2}$ ices in the two models  tend to be anticorrelated. In the UV-illuminated case,
the abundance ratios for the CH$_{2}$ ice become negligible at higher densities, $n_{\rm H} \gtrsim 10^{8}$~cm$^{-3}$
compared to the C ice. At these high densities, the relative abundances
of CH$_2$ ice become higher than those of the C ice and hence are less prone to abundance variations.
Abundance ratios for both C and CH$_{2}$ ices are much lower in the UV-dark case, where most of the volatiles
are depleted and surface chemistry converts a substantial fraction of C into complex organic ices,
lowering the role of the C $+$ H$_{2}$ reaction and the C and H$_{2}$ ice abundances at 1~Myr.

Time-dependent abundances relative to the total amount of hydrogen nuclei for C and CH$_{2}$ ices are shown in
Fig.~\ref{fig:gc_gch2_1D}. At the chosen physical conditions
$T = 12$~K, $n_{\rm H} = 3.79 \times 10^{10}$~cm$^{-3}$ , the abundance changes are significant for most of the species in the chemical model. These physical conditions are representative of the densest regions of low-mass starless cores and protoplanetary disk mid-planes. The freeze-out timescale onto $0.1\mu$m grains at this density is only
a few years.

The abundance variations between the two chemical models for both C and CH$_2$ ices are initially very high,
$\sim 5-8$ and $\sim 1$ orders of magnitude, respectively, and decrease with time.
The relative abundances of the C ice are lower in the modified model with the fast C $+$ H$_2$ reaction.
In contrast, the  abundances of the CH$_2$ ice are higher in the modified model.
The pace of abundance convergence between the two models
for the C ice is rapid until $\sim 1-3 \times 10^3$~years, and becomes slower
after that. For the CH$_2$ ice the pace of the abundance convergence is slower than for the C ice.
Both C and  CH$_2$ ices experience a transition to another chemical state
at $\sim 10^5$~years, where their abundance further decreases.
These two timescales, $\sim 1-3 \times 10^3$ and $\sim 10^5$~years, correspond to the
final conversion of O ice into OH ice and OH ice into water ice, respectively.
After this, more surface hydrogen atoms and molecules become
available for competitive reactions, for instance, involving C and hydrocarbon ices.
In the end, the elemental carbon is reshuffled mainly into methane and CO ices, with
the addition of the COM ices.

\begin{figure*}[!ht]
\centering
\setlength\tabcolsep{-0.5pt}
\renewcommand{\arraystretch}{0}
        \begin{tabular}{@{}R{3.6cm}C{3cm}C{3cm}C{3cm}C{3cm}L{3.7cm}@{}}
        \includegraphics[width=0.159\textwidth]{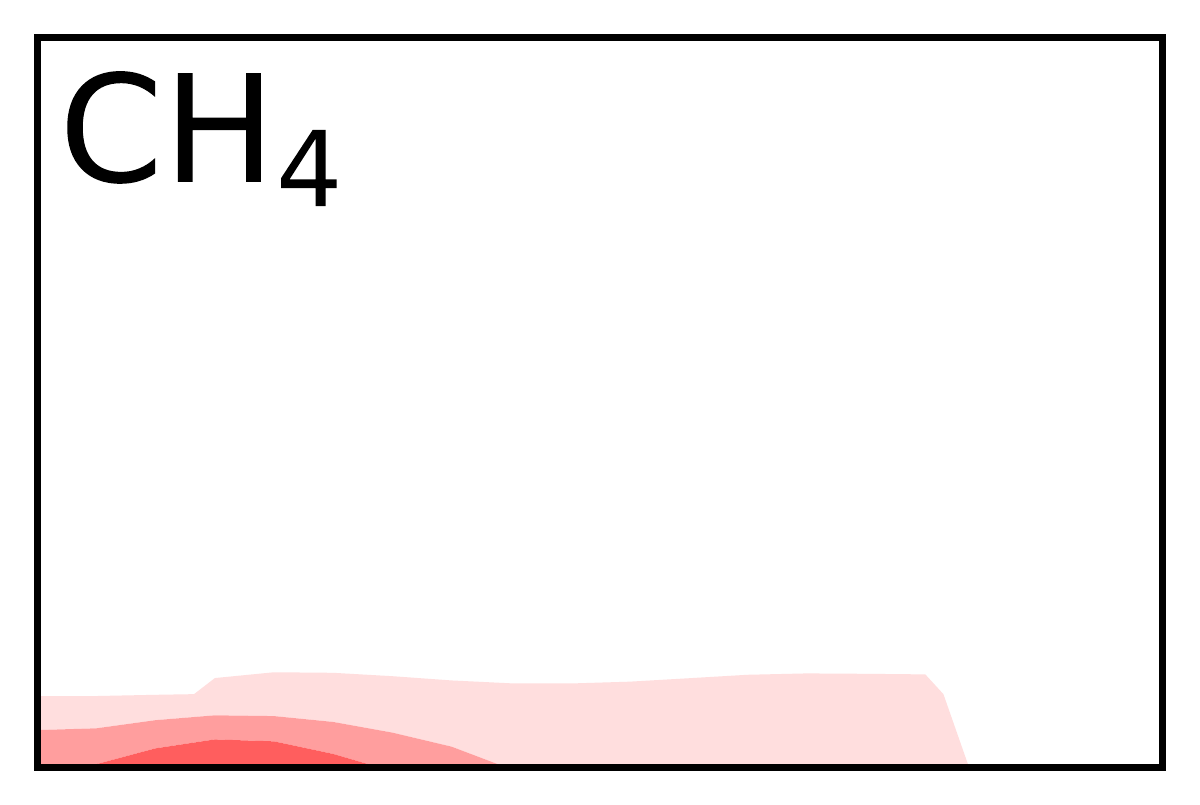} &
        \includegraphics[width=0.159\textwidth]{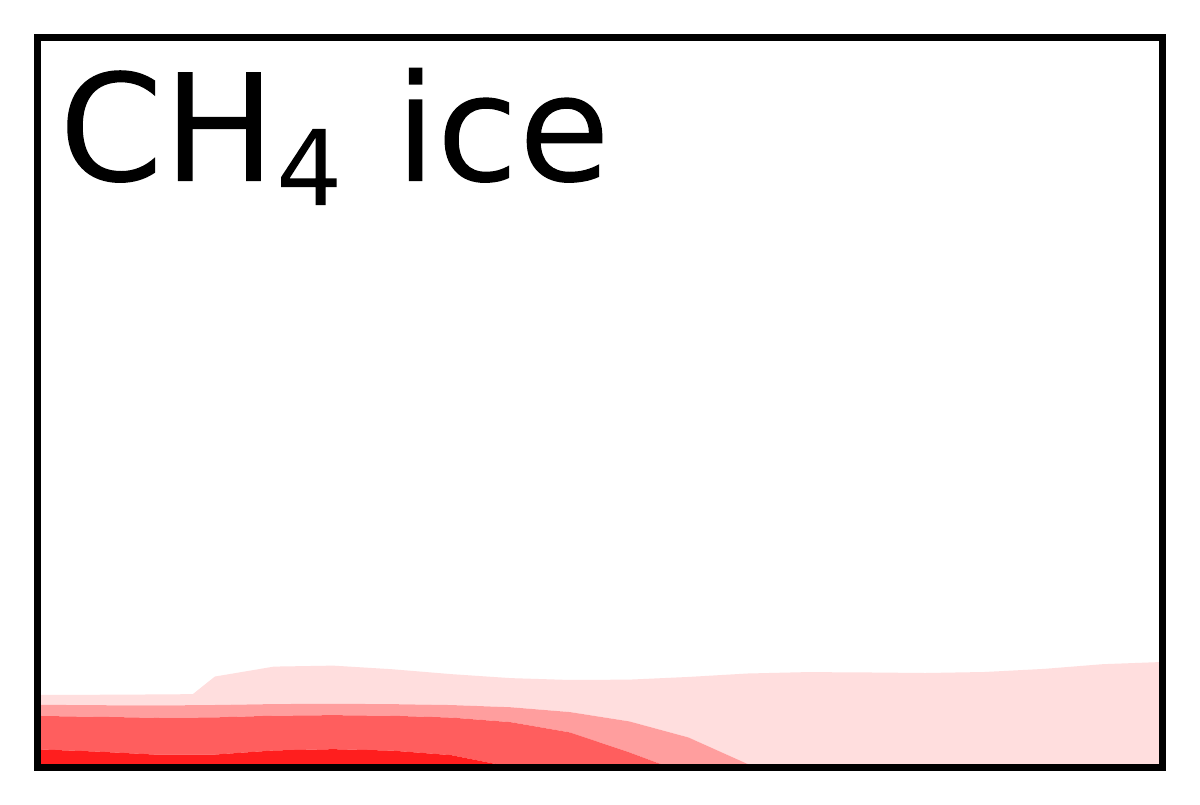} &
        \includegraphics[width=0.159\textwidth]{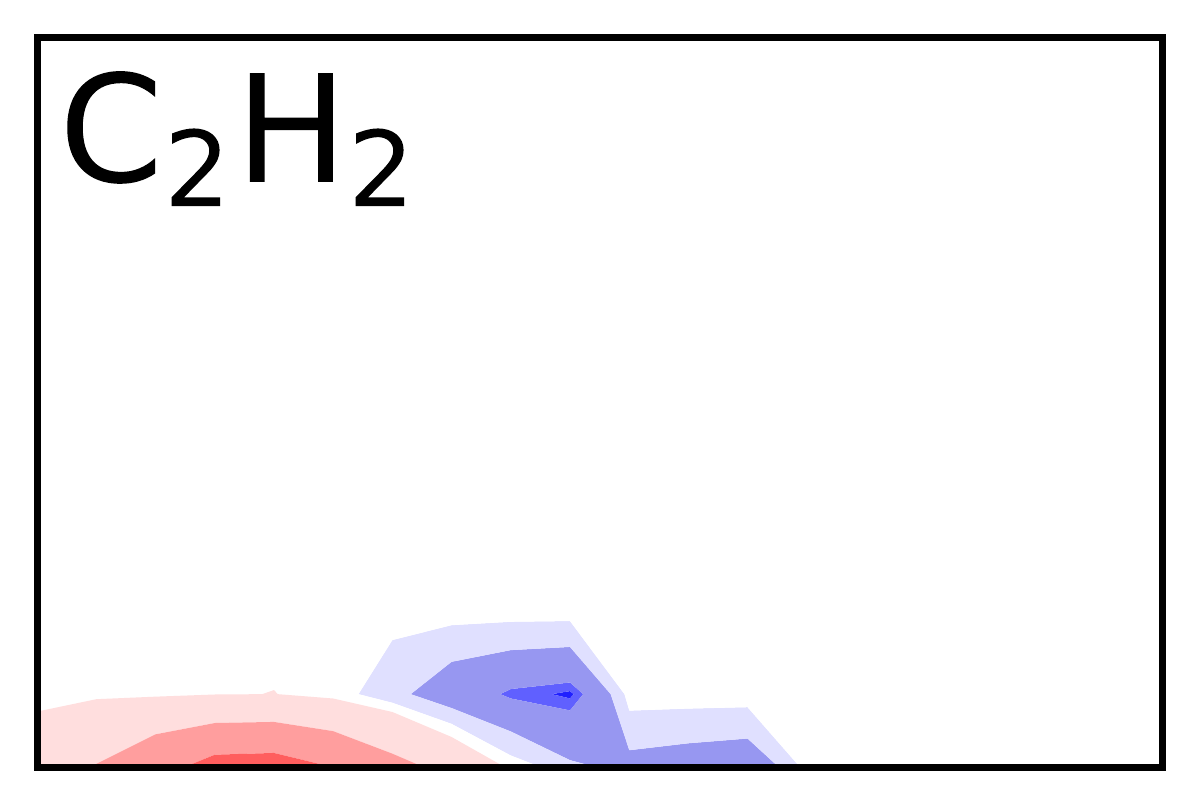} &
        \includegraphics[width=0.159\textwidth]{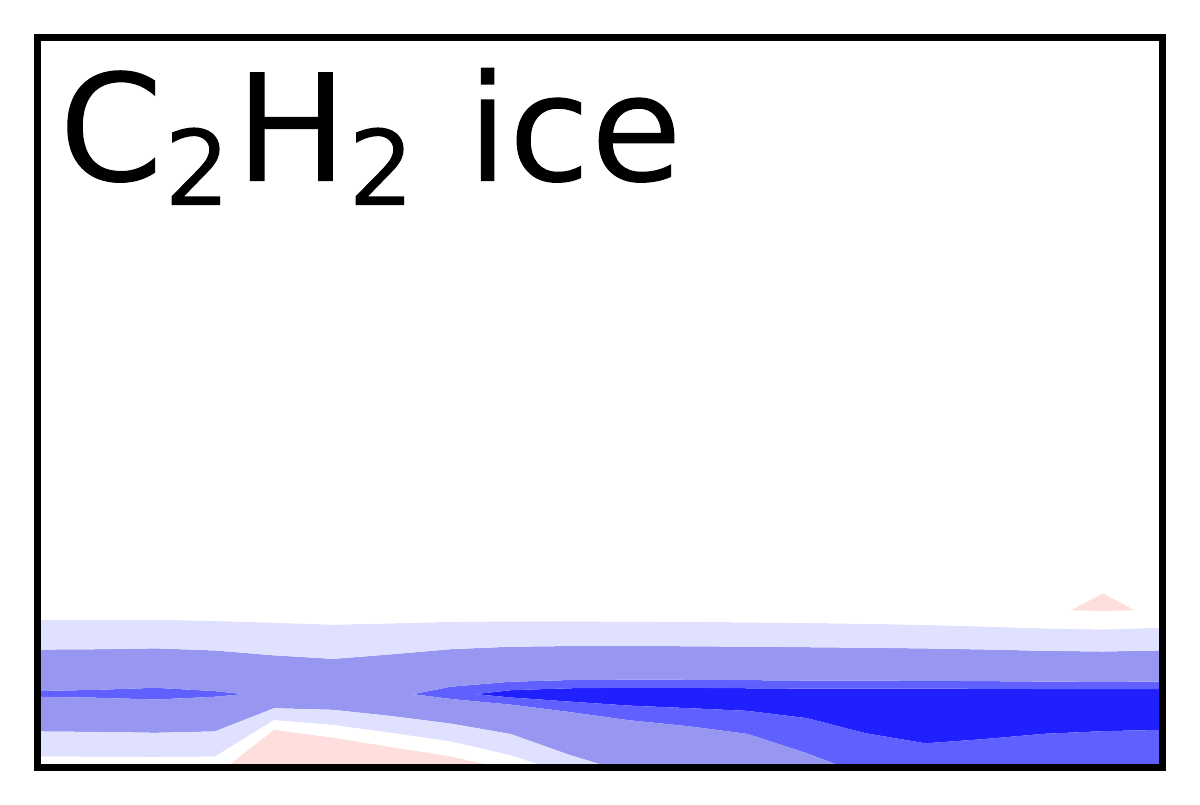} &
        \includegraphics[width=0.159\textwidth]{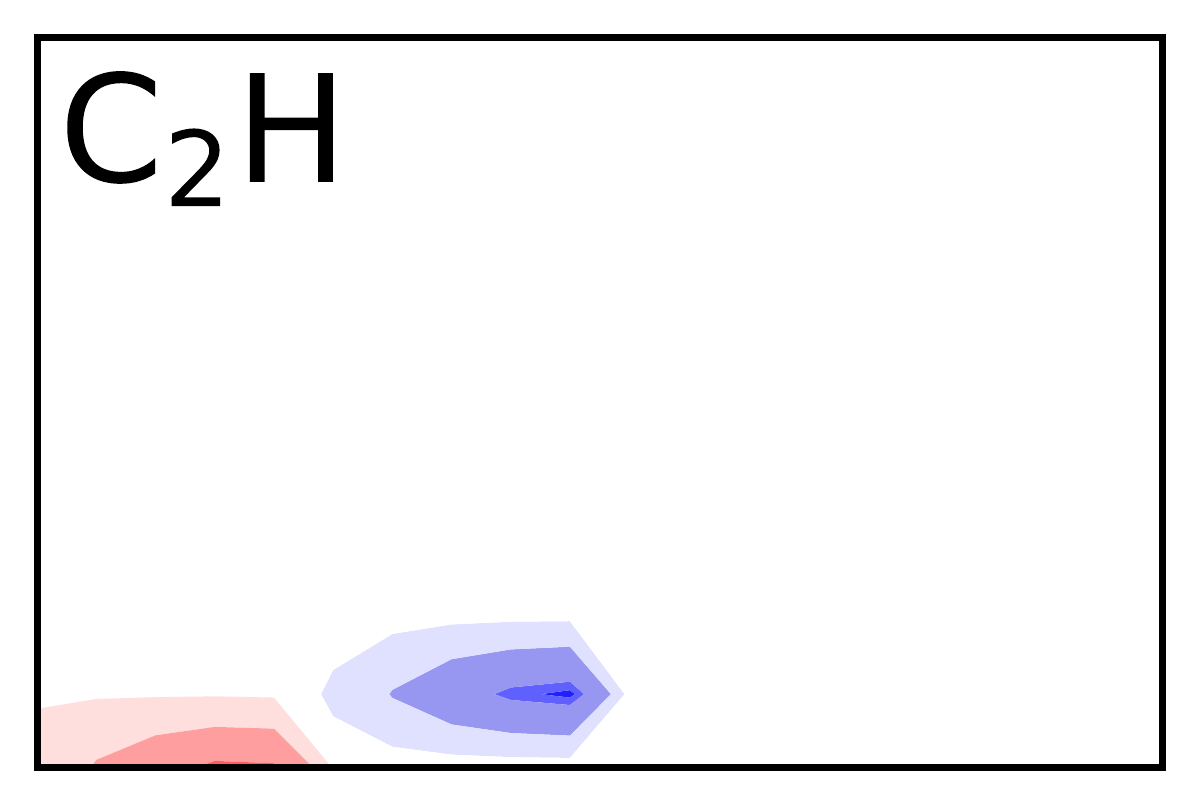} &
        \includegraphics[width=0.159\textwidth]{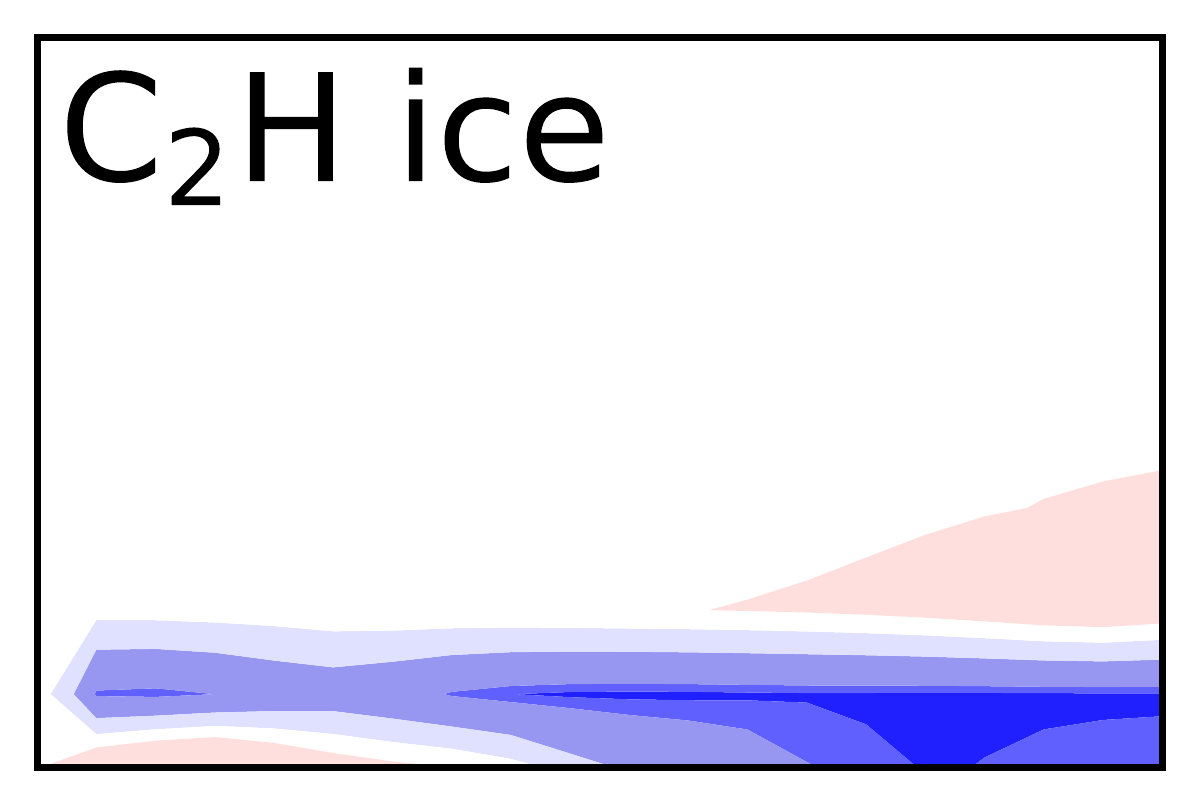} \\
        \includegraphics[width=0.159\textwidth]{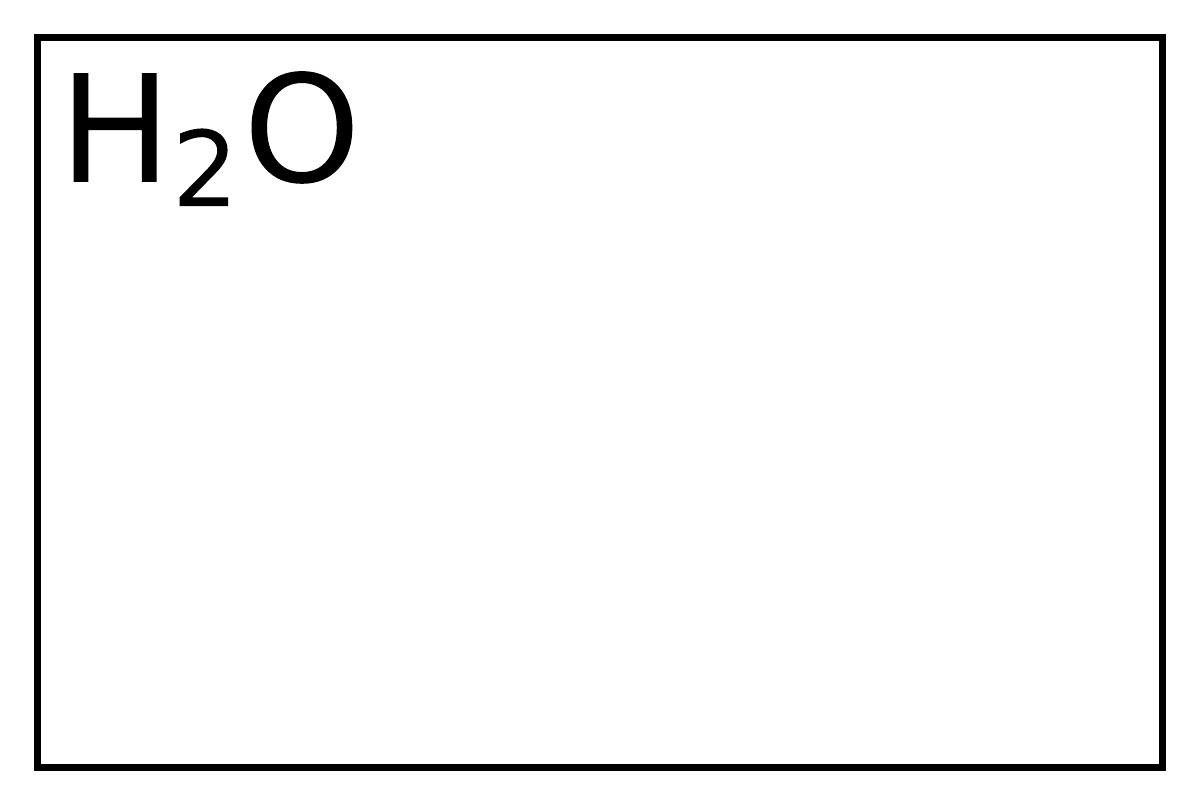} &
        \includegraphics[width=0.159\textwidth]{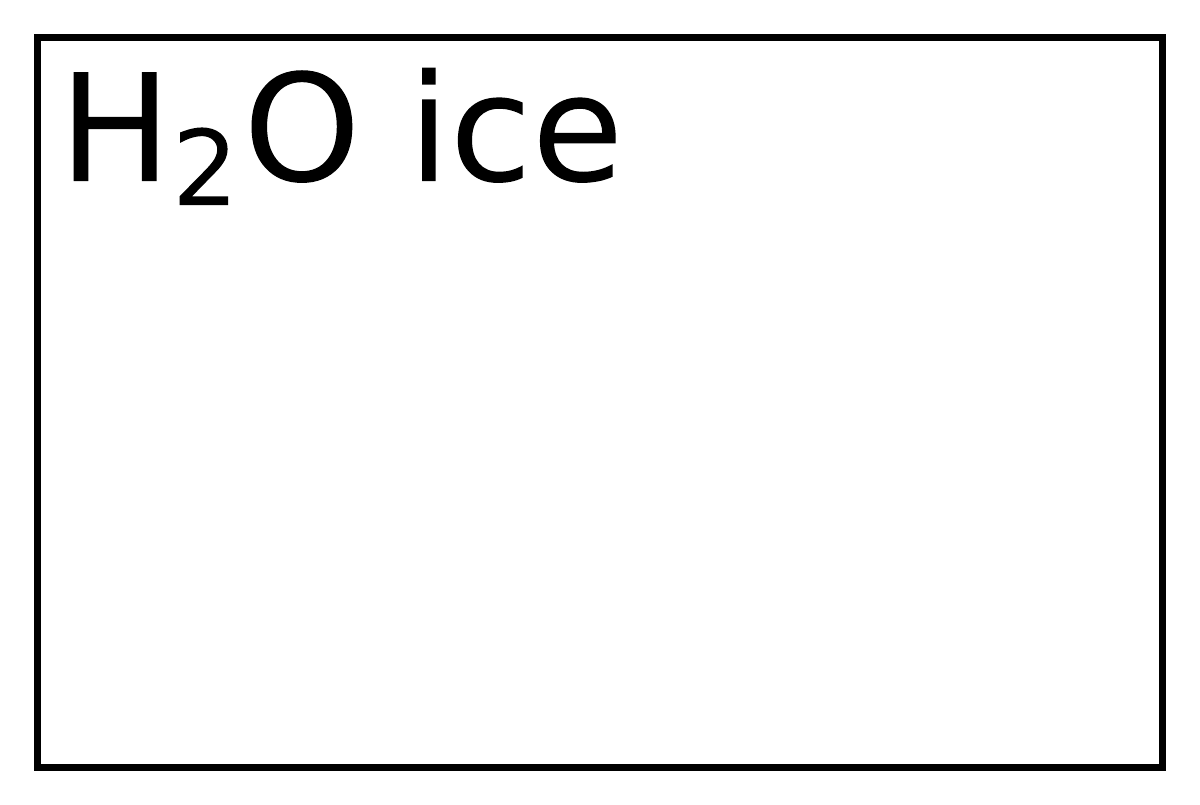} &
        \includegraphics[width=0.159\textwidth]{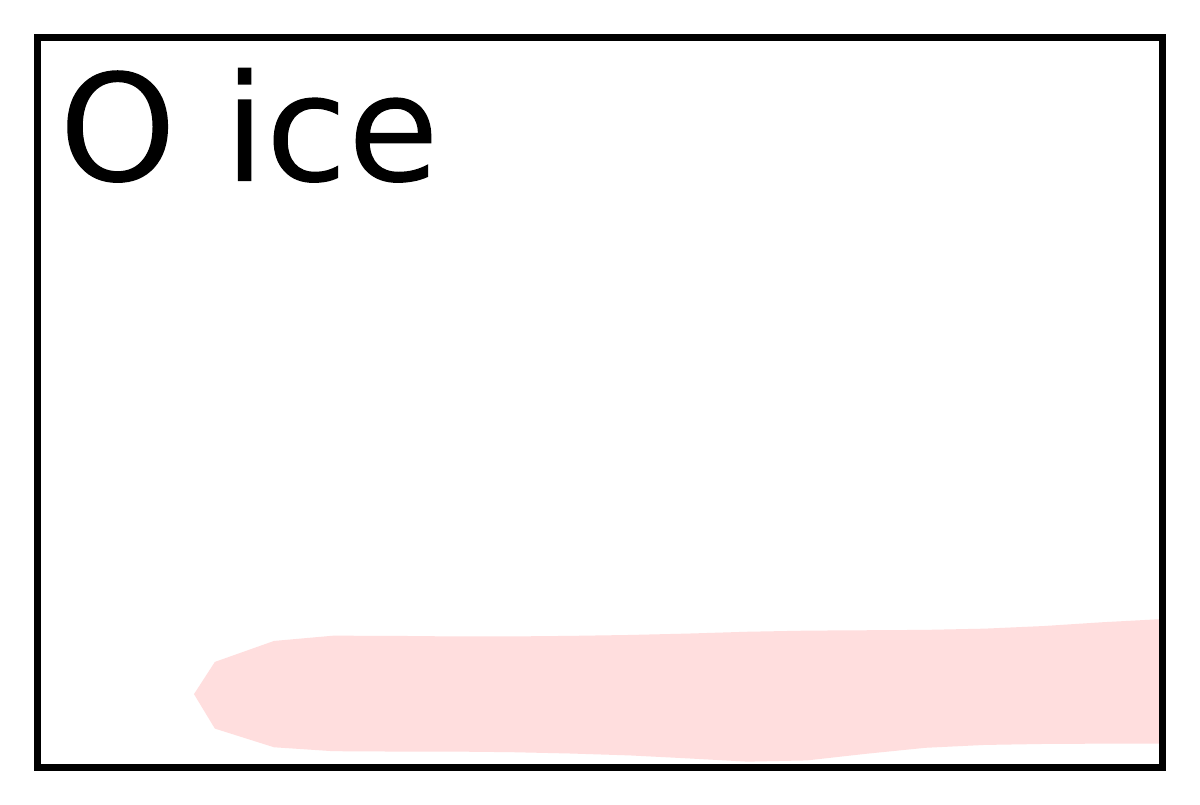} &
        \includegraphics[width=0.159\textwidth]{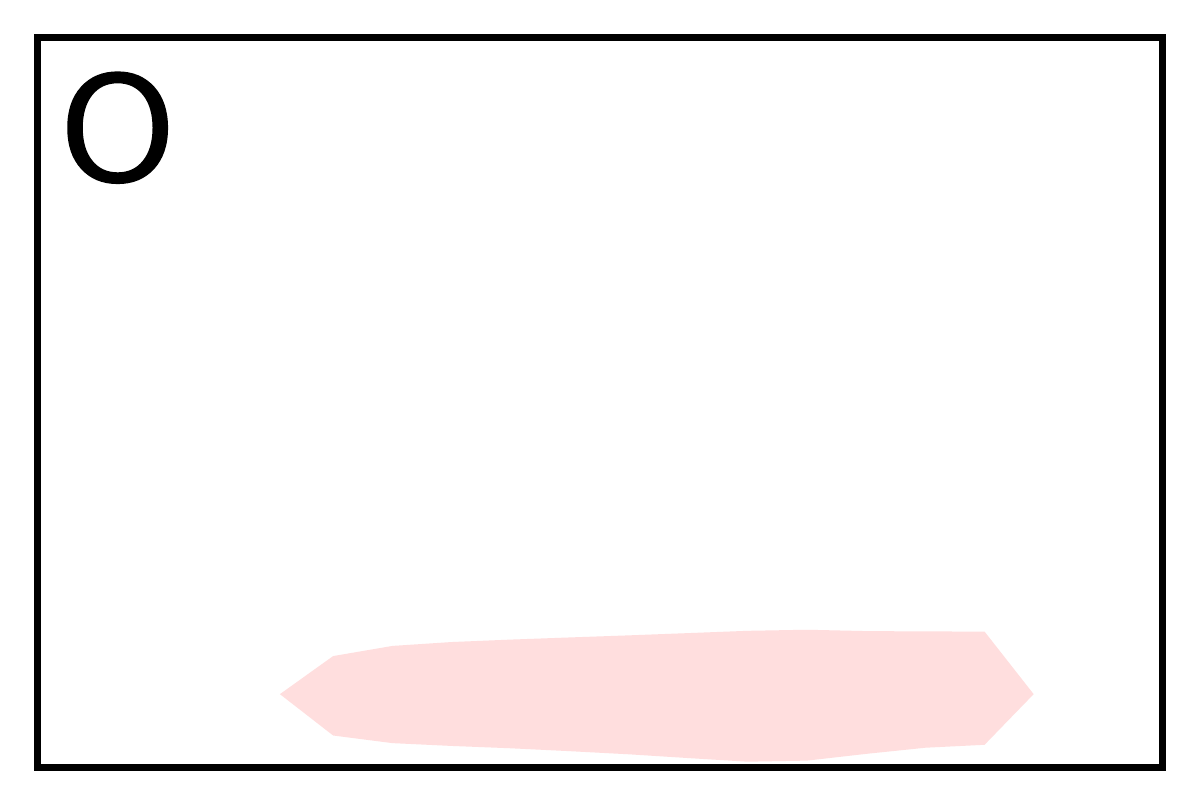} &
        \includegraphics[width=0.159\textwidth]{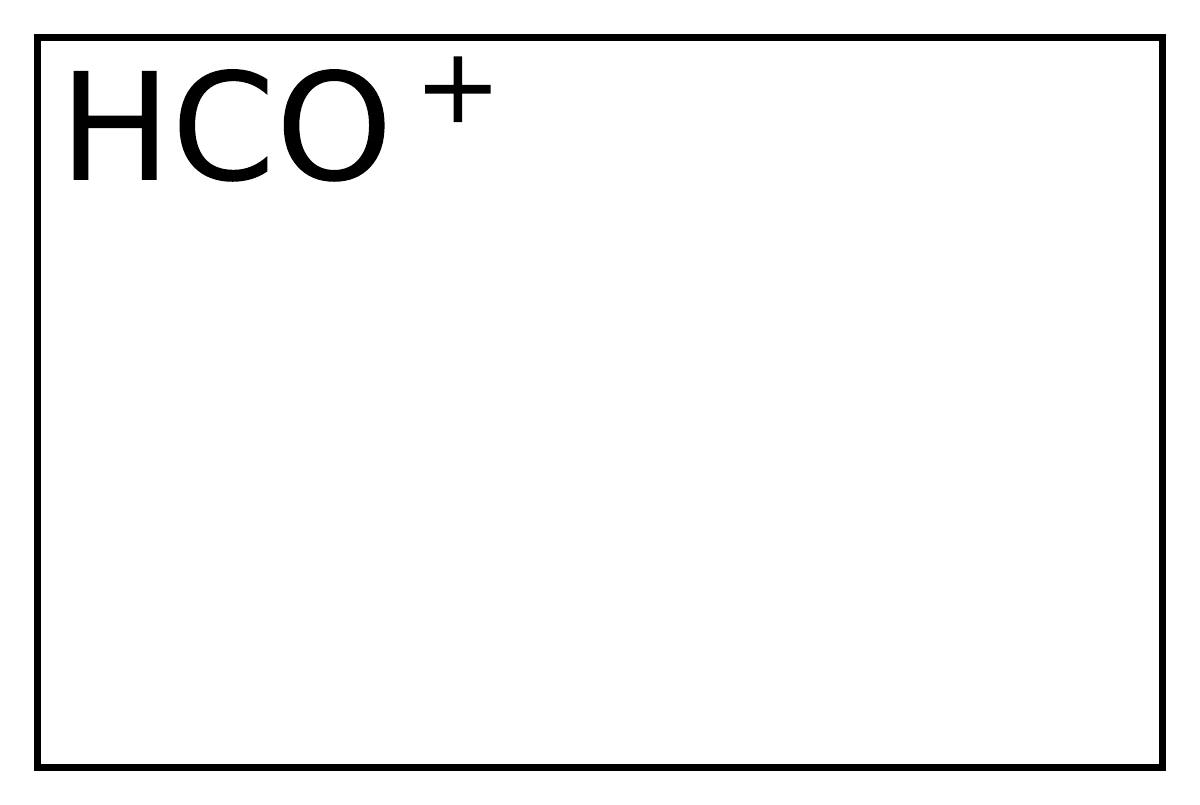} &
        \includegraphics[width=0.159\textwidth]{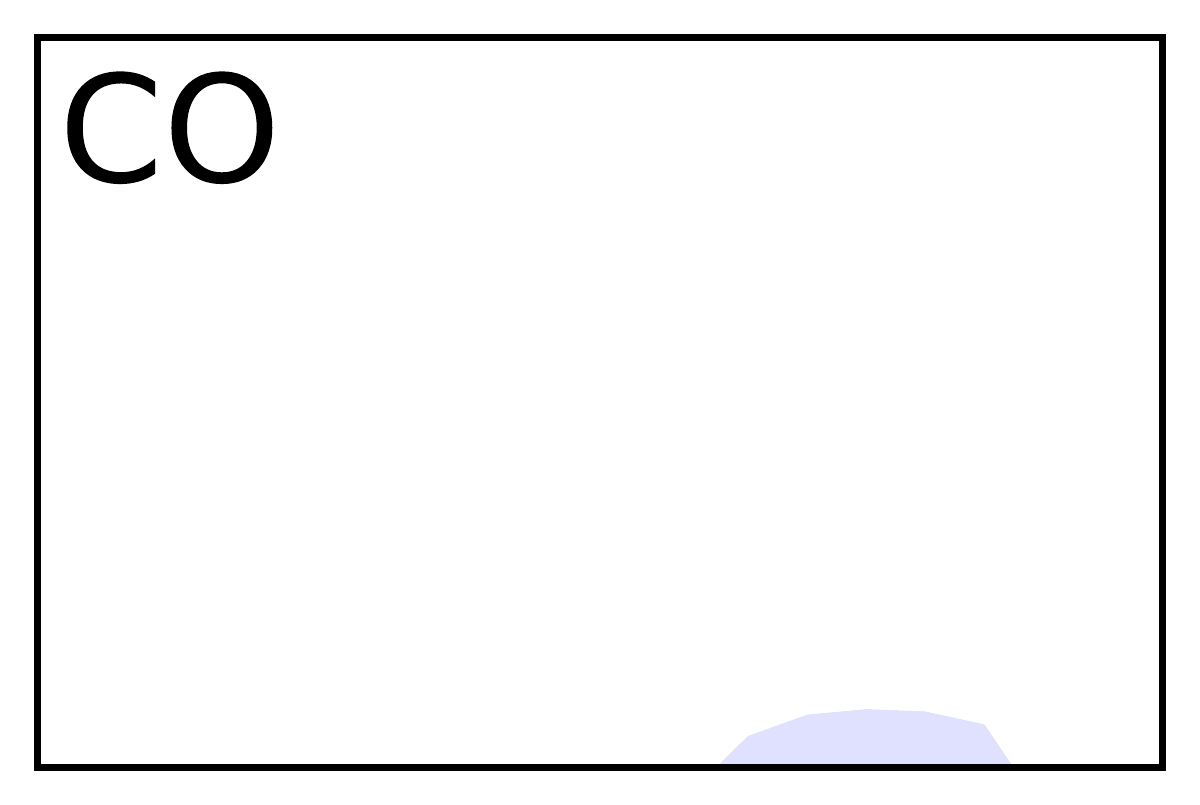} \\
        \includegraphics[width=0.159\textwidth]{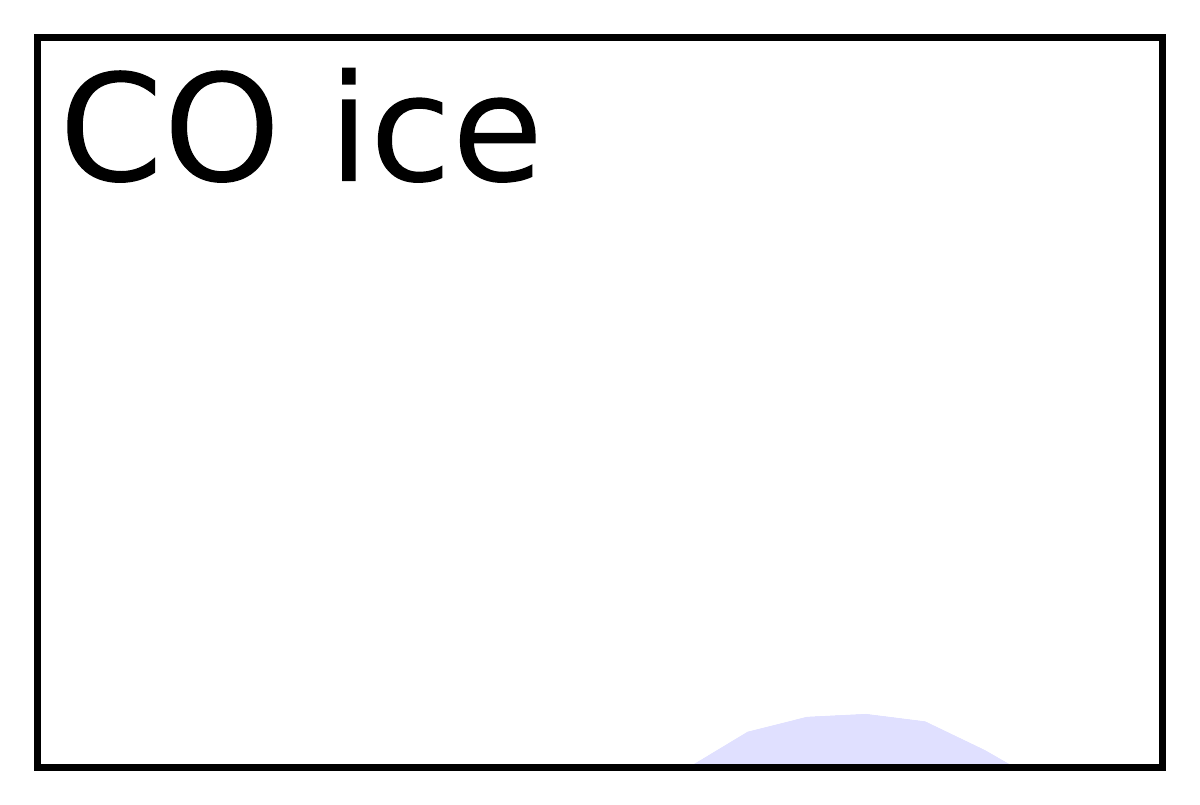} &
        \includegraphics[width=0.159\textwidth]{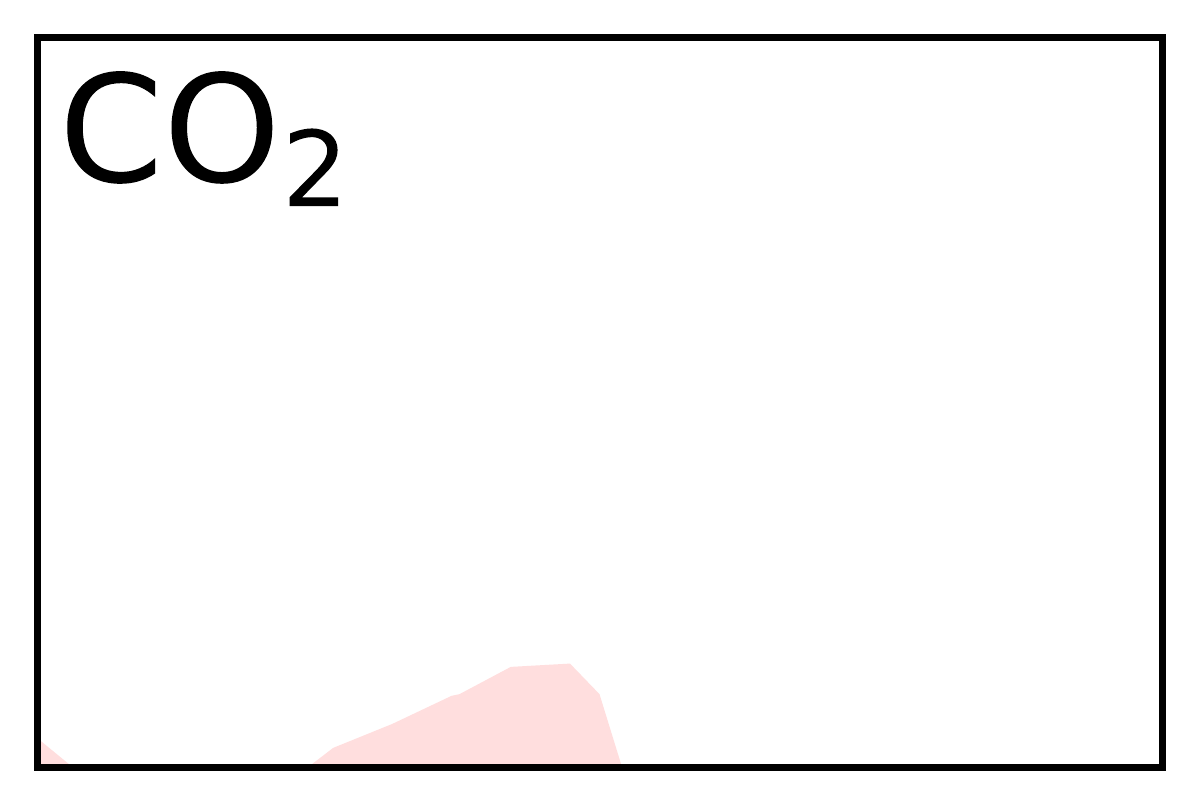} &
        \includegraphics[width=0.159\textwidth]{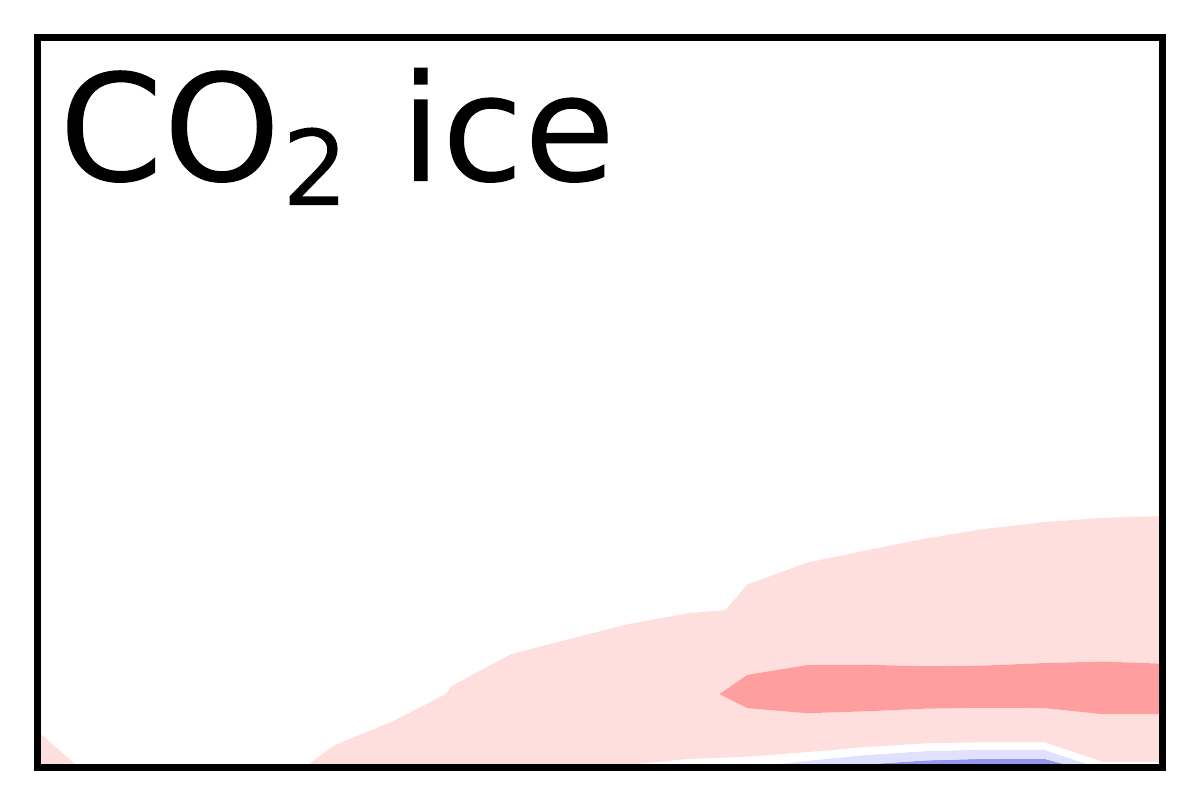} &
        \includegraphics[width=0.159\textwidth]{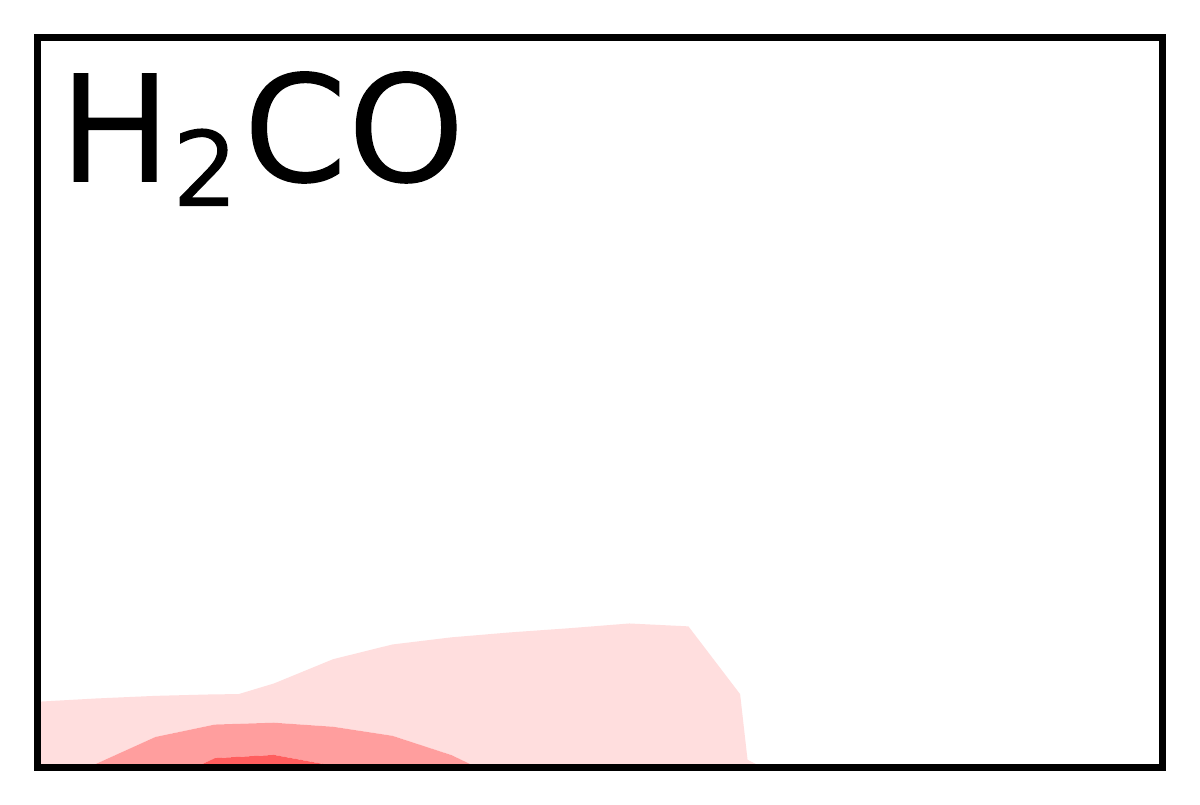} &
        \includegraphics[width=0.159\textwidth]{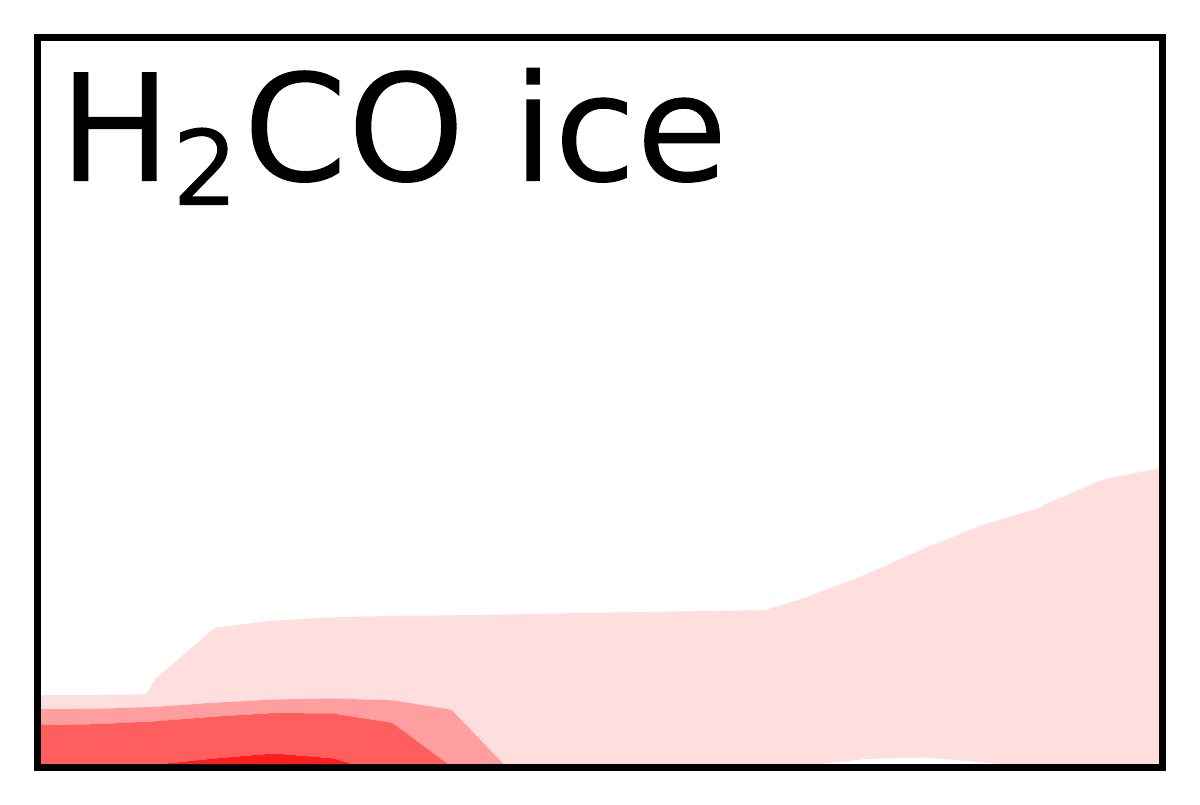} &
        \includegraphics[width=0.159\textwidth]{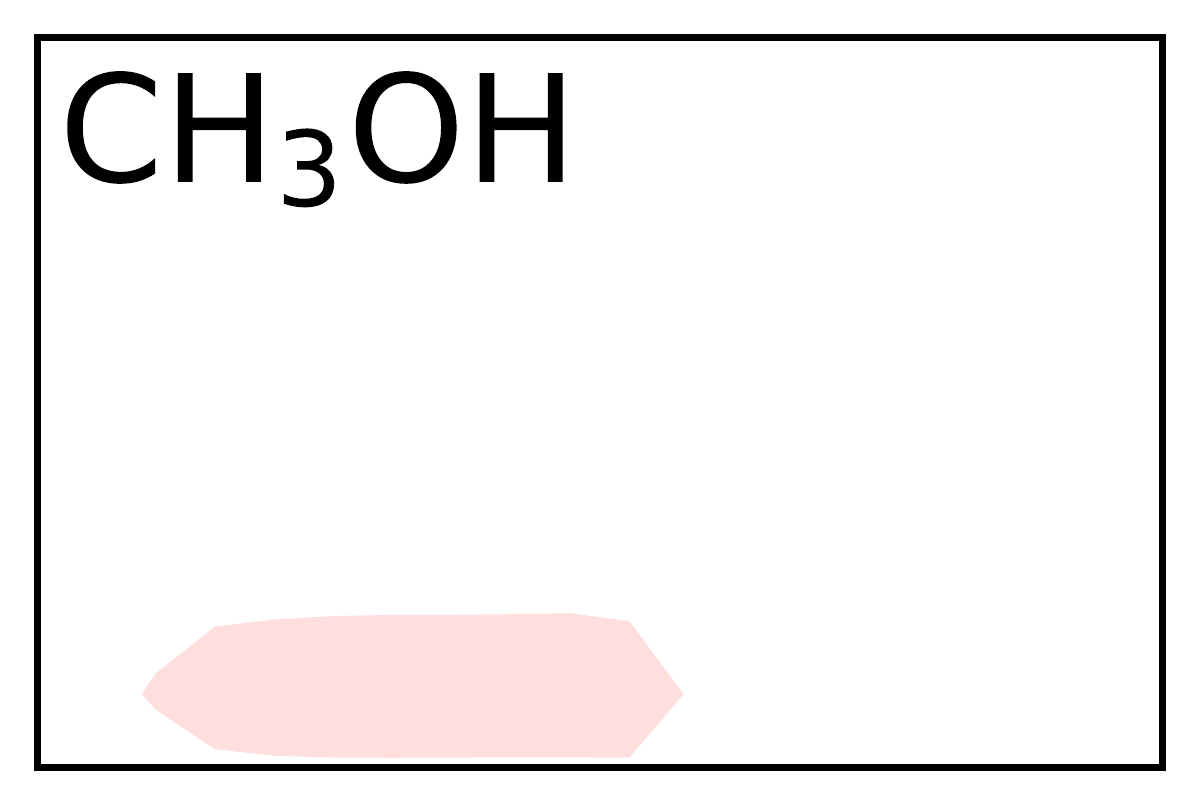} \\
        \includegraphics[width=0.159\textwidth]{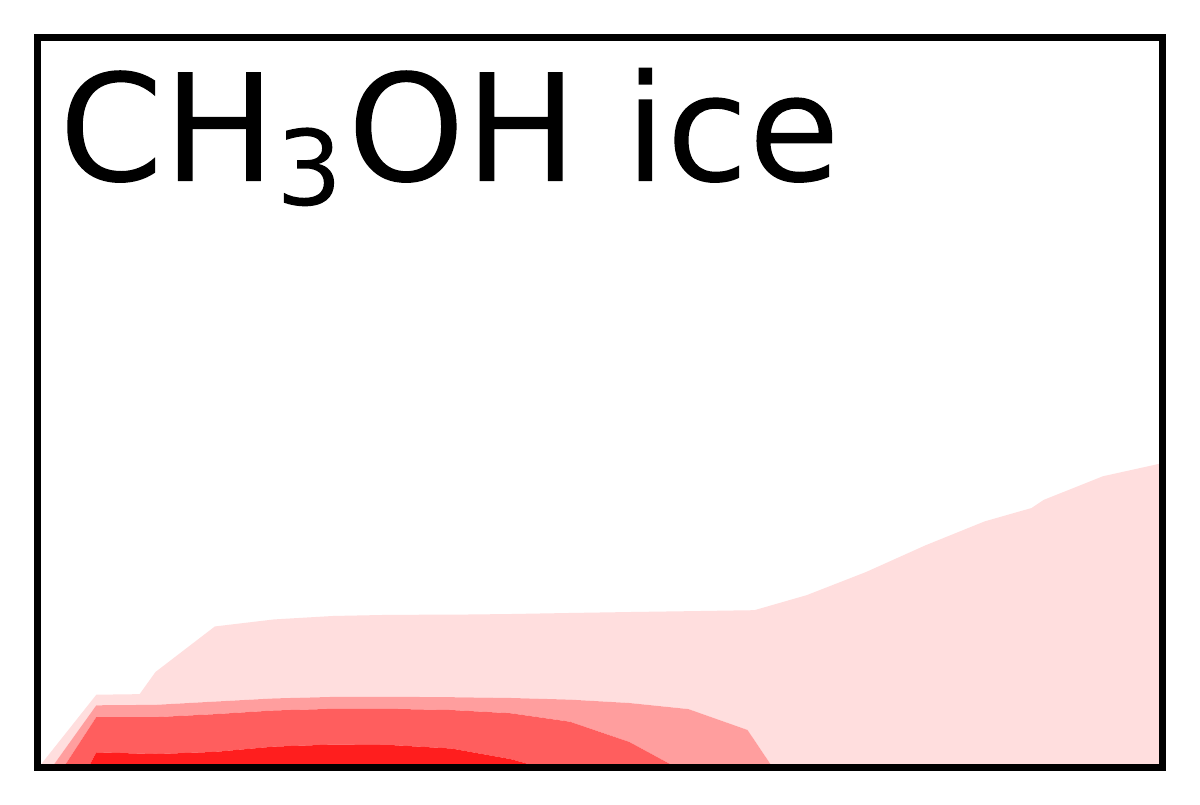} &
        \includegraphics[width=0.159\textwidth]{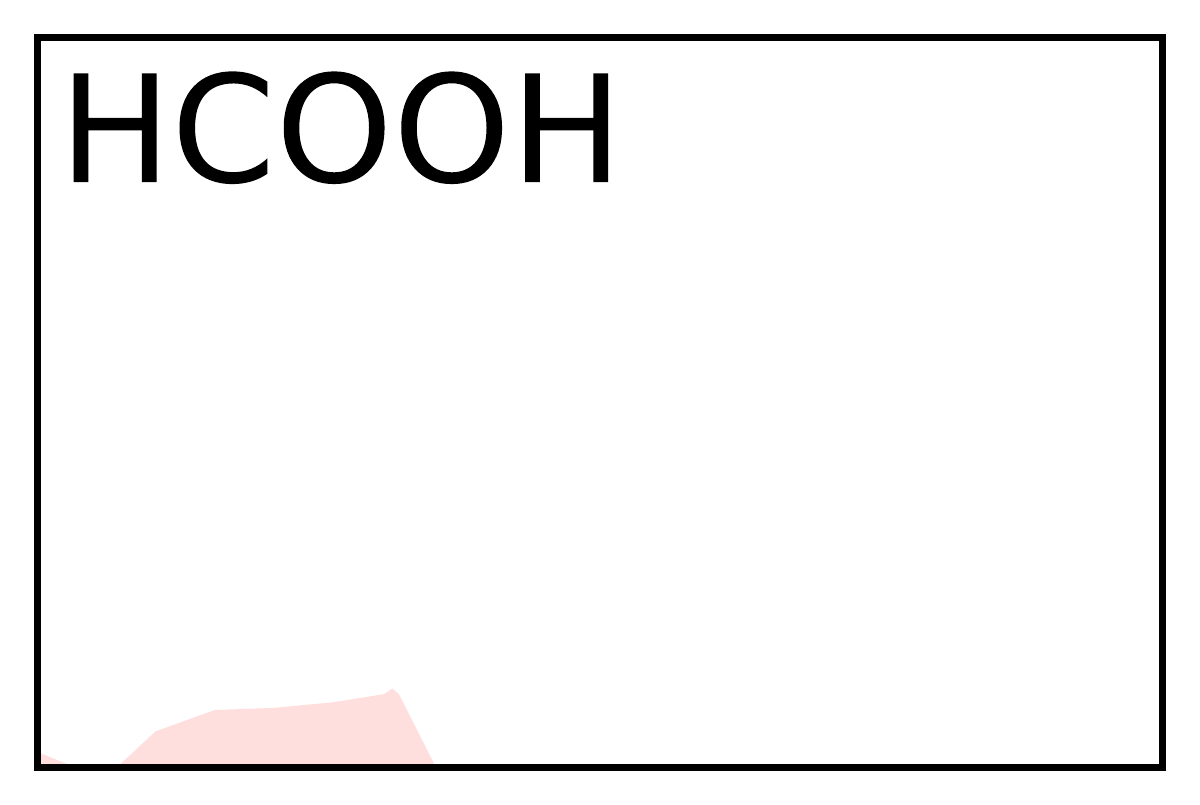} &
        \includegraphics[width=0.159\textwidth]{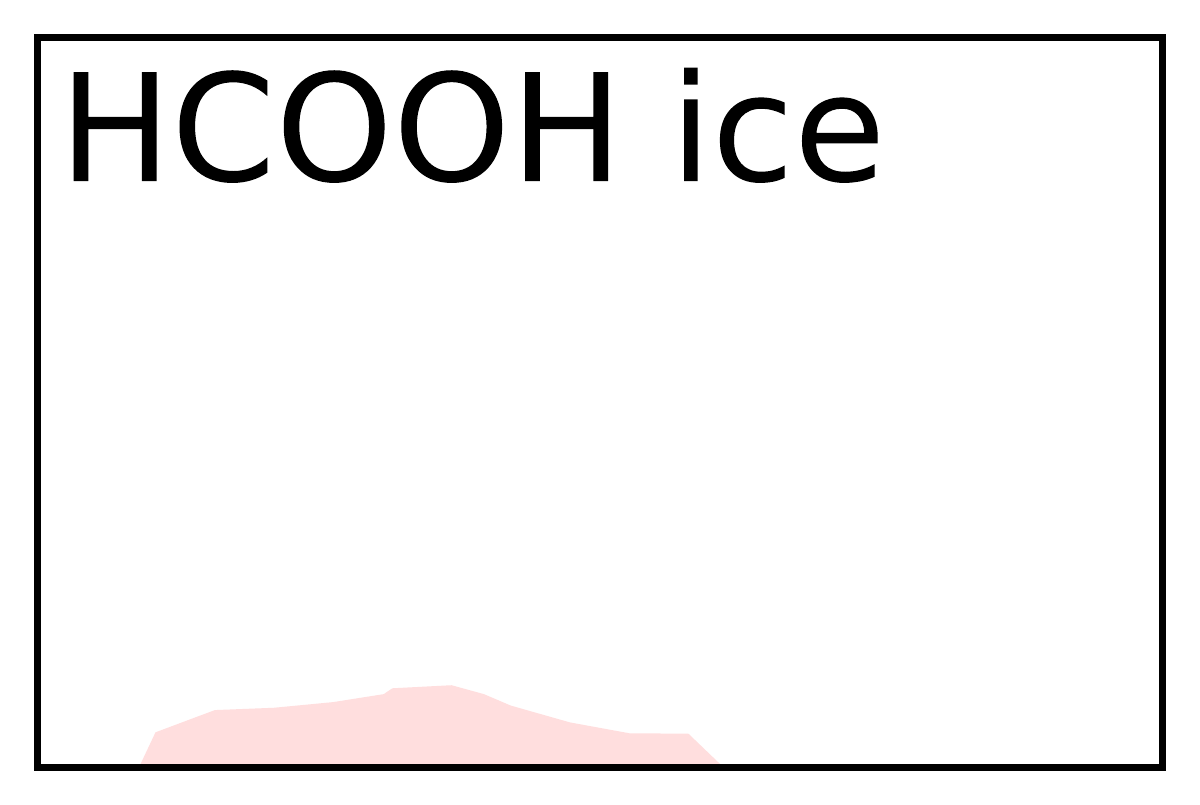} &
        \includegraphics[width=0.159\textwidth]{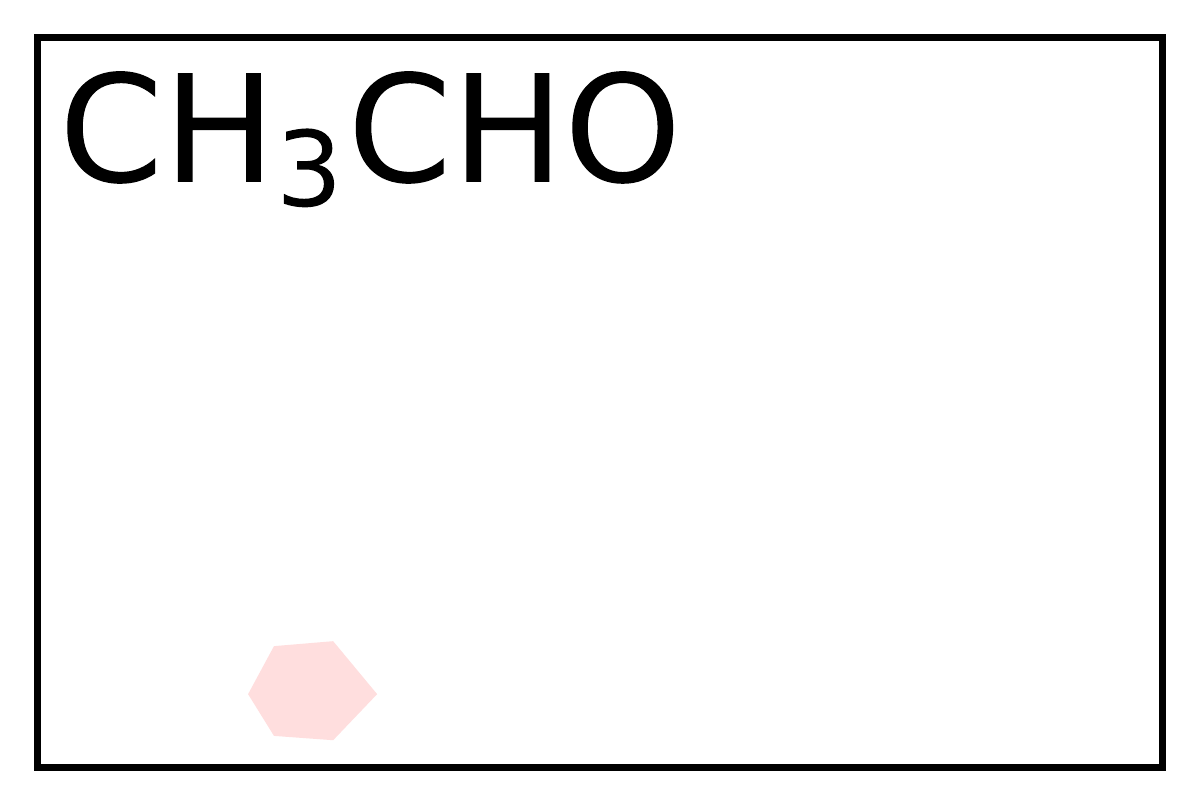} &
        \includegraphics[width=0.159\textwidth]{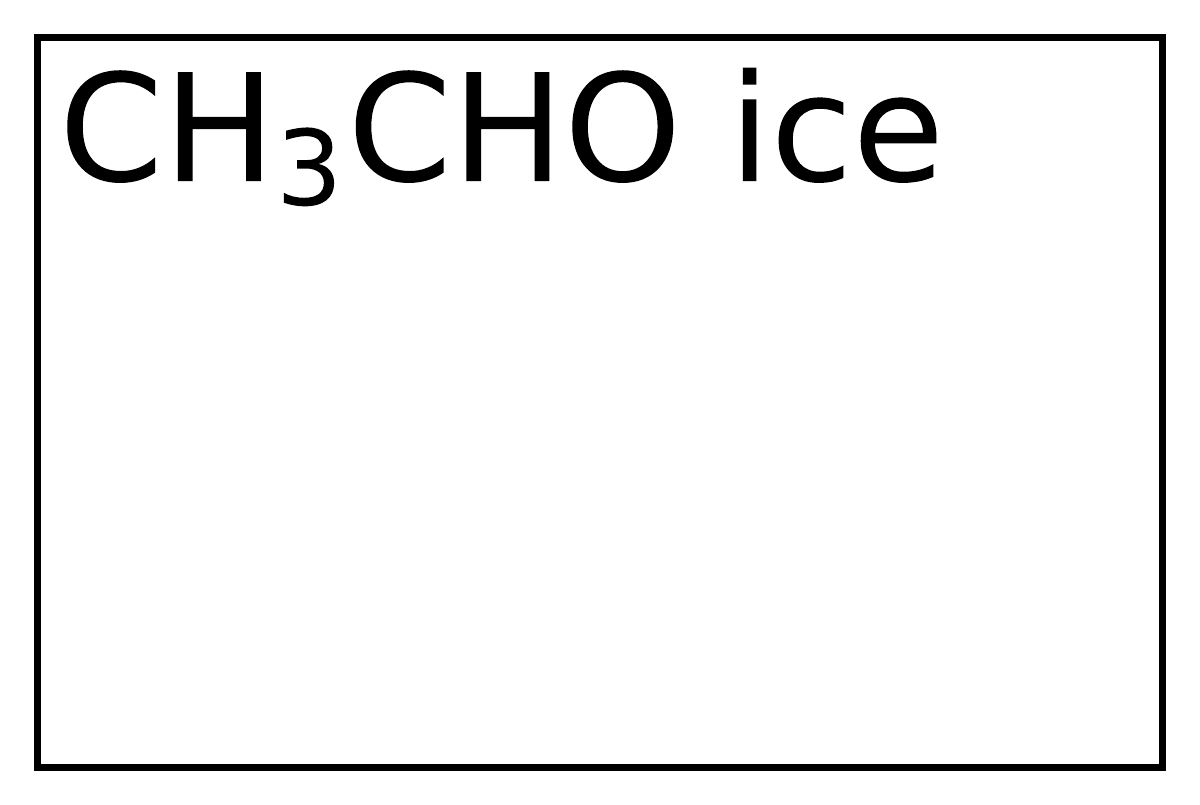} &
        \includegraphics[width=0.159\textwidth]{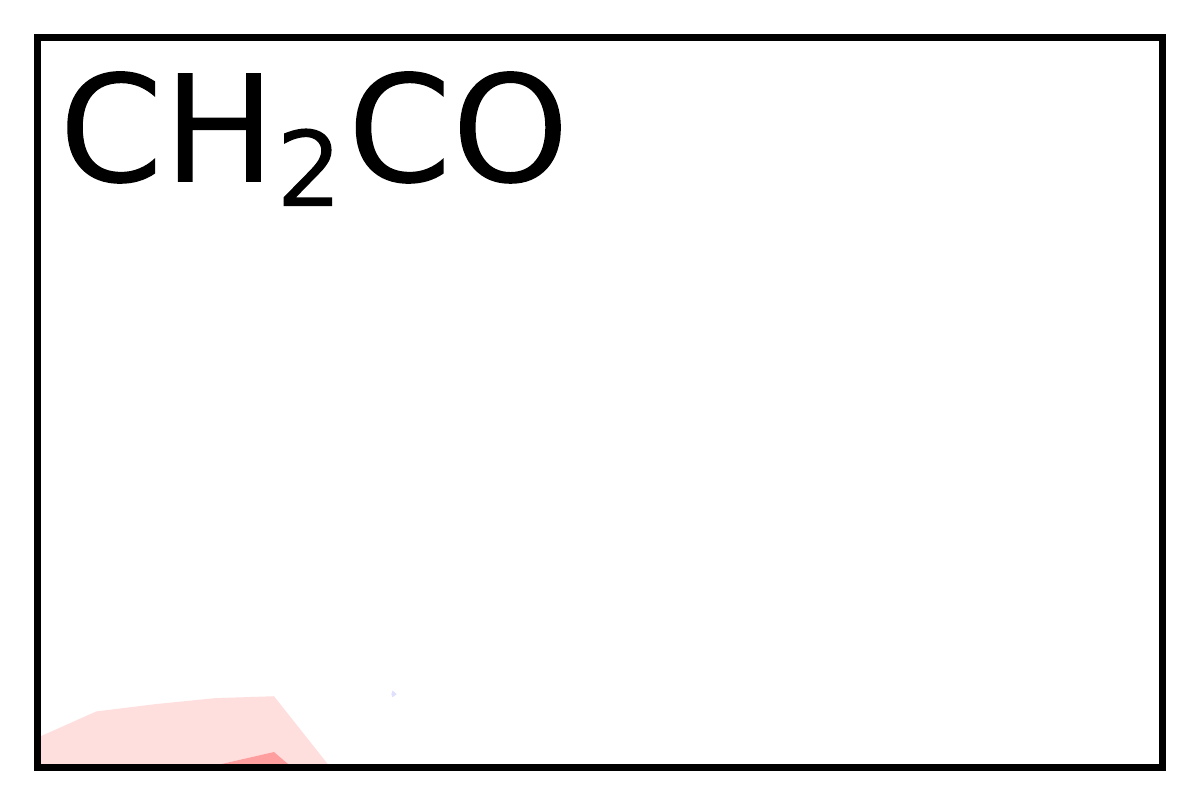} \\
        \includegraphics[width=0.159\textwidth]{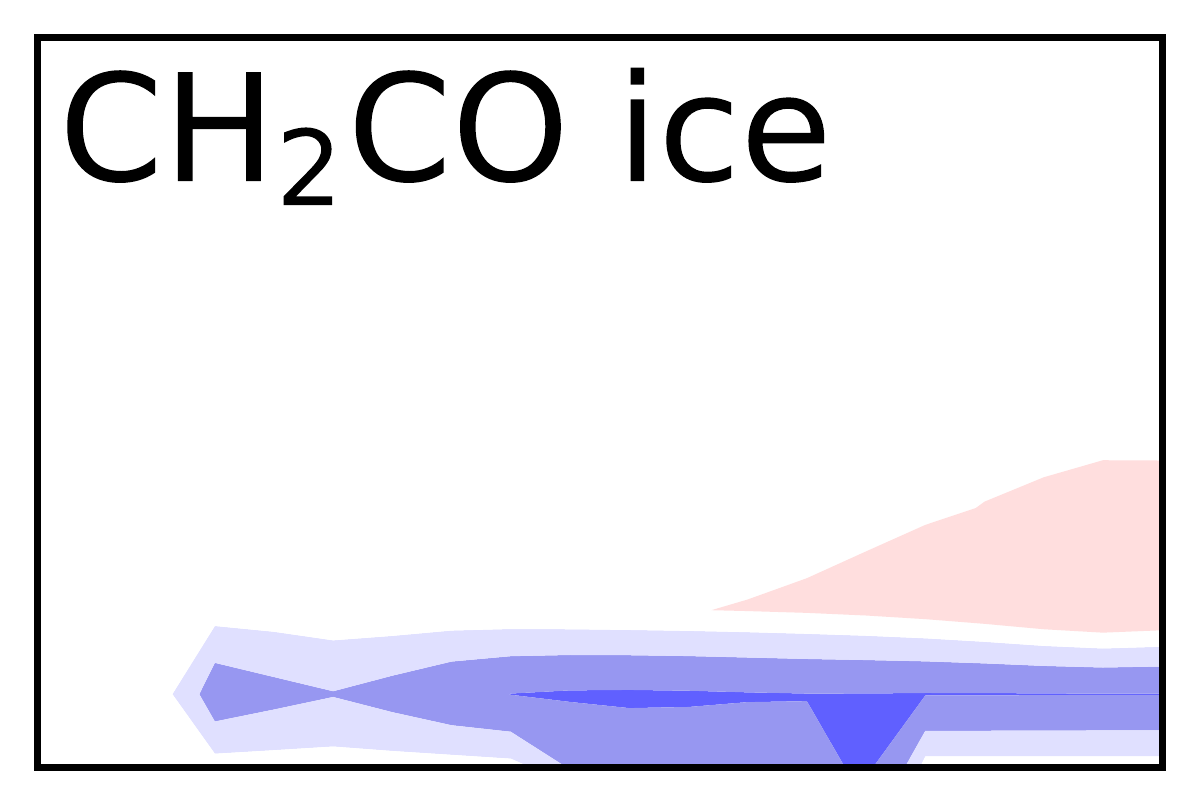} &
        \includegraphics[width=0.159\textwidth]{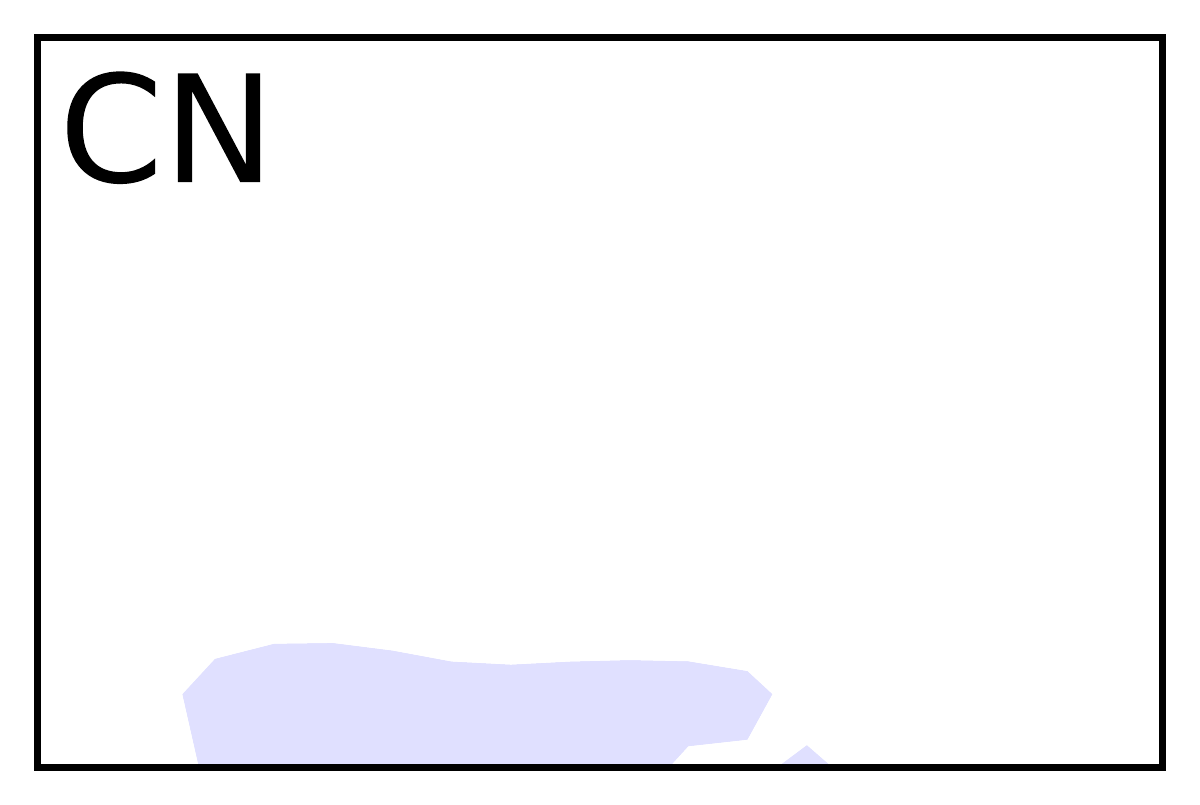} &
        \includegraphics[width=0.159\textwidth]{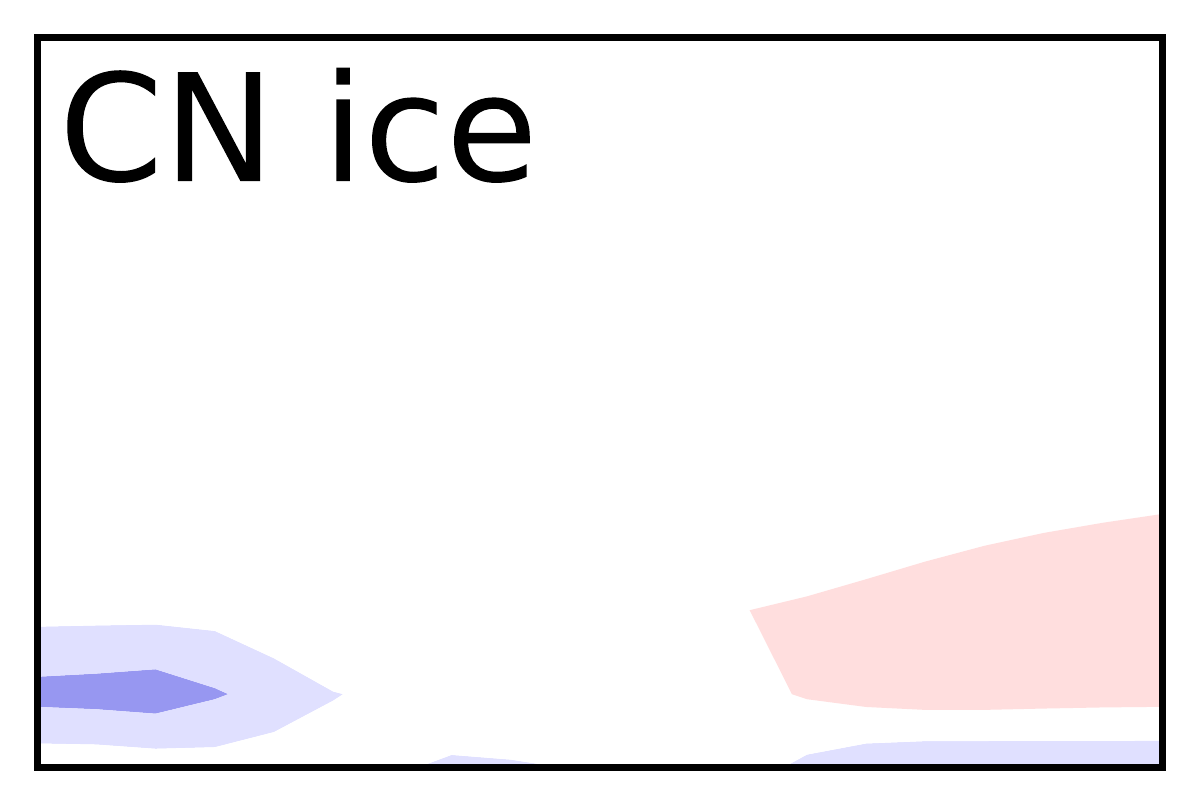} &
        \includegraphics[width=0.159\textwidth]{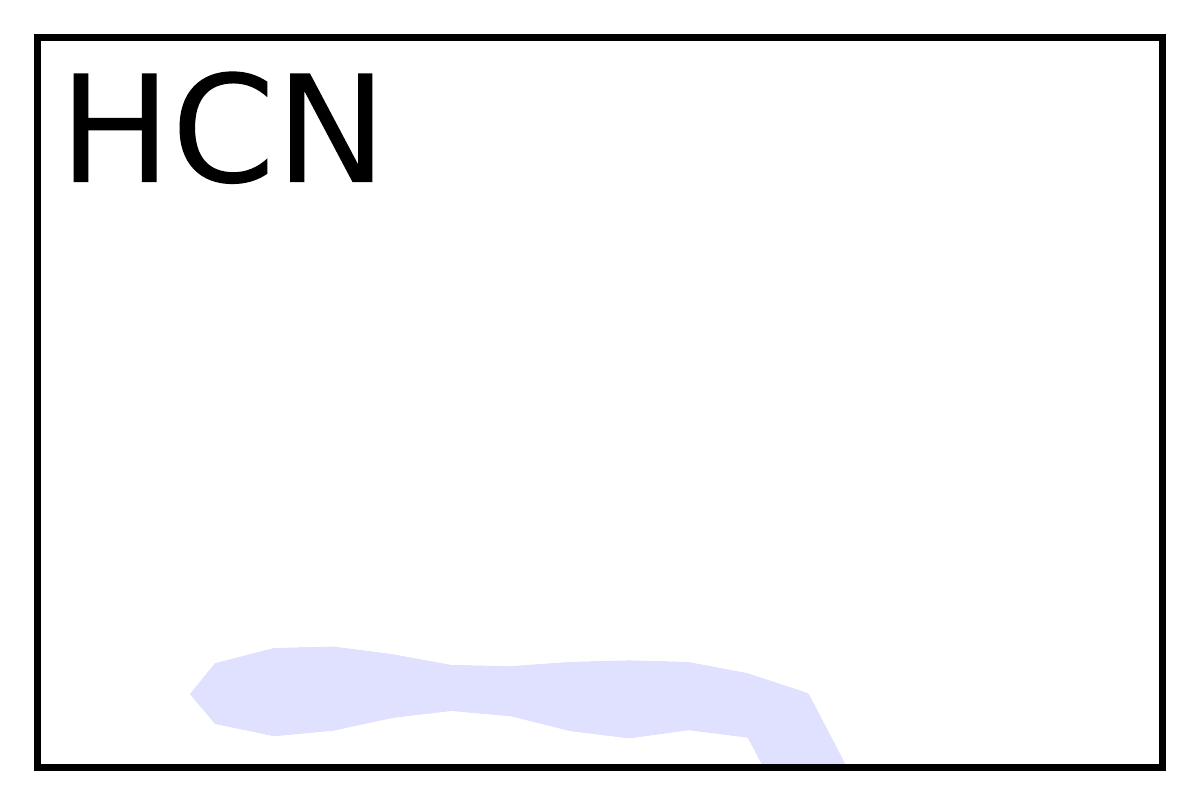} &
        \includegraphics[width=0.159\textwidth]{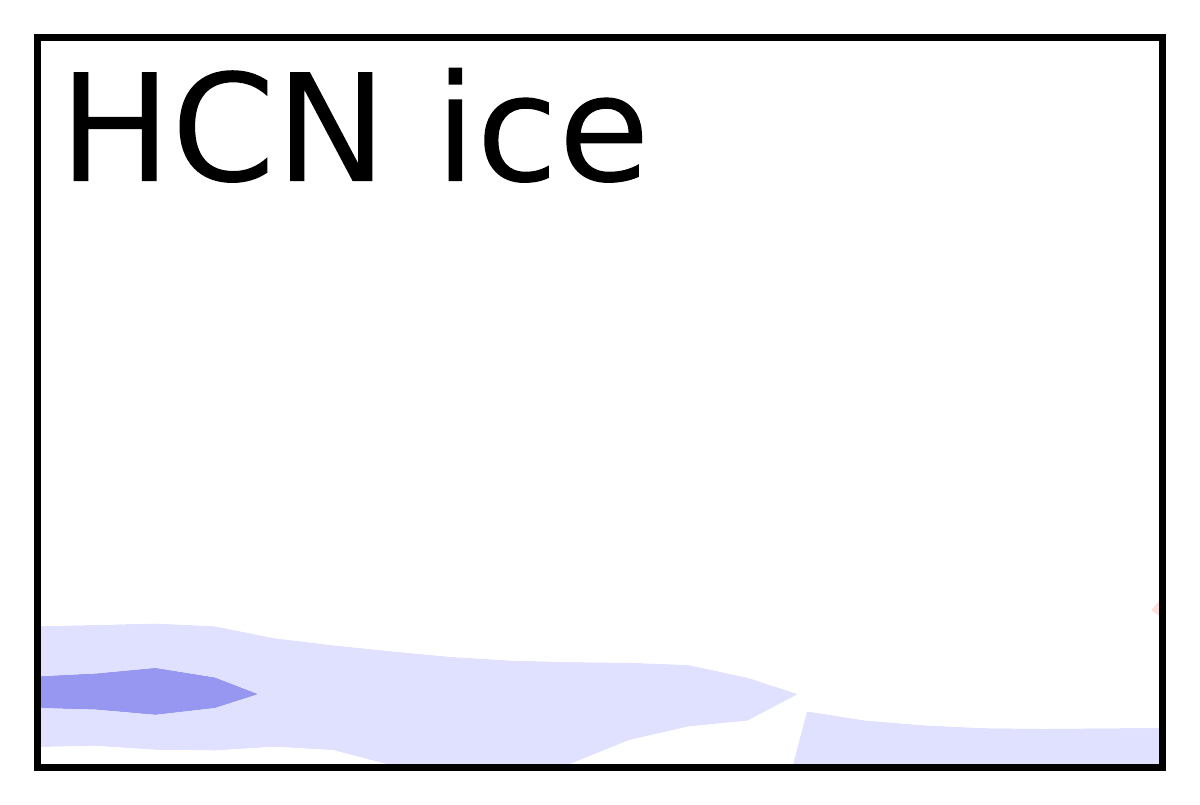} &
        \includegraphics[width=0.159\textwidth]{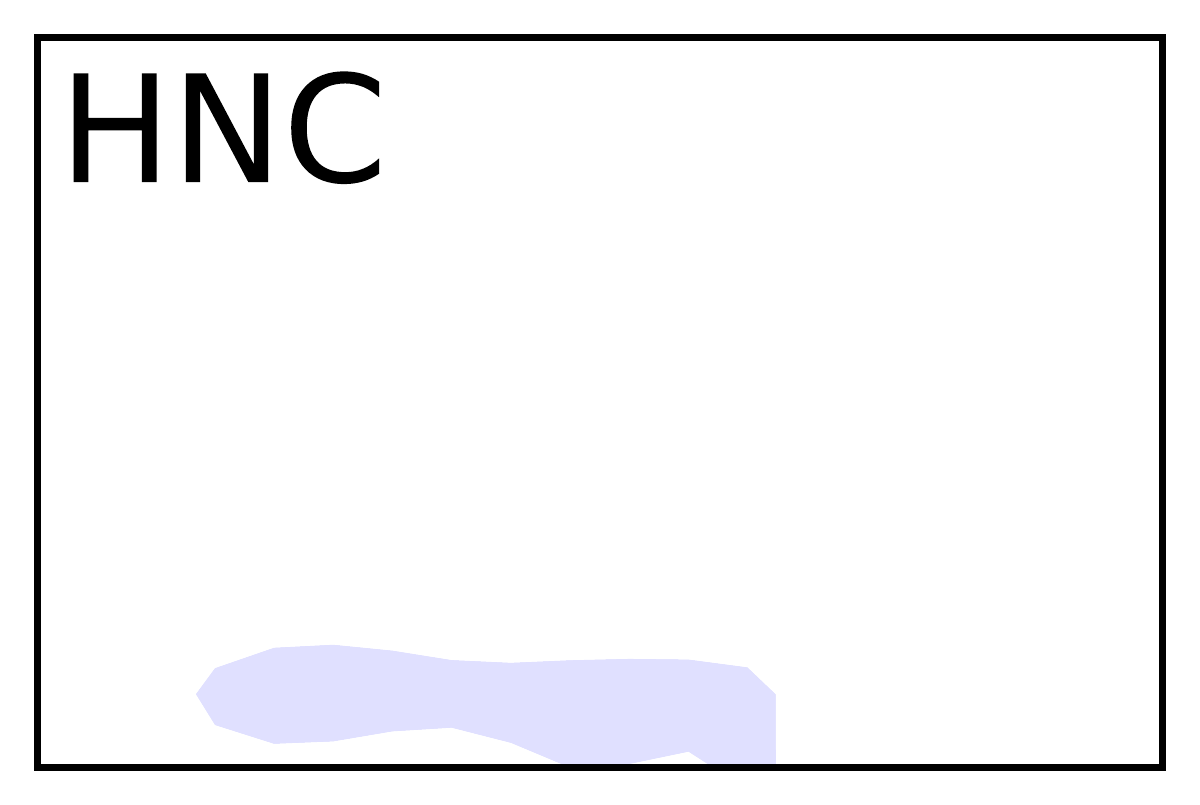} \\
        \includegraphics[width=0.159\textwidth]{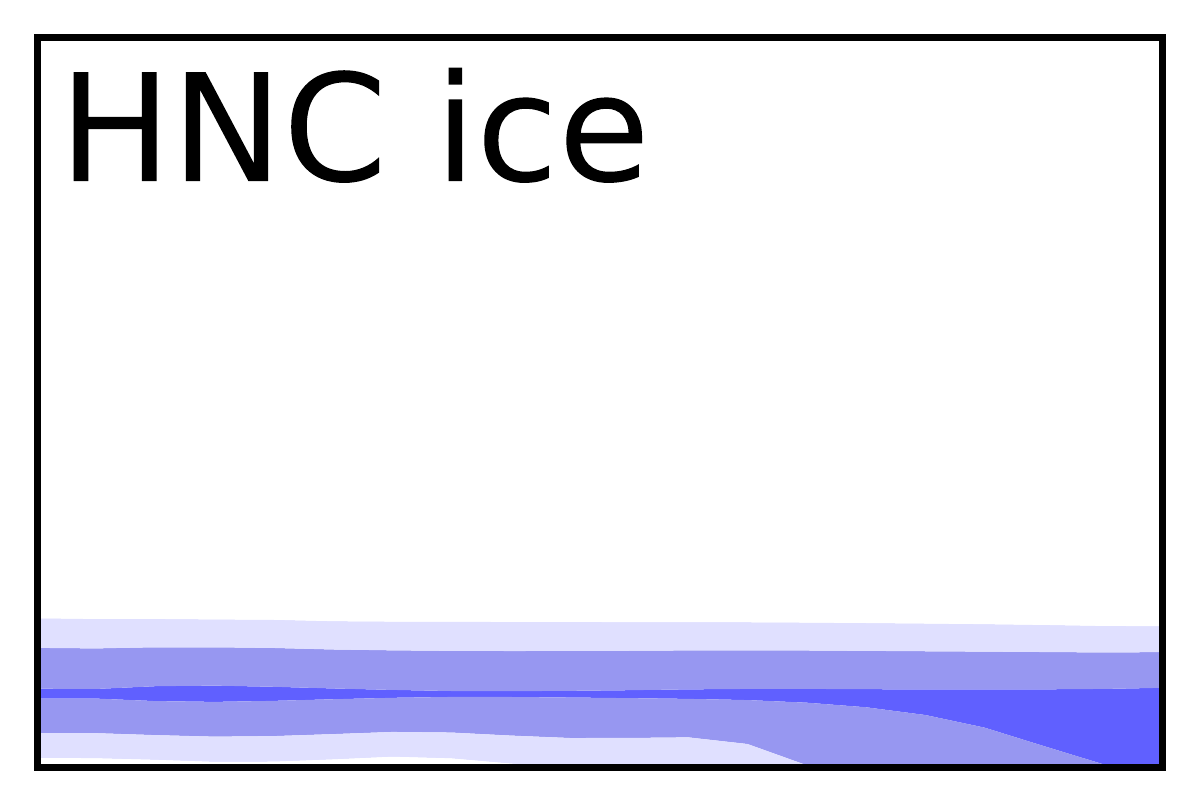} &
        \includegraphics[width=0.159\textwidth]{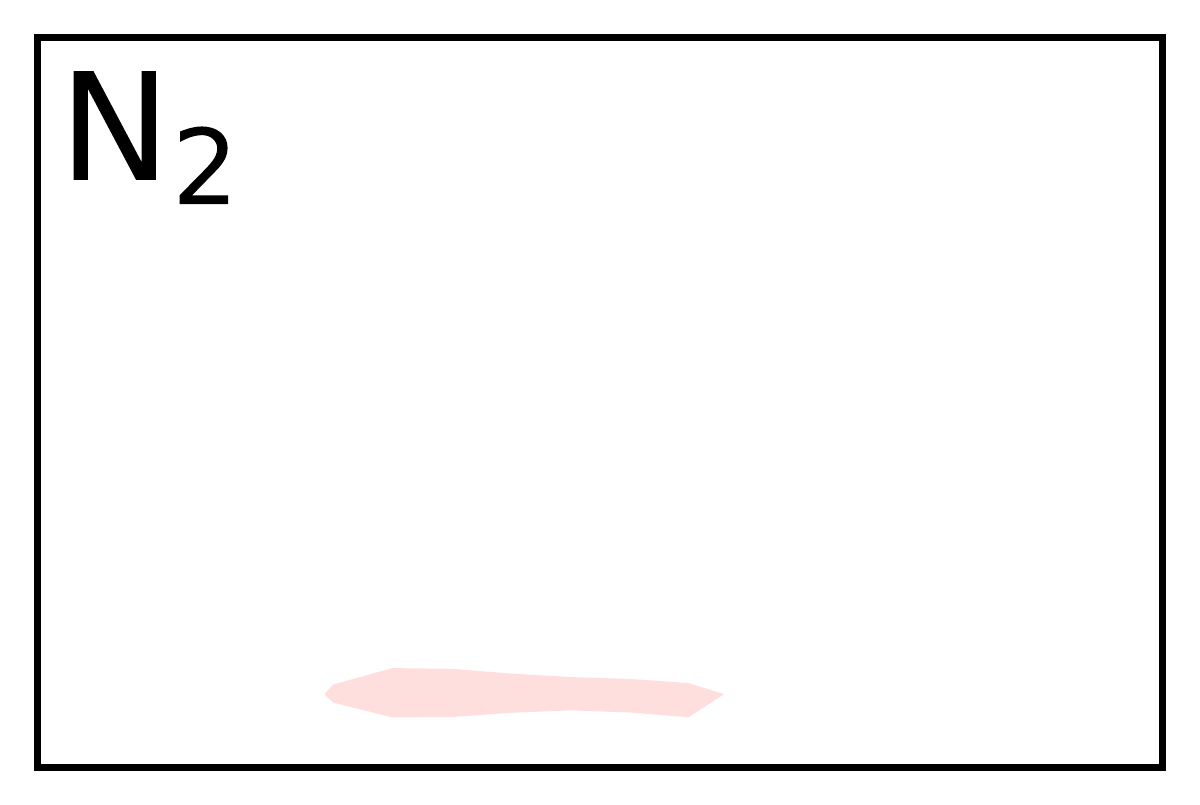} &
        \includegraphics[width=0.159\textwidth]{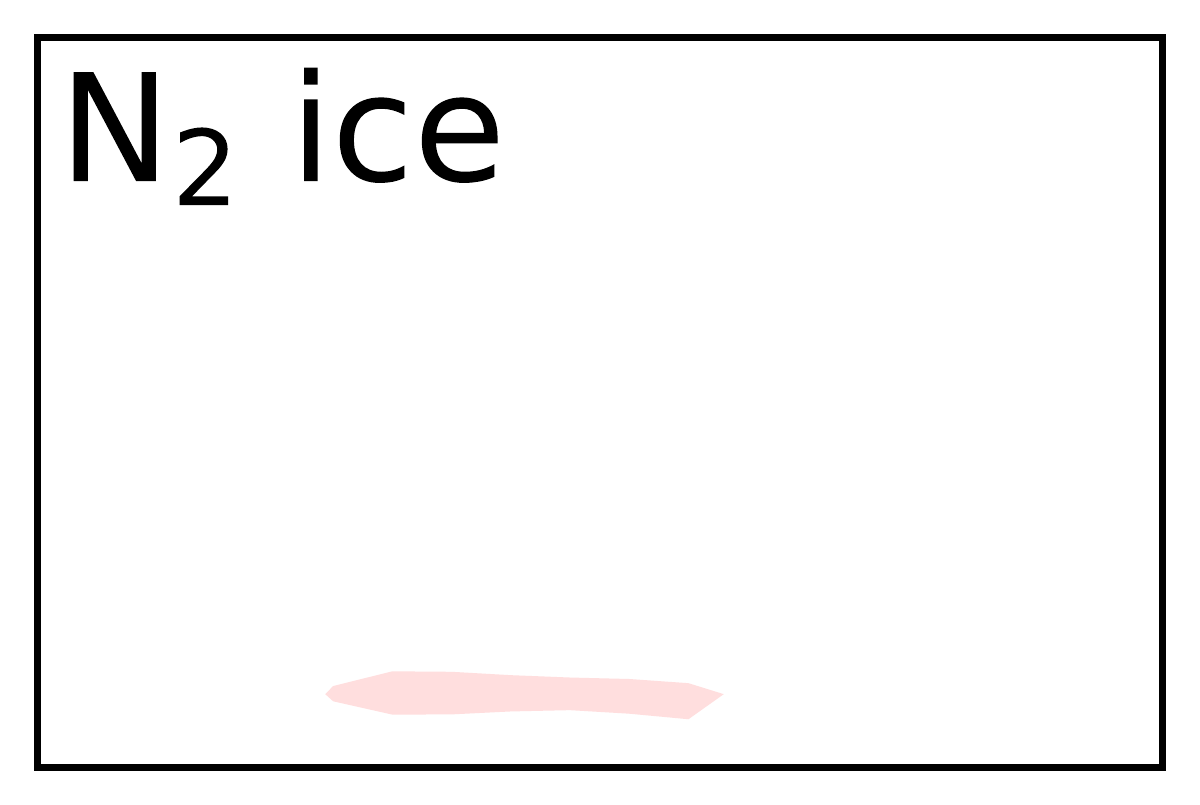} &
        \includegraphics[width=0.159\textwidth]{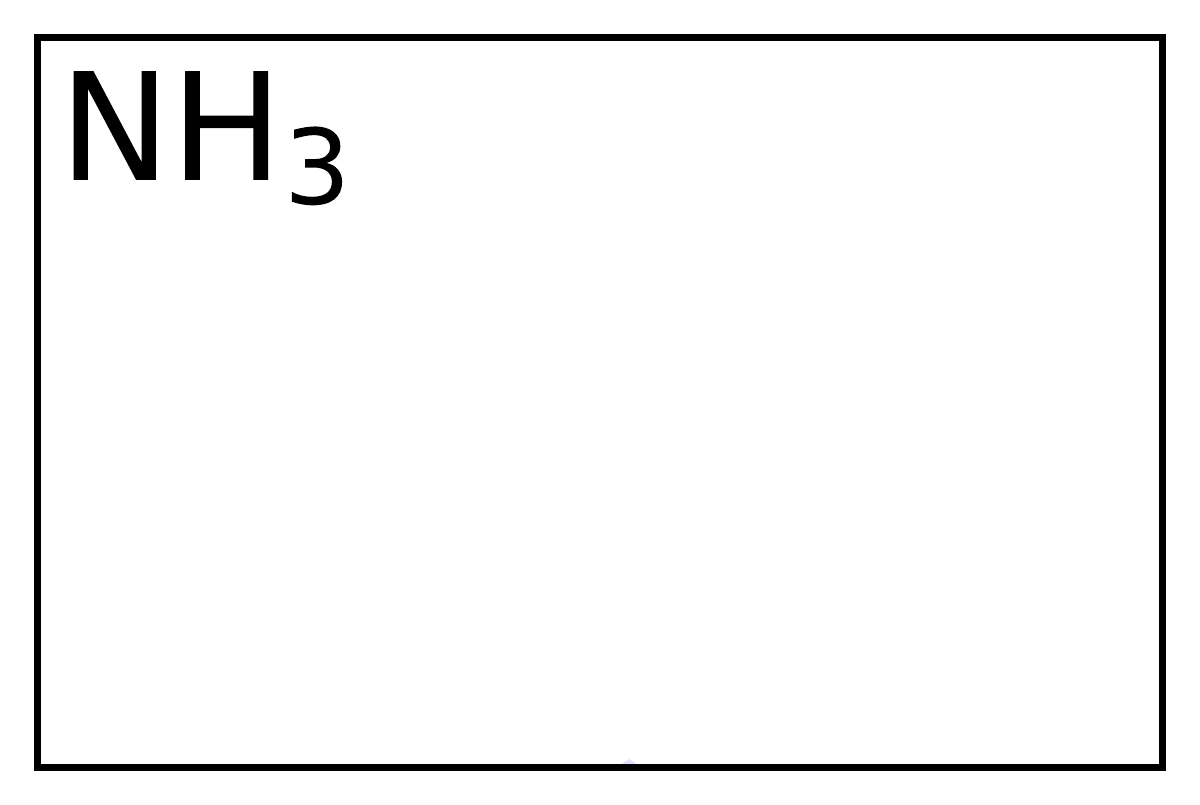} &
        \includegraphics[width=0.159\textwidth]{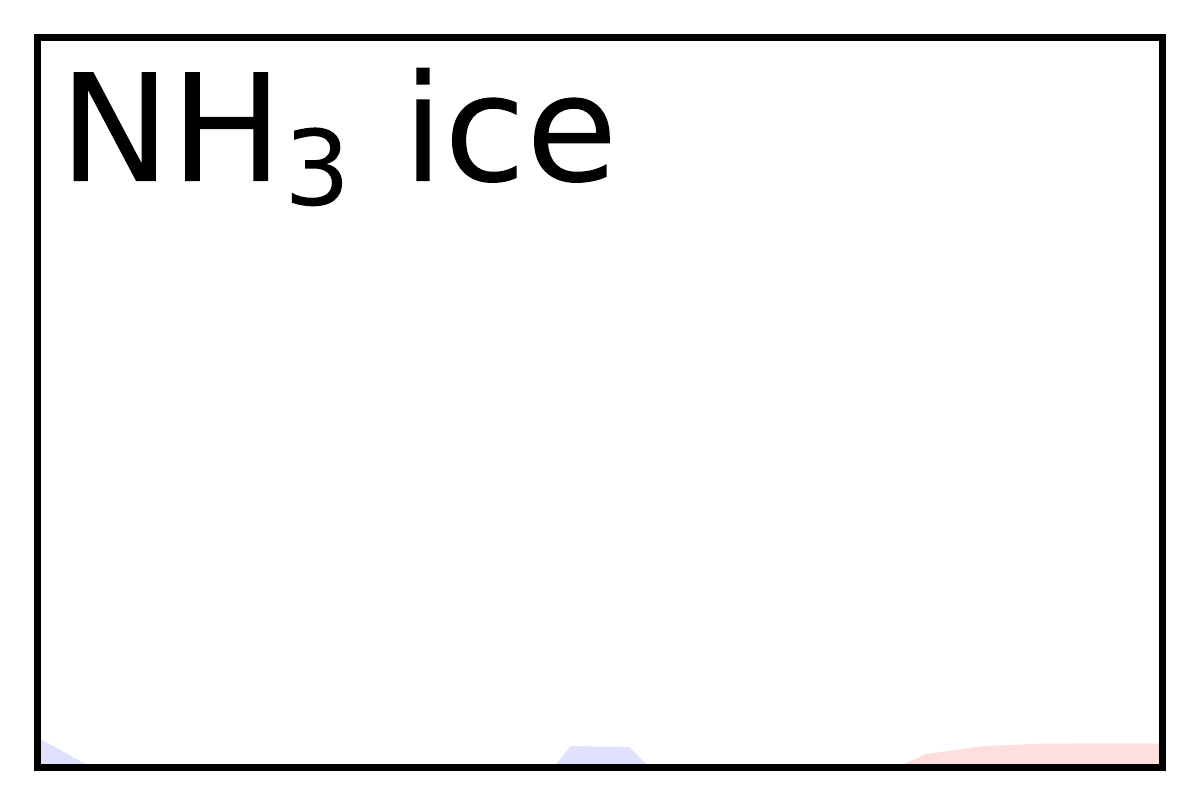} &
        \includegraphics[width=0.159\textwidth]{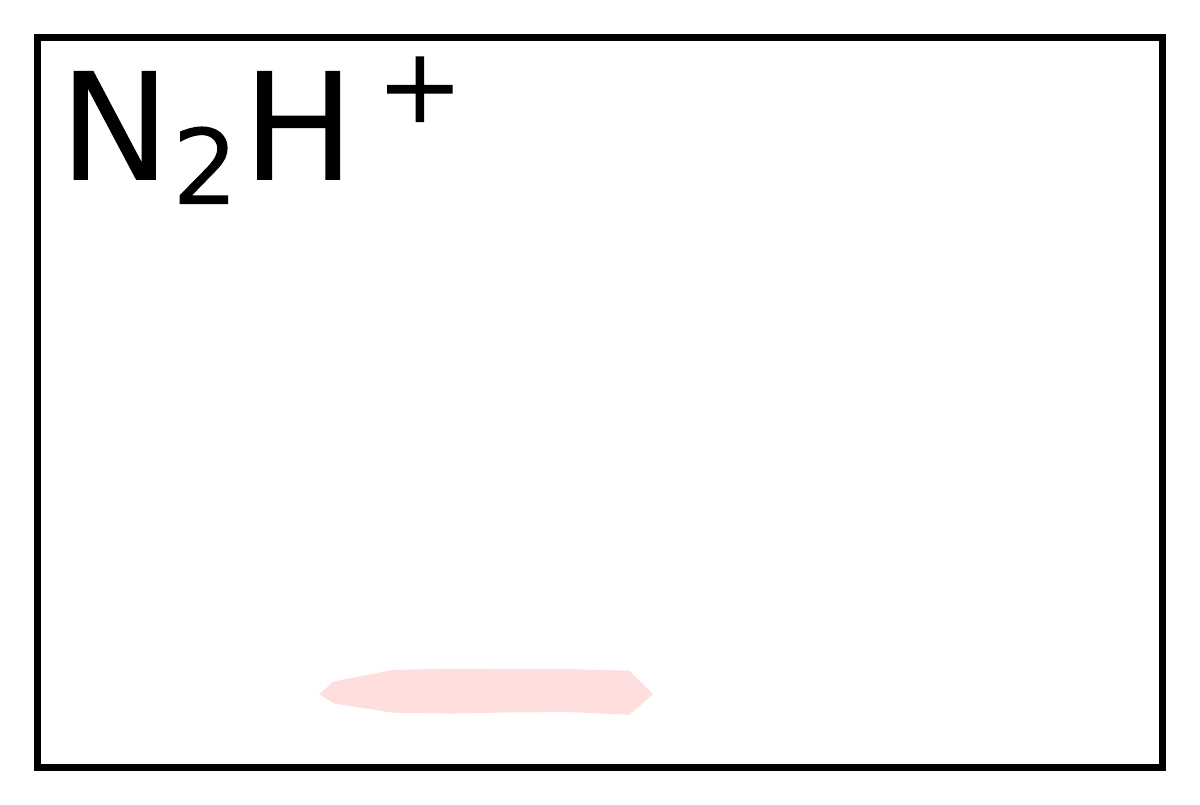} \\
        \includegraphics[width=0.159\textwidth]{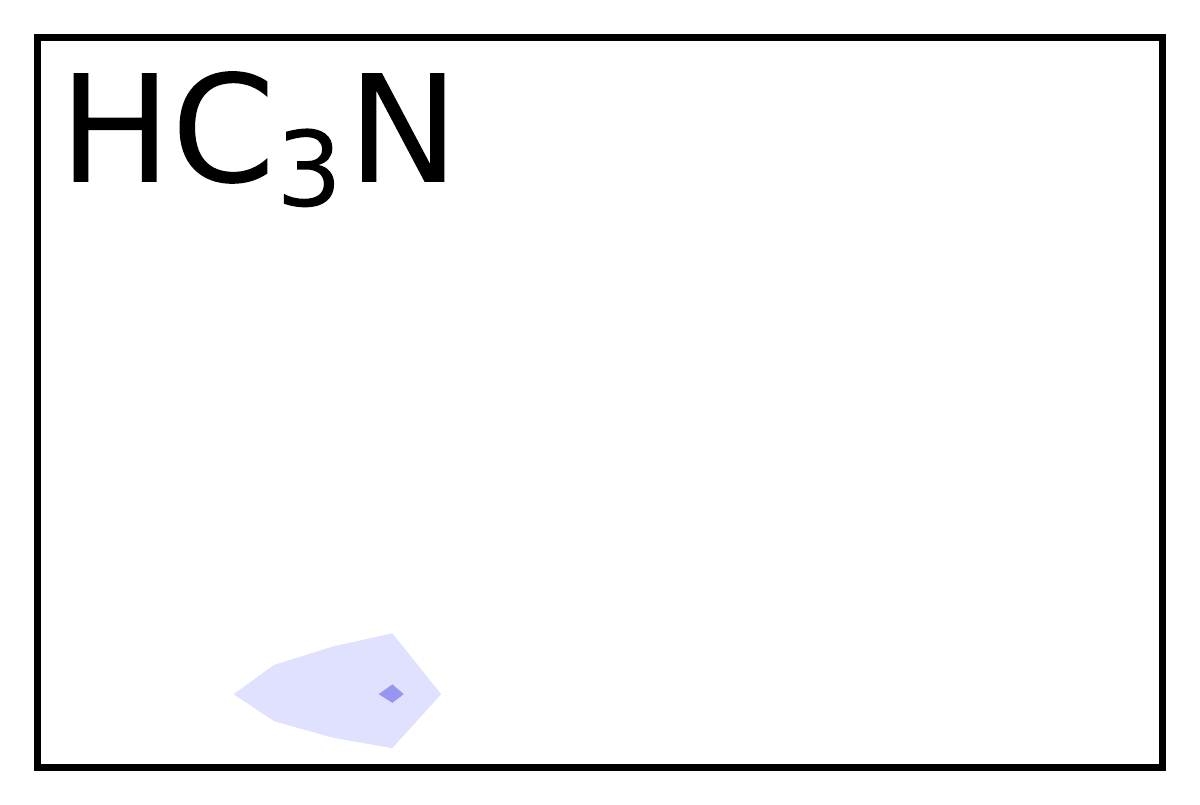} &
        \includegraphics[width=0.159\textwidth]{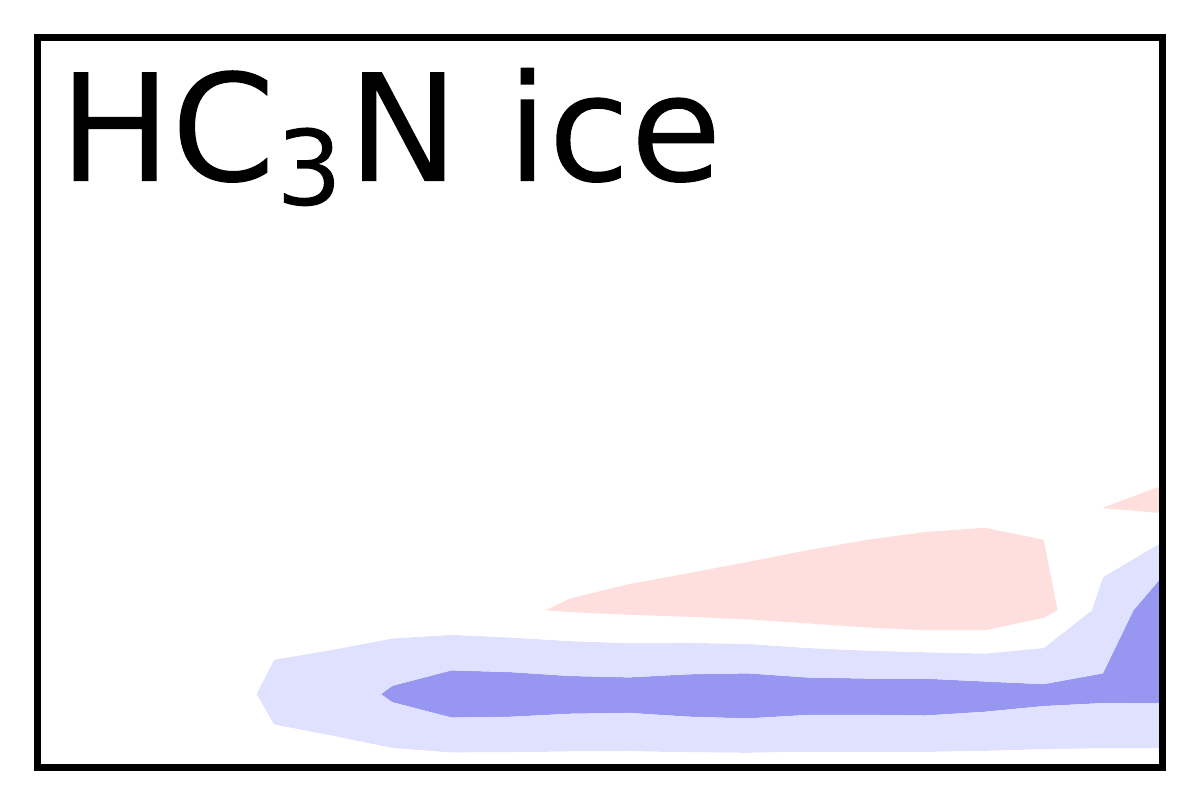} &
        \includegraphics[width=0.159\textwidth]{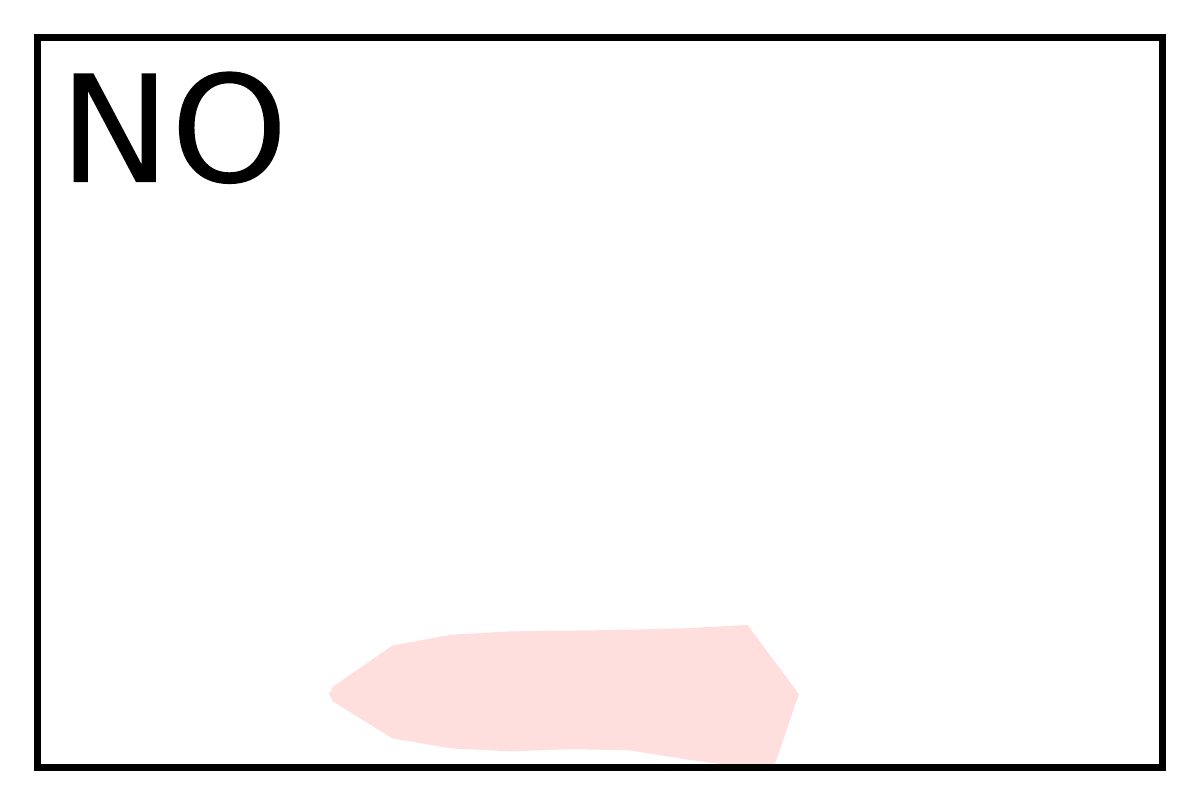} &
        \includegraphics[width=0.159\textwidth]{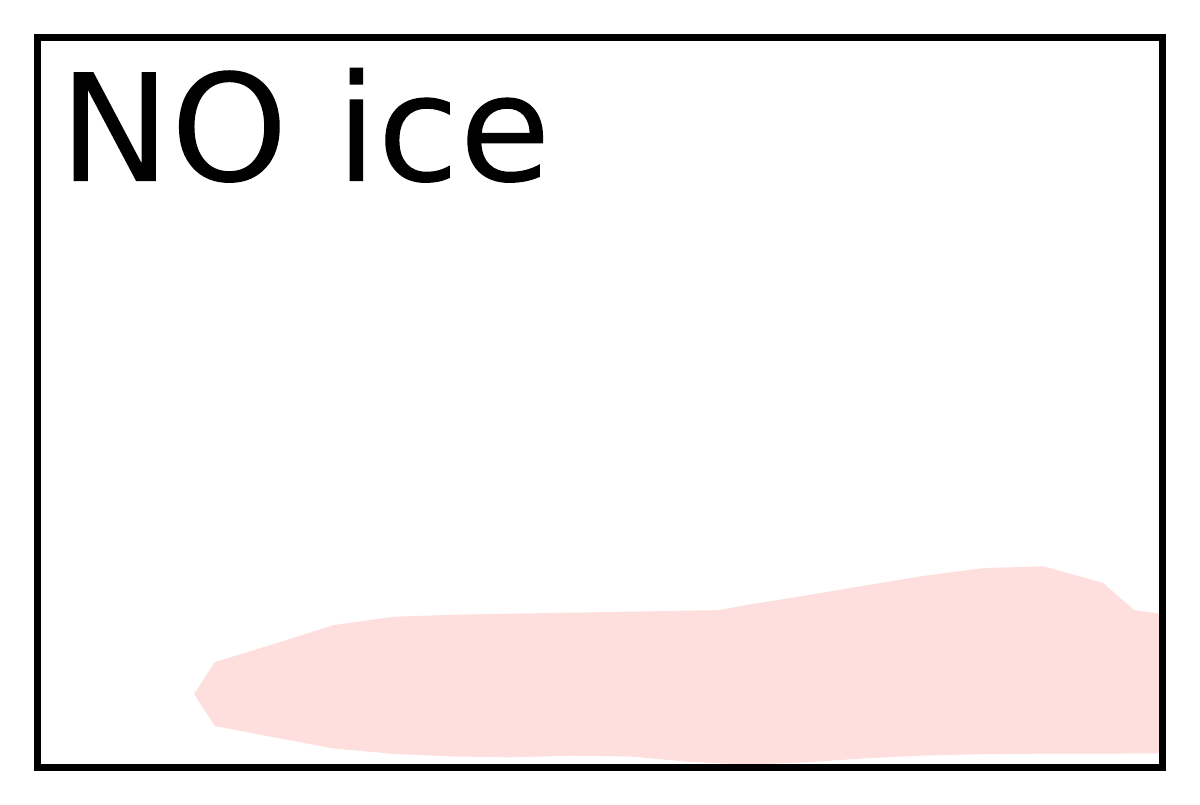} &
        \includegraphics[width=0.159\textwidth]{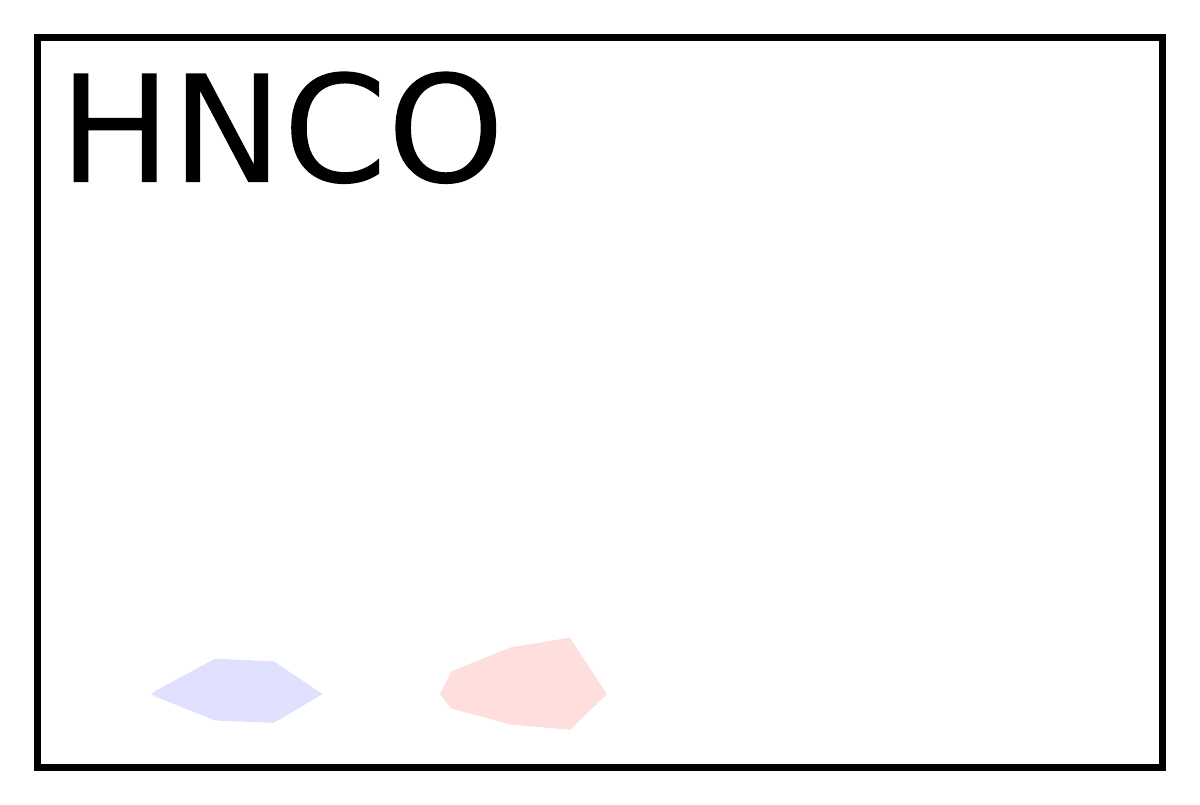} &
        \includegraphics[width=0.159\textwidth]{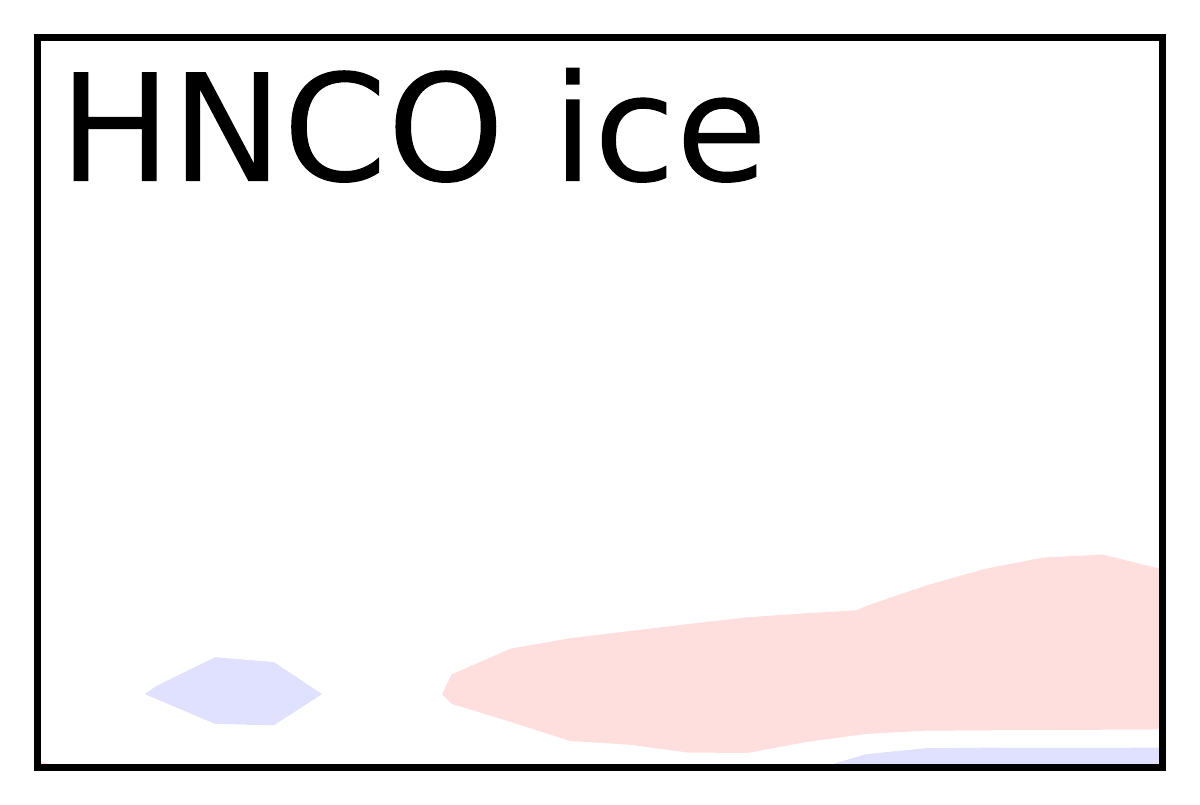} \\
        \includegraphics[width=0.159\textwidth]{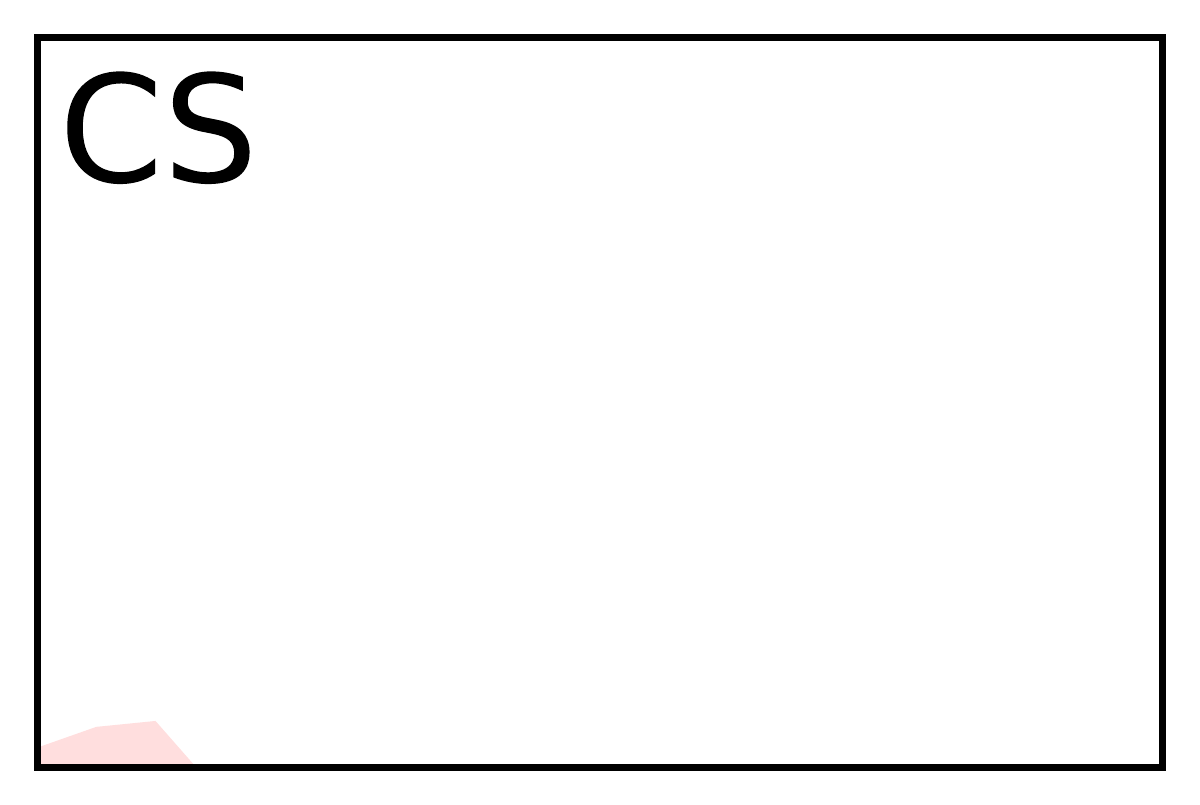} &
        \includegraphics[width=0.159\textwidth]{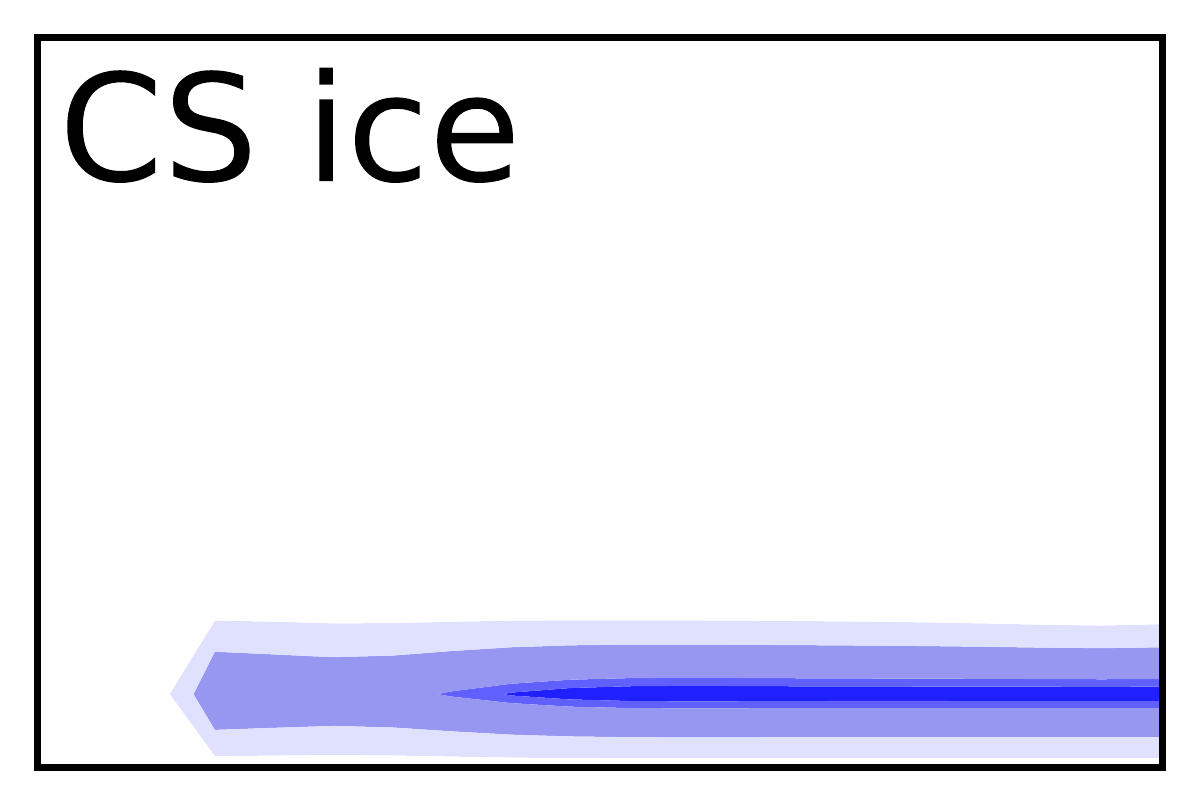} &
        \includegraphics[width=0.159\textwidth]{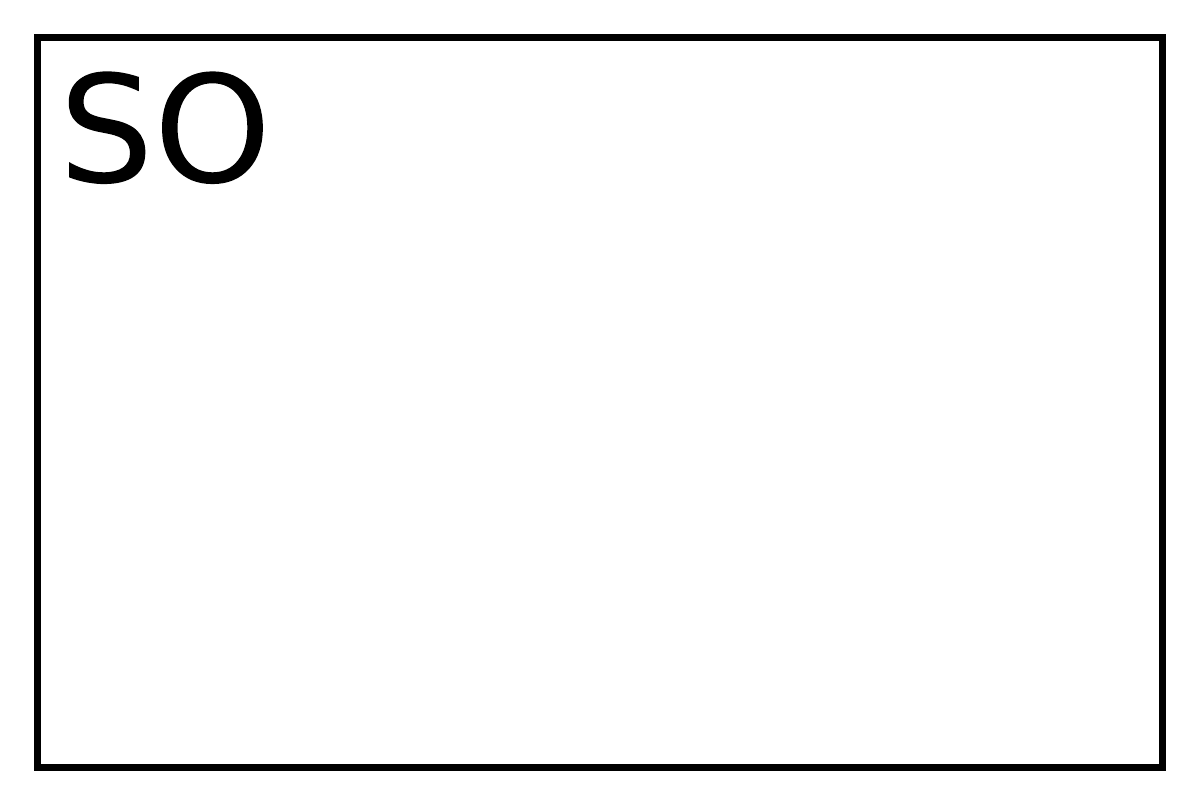} &
        \includegraphics[width=0.159\textwidth]{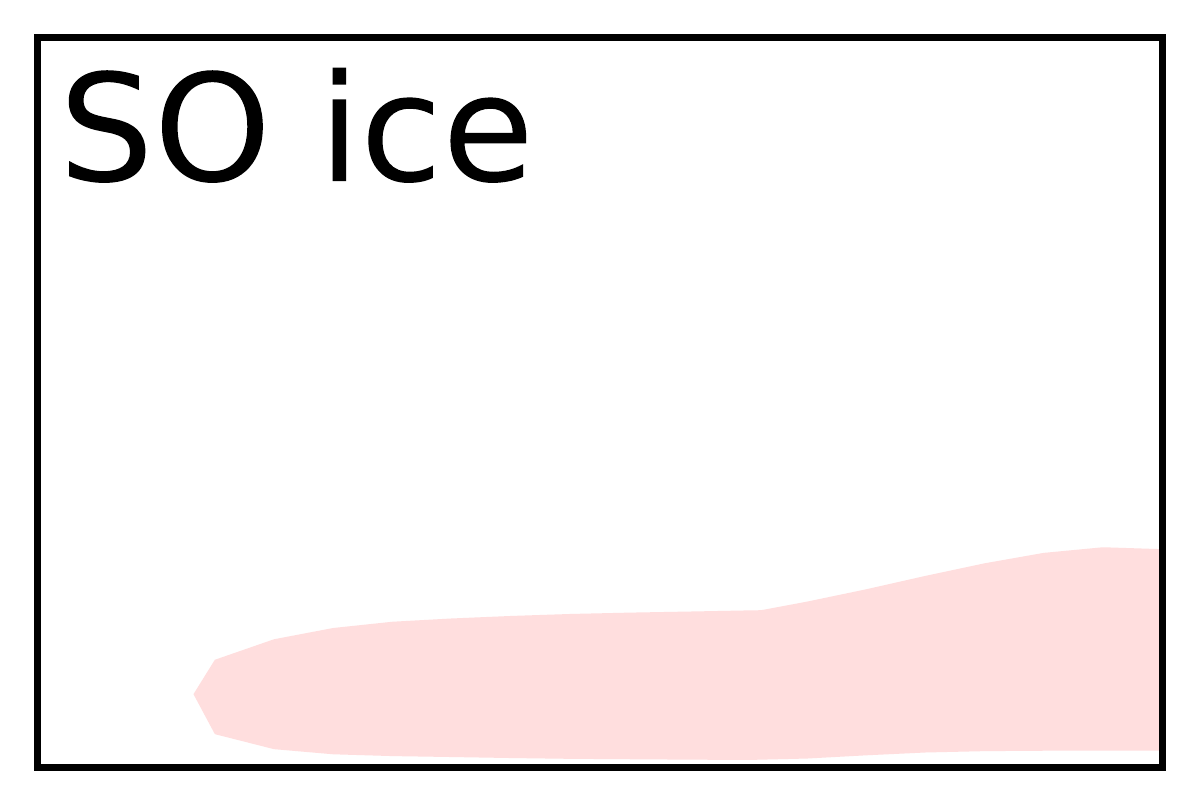} &
        \includegraphics[width=0.159\textwidth]{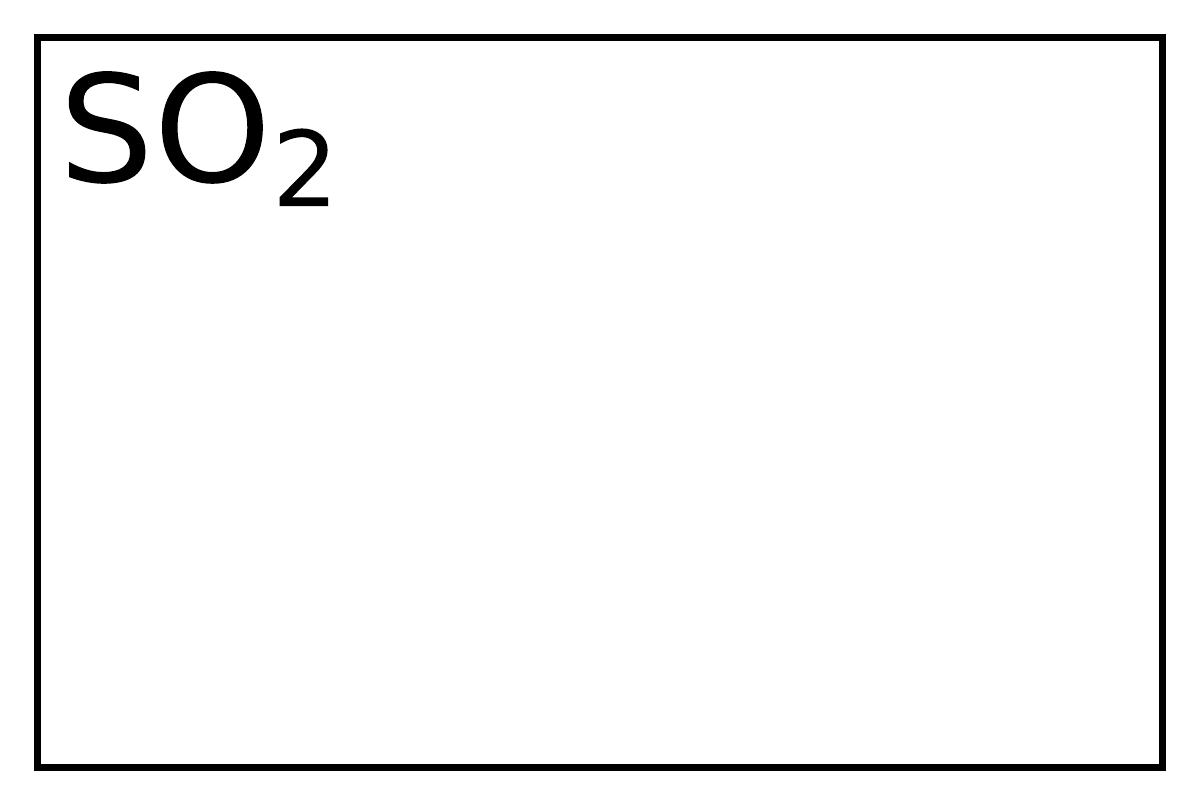} &
        \includegraphics[width=0.159\textwidth]{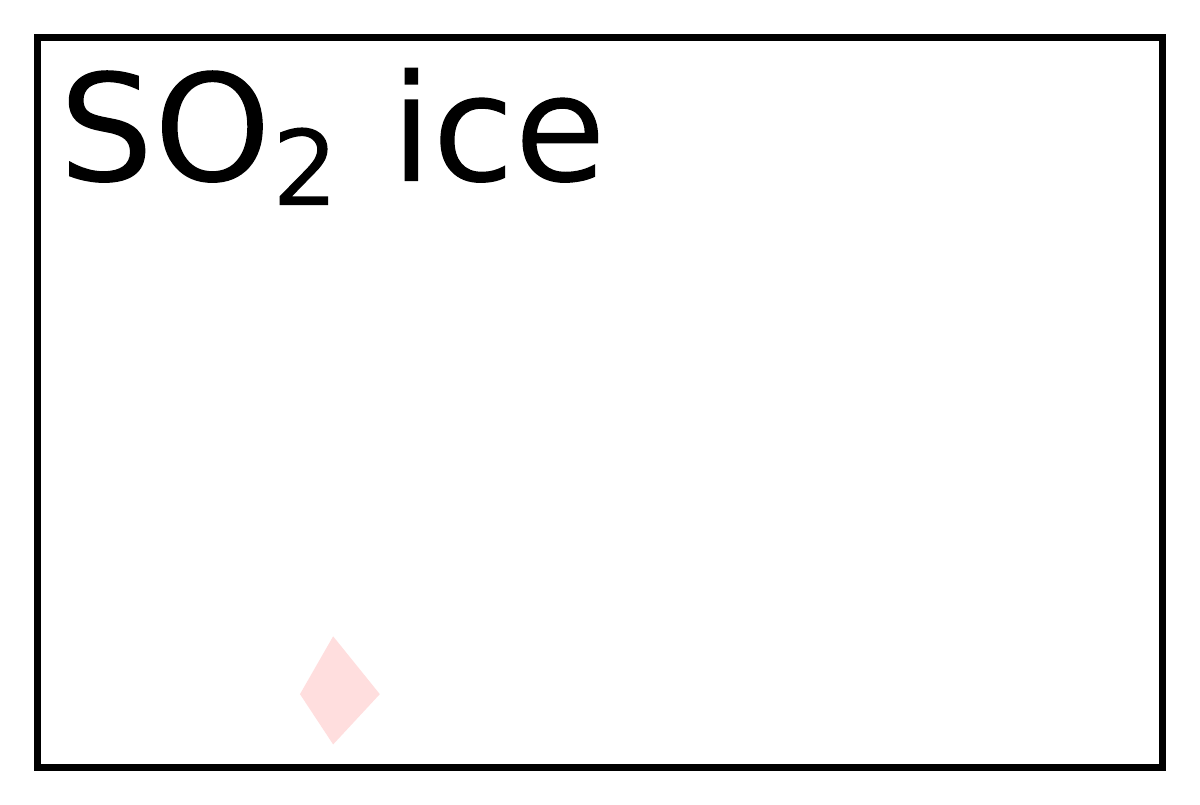} \\
        \includegraphics[width=0.189\textwidth]{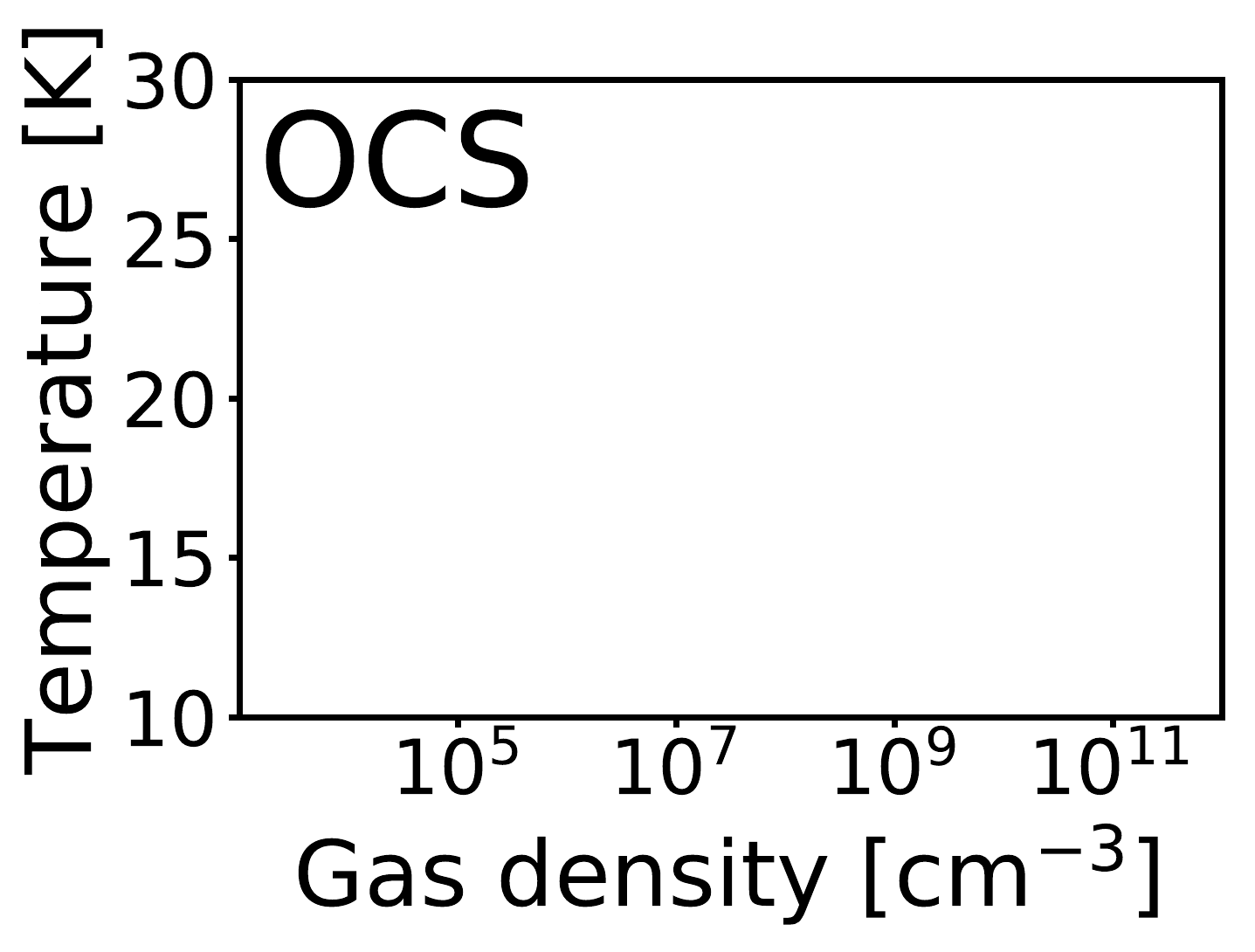} &
        \includegraphics[width=0.159\textwidth]{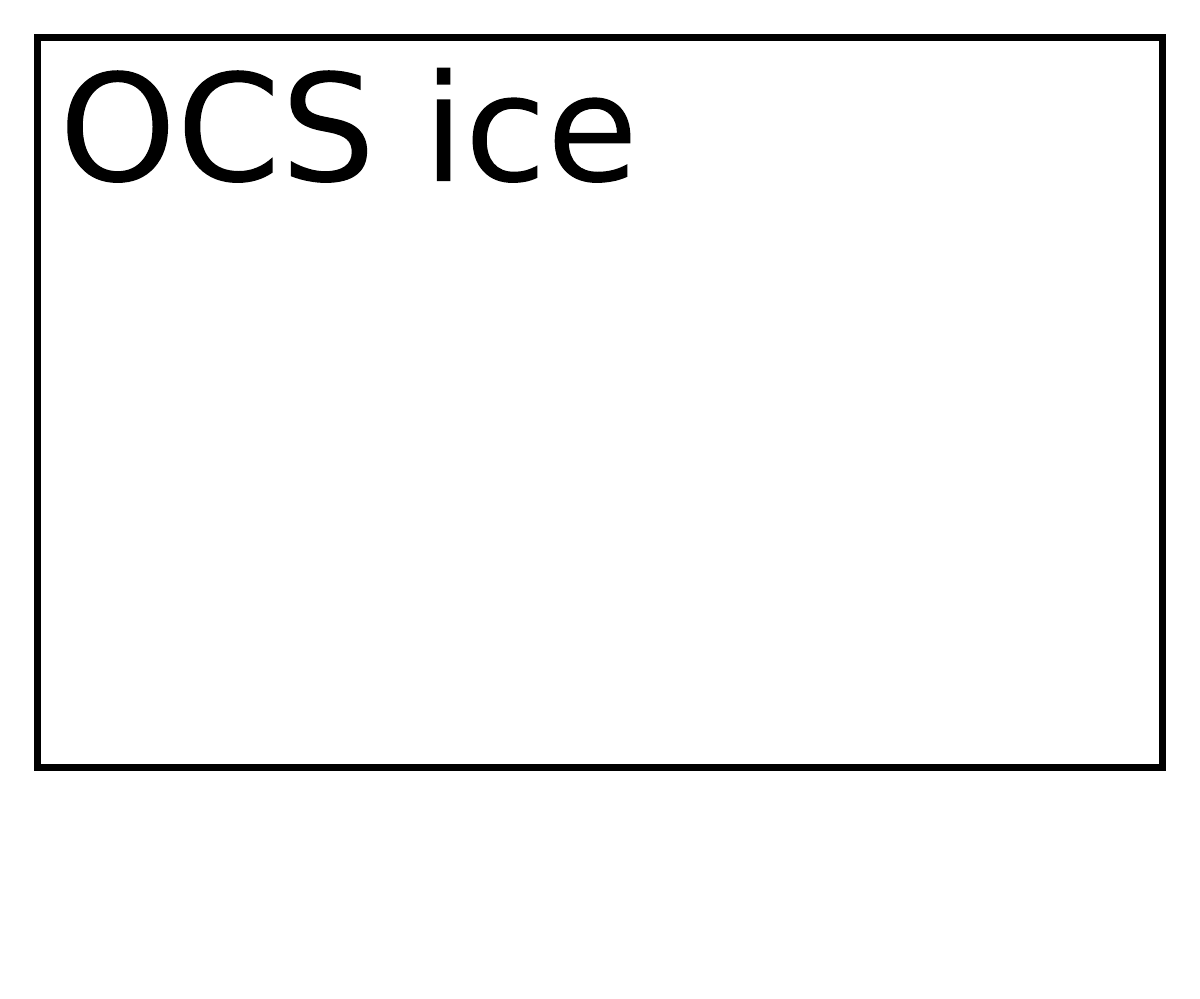} &
        \includegraphics[width=0.159\textwidth]{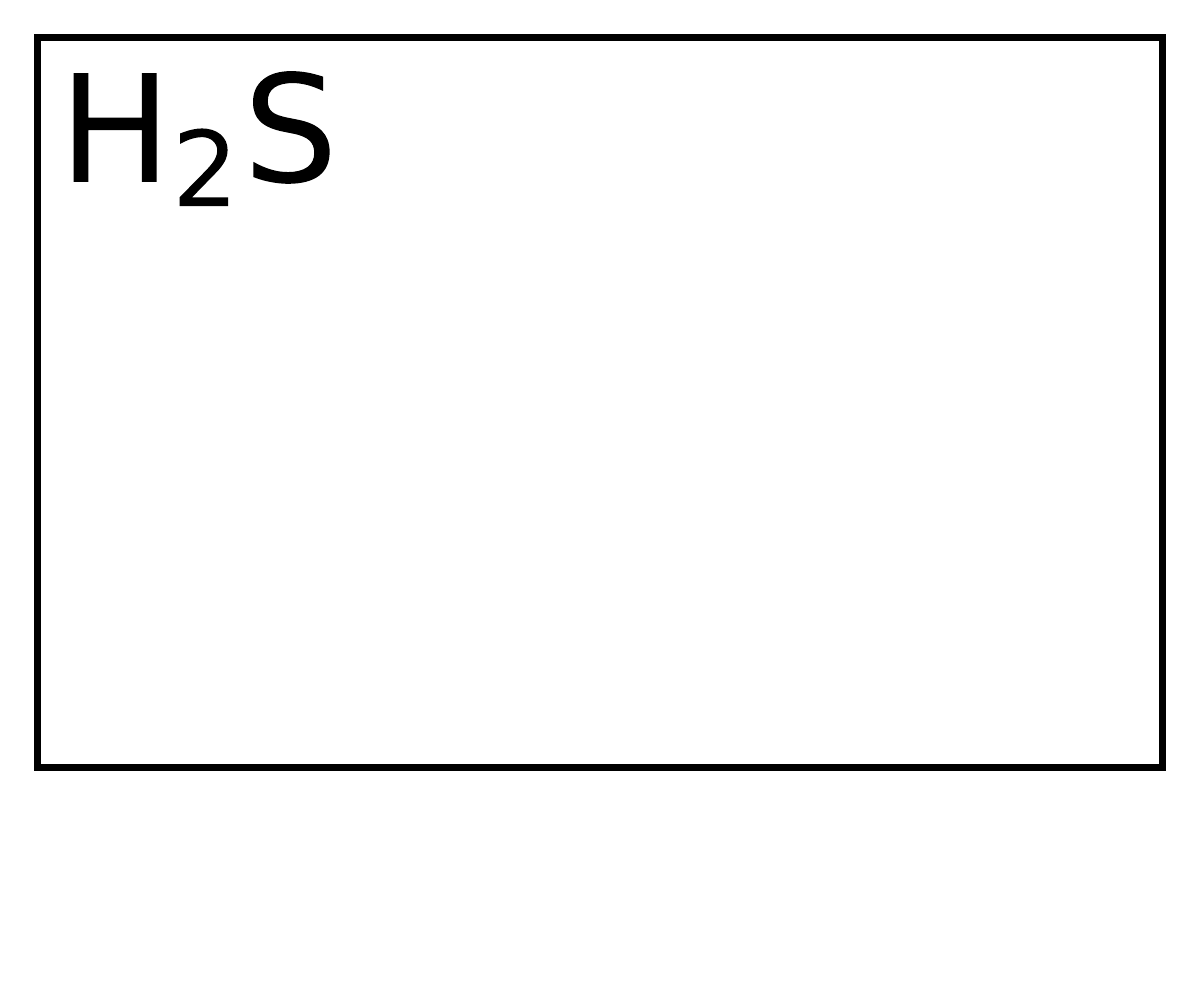} &
        \includegraphics[width=0.159\textwidth]{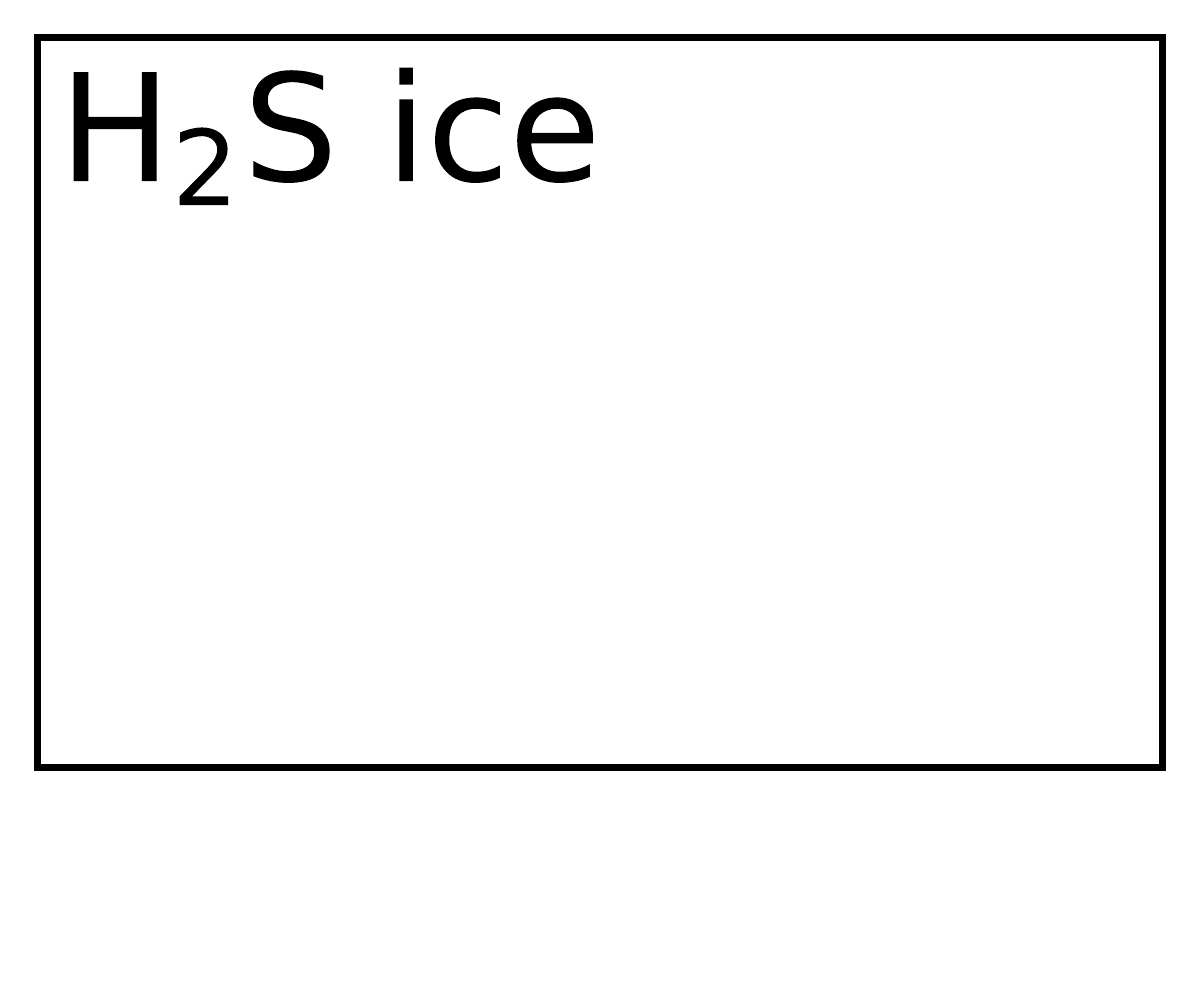} &
        \includegraphics[width=0.159\textwidth]{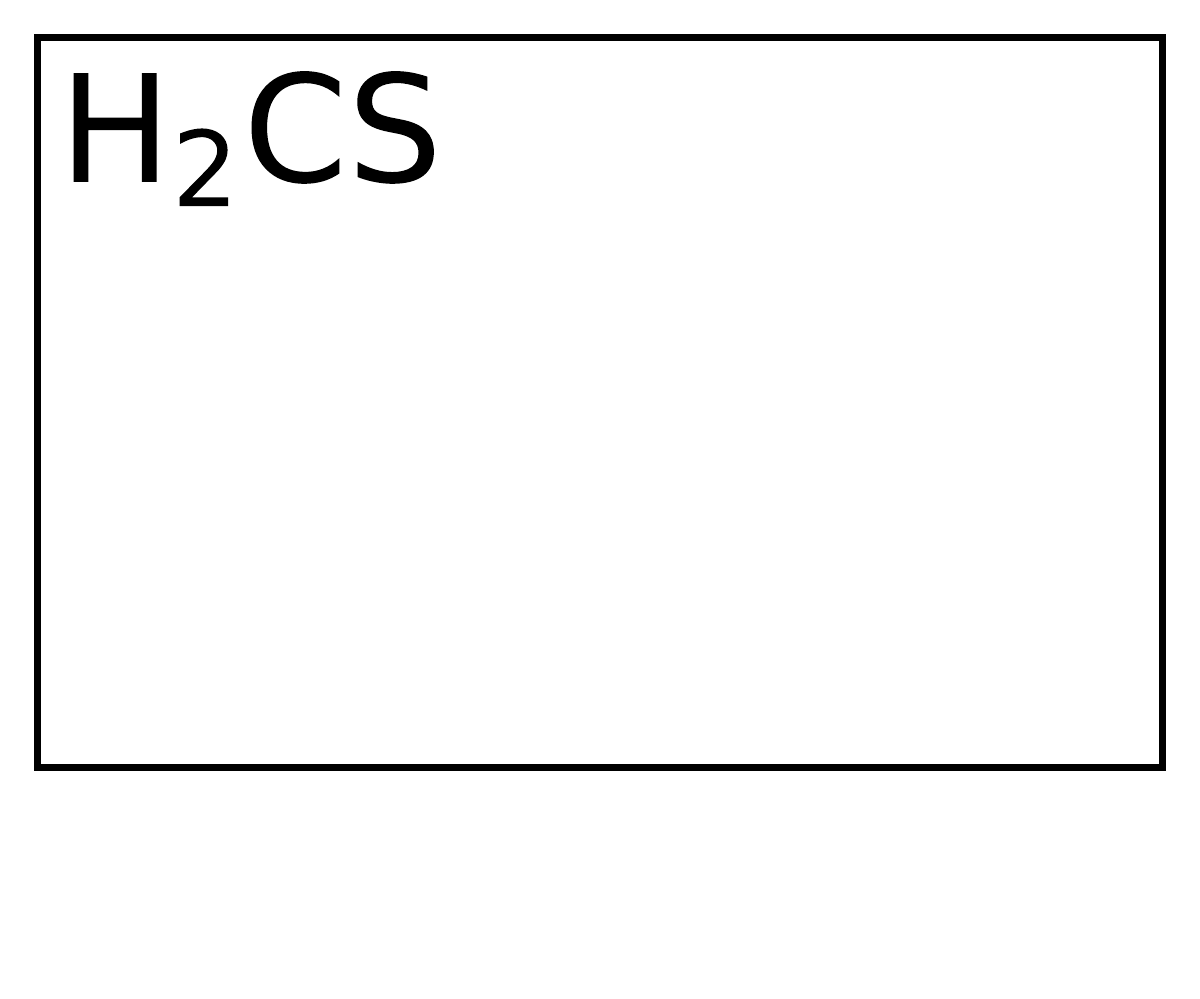} &
        \includegraphics[width=0.211\textwidth]{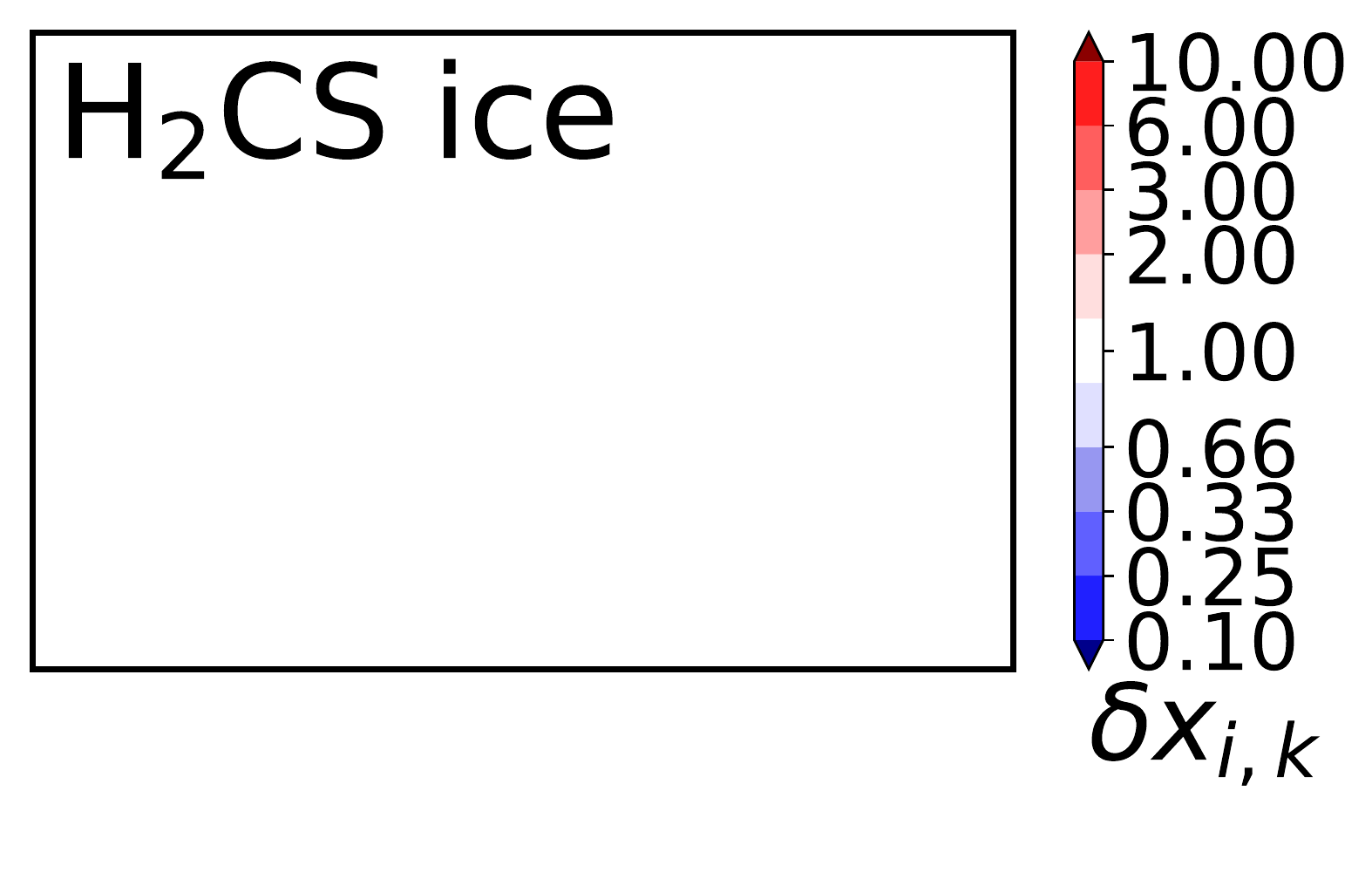} \\
        \end{tabular}\\
\caption{Abundance ratios $\delta x$ between the modified and original networks
        are shown for the entire 2D physical grid at $t_{j}=10^{6}$~years for selected
        observed species. The results are depicted for the $A_{\rm V}=3$~mag case.
        }
        \label{fig:1d2d_Av_3}
\end{figure*}

\begin{figure*}[!ht]
\centering
\setlength\tabcolsep{-0.5pt}
\renewcommand{\arraystretch}{0}
        \begin{tabular}{@{}R{3.6cm}C{3cm}C{3cm}C{3cm}C{3cm}L{3.7cm}@{}}
        \includegraphics[width=0.159\textwidth]{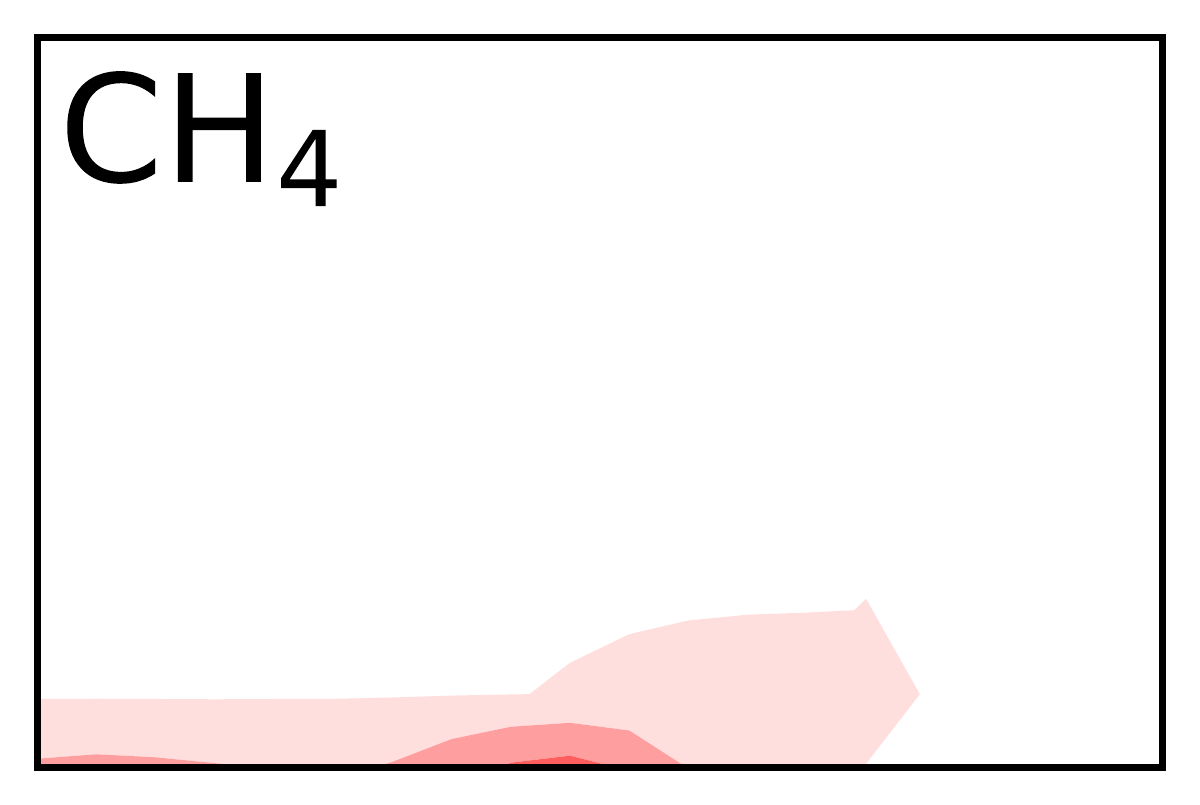} &
        \includegraphics[width=0.159\textwidth]{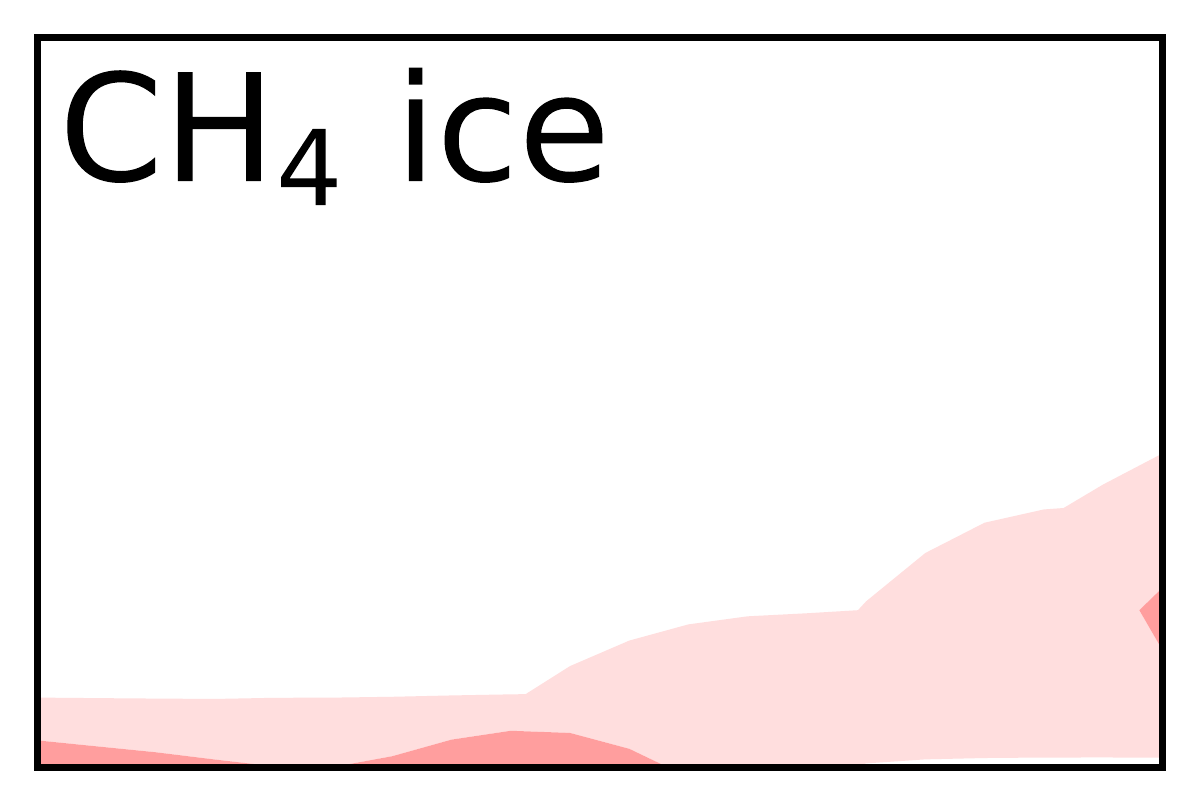} &
        \includegraphics[width=0.159\textwidth]{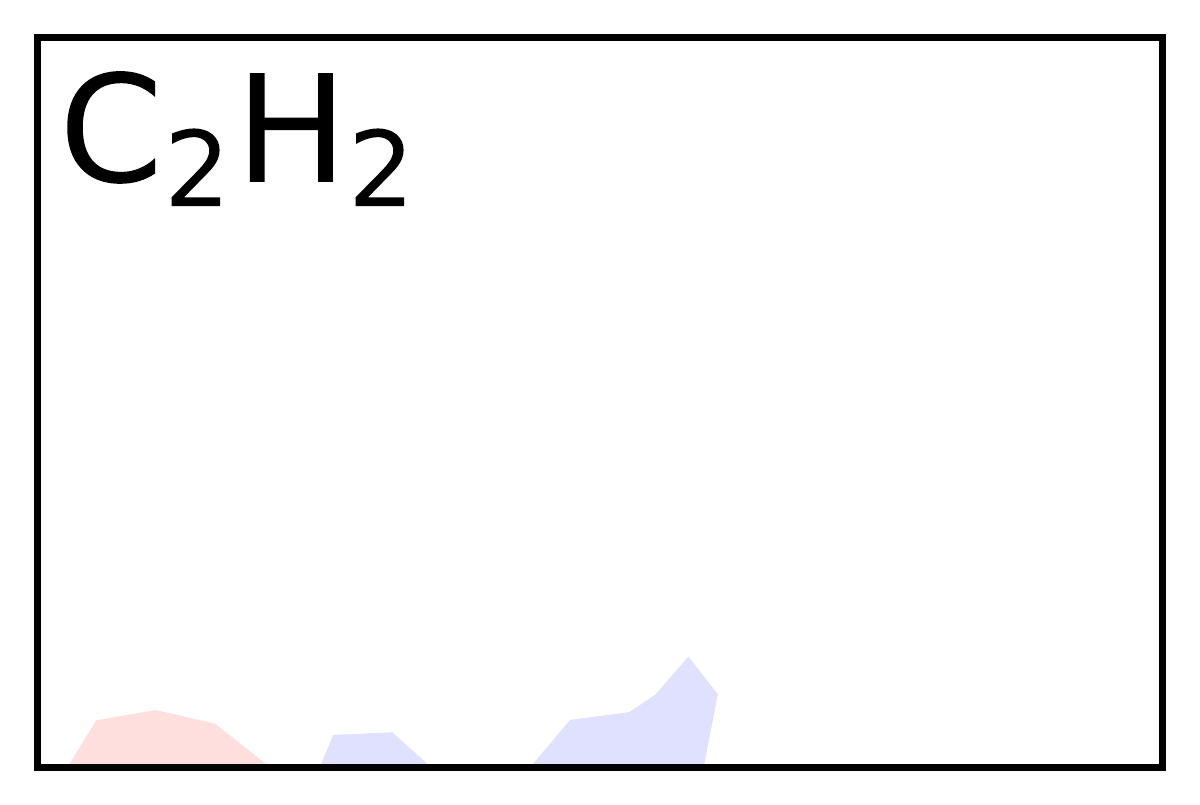} &
        \includegraphics[width=0.159\textwidth]{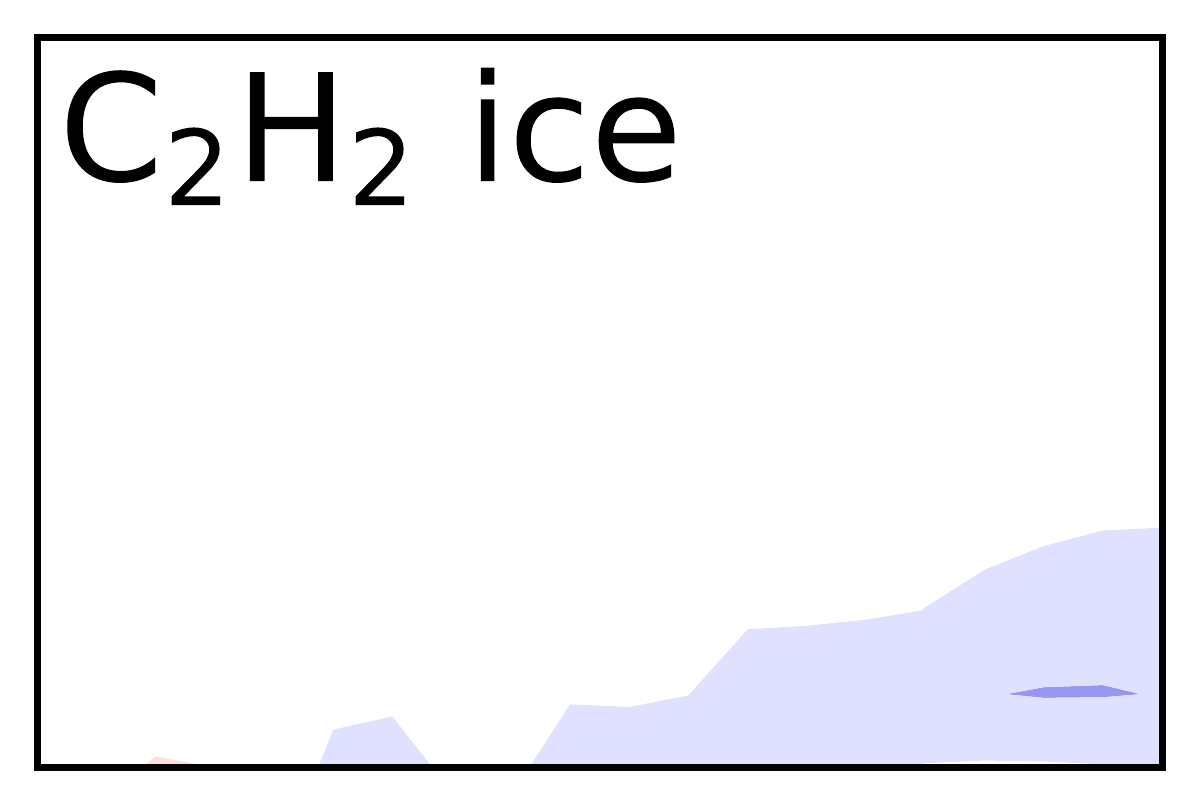} &
        \includegraphics[width=0.159\textwidth]{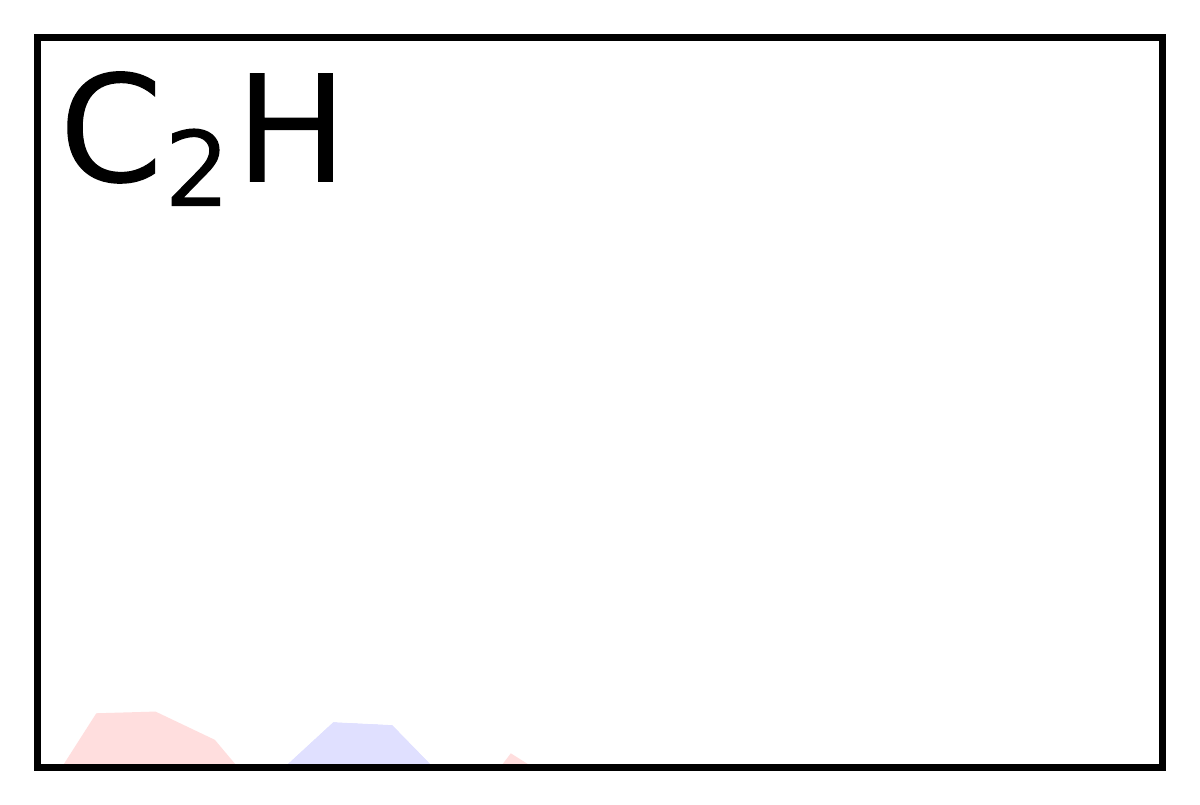} &
        \includegraphics[width=0.159\textwidth]{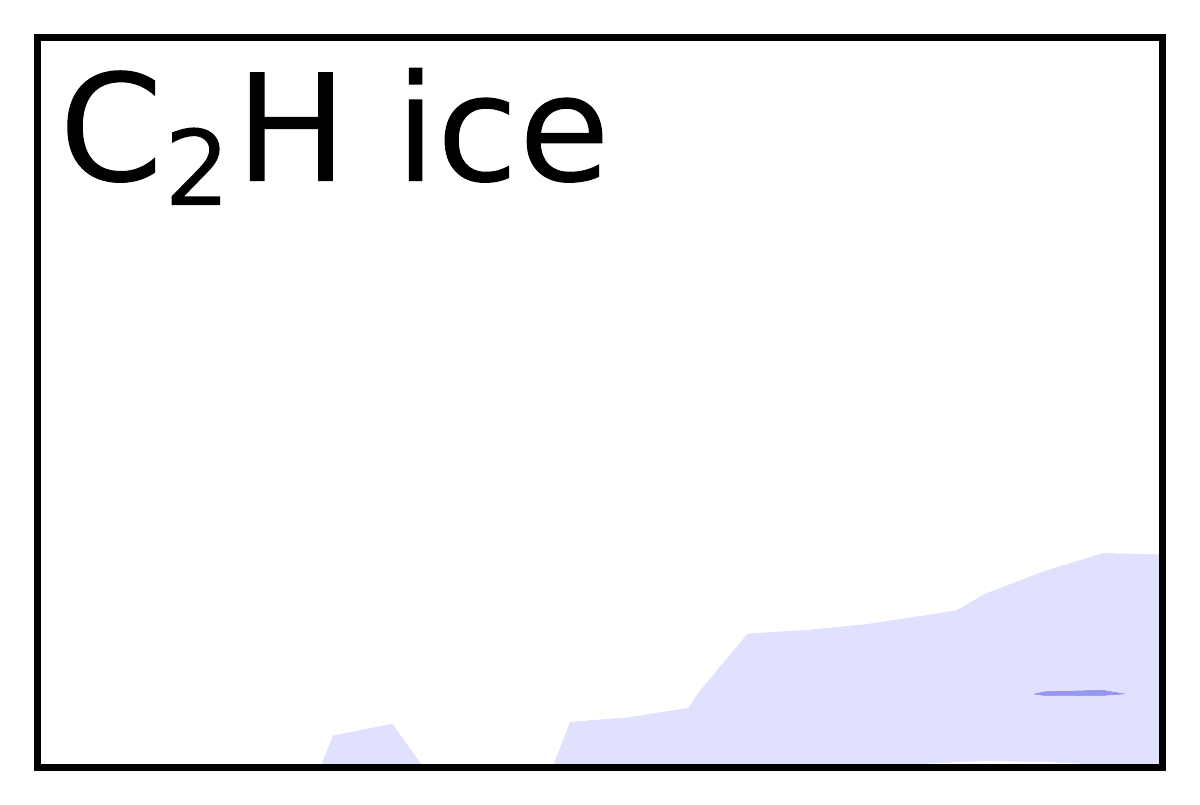} \\
        \includegraphics[width=0.159\textwidth]{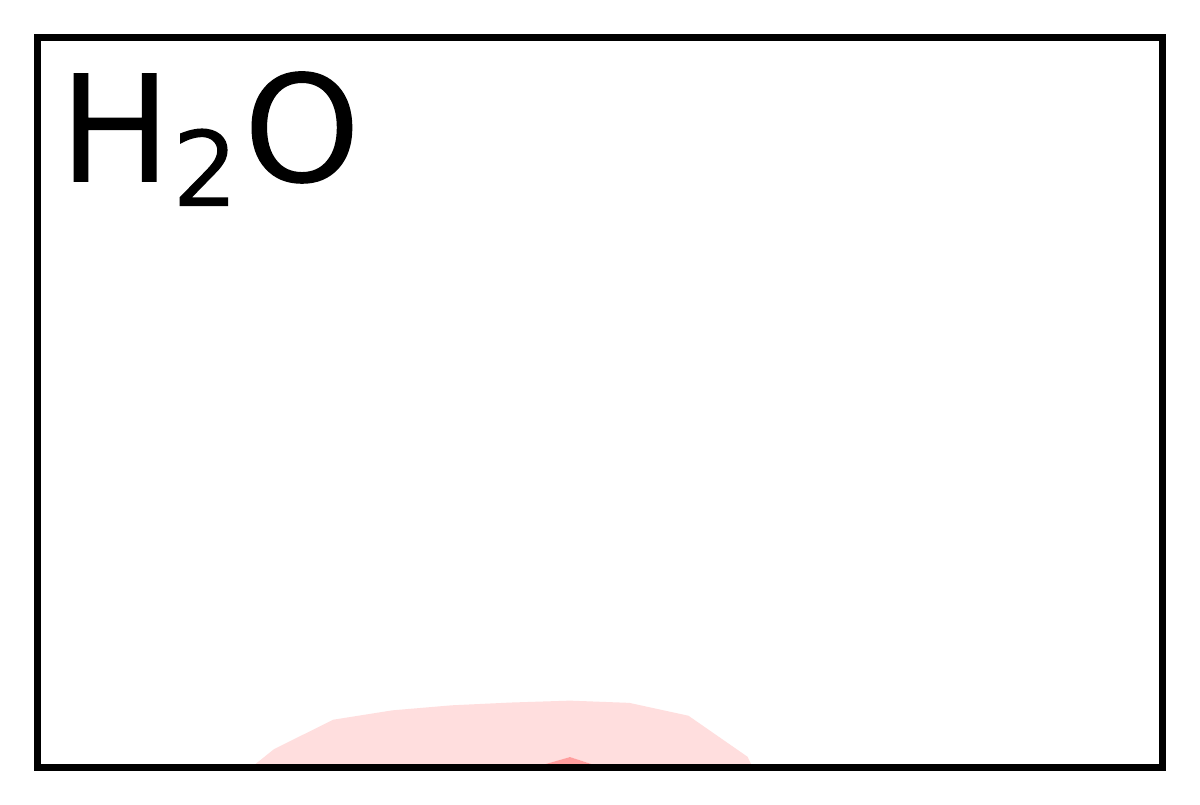} &
        \includegraphics[width=0.159\textwidth]{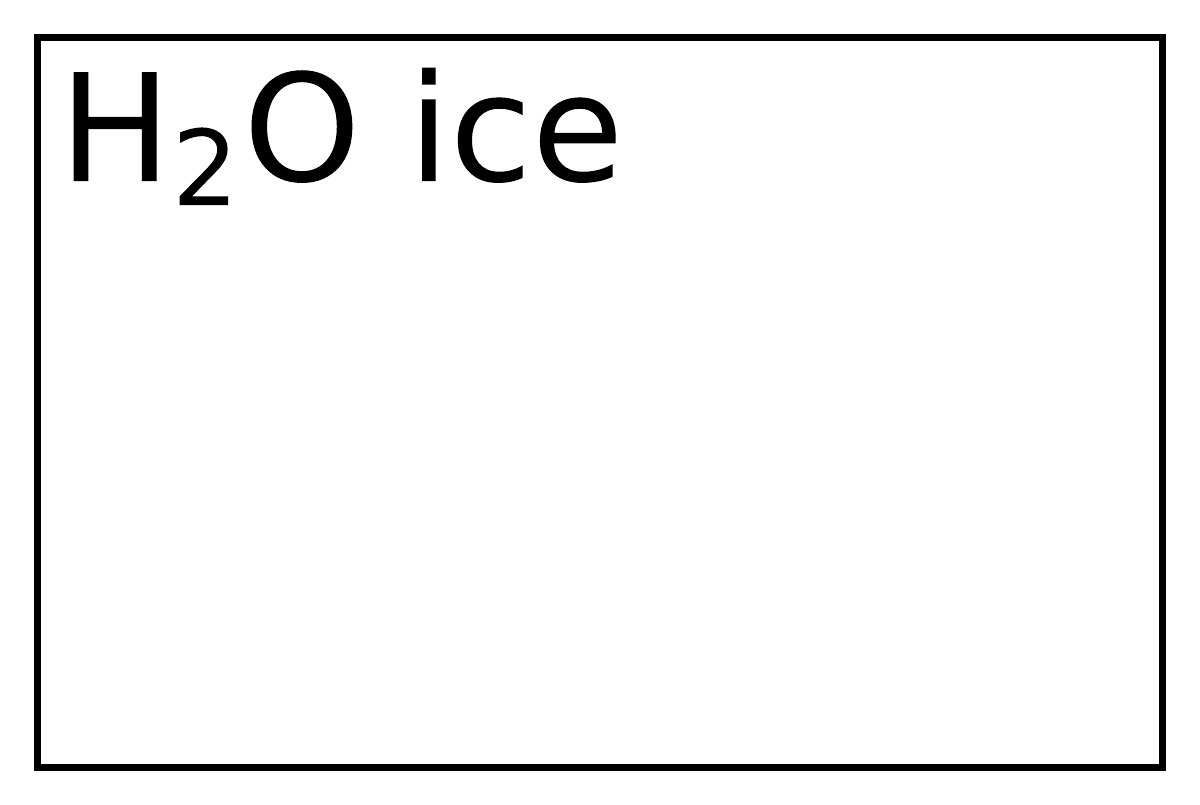} &
        \includegraphics[width=0.159\textwidth]{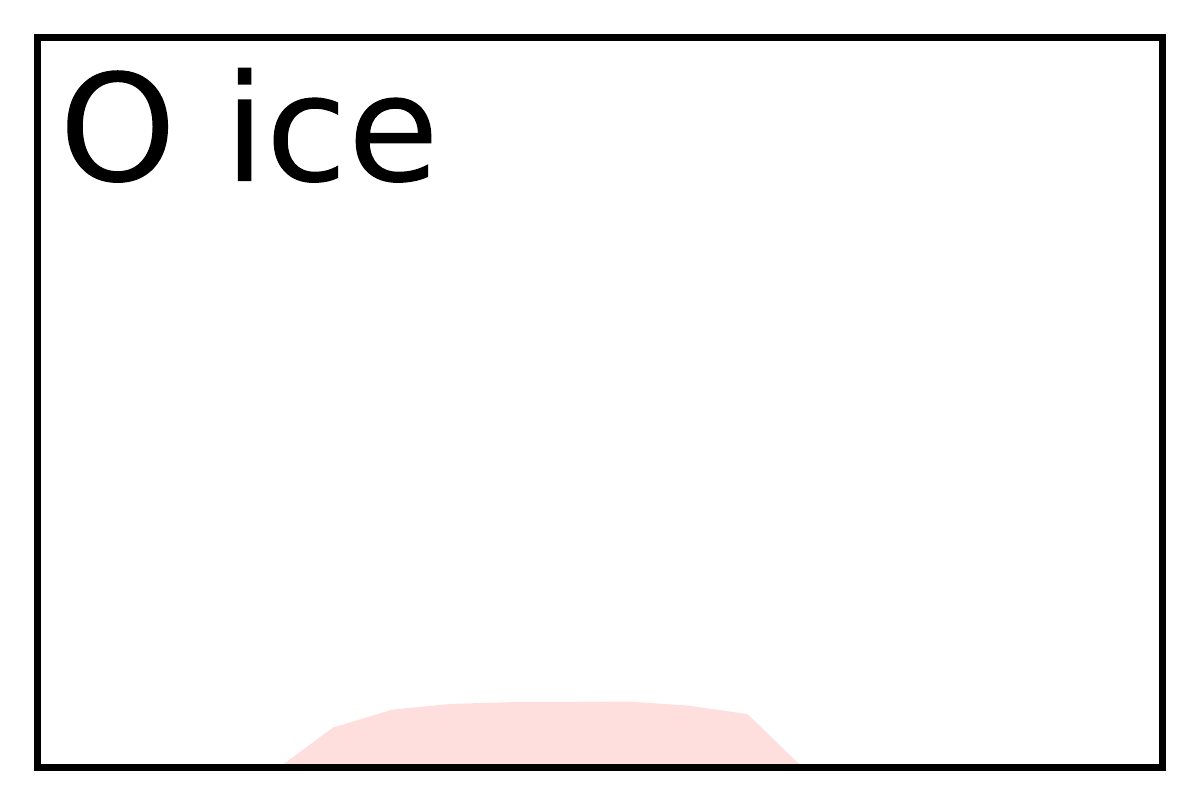} &
        \includegraphics[width=0.159\textwidth]{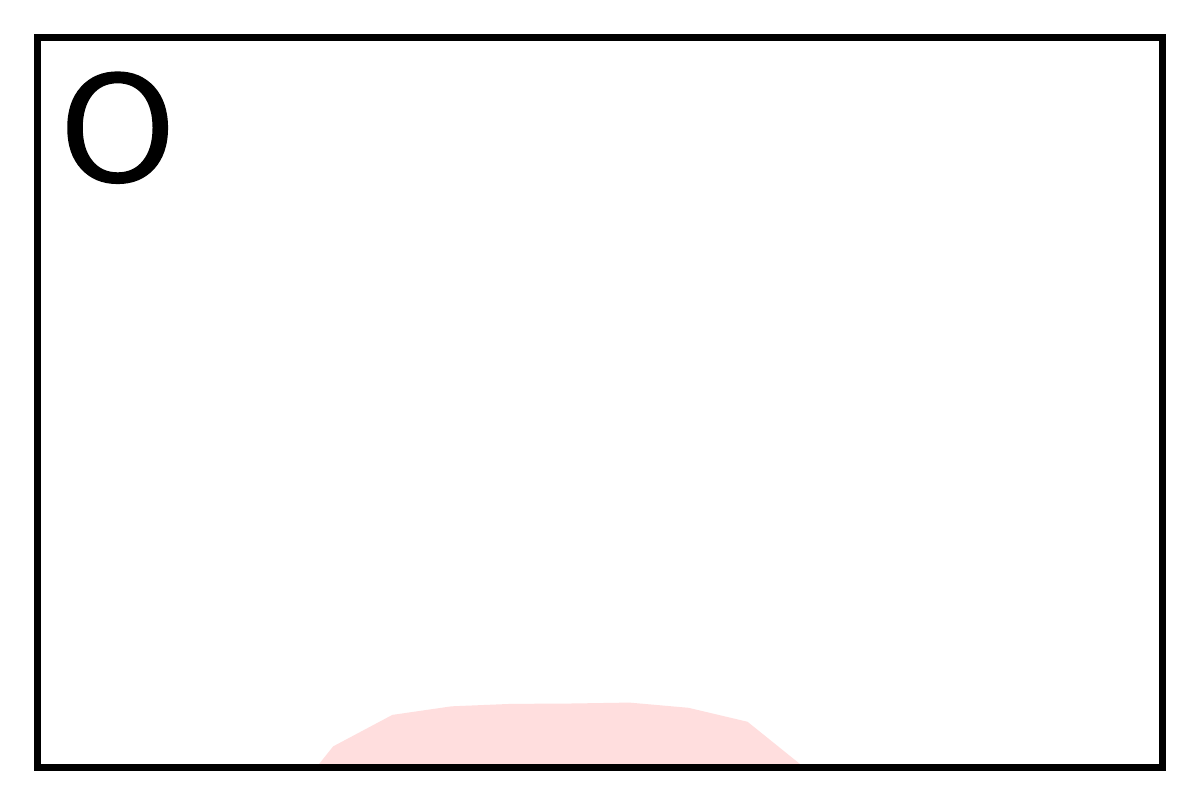} &
        \includegraphics[width=0.159\textwidth]{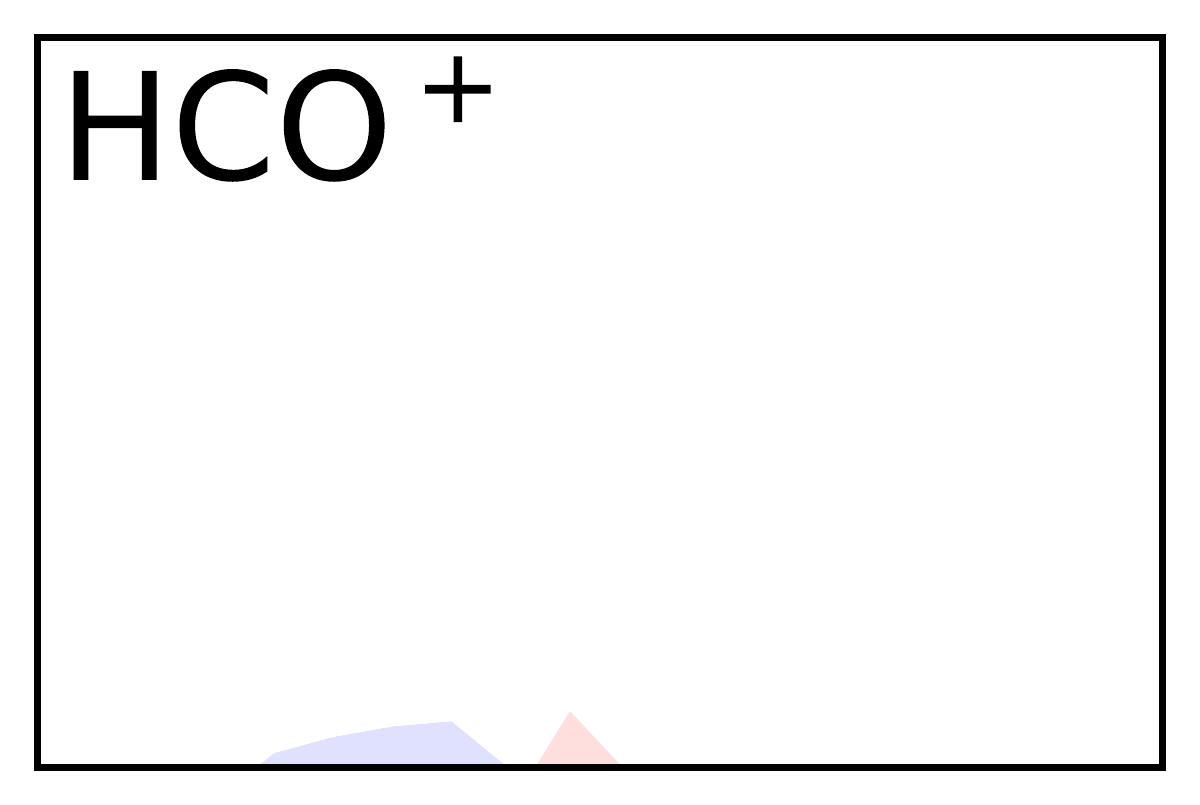} &
        \includegraphics[width=0.159\textwidth]{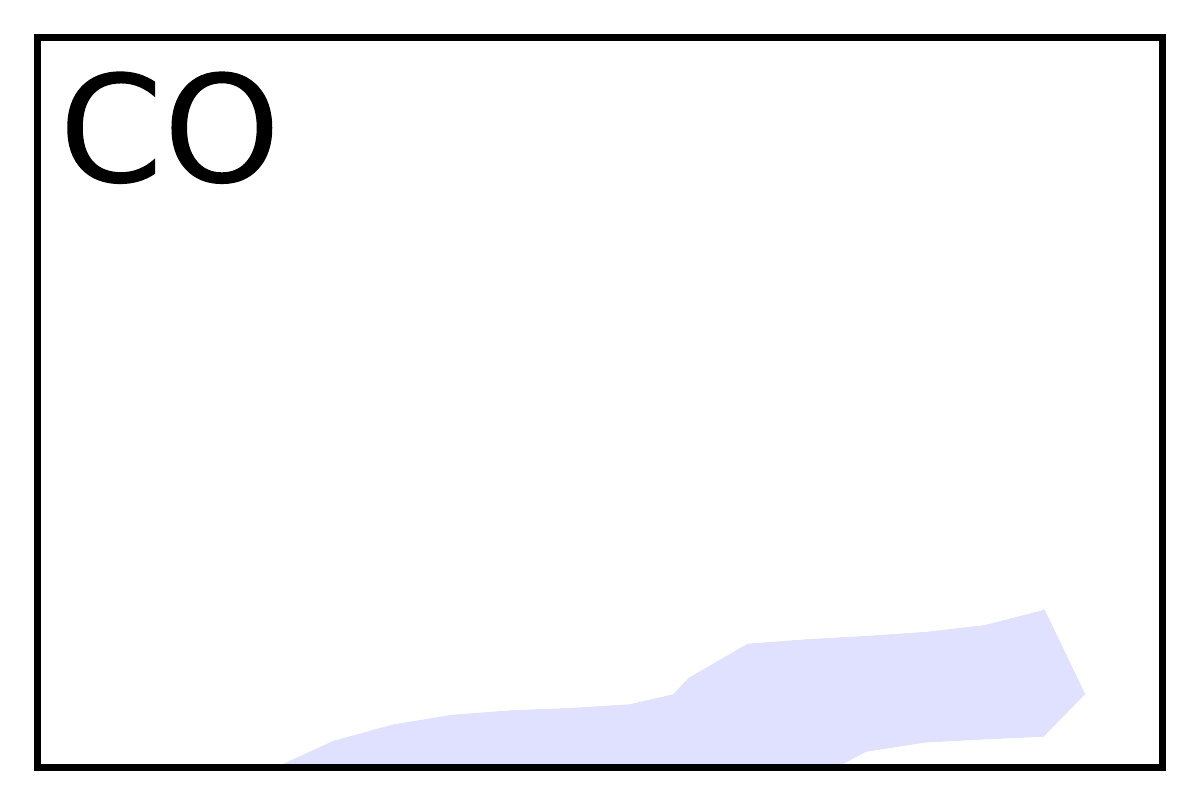} \\
        \includegraphics[width=0.159\textwidth]{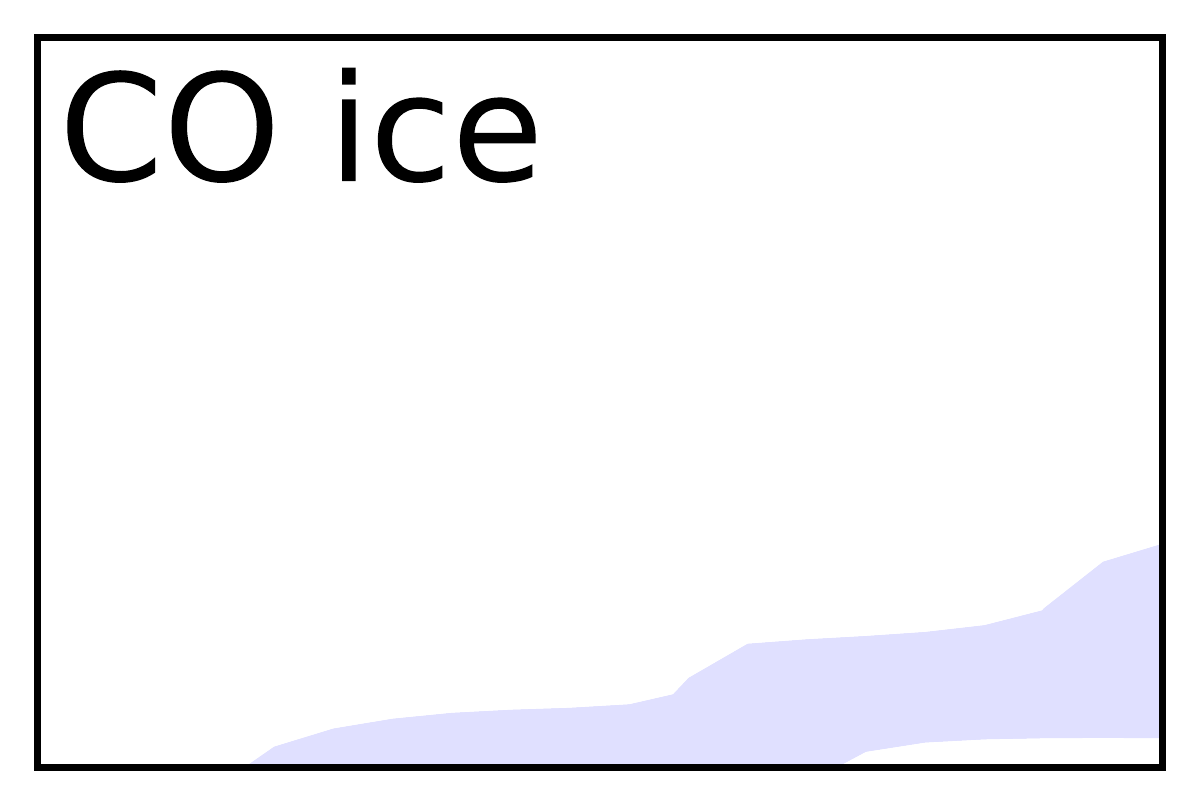} &
        \includegraphics[width=0.159\textwidth]{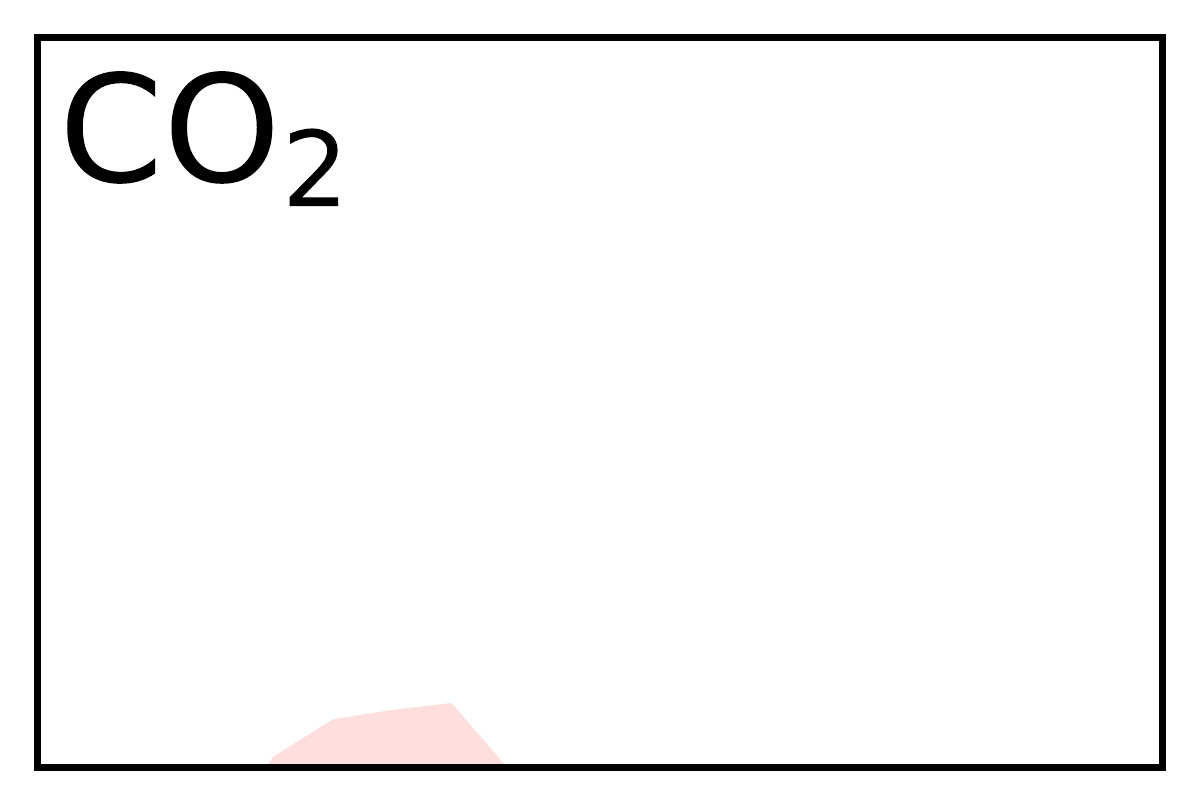} &
        \includegraphics[width=0.159\textwidth]{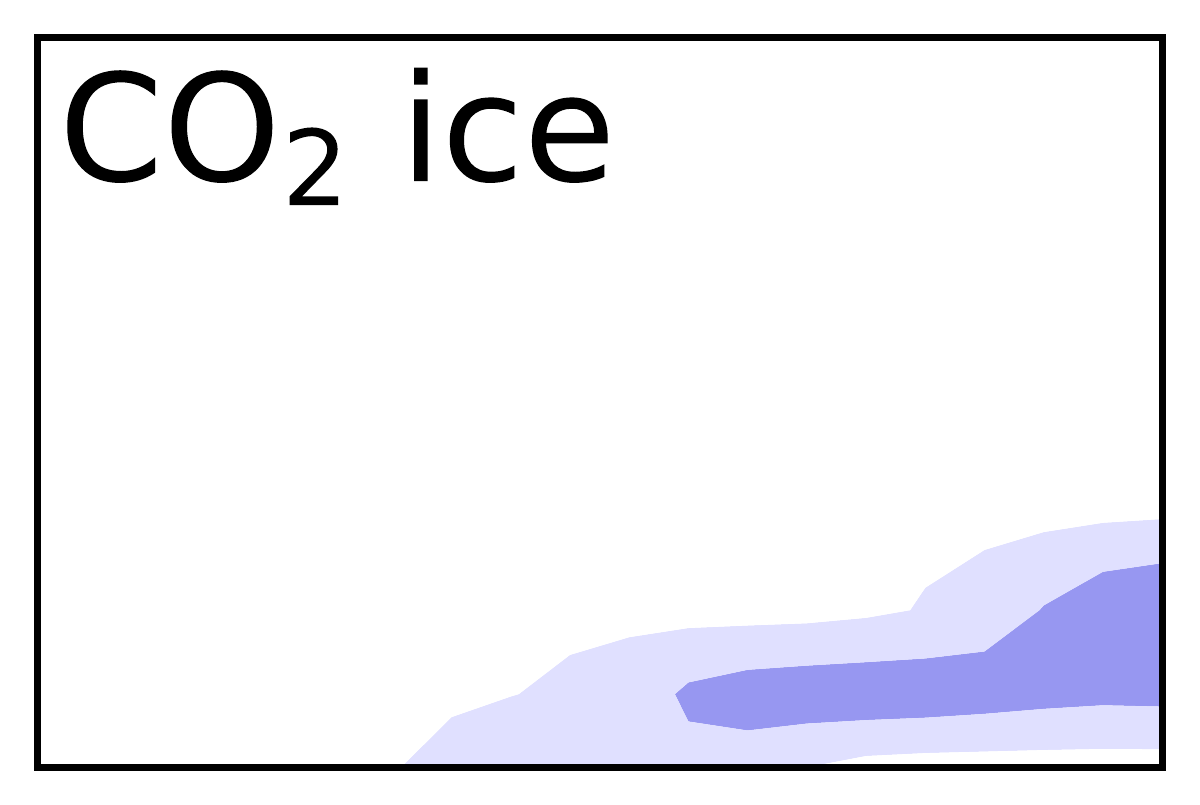} &
        \includegraphics[width=0.159\textwidth]{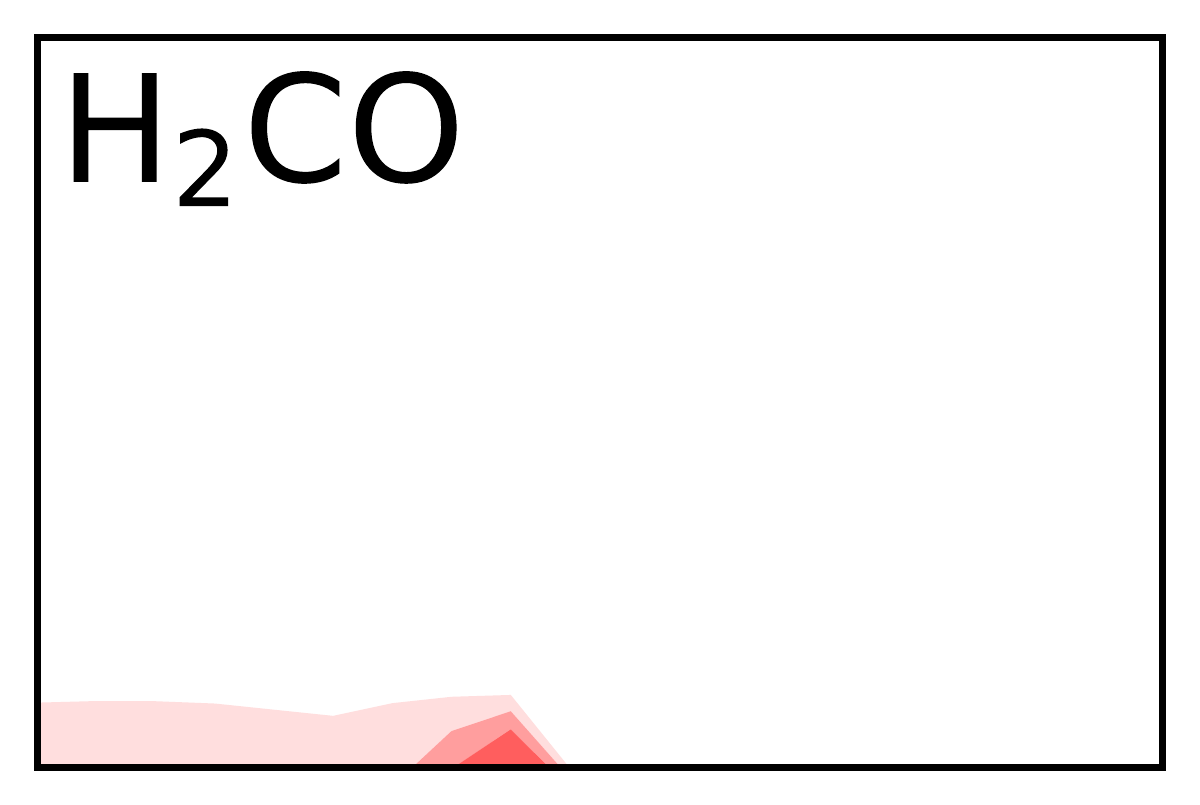} &
        \includegraphics[width=0.159\textwidth]{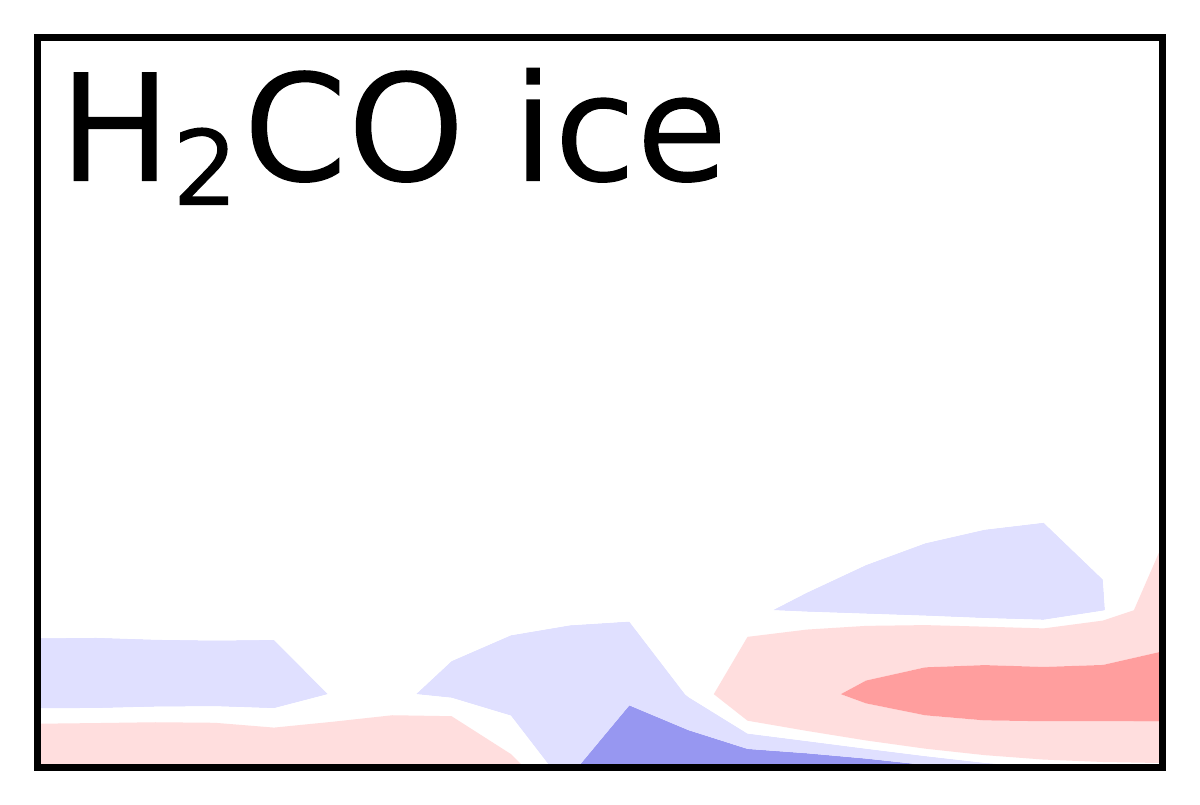} &
        \includegraphics[width=0.159\textwidth]{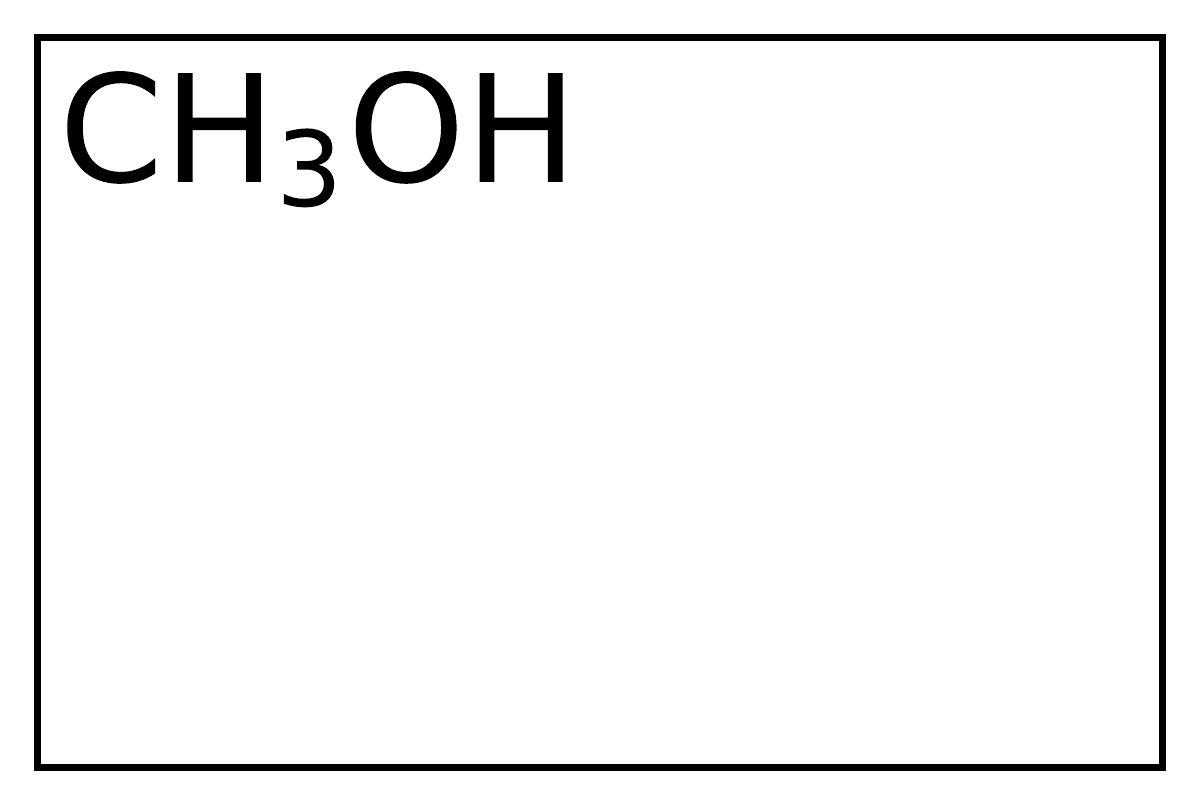} \\
        \includegraphics[width=0.159\textwidth]{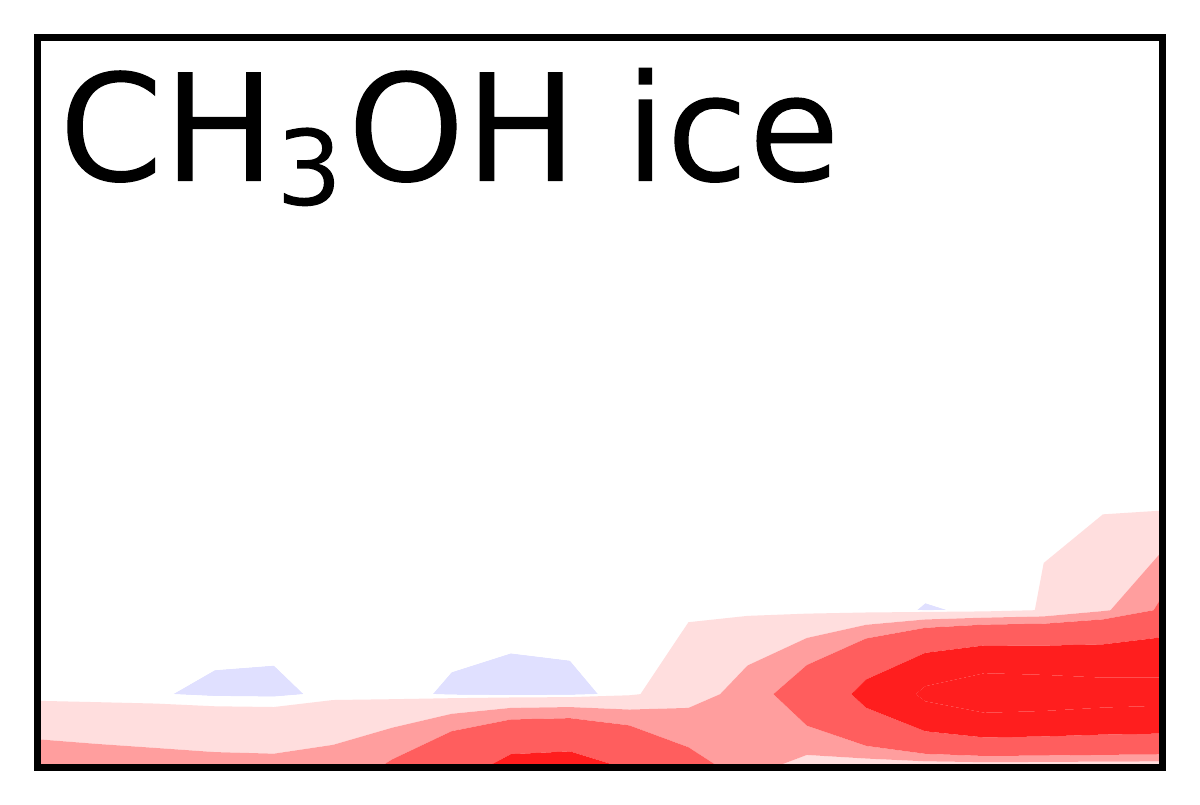} &
        \includegraphics[width=0.159\textwidth]{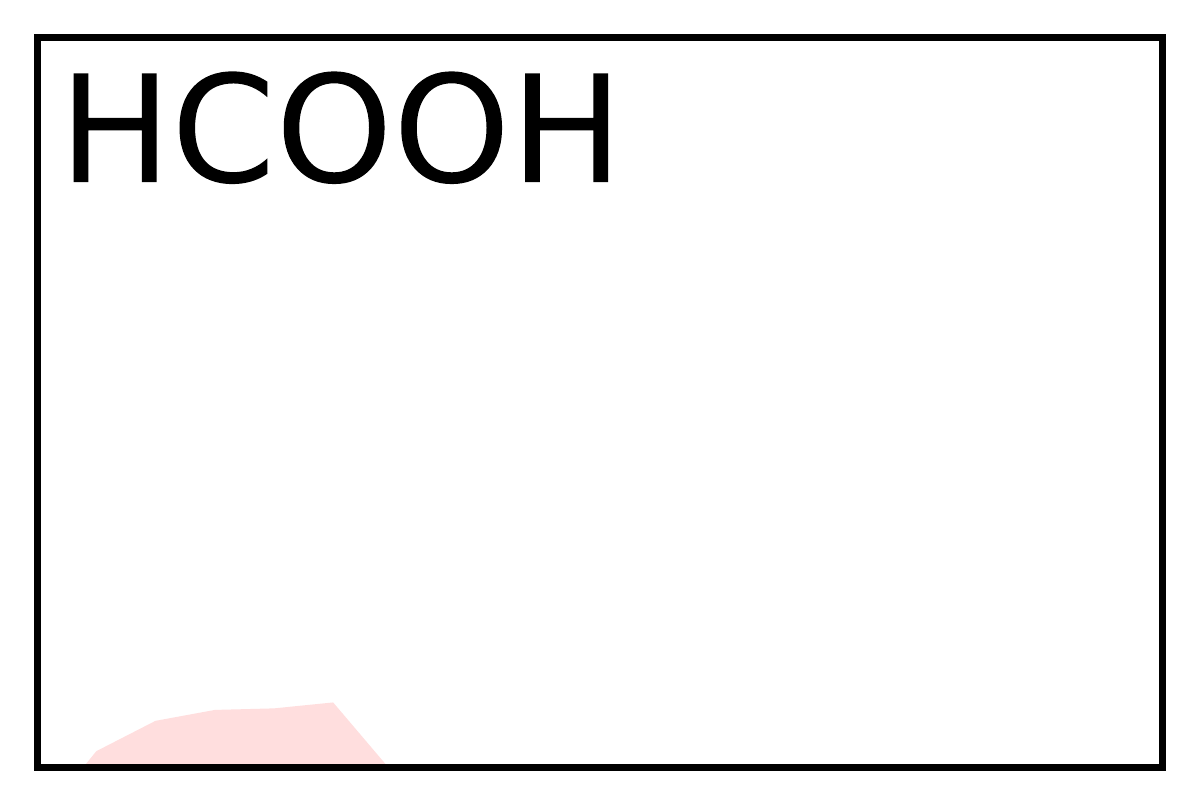} &
        \includegraphics[width=0.159\textwidth]{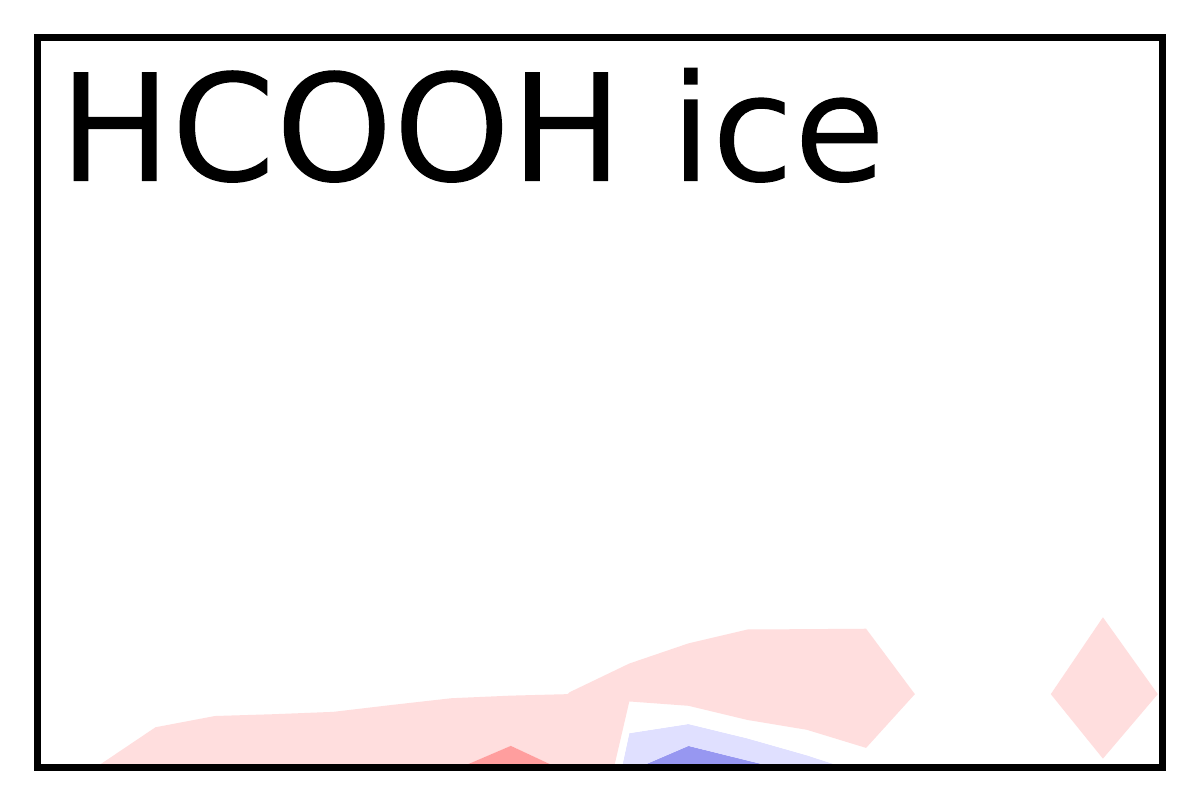} &
        \includegraphics[width=0.159\textwidth]{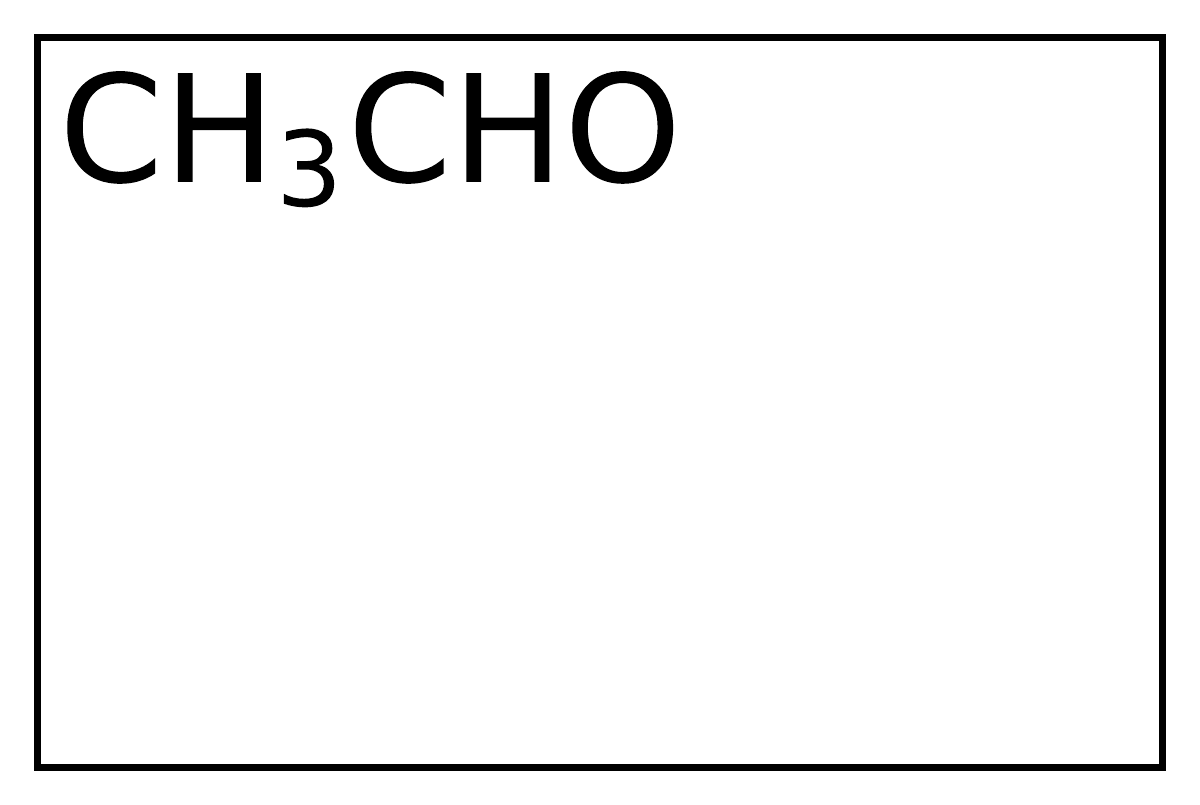} &
        \includegraphics[width=0.159\textwidth]{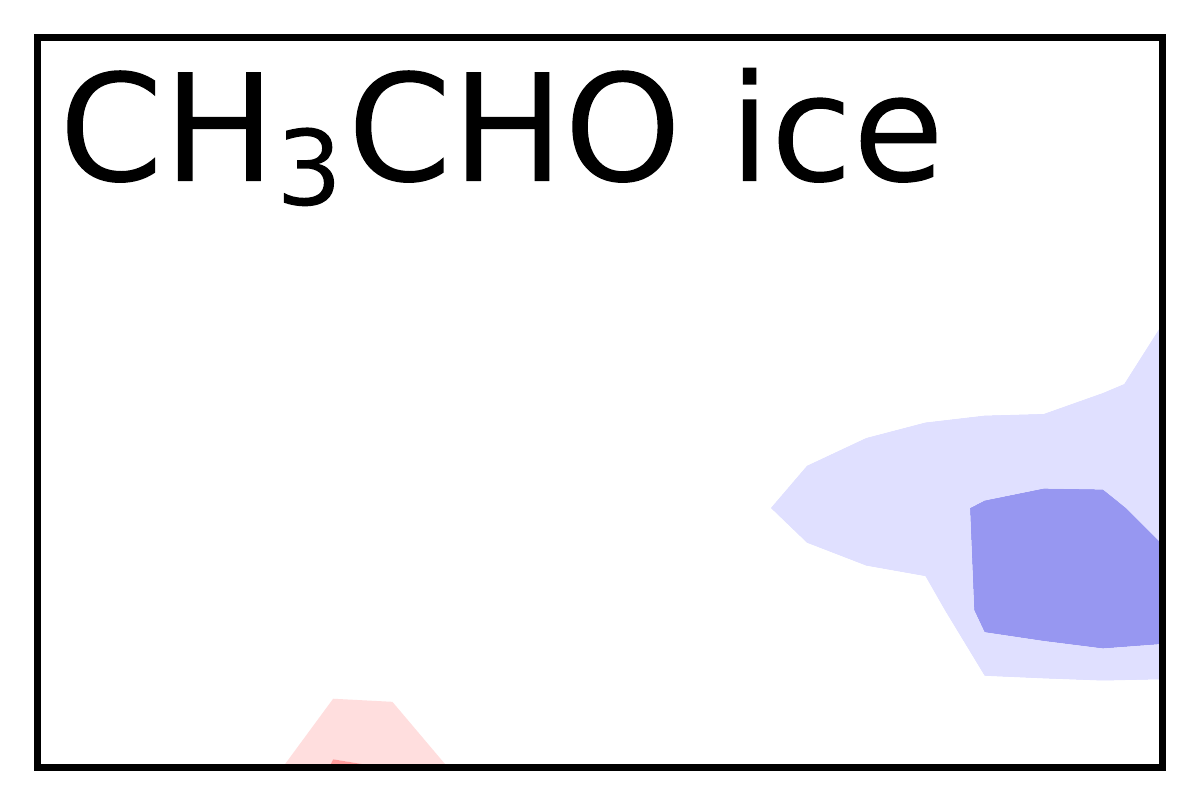} &
        \includegraphics[width=0.159\textwidth]{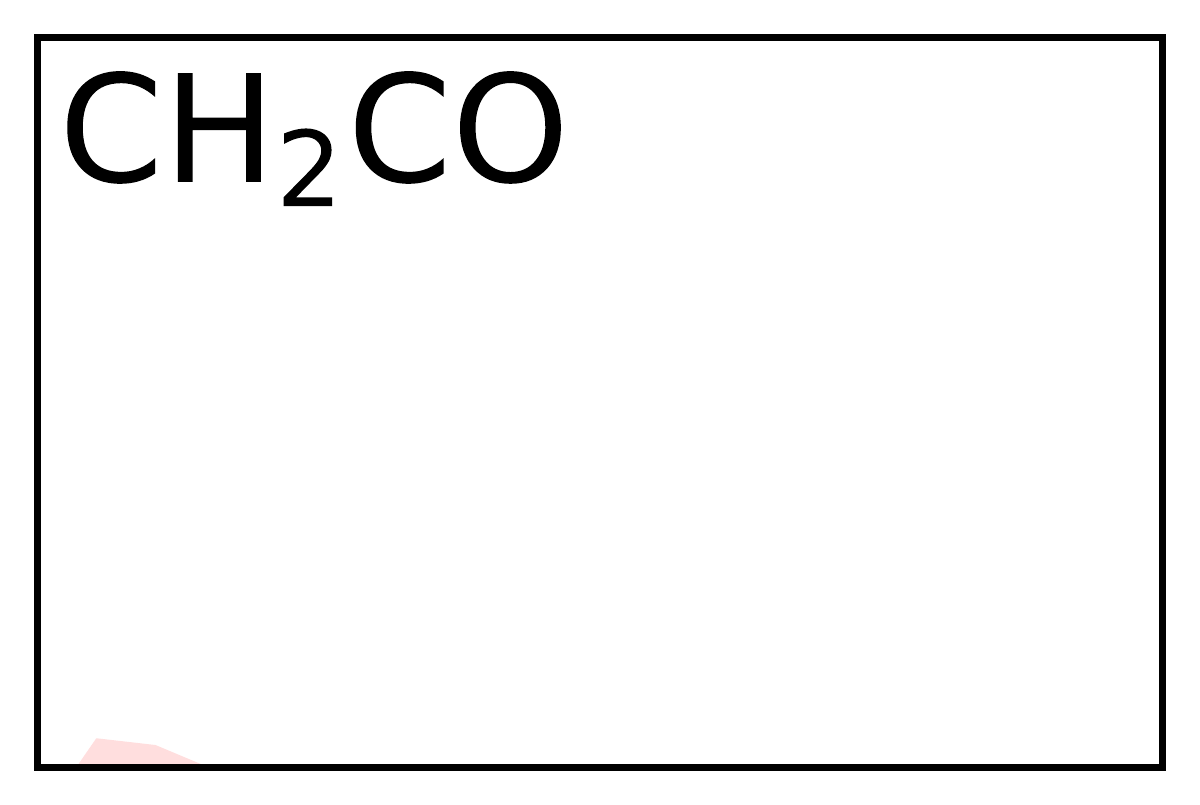} \\
        \includegraphics[width=0.159\textwidth]{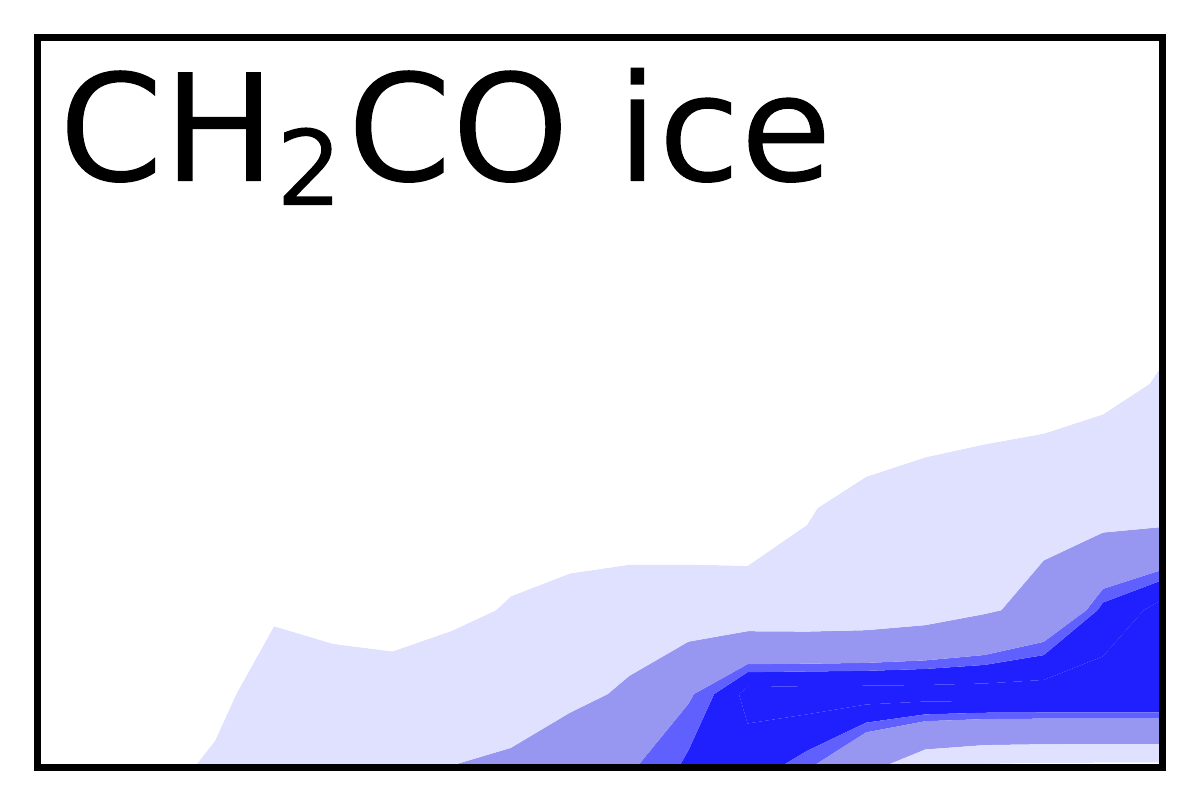} &
        \includegraphics[width=0.159\textwidth]{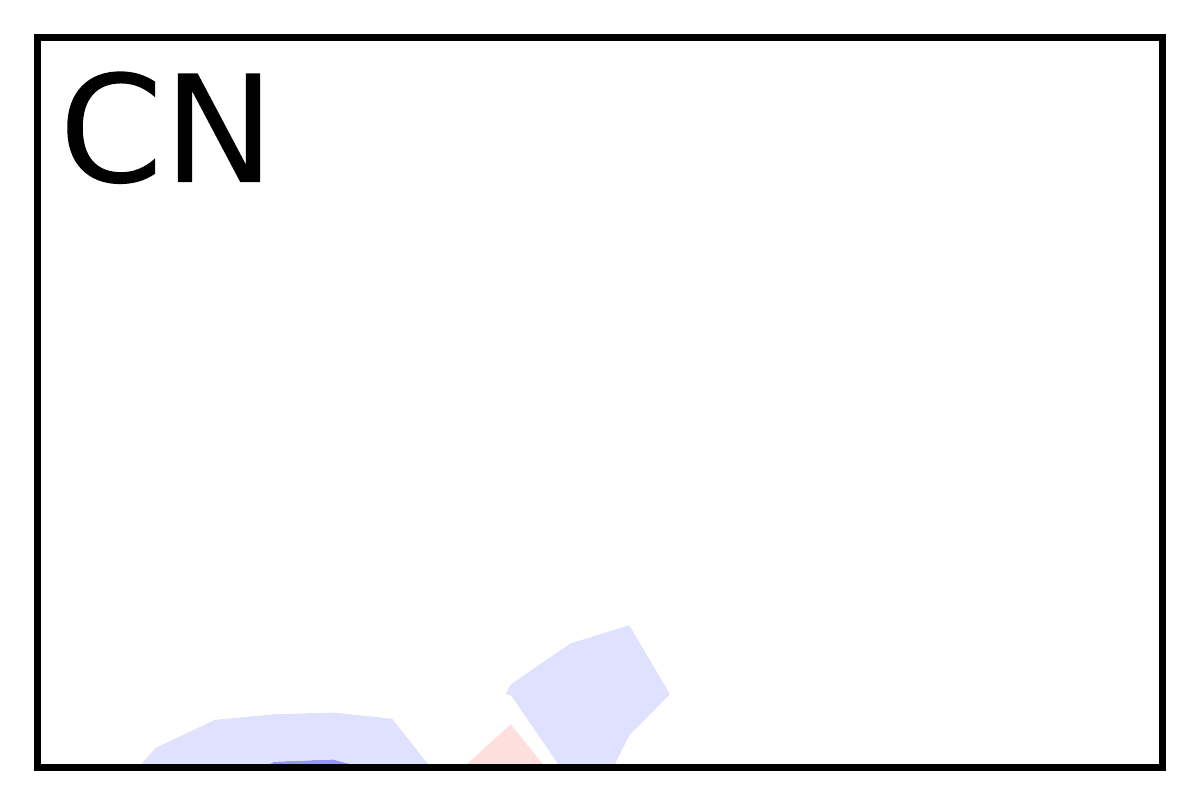} &
        \includegraphics[width=0.159\textwidth]{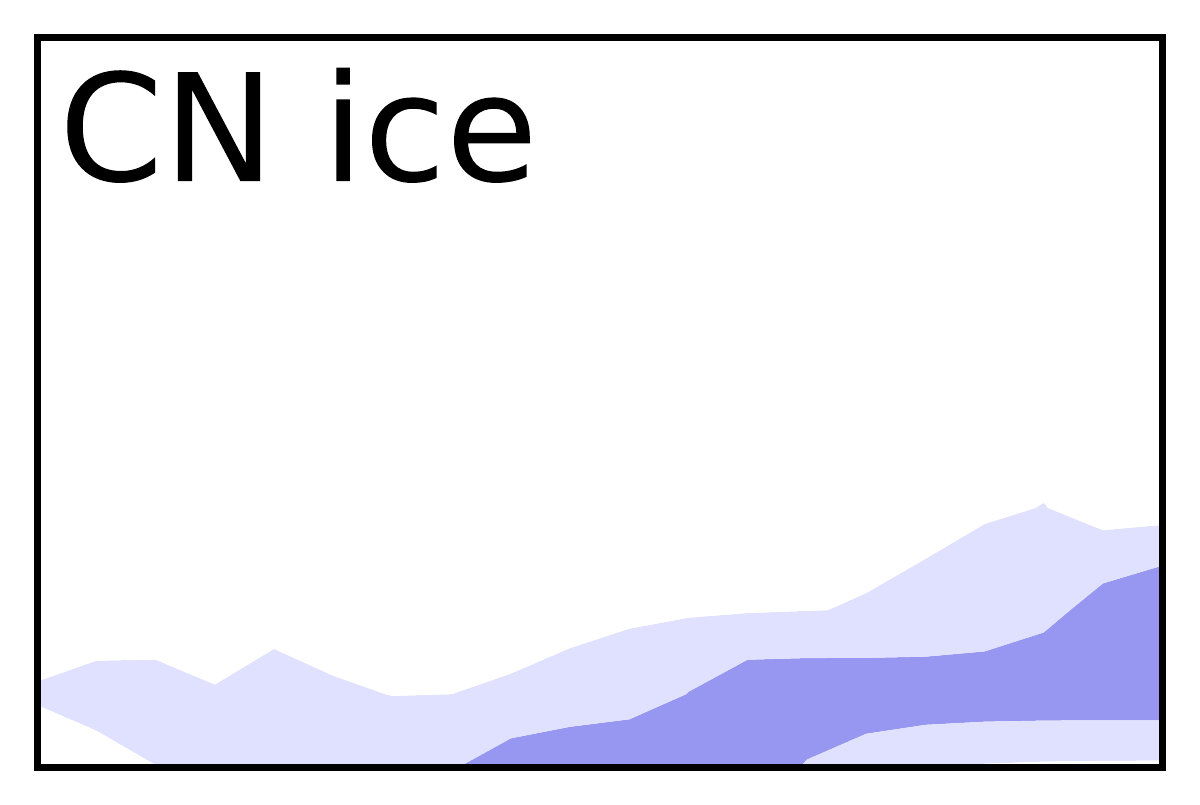} &
        \includegraphics[width=0.159\textwidth]{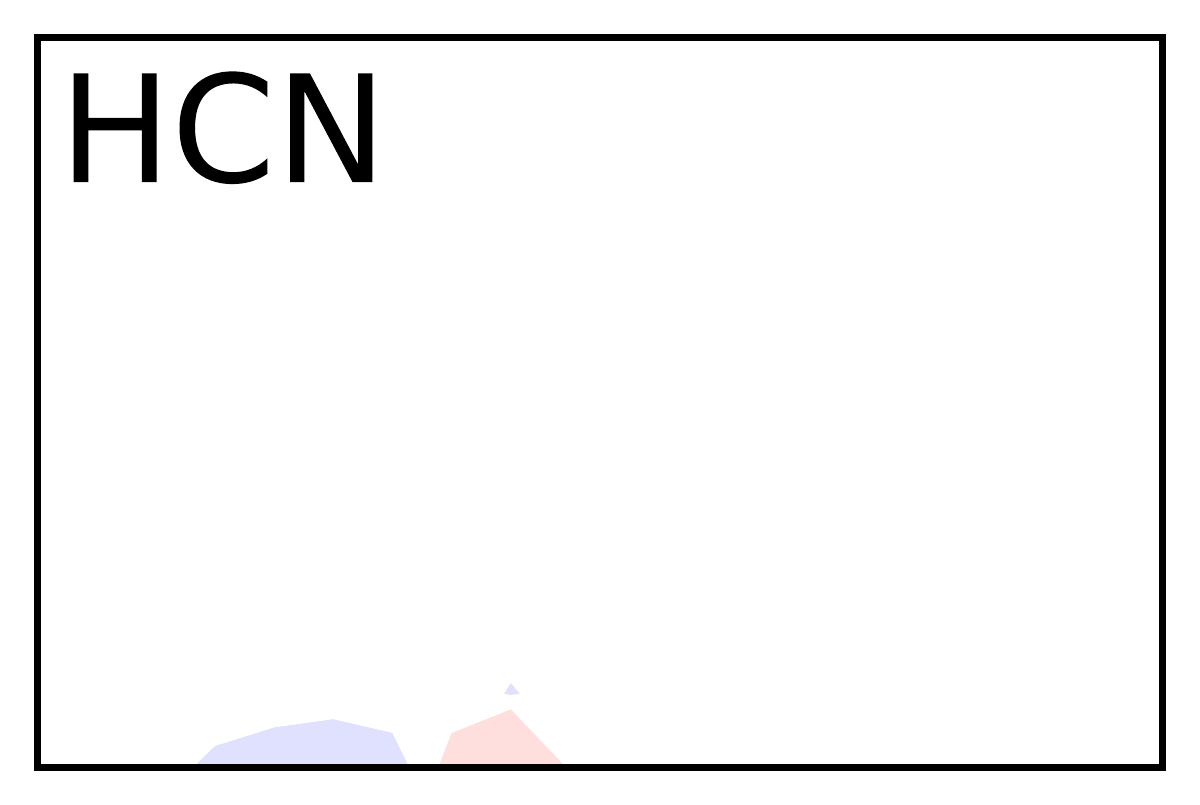} &
        \includegraphics[width=0.159\textwidth]{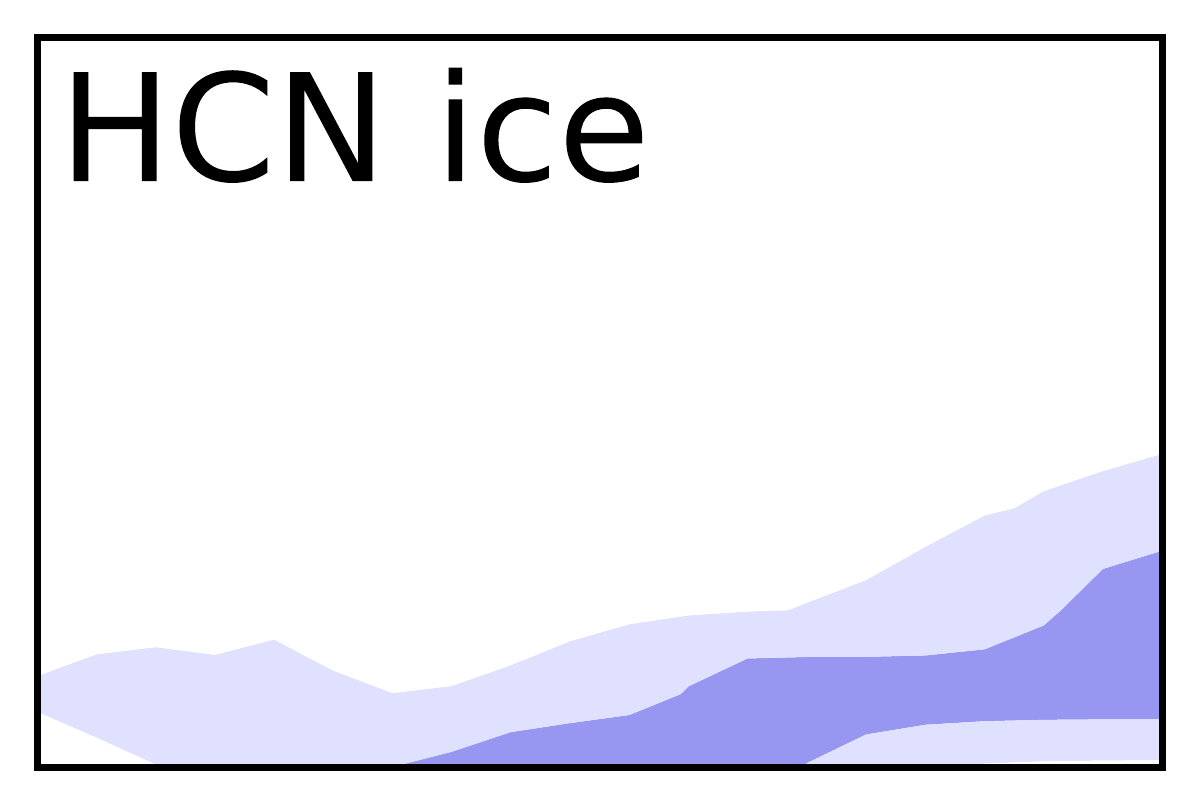} &
        \includegraphics[width=0.159\textwidth]{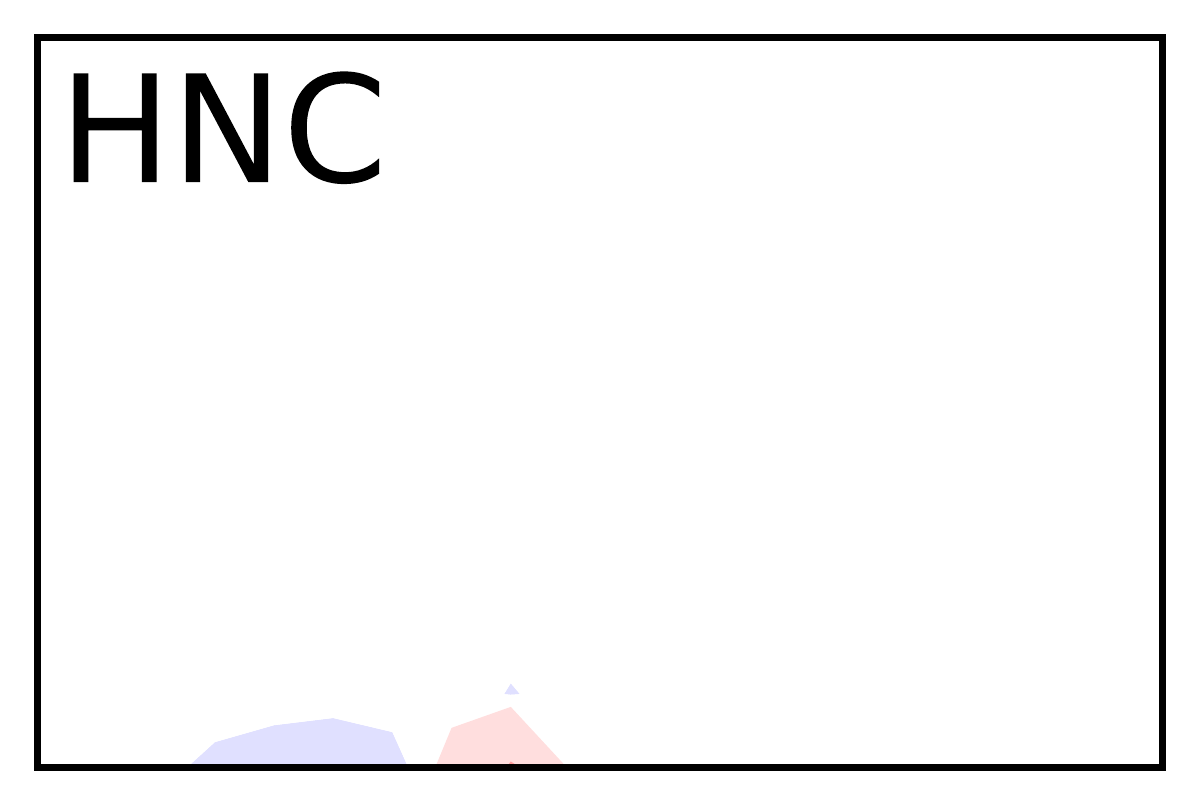} \\
        \includegraphics[width=0.159\textwidth]{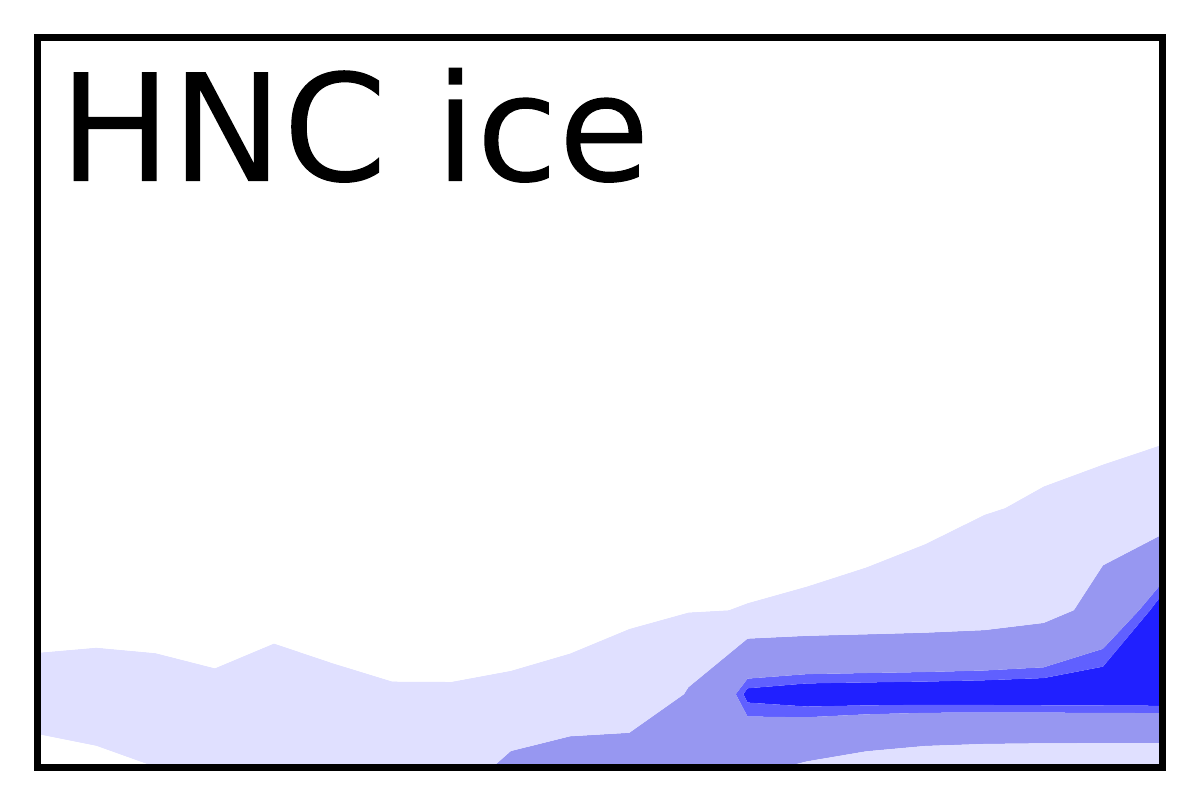} &
        \includegraphics[width=0.159\textwidth]{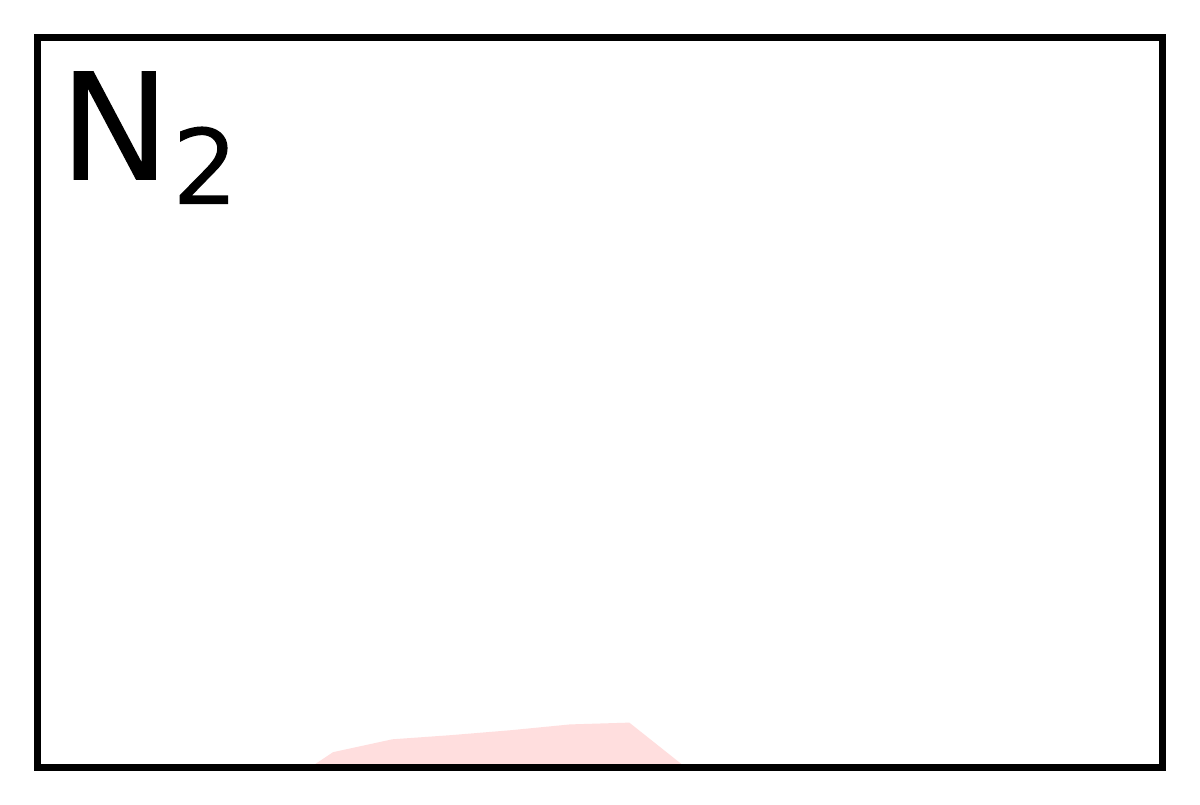} &
        \includegraphics[width=0.159\textwidth]{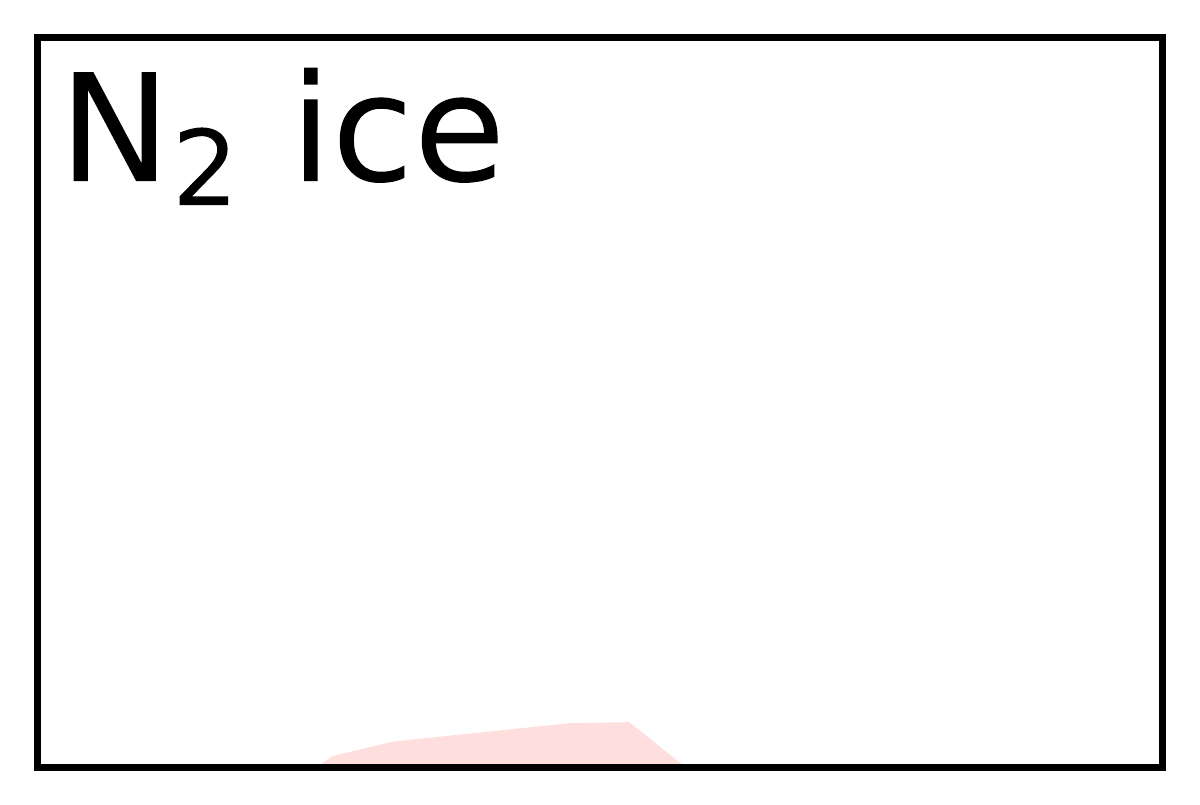} &
        \includegraphics[width=0.159\textwidth]{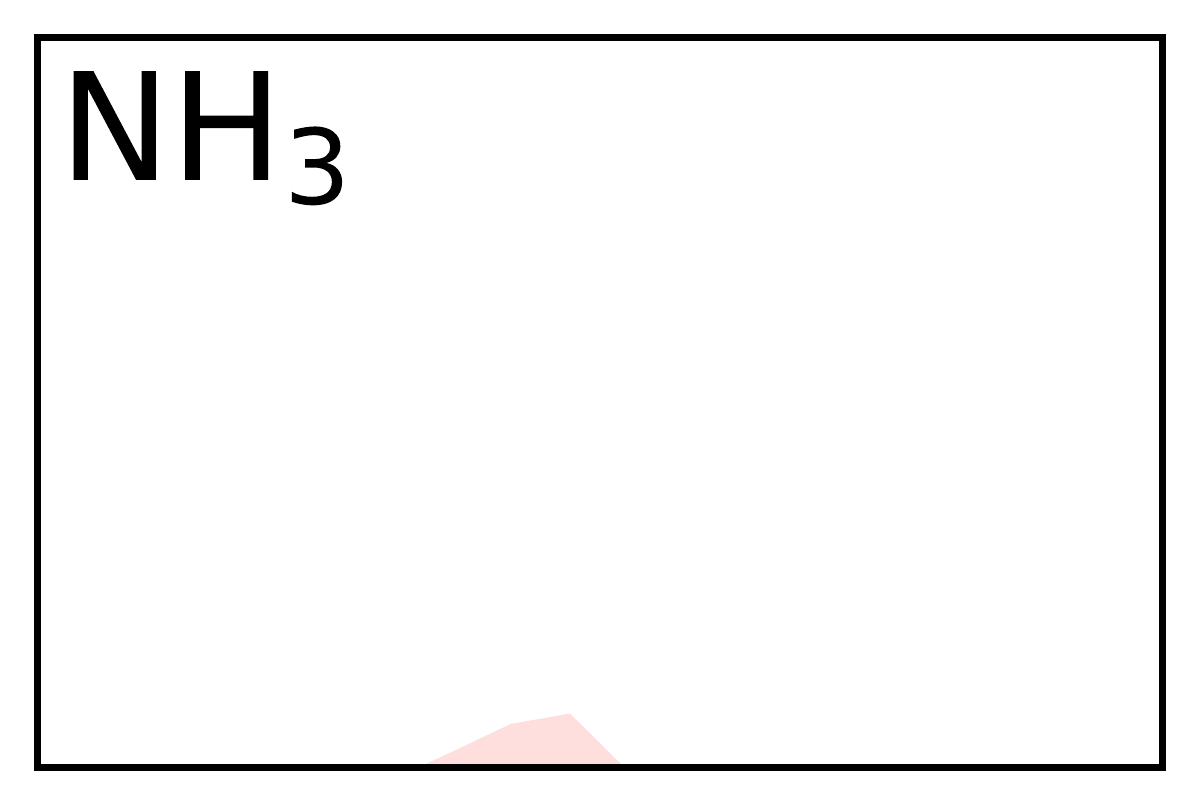} &
        \includegraphics[width=0.159\textwidth]{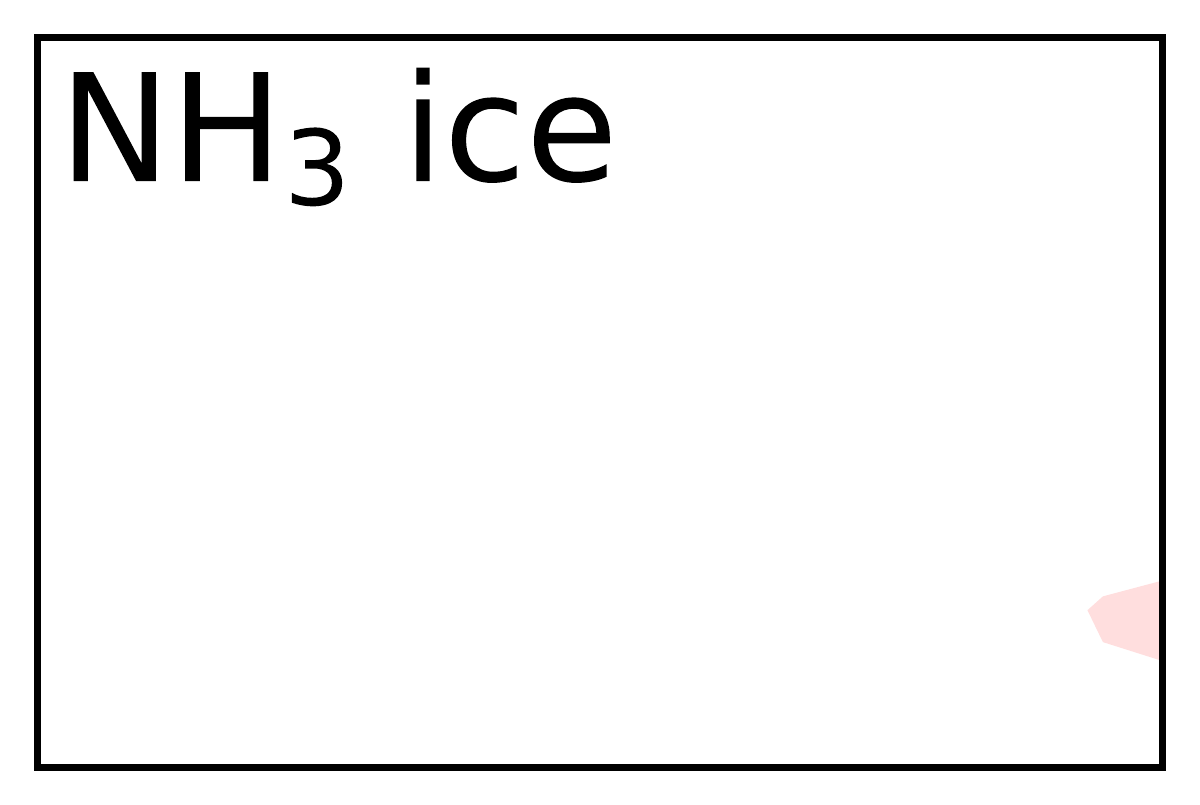} &
        \includegraphics[width=0.159\textwidth]{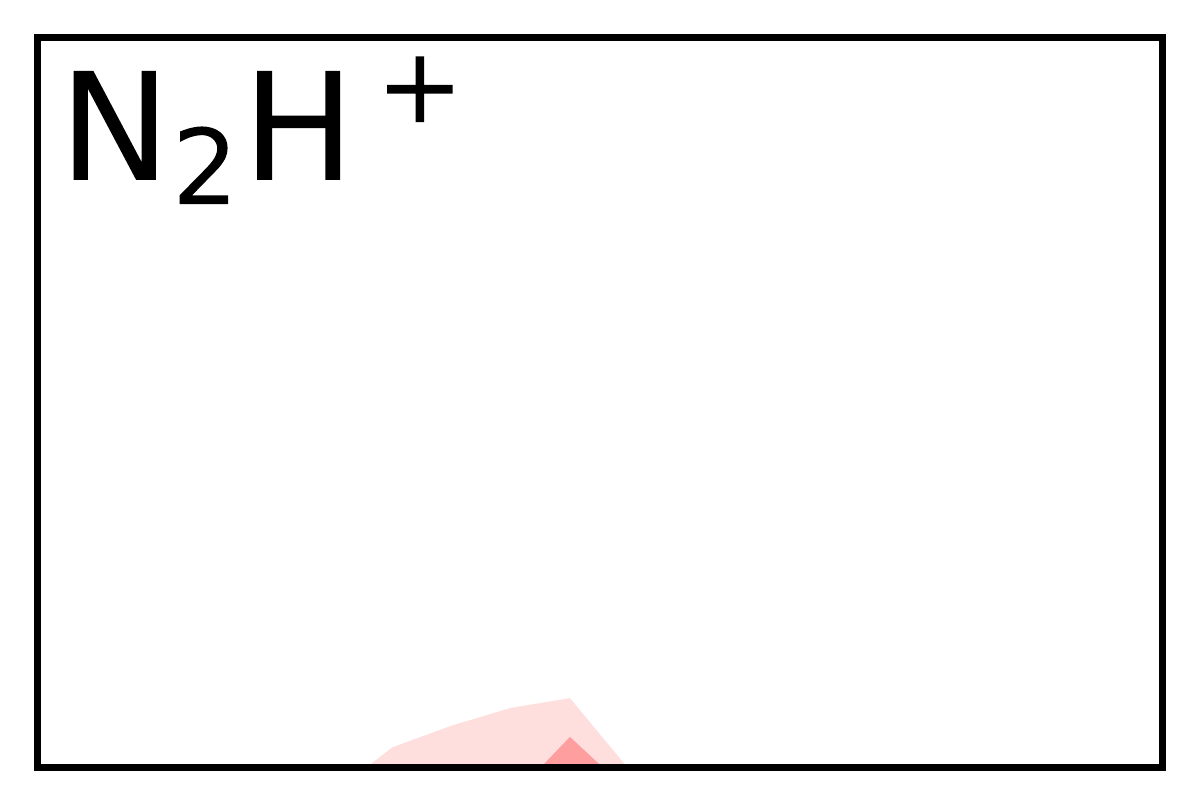} \\
        \includegraphics[width=0.159\textwidth]{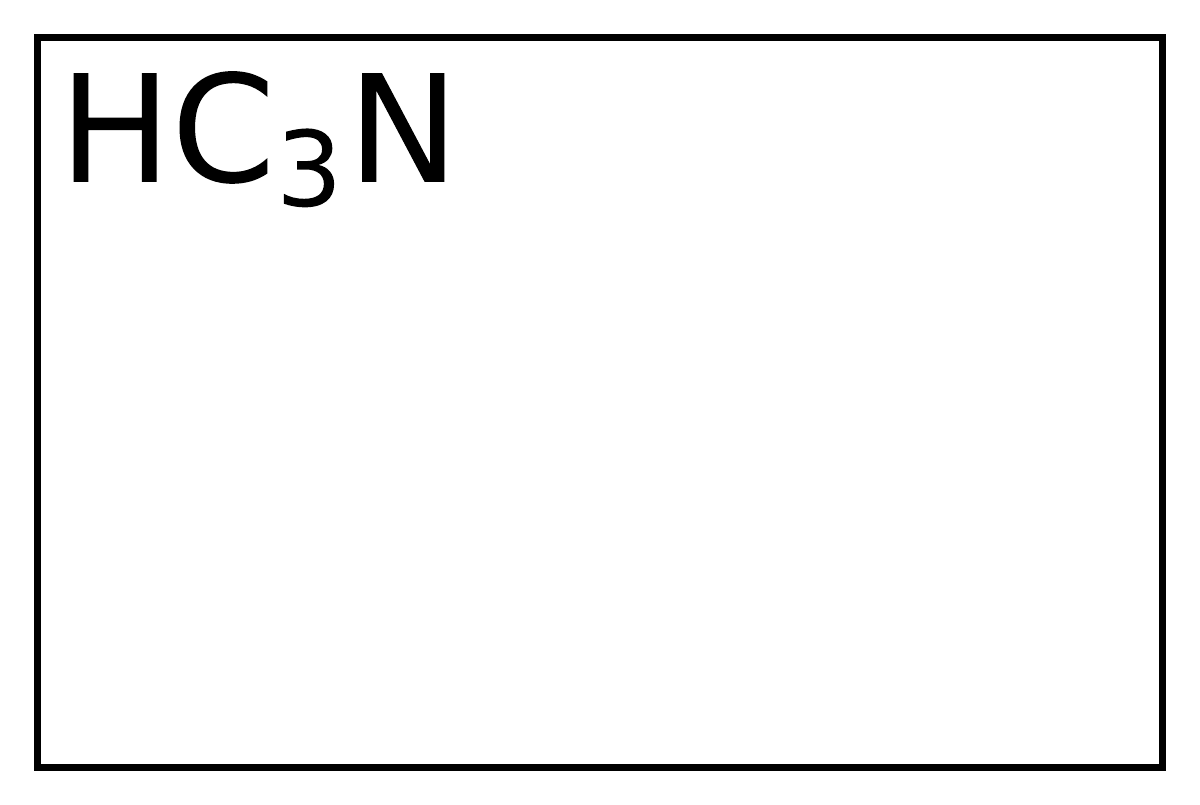} &
        \includegraphics[width=0.159\textwidth]{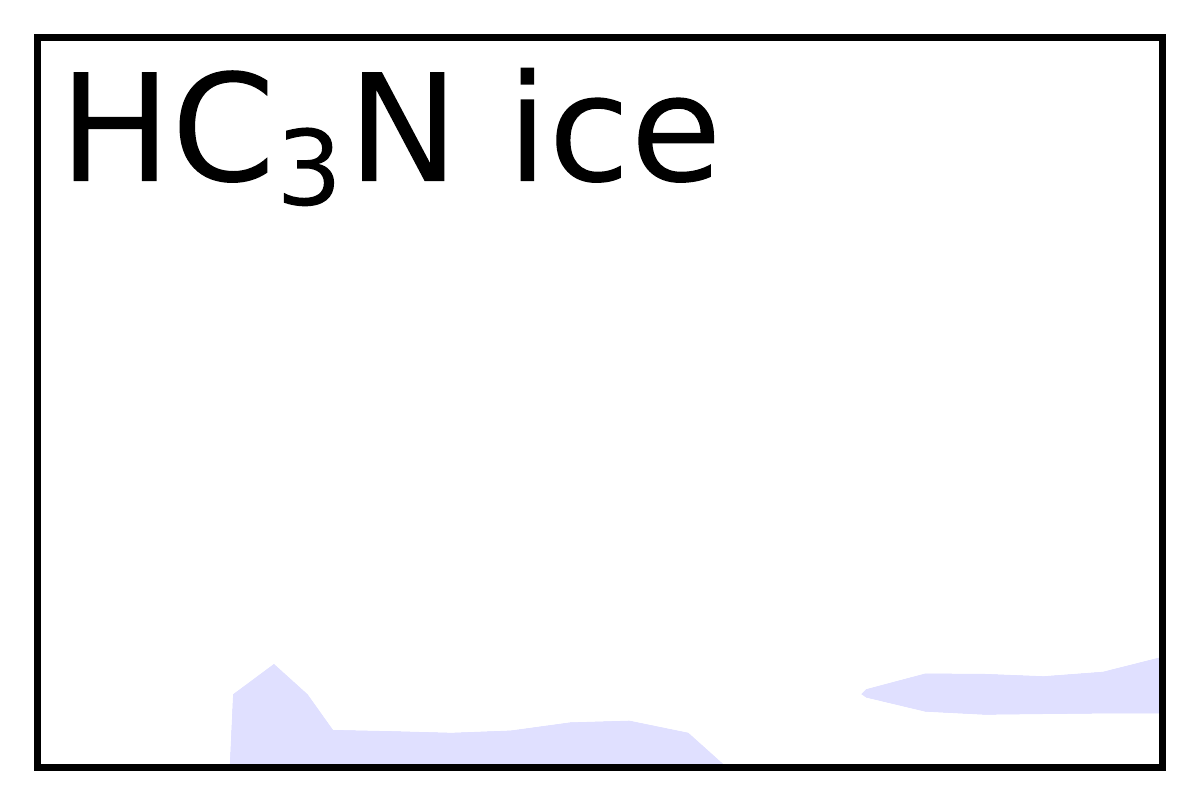} &
        \includegraphics[width=0.159\textwidth]{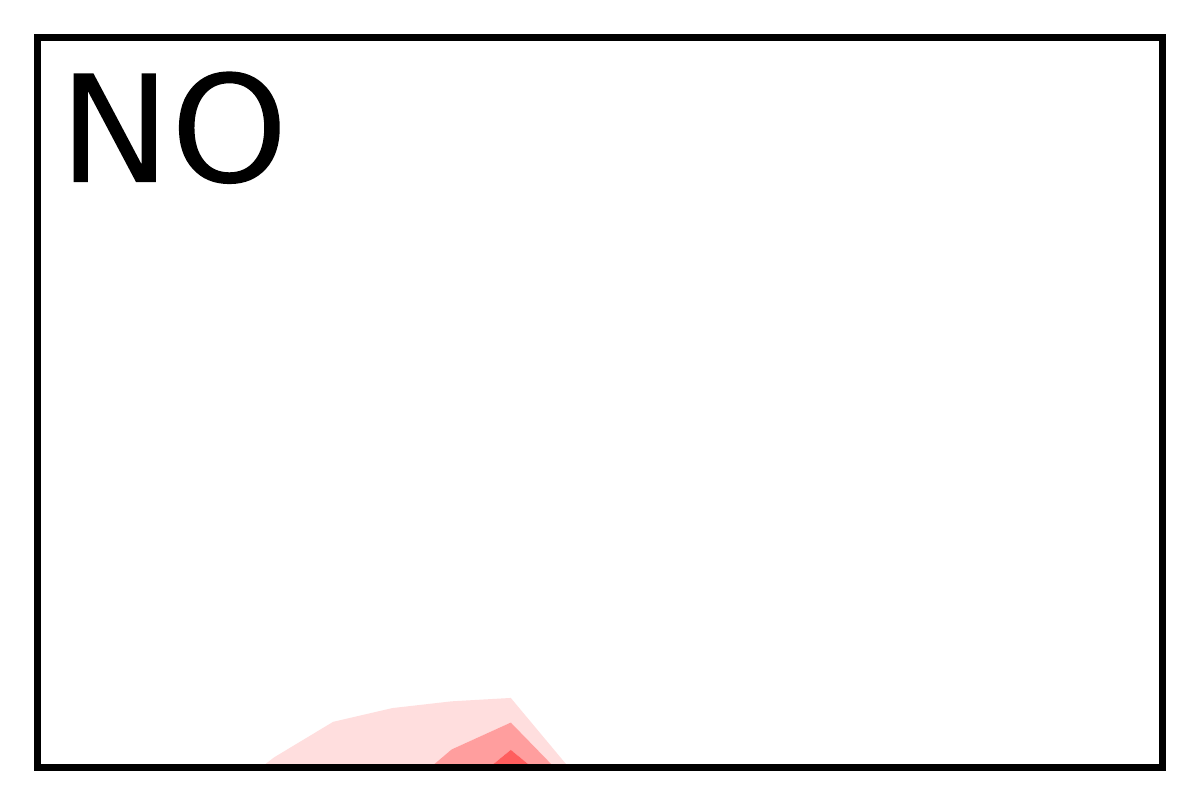} &
        \includegraphics[width=0.159\textwidth]{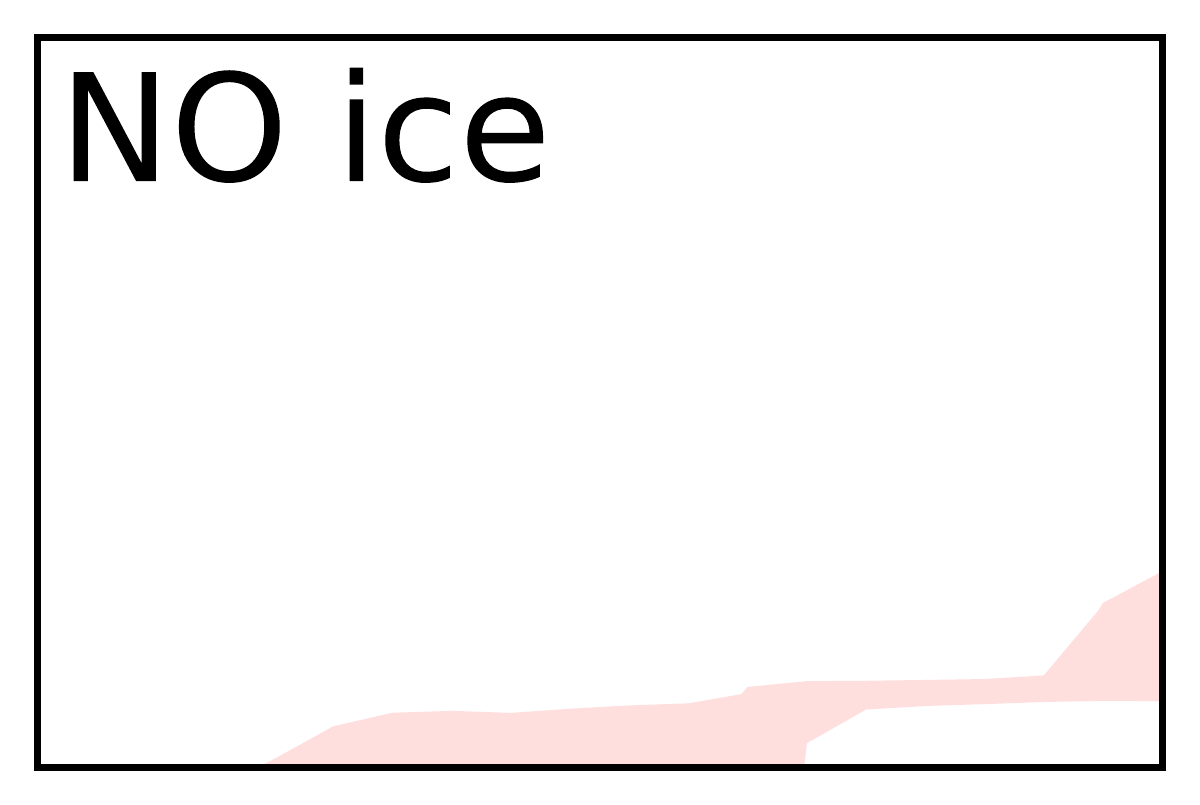} &
        \includegraphics[width=0.159\textwidth]{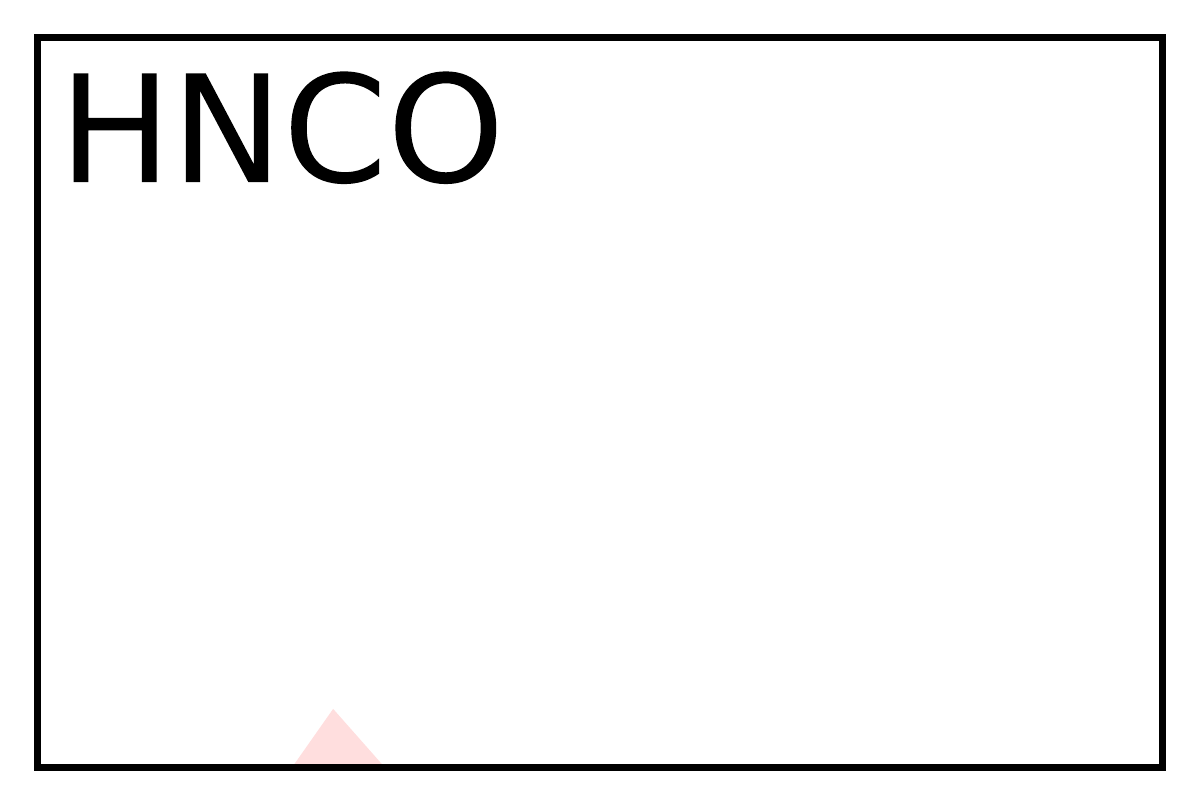} &
        \includegraphics[width=0.159\textwidth]{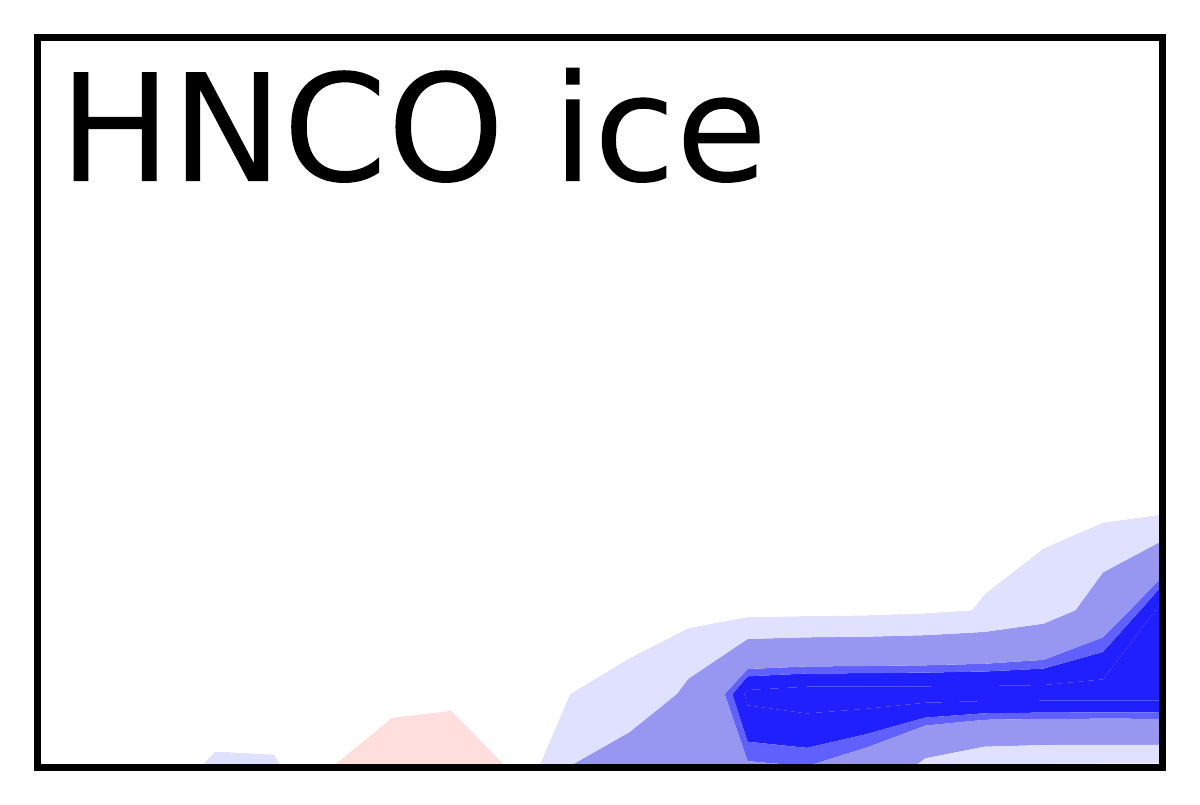} \\
        \includegraphics[width=0.159\textwidth]{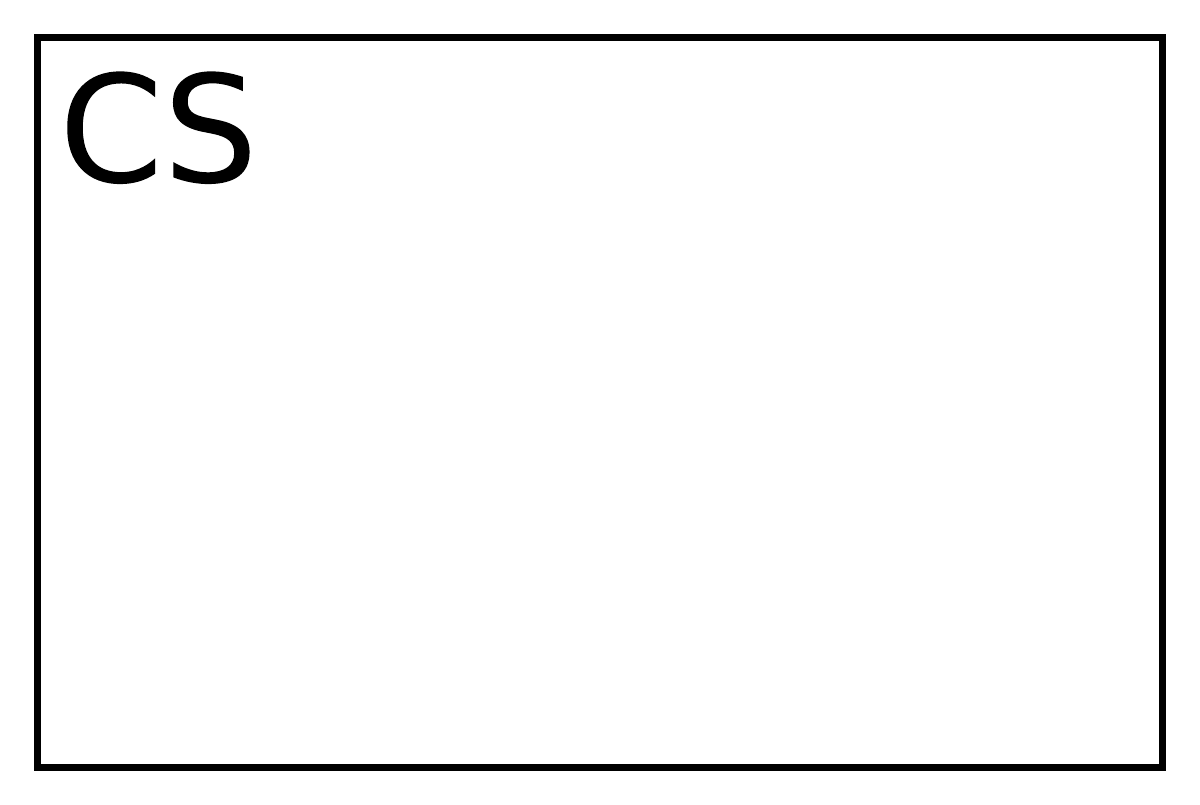} &
        \includegraphics[width=0.159\textwidth]{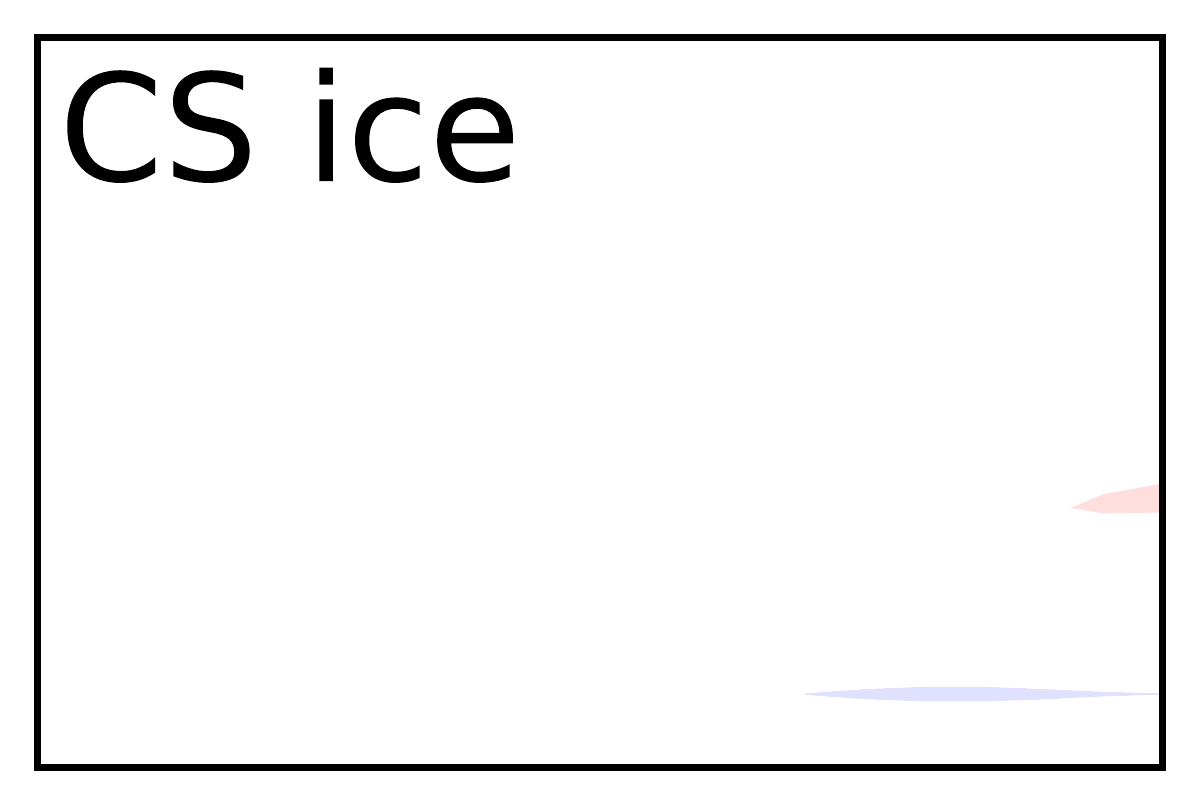} &
        \includegraphics[width=0.159\textwidth]{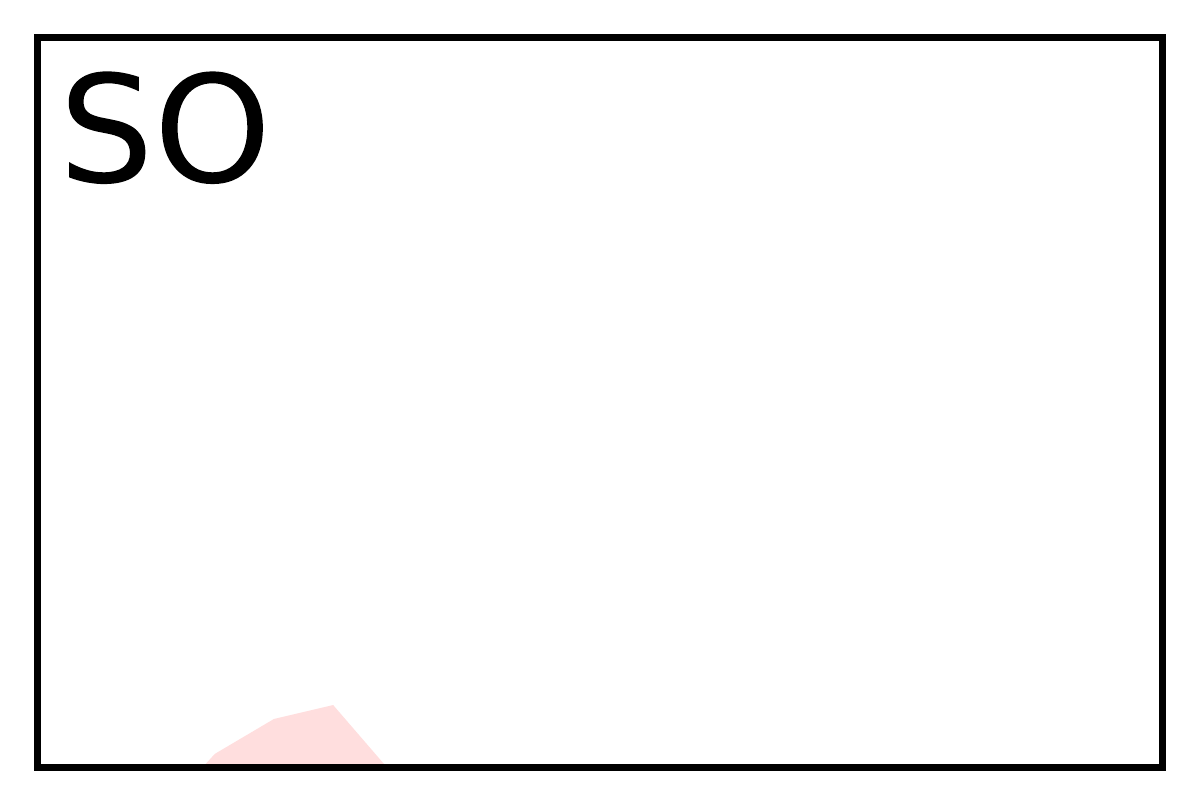} &
        \includegraphics[width=0.159\textwidth]{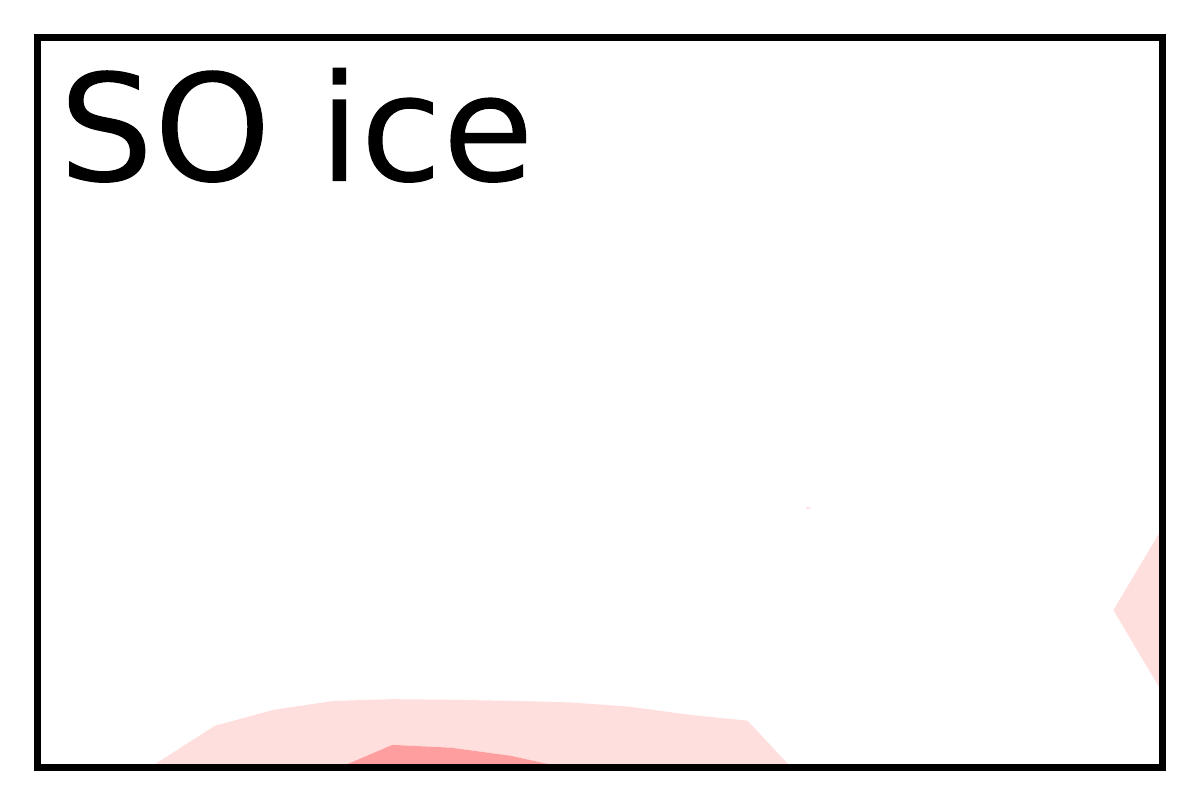} &
        \includegraphics[width=0.159\textwidth]{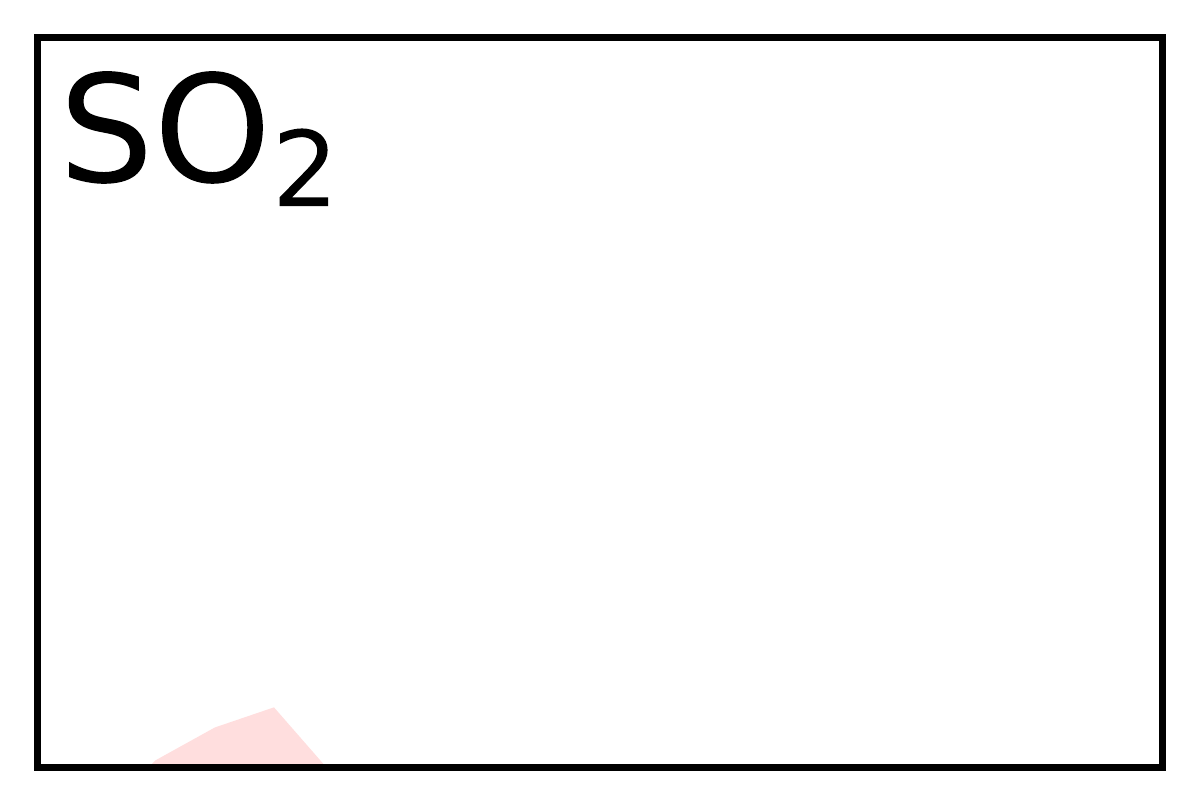} &
        \includegraphics[width=0.159\textwidth]{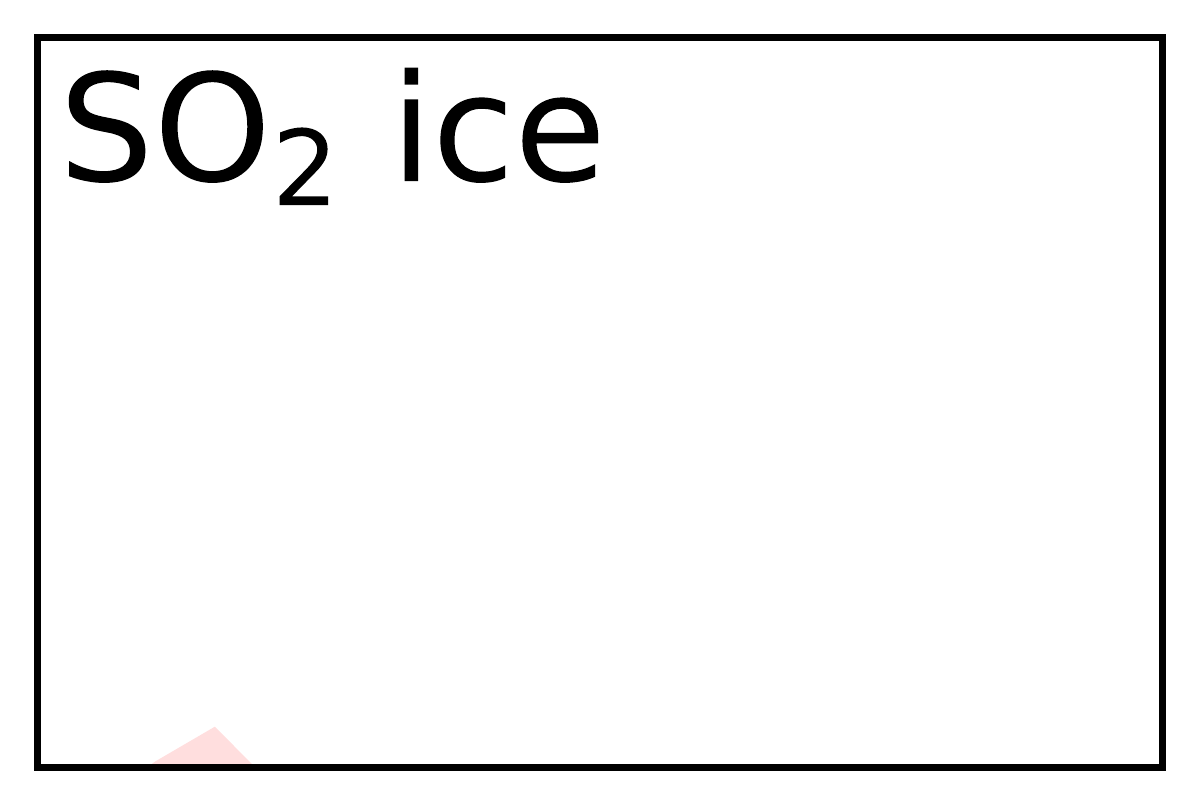} \\
        \includegraphics[width=0.189\textwidth]{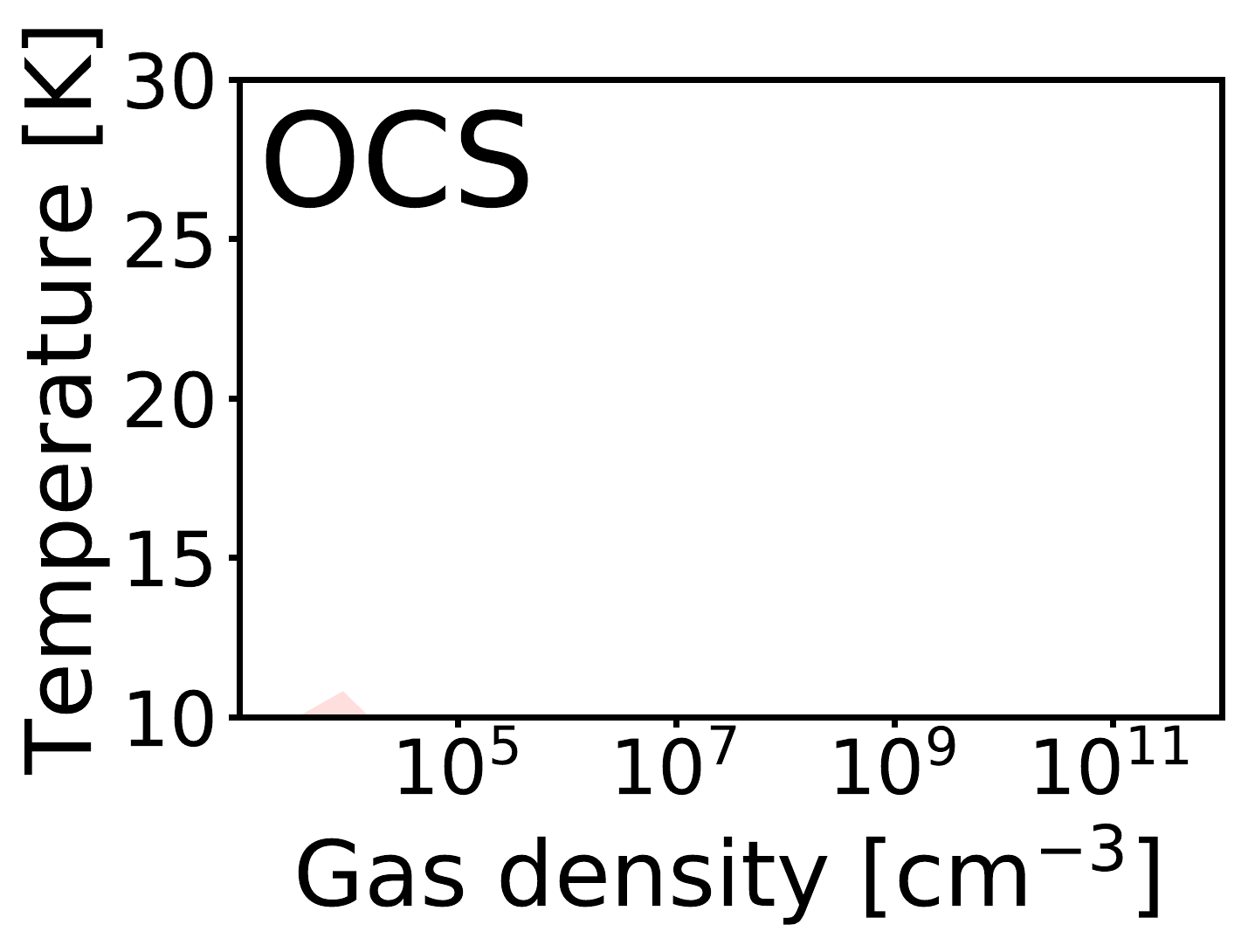} &
        \includegraphics[width=0.159\textwidth]{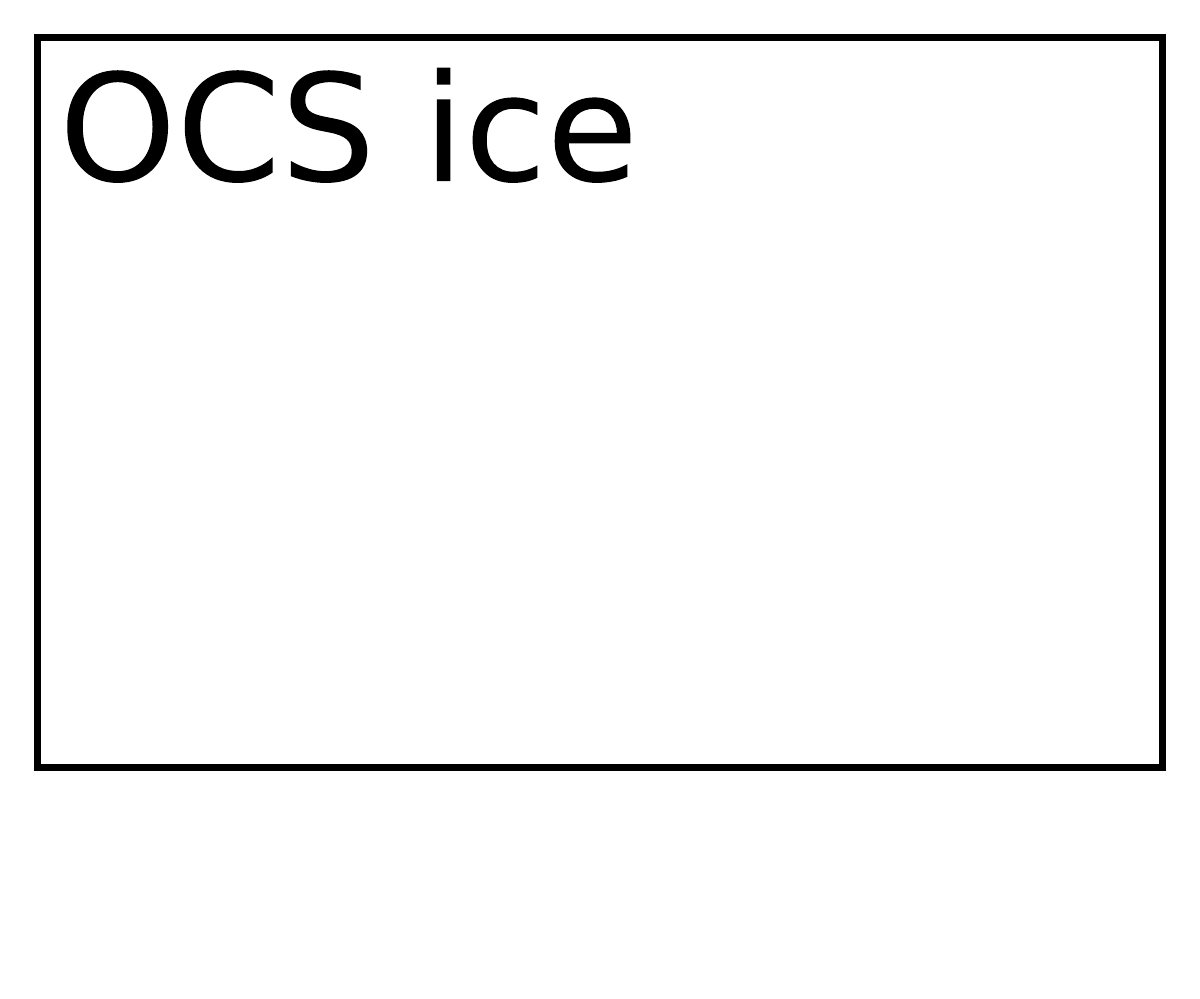} &
        \includegraphics[width=0.159\textwidth]{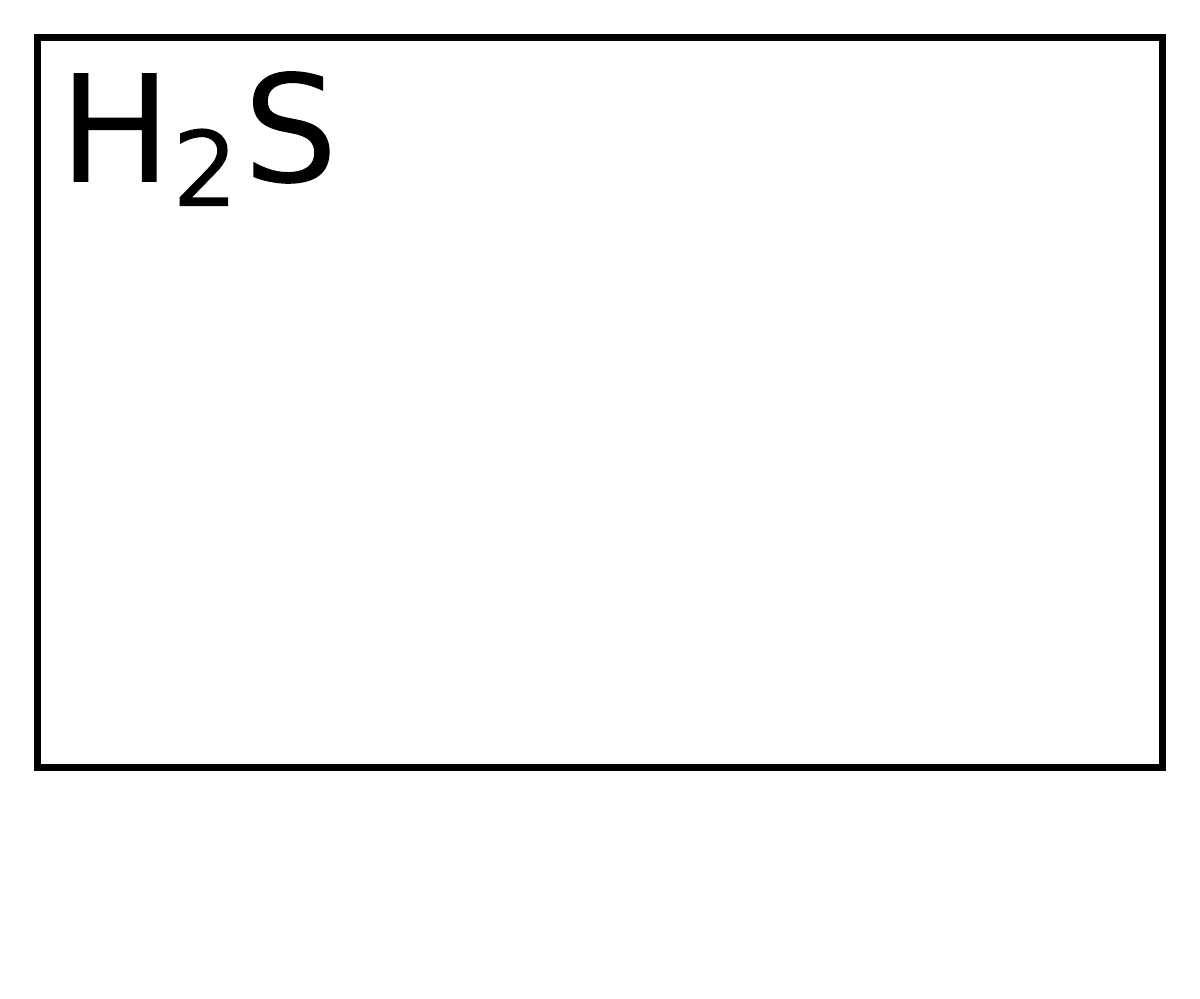} &
        \includegraphics[width=0.159\textwidth]{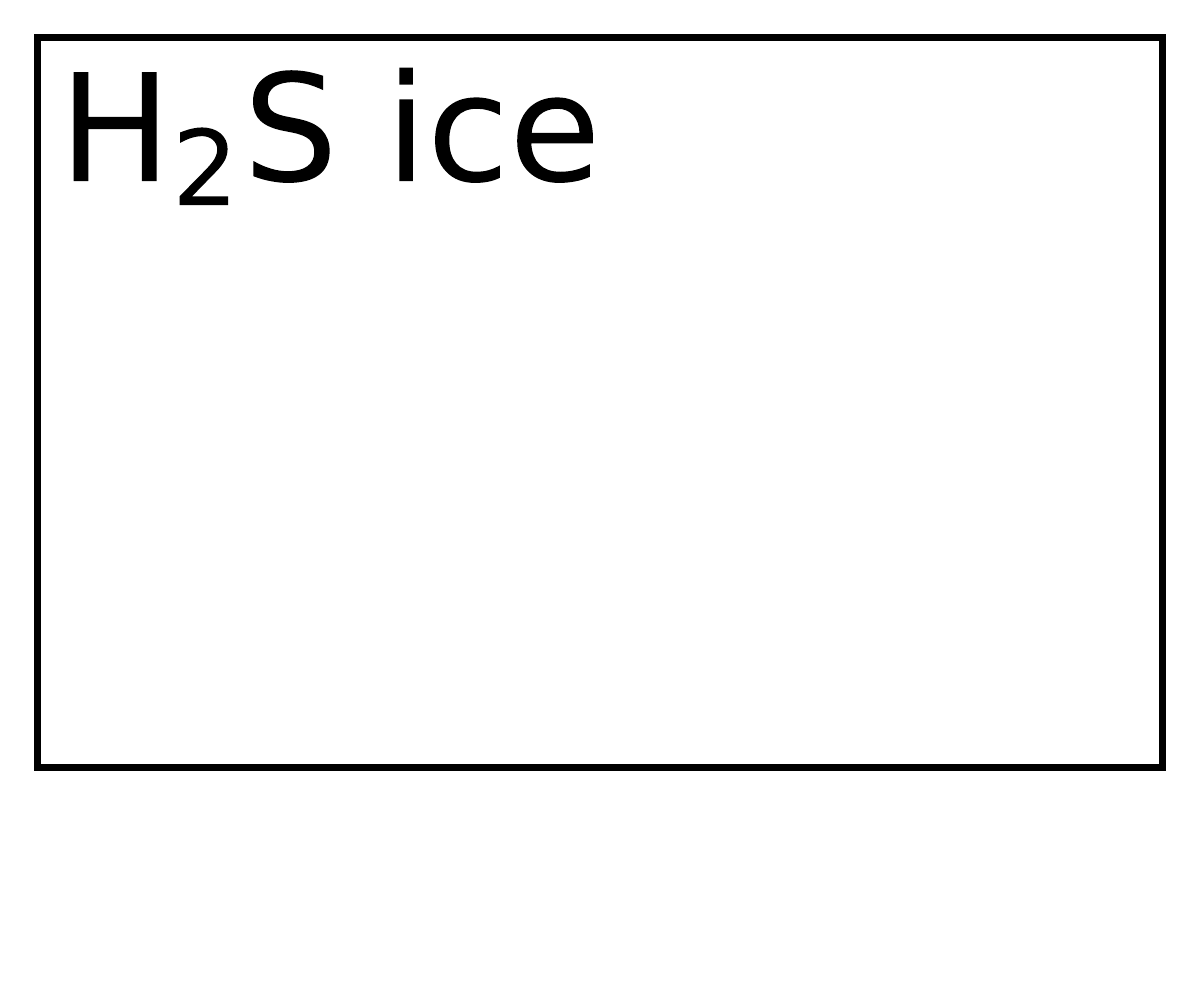} &
        \includegraphics[width=0.159\textwidth]{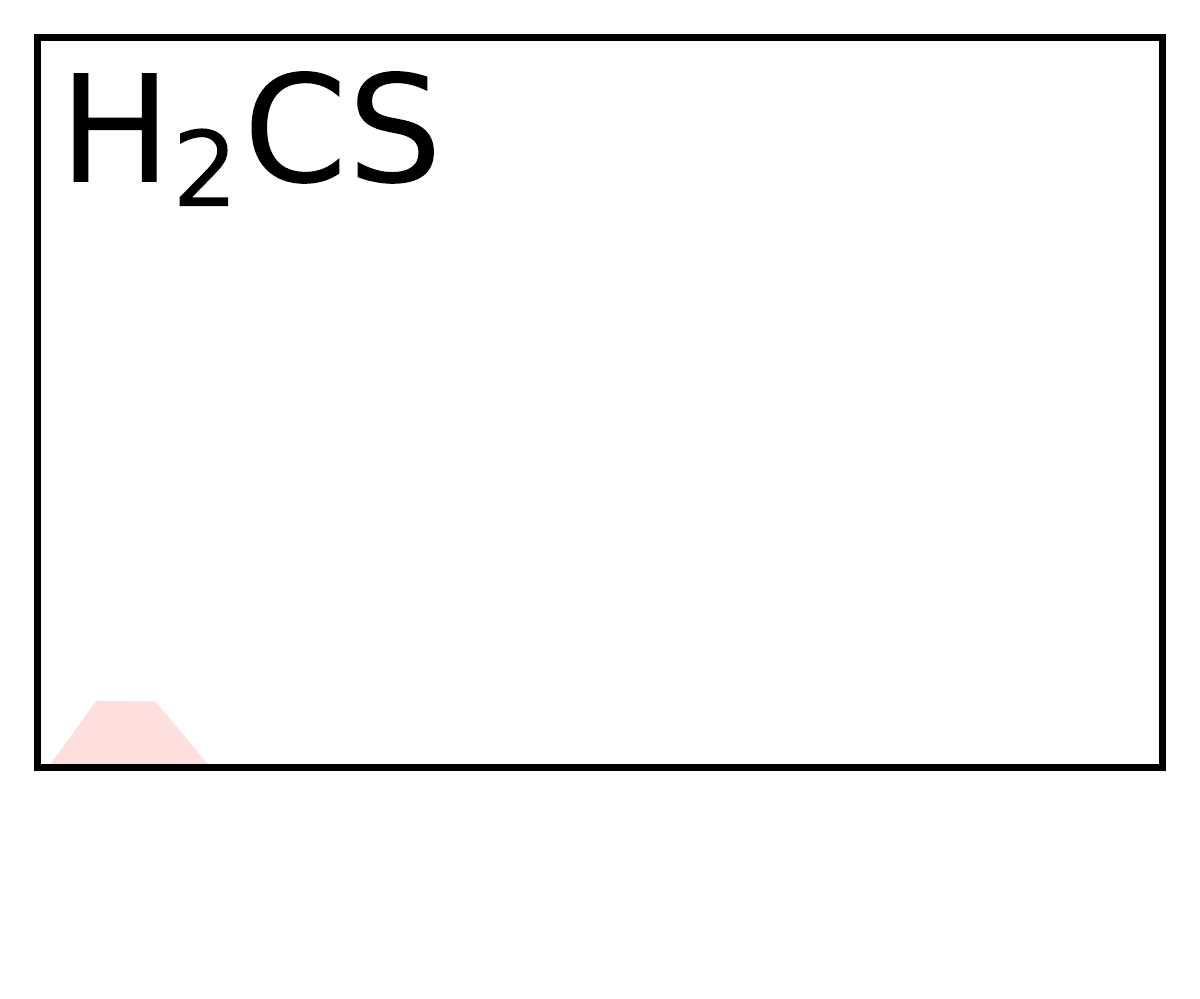} &
        \includegraphics[width=0.211\textwidth]{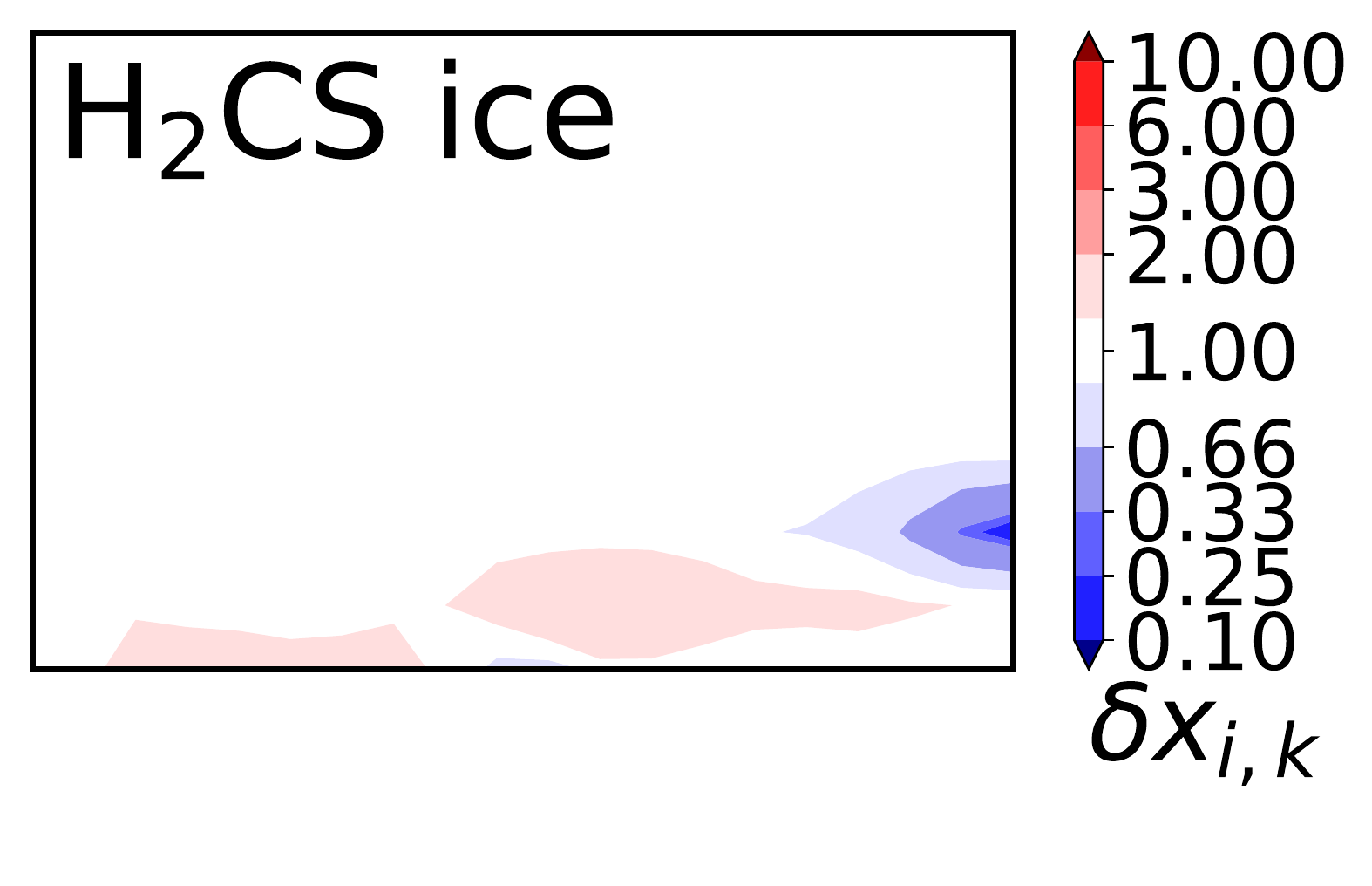} \\
        \end{tabular}\\
        \caption{Same as Fig.~\ref{fig:1d2d_Av_3}, but for the $A_{\rm V}=20$~mag case.}
        \label{fig:1d2d_Av_20}
\end{figure*}

To place these findings into a more general context, we show similar abundance ratio plots
for a range of selected gaseous species observed in the ISM and their ices for the two models with different $A_{\rm V}$ in
Figs.~\ref{fig:1d2d_Av_3} and \ref{fig:1d2d_Av_20}.
Highly abundant CO, H$_2$O, and N$_2$ as well as other
simple species such as O, HCO$^+$, N$_2$H$^+$, NH$_3$, NO and most of S-bearing molecules clearly remain relatively unaffected by the surface C + H$_2$ barrier change.
In contrast, abundances of CO$_2$ ice, H$_2$CO, CN, HCN, HNC, CS, H$_2$CS, hydrocarbons, and organic molecules
such as H$_2$CO, CH$_3$OH, and HNCO (either gas and/or ice) vary by more than a factor of two between the original and modified chemical models.

%-------------------------------------------------------------------

%-------------------------------------------------------------------

The abundances of key species, CO, H$_2$O, and N$_2$, do change between
the original and modified chemical models, but these deviations are small compared to their relative abundances.
For instance, the relative abundance differences for the CO ice in the original and modified chemical models are not negligible, $\sim 10^{-6}-10^{-5}$, but these values are at least several times lower than the  total CO relative abundance $\gtrsim 5\times 10^{-5}$. This leads to a small difference in the CO relative abundance ratios in both models. The relative abundance differences for the CO ice between the original and modified chemical models are $\Delta x({\rm CO}) \lesssim 10^{-6}-10^{-5}$, while its relative abundances $x({\rm CO})$ at 1~Myr are at least a few times $10^{-5}$.

In \Cref{fig:1d2d_Av_3,fig:1d2d_Av_20}, the rise in abundance of CH$_4$ at low temperatures $\lesssim 15$~K in the modified model is related to more efficient surface hydrogenation of CH$_2$. CH$_2$ is rapidly produced through the C $+$ H$_2$ surface reaction when it has no barrier.
The abundances of larger hydrocarbons with two or more C atoms in general decrease as a result of competition,
as more reactive carbon becomes locked in methane ice.
This also affects
the abundances of CN, HCN, and HNC (gas/ice) as well as CS and H$_2$ ice in a similar manner because their formation pathways directly involve light hydrocarbons.

The abundances of simple organic molecules such as H$_{2}$CO and CH$_{3}$OH mostly
increase in the modified model. This is due to a similar mechanism of water ice
formation that we discussed before for the evolution of C and CH$_2$ ices.
The increase in surface H at late times allows more surface CO to become hydrogenated
and thus more H$_2$CO and CH$_3$OH ices can be synthesized. In the modified model CH$_4$ is formed as a consequence of surface hydrogenation of mono-carbonic ices as well as by our evaluated reaction. The increase in the efficiency of the latter keeps some of the atomic H free to form organic molecules.

The abundances of the CH$_2$CO ice decreases in the modified chemical model
due to the competition between formation of the CH$_n$ and C$_2$H$_m$ ices. CH$_2$CO ice
in our chemical network mainly forms from the C$_2$-bearing ices.
A similar abundance change occurs with the HNCO and HC$_3$N ices, whose formation is related
to the chemistry of nitriles.

\section{Discussion}
\label{sec:diss}

\subsection{Reactivity of carbon atoms on dust grain surfaces: missing processes from chemical networks}
\label{sec:diss_reactivity}
One of the main results of this study is that the high reactivity of atomic carbon when the surface reaction C $+$ H$_2$ is barrierless increases abundances of complex organic ices. This influence might be particularly important
during the early stages of molecular cloud formation, when the CII $\rightarrow$ CI $\rightarrow$ CO conversion
is not yet completed. In our modeling, the initial form of carbon in the ISM is expected to be CII.
%The carbonaceous dust injected into the ISM by dying stars is reprocessed by supernovae shocks and gets partly atomized \citep{Draine03}.
During formation of molecular clouds, the atomic carbon undergoes a series of chemical transformations in the gas phase, leading to the formation of CO. If the cloud cools to low, $\lesssim 30$~K temperatures,
C atoms will start to accrete onto dust surfaces, where more complex species may form, including COMs.

Despite the importance of the surface chemistry involving atomic carbon,  contemporary chemical databases do not include many of these reactions. Our chemical network and the KIDA database contains only about 35 surface reactions involving C atoms. Most of these reactions are related to the growth of the carbon chains C$_{n}$X $+$ C $\rightarrow$ C$_{n+1}$X, where X could be H, O, N, S, or none. This is in strong contrast to numerous laboratory studies that demonstrated
that C atoms can react without a  barrier with almost all species present in the ice mantles \citep{Chastaing00,Chastaing01,Kaiser02,Kaiser99b,Krasnokutski17,Shannon14,Krasnokutski19a,Krasnokutski19b,Krasnokutski20,Hickson16}.

For example, the reaction C $+$ NH$_{3}$ $\rightarrow$ CH$_{2}$NH has recently been found to be barrierless and therefore is expected to be important for the formation of glycine and other amino acids and COM ices \citep{Krasnokutski20}.
Furthermore, the product of this reaction, CH$_{2}$NH, is formed in a long-living excited state, allowing it to participate in other chemical reactions with considerable energy barriers. The low-temperature reaction
C $+$ H$_{2}$O $\rightarrow$ H$_{2}$CO is expected to proceed fast on the ice surfaces due to quantum tunnelling \citep{Hickson16}.
It might therefore be expected that by extending a set of surface reactions with atomic carbon in modern chemical networks,
complexity and predicted abundances of the COM ices increase.

%Furthermore, some of these reactions may proceed differently from the standard Langmuir-Hinshelwood diffusive
%approach, e.g., via the Eley–Rideal mechanism. Also, modelling of the chain reactions, when the energy of the first reaction triggers the subsequent reactions, is often overlooked in the models. On top of that, there is a limited number of experimental studies of surface processes that limits the global accuracy and predictive power of astrochemical models.

%However, in many cases the results could be adopted from quantum chemical computations or from the gas phase studies. In the last case, the thermalization and stabilization of the products due to the presence of the third body should be taken into account.

\subsection{TMC1 benchmark}
\label{sec:diss_TMC1}

The question arises whether the theoretical chemical modeling
we presented might be applied to real astronomical objects
and whether the adopted chemical model and networks are feasible.
To test the feasibility of the original and modified chemical models, we compared computed and observed abundances
in the prototypical TMC1 molecular cloud. We used the observed relative abundances from \citet{Agundez2013},
as well as their initial abundances and 0D physical conditions: uniform gas and dust temperature
$T=10$~K, $n_{\rm H} = 2\times 10^4$~cm$^{-3}$, $A_{\rm V} = 10$ mag,
a dust-to-gas-mass ratio of 0.01, and a grain radius of $0.1\mu$m. The agreement $a_i$ between the modeled and observed abundances was calculated for each of the observed species as follows:
\begin{gather*} a_i =
        \begin{cases}
                \: 1 \: \quad ; \qquad 0.1 \leq x_{i}(t_j)/x_{i,obs} \leq 10 \\
                \: 0 \: \quad ; \qquad otherwise,
        \end{cases}
\end{gather*}
where $x_{i}(t_j)$ is the modeled relative abundance of a species $i$ at a particular time moment ($t_j$) for the original (ORG) or modified (MOD) model, and $x_{i,obs}$ is the relative observed abundance
(with respect to the amount of molecular hydrogen). The agreement values $a_i$ were summed over all observed species.

\begin{figure}[h]
\centering
\setlength\tabcolsep{-0.5pt}
\renewcommand{\arraystretch}{0}
        \includegraphics[width=0.40\textwidth]{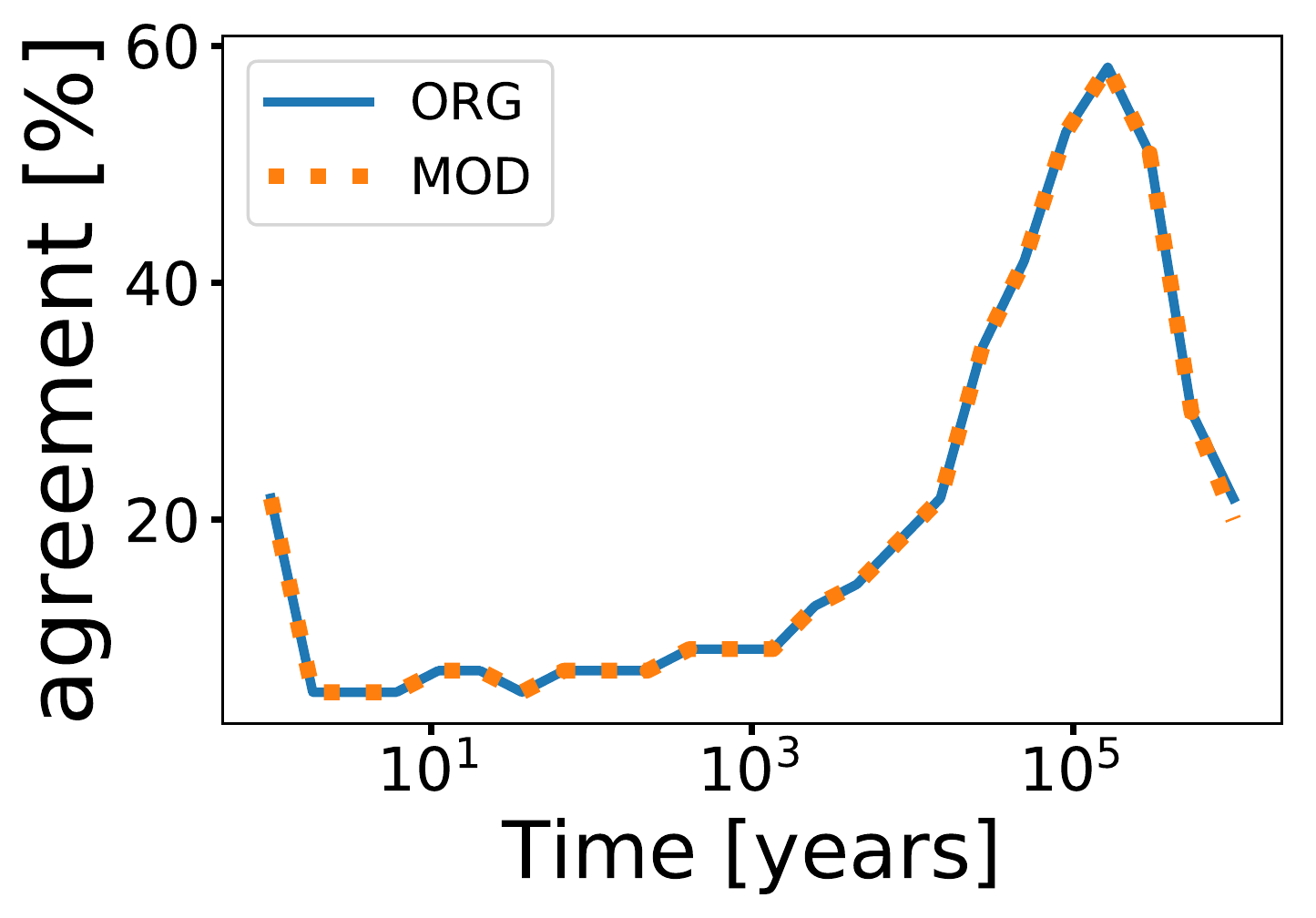}
        \caption{Percentage of the observed species in the TMC1 cloud reproduced by our two chemical models as a function of time.}
        \label{fig:agreemnent_TMC1}
\end{figure}

The percentage of the species reproduced by the two chemical models plotted as a function of time is shown in
Fig. \ref{fig:agreemnent_TMC1}.
Both models clearly show the best agreement of about $60\%$ at a late time, $\sim 10^5$~years,
similar to what was found by \citet{Agundez_ea08} and other previous studies.
There is no significant difference between the two chemical models because the observed species
in TMC1 are mainly simple gas-phase species, not that many hydrocarbons that are
sensitive to the barrier change in the surface C $+$ H$_2$ reaction were detected there.

Other environments were the effect of the barrier change in the 
surface reaction between C and H$_2$ on the chemical composition
could be more pronounced  are protoplanetary disks with their cold 
outer mid-plane regions and warmer molecular layers. However, 
for the sake of brevity, we  leave this investigation for 
another paper.

\section{Conclusion}
%-------------------------------------------------------------------
We performed a thorough theoretical study of the effect of the barrier change in the single
surface reaction C + H$_2$ $\rightarrow$ CH$_2$ on the time-dependent chemical evolution
in a wide range of physical conditions typical for the interstellar medium. We found that the most affected species
are hydrocarbon ices and their gas-phase counterparts. The absence of the barrier in the C + H$_2$
reaction leads to a faster conversion of the surface carbon atoms into methane ice.
More efficient synthesis of mono-carbonic hydrocarbon ices (CH$_2$, CH$_3$ , and CH$_4$) lowers the efficiency of formation for
larger hydrocarbons, cyanides, and nitriles, and CS-bearing species at low temperatures $\lesssim 10-30$~K.
The absence of the barrier in the C + H$_2$ reaction increases the efficiency of the surface hydrogenation of CO and hence the synthesis of complex organic ices.
However, abundances of major species such as CO, H$_2$O, and N$_2$ as well as O, HCO$^+$, N$_2$H$^+$, NH$_3$, NO,
and most of the S-bearing molecules remain almost unaffected.
%A comparison between the models with
%and without a barrier in the surface reaction C + H$_2$ $\rightarrow$ CH$_2$ showed that it
%has almost no effect on abundances of detected gaseous species in the TMC1 molecular cloud.
These global chemical trends are similar in the UV-irradiated ($A_{\rm V}=3$~mag) and UV-dark environments
($A_{\rm V}=20$~mag). The effect of the C + H$_2$ surface reaction barrier value on the gas-phase chemistry is limited by the low efficiency of the desorption mechanism in our molecular cloud model. In contrast, in less UV-dark  astrophysical environments,  this higher desorption would produce a larger effect on the gas-phase abundances.

We conclude that the fast reactivity of atomic carbon on dust grain surfaces, which has recently been
measured in laboratory experiments and confirmed by quantum chemical calculations, is of considerable
importance for contemporary astrochemical models. These processes, which are currently not fully
included in public astrochemical databases such as KIDA,  need to be taken into account
to better understand the chemical cycle of carbon and the synthesis of hydrocarbon and
organic ices in the ISM. Further laboratory measurements and/or quantum chemical computations
are required to determine other important surface reactions with carbon atoms.
%Moreover, for correct accounting of these reactions the inclusion of new types of the chemical mechanisms such as Eley–Rideal and chain reactions is required.

\section{Acknowledgements}
D.S. acknowledges support by the Deutsche Forschungsgemeinschaft through SPP 1833:
``Building a Habitable Earth'' (SE 1962/6-1). S.K. is grateful for the support by
the Max Planck Institute for Astronomy and the Deutsche Forschungsgemeinschaft
DFG (grants No. KR 3995/4-1). M.S. is thankful for the support from the Erasmus+ Programme and the Slovenian Research Agency (ARRS) through research core fundings No. P1-0201.
T.H. acknowledges support from the European Research Council under the
Horizon~2020 Framework Program via the ERC Advanced Grant ``Origins'' 83 24 28.

%-------------------------------------------------------------------
\bibliography{refs}

\begin{thebibliography}{45}
\expandafter\ifx\csname natexlab\endcsname\relax\def\natexlab#1{#1}\fi

\bibitem[{{Ag{\'u}ndez} {et~al.}(2008){Ag{\'u}ndez}, {Cernicharo}, \&
  {Goicoechea}}]{Agundez_ea08}
{Ag{\'u}ndez}, M., {Cernicharo}, J., \& {Goicoechea}, J.~R. 2008, A\&A, 483,
  831

\bibitem[{{Ag{\'u}ndez} \& {Wakelam}(2013)}]{Agundez2013}
{Ag{\'u}ndez}, M. \& {Wakelam}, V. 2013, Chemical Reviews, 113, 8710

\bibitem[{{Becker} {et~al.}(1989){Becker}, {Engelhardt}, {Wiesen}, \&
  {Bayes}}]{Becker89}
{Becker}, K.~H., {Engelhardt}, B., {Wiesen}, P., \& {Bayes}, K.~D. 1989, cpl,
  154, 342

\bibitem[{{Belloche} {et~al.}(2014){Belloche}, {Garrod}, {M{\"u}ller}, \&
  {Menten}}]{Belloche_ea14}
{Belloche}, A., {Garrod}, R.~T., {M{\"u}ller}, H.~S.~P., \& {Menten}, K.~M.
  2014, Science, 345, 1584

\bibitem[{{Bertin} {et~al.}(2016){Bertin}, {Romanzin}, {Doronin}, {Philippe},
  {Jeseck}, {Ligterink}, {Linnartz}, {Michaut}, \& {Fillion}}]{Bertin2016}
{Bertin}, M., {Romanzin}, C., {Doronin}, M., {et~al.} 2016, ApJL, 817, L12

\bibitem[{{Biham} {et~al.}(2001){Biham}, {Furman}, {Pirronello}, \&
  {Vidali}}]{Bihamea01}
{Biham}, O., {Furman}, I., {Pirronello}, V., \& {Vidali}, G. 2001, \apj, 553,
  595

\bibitem[{{Bussery-Honvault} {et~al.}(2005){Bussery-Honvault}, {Julien},
  {Honvault}, \& {Launay}}]{Bussery-Honvault2005}
{Bussery-Honvault}, B., {Julien}, J., {Honvault}, P., \& {Launay}, J.~M. 2005,
  pccp, 7, 1476

\bibitem[{{Chastaing} {et~al.}(2000){Chastaing}, {Le Picard}, \&
  {Sims}}]{Chastaing00}
{Chastaing}, D., {Le Picard}, S.~D., \& {Sims}, I.~R. 2000, Journal of Chemical
  Physics, 112, 8466

\bibitem[{{Chastaing} {et~al.}(2001){Chastaing}, {Le Picard}, {Sims}, \&
  {Smith}}]{Chastaing01}
{Chastaing}, D., {Le Picard}, S.~D., {Sims}, I.~R., \& {Smith}, I.~W.~M. 2001,
  A\&A, 365, 241

\bibitem[{{Cruz-Diaz} {et~al.}(2016){Cruz-Diaz}, {Mart{\'{\i}}n-Dom{\'e}nech},
  {Mu{\~n}oz Caro}, \& {Chen}}]{Cruz_Diaz2016}
{Cruz-Diaz}, G.~A., {Mart{\'{\i}}n-Dom{\'e}nech}, R., {Mu{\~n}oz Caro}, G.~M.,
  \& {Chen}, Y.-J. 2016, A\&A, 592, A68

\bibitem[{{Cuppen} {et~al.}(2017){Cuppen}, {Walsh}, {Lamberts}, {Semenov},
  {Garrod}, {Penteado}, \& {Ioppolo}}]{Cuppen2017}
{Cuppen}, H.~M., {Walsh}, C., {Lamberts}, T., {et~al.} 2017, \ssr

\bibitem[{{Dean} {et~al.}(1991){Dean}, {Davidson}, \& {Hanson}}]{Dean1991}
{Dean}, A.~J., {Davidson}, D.~F., \& {Hanson}, R.~K. 1991, jpc, 95, 183

\bibitem[{{Draine} \& {Bertoldi}(1996)}]{DB96}
{Draine}, B.~T. \& {Bertoldi}, F. 1996, Astrophys.~J, 468, 269

\bibitem[{{Garrod} \& {Herbst}(2006)}]{Garrod_Herbst06}
{Garrod}, R.~T. \& {Herbst}, E. 2006, \aap, 457, 927

\bibitem[{{Garrod} {et~al.}(2007){Garrod}, {Wakelam}, \&
  {Herbst}}]{2007A&A...467.1103G}
{Garrod}, R.~T., {Wakelam}, V., \& {Herbst}, E. 2007, \aap, 467, 1103

\bibitem[{{Graedel} {et~al.}(1982){Graedel}, {Langer}, \&
  {Frerking}}]{1982ApJS...48..321G}
{Graedel}, T.~E., {Langer}, W.~D., \& {Frerking}, M.~A. 1982, \apjs, 48, 321

\bibitem[{{Harada} {et~al.}(2010){Harada}, {Herbst}, \&
  {Wakelam}}]{2010ApJ...721.1570H}
{Harada}, N., {Herbst}, E., \& {Wakelam}, V. 2010, \apj, 721, 1570

\bibitem[{{Harding} {et~al.}(1993){Harding}, {Guadagnini}, \&
  {Schatz}}]{Harding1993}
{Harding}, L.~B., {Guadagnini}, R., \& {Schatz}, G.~C. 1993, jpc, 97, 5472

\bibitem[{{Hasegawa} {et~al.}(1992){Hasegawa}, {Herbst}, \& {Leung}}]{HHL92}
{Hasegawa}, T.~I., {Herbst}, E., \& {Leung}, C.~M. 1992, \apjs, 82, 167

\bibitem[{{Henning} \& {Semenov}(2013)}]{2013ChRv..113.9016H}
{Henning}, T. \& {Semenov}, D. 2013, Chemical Reviews, 113, 9016

\bibitem[{{Henning} \& {Krasnokutski}(2019)}]{Krasnokutski19b}
{Henning}, T.~K. \& {Krasnokutski}, S.~A. 2019, Nat. Astron., 3, 568

\bibitem[{{Herbst} \& {van Dishoeck}(2009)}]{Herbst_vDishoeck09}
{Herbst}, E. \& {van Dishoeck}, E.~F. 2009, ARA\&A, 47, 427

\bibitem[{{Herbst} \& {Yates}(2013)}]{herbst2013}
{Herbst}, E. \& {Yates}, J.~T. 2013, Chemical Reviews, 113, 8707

\bibitem[{{Hickson} {et~al.}(2016){Hickson}, {Loison}, {Nunez-Reyes}, \&
  {Mereau}}]{Hickson16}
{Hickson}, K.~M., {Loison}, J.~C., {Nunez-Reyes}, D., \& {Mereau}, R. 2016,
  JPCL, 7, 3641

\bibitem[{{Hollenbach} \& {Salpeter}(1971)}]{Hollenbach_Salpeter71}
{Hollenbach}, D. \& {Salpeter}, E.~E. 1971, \apj, 163, 155

\bibitem[{{Husain} \& {Young}(1975)}]{Husain1975}
{Husain}, D. \& {Young}, A.~N. 1975, J. Chem. Soc., Faraday Trans. 2, 71, 525

\bibitem[{{Kaiser} \& {Mebel}(2002)}]{Kaiser02}
{Kaiser}, R.~I. \& {Mebel}, A.~M. 2002, International Reviews in Physical
  Chemistry, 21, 307

\bibitem[{{Kaiser} {et~al.}(1999){Kaiser}, {Mebel}, {Chang}, {Lin}, \&
  {Lee}}]{Kaiser99b}
{Kaiser}, R.~I., {Mebel}, A.~M., {Chang}, A.~H.~H., {Lin}, S.~H., \& {Lee},
  Y.~T. 1999, Journal of Chemical Physics, 110, 10330

\bibitem[{{Krasnokutski} {et~al.}(2017{\natexlab{a}}){Krasnokutski}, {Goulart},
  {Gordon}, A., C., M., {Salvenmoser}, Th, \& {Scheier}}]{Krasnokutski2017}
{Krasnokutski}, S.~A., {Goulart}, M., {Gordon}, E.~B., {et~al.}
  2017{\natexlab{a}}, ApJ, 847, 89

\bibitem[{{Krasnokutski} \& {Huisken}(2014)}]{Krasnokutski14}
{Krasnokutski}, S.~A. \& {Huisken}, F. 2014, Applied Physics Letters, 105,
  113506

\bibitem[{{Krasnokutski} {et~al.}(2017{\natexlab{b}}){Krasnokutski}, {Huisken},
  {J{\"a}ger}, \& {Henning}}]{Krasnokutski17}
{Krasnokutski}, S.~A., {Huisken}, F., {J{\"a}ger}, C., \& {Henning}, T.
  2017{\natexlab{b}}, ApJ, 836, 32

\bibitem[{Krasnokutski {et~al.}(2020)Krasnokutski, Jäger, \&
  Henning}]{Krasnokutski20}
Krasnokutski, S.~A., Jäger, C., \& Henning, T. 2020, ApJ, 889, 67

\bibitem[{{Krasnokutski} {et~al.}(2016){Krasnokutski}, {Kuhn}, {Renzler},
  {J{\"a}ger}, {Henning}, \& {Scheier}}]{Krasnokutski2016}
{Krasnokutski}, S.~A., {Kuhn}, M., {Renzler}, M., {et~al.} 2016, ApJL, 818, L31

\bibitem[{{Krasnokutski} {et~al.}(2019){Krasnokutski}, {Tkachenko}, C., \&
  Th}]{Krasnokutski19a}
{Krasnokutski}, S.~A., {Tkachenko}, O., C., J., \& Th, H. 2019, PCCP, 21, 12986

\bibitem[{{Lee} {et~al.}(1996){Lee}, {Herbst}, {Pineau des For\^ets}, {Roueff},
  \& {Le Bourlot}}]{1996A&A...311..690L}
{Lee}, H.-H., {Herbst}, E., {Pineau des For\^ets}, G., {Roueff}, E., \& {Le
  Bourlot}, J. 1996, A\&A, 311, 690

\bibitem[{{Lee} {et~al.}(1998){Lee}, {Roueff}, {Pineau des Forets},
  {Shalabiea}, {Terzieva}, \& {Herbst}}]{Lea98}
{Lee}, H.-H., {Roueff}, E., {Pineau des Forets}, G., {et~al.} 1998, \aap, 334,
  1047

\bibitem[{{Lin} \& {Guo}(2004)}]{Lin2004}
{Lin}, S.~Y. \& {Guo}, H. 2004, JPCa, 108, 10066

\bibitem[{{McGuire}(2018)}]{McGuire18}
{McGuire}, B.~A. 2018, \apjs, 239, 17

\bibitem[{{Rimmer} \& {Helling}(2016)}]{Rimmer_Helling16}
{Rimmer}, P.~B. \& {Helling}, C. 2016, \apjs, 224, 9

\bibitem[{{Semenov} {et~al.}(2010){Semenov}, {Hersant}, {Wakelam}, {Dutrey},
  {Chapillon}, {Guilloteau}, {Henning}, {Launhardt}, {Pi{\'e}tu}, \&
  {Schreyer}}]{Semenov_ea10}
{Semenov}, D., {Hersant}, F., {Wakelam}, V., {et~al.} 2010, Astron.~Astrophys,
  522, A42

\bibitem[{{Semenov} \& {Wiebe}(2011)}]{SW2011}
{Semenov}, D. \& {Wiebe}, D. 2011, ApJS, 196, 25

\bibitem[{{Shannon} {et~al.}(2014){Shannon}, {Cossou}, {Loison}, {Caubet},
  {Balucani}, {Seakins}, {Wakelam}, \& {Hickson}}]{Shannon14}
{Shannon}, R.~J., {Cossou}, C., {Loison}, J.~C., {et~al.} 2014, Rsc Advances,
  4, 26342

\bibitem[{{Tielens}(2010)}]{Tielens_10}
{Tielens}, A.~G.~G.~M. 2010, {The Physics and Chemistry of the Interstellar
  Medium}

\bibitem[{{Vasyunin} \& {Herbst}(2013)}]{2013ApJ...769...34V}
{Vasyunin}, A.~I. \& {Herbst}, E. 2013, \apj, 769, 34

\bibitem[{{Wakelam} {et~al.}(2012){Wakelam}, {Herbst}, {Loison}, {Smith},
  {Chandrasekaran}, {Pavone}, {Adams}, {Bacchus-Montabonel}, {Bergeat},
  {B{\'e}roff}, {Bierbaum}, {Chabot}, {Dalgarno}, {van Dishoeck}, {Faure},
  {Geppert}, {Gerlich}, {Galli}, {H{\'e}brard}, {Hersant}, {Hickson},
  {Honvault}, {Klippenstein}, {Le Picard}, {Nyman}, {Pernot}, {Schlemmer},
  {Selsis}, {Sims}, {Talbi}, {Tennyson}, {Troe}, {Wester}, \&
  {Wiesenfeld}}]{KIDA}
{Wakelam}, V., {Herbst}, E., {Loison}, J.-C., {et~al.} 2012, \apjs, 199, 21

\end{thebibliography}
%-------------------------------------------------------------------

\end{document}